\documentstyle[12pt,epsf,amssymb]{book}  
\newcommand{\ncm}{\newcommand}
\renewcommand{\theequation}{\thesection.\arabic{equation}}
\newcommand{\sectiona}[1]{\setcounter{equation}{0}\section{#1}}
\newcommand{\sectie}{\def\thesection{\thechapter.\arabic{section}}}
\newcommand{\aanhangsel}{\def\thesection{\thechapter.\Alph{section}}
\setcounter{section}{0}}
\ncm{\oH}{\bar{H}}
\ncm{\us}{\quad\mbox{using}\quad}
\ncm{\ra}{\rightarrow}
\ncm{\ot}{\otimes}
\ncm{\DH}{D(H)}
\ncm{\DW}{D^{\omega}(H)}
\ncm{\TH}{T(\oH)}
\ncm{\ssc}{\displaystyle}
\ncm{\oq}{\eta} 
\ncm{\im}{\imath}
\ncm{\ba}{\begin{array}}
\ncm{\ea}{\end{array}}
\ncm{\ul}{\underline}
\ncm{\ol}{\overline}

\def\hoek{\hbox{\vrule height 2.5ex depth 0pt \vrule width 2.5ex height .4pt
 depth 0pt}}
\def\haak#1#2{
\mathop{\hoek\llap{\vbox to 2.5ex{ \vfil
\hbox{$\scriptstyle#1$\hskip 2.8ex} \vfil}}}
\limits_{#2} }
\def\hook#1#2{\setbox0=\hbox{$\scriptstyle#1$}
\hskip\wd0\haak{\box0}{#2}}
\ncm{\str}{\rule{0cm}{3.5mm}}
\ncm{\om}{\omega}
\ncm{\ep}{\epsilon}
\newlength{\extraspace}
\newlength{\extraspaces}
\newcommand{\be}{\begin{equation}
\addtolength{\abovedisplayskip}{\extraspaces}
\addtolength{\belowdisplayskip}{\extraspaces}
\addtolength{\abovedisplayshortskip}{\extraspace}
\addtolength{\belowdisplayshortskip}{\extraspace}}
\newcommand{\ee}{\end{equation}}

\makeatletter

\def\eqnfourarray{\stepcounter{equation}%
  \def\@currentlabel{\p@equation\theequation}\global\@eqnswtrue \m@th
  \global\@eqcnt\z@ \tabskip\@centering
  \let\\\@eqncr \let\@@eqncr\@@eqnfourcr
  $$\everycr{}\halign to\displaywidth \bgroup
  \hskip\@centering $\displaystyle \tabskip\z@skip {##}$\@eqnsel &%
  \global\@eqcnt\@ne \hskip\tw@\arraycolsep \hfil ${##}$\hfil &%
  \global\@eqcnt\tw@ \hskip\tw@\arraycolsep $\displaystyle {##}$\hfil &%
  \global\@eqcnt\thr@@ \hskip\tw@\arraycolsep \hfil \hbox{##}%
    \tabskip\@centering &%
  \global\@eqcnt 4 \hbox\bgroup \hss ##\egroup
    \tabskip\z@skip \cr}

\def\endeqnfourarray{\@@eqncr \egroup
  \global\advance\c@equation\m@ne $$\global\@ignoretrue}

\def\@@eqnfourcr{\let\reserved@a\relax
  \ifcase\@eqcnt \def\reserved@a{& & & &}%
    \or \def\reserved@a{& & &}%
    \or \def\reserved@a{& &}%
    \or \def\reserved@a{&}%
    \else \let\reserved@a\@empty
      \@latex@error{Too many columns in eqnfourarray environment}\@ehc
  \fi
  \reserved@a \if@eqnsw\@eqnnum \stepcounter{equation}\fi
  \global\@eqnswtrue \global\@eqcnt \z@\cr}

\makeatother

\newcommand{\bea}{\begin{eqnarray}
\addtolength{\abovedisplayskip}{\extraspaces}
\addtolength{\belowdisplayskip}{\extraspaces}
\addtolength{\abovedisplayshortskip}{\extraspace}
\addtolength{\belowdisplayshortskip}{\extraspace}}
\newcommand{\eea}{\end{eqnarray}}
\newcommand{\beas}{\begin{eqnarray*}
\addtolength{\abovedisplayskip}{\extraspaces}
\addtolength{\belowdisplayskip}{\extraspaces}
\addtolength{\abovedisplayshortskip}{\extraspace}
\addtolength{\belowdisplayshortskip}{\extraspace}}
\newcommand{\eeas}{\end{eqnarray*}}
\ncm{\Z}{{\mbox{\bf Z}}}
\ncm{\al}{\alpha}
\ncm{\bt}{\beta}
\ncm{\gm}{\gamma}
\ncm{\dl}{\delta}
\ncm{\varep}{\varepsilon}
\ncm{\zt}{\zeta}
\ncm{\et}{\eta}
\ncm{\th}{\theta}
\ncm{\kp}{\kappa}
\ncm{\lm}{\lambda}
\ncm{\rh}{\rho}
\ncm{\hl}{\hline}
\ncm{\sg}{\sigma}
\ncm{\ta}{\tau}
\ncm{\ph}{\phi}
\ncm{\phv}{\varphi}
\ncm{\ch}{\chi}
\ncm{\ps}{\Phi}
\ncm{\nn}{\nonumber}

\begin{document}

\vspace*{2cm}
\begin{center}
{\Huge \bf  Topological  Interactions in \\
Broken  Gauge Theories}\\
\vspace{5cm}
{\large ACADEMISCH PROEFSCHRIFT}\\
\vspace{1cm}
\end{center}
\bgroup\parindent0pt%\parfillskip0pt
\leftskip3.9cm\rightskip\leftskip
\let\\\relax\hyphenpenalty100000
ter verkrijging van de graad van doctor\\
aan de Universiteit van Amsterdam\\
op gezag van de Rector Magnificus\\
Prof.\ dr.\ P.W.M.\ de Meijer\\
ten overstaan van een door het college van dekanen ingestelde commissie\\
in het openbaar te verdedigen in de Aula der Universiteit\\
op dinsdag 19 september 1995  te  13.30 uur\\
\par\egroup
\begin{center}
\vspace{.5cm}
door\\
\vspace{1cm}
{\bf\large Mark Dirk Frederik de Wild Propitius}\\
\vspace{1cm}
\rm geboren te Amsterdam
\end{center}
\pagestyle{empty}
\newpage
\vspace*{1cm}
\begin{flushleft}
\begin{tabbing}
Promotiecommisie: bl \=         \kill
Promotor: \> Prof.\ dr.\ ir.\ F.A.\ Bais\\
         \>     \\
Promotiecommisie: \>Prof.\ dr.\ P.J.\ van Baal  \\
                  \>Prof.\ dr.\ R.H.\ Dijkgraaf \\
                  \>Prof.\ dr.\ G.\ 't Hooft \\
                  \>Prof.\ dr.\ A.M.M.\ Pruisken    \\
                  \>Prof.\ dr.\ J. Smit \\
                  \>Prof.\ dr.\ H.L.\ Verlinde
\end{tabbing}
\vspace{2cm}
Faculteit der Wiskunde, Informatica, Natuur- en Sterrenkunde\\
\vspace{.5cm}
Instituut voor Theoretische Fysica \\
\vspace{3cm}
This thesis is based on the following papers:
\begin{itemize}
\item  F.A. Bais, P. van Driel and M. de Wild Propitius,
       Phys.~Lett.~{\bf B280} (1992) 63. 
\item  F.A. Bais, P. van Driel and M. de Wild Propitius, 
       Nucl. Phys. {\bf B393} (1993) 547. 
\item  F.A. Bais, A. Morozov and M. de Wild Propitius,
                Phys. Rev. Lett. {\bf 71}  (1993) 2383.
            
\item  F.A. Bais and M. de Wild Propitius, 
       in {\sl The III International
       Conference on Mathematical Physics, String Theory and Quantum Gravity}, 
       Proceedings of the Conference, Alushta, 1993, 
       Theor. Math. Phys. {\bf 98} (1994) 509.
\item  M. de Wild Propitius and F.A. Bais, {\sl Discrete gauge theories}, 
       to appear in the proceedings of the CRM-CAP Summer School 
`Particles and Fields 94', Bannf, Alberta, Canada, August 16-24, 1994.
\end{itemize}
\end{flushleft}
\newpage
\vspace*{15cm}
\begin{flushright}
{\sl Voor mijn moeder}
\end{flushright}
\newpage
\leavevmode
\newpage
\vspace*{8cm}
\begin{flushright}
{\bf Who  needs physics when we've got chemistry?}\\
\vspace{.5cm}
{\bf -}Peggy Sue Got Married, Francis Coppola
\end{flushright}
\newpage

\pagestyle{headings}
\setcounter{page}{1}
\pagenumbering{roman}

\tableofcontents

\newpage
\pagestyle{headings}
\pagenumbering{arabic}
\setcounter{page}{1}

\chapter*{Preface}
\addcontentsline{toc}{chapter}{Preface}
\markboth{PREFACE}{PREFACE}

Symmetry has become one of the main  
guiding principles in  physics during the twentieth century.
Over the last ten decades,
we have progressed from external to internal, 
from global to local, from finite to infinite,
from ordinary to supersymmetry and recently arrived at the notion
of quantum groups.

In general, a physical system  consists of 
a finite or infinite number of degrees of freedom which may or may not 
interact. The dynamics is prescribed by a set of 
evolution equations which follow from varying the action 
with respect to the different degrees of freedom.
A symmetry then corresponds to  a group of 
transformations on the space time coordinates and/or the 
degrees of freedom that leave the action 
and therefore also the evolution equations invariant.
External symmetries have to do  with invariances (e.g.\ Lorentz invariance)
under transformations on the space time  coordinates.
Symmetries not related with transformations of space time coordinates
are  called internal symmetries. We also discriminate 
between global symmetries and local symmetries. 
A global or rigid symmetry transformation is the same 
throughout space time and usually leads to a conserved quantity.
Turning a global symmetry into a local symmetry, i.e.\ 
allowing the symmetry transformations to vary continuously 
from one point in space time to another,
requires the introduction of additional gauge degrees 
of freedom mediating a force. 
This so-called gauge principle has eventually led to 
the extremely successful standard model 
of the strong and electro-weak interactions between the 
elementary particles based on the local gauge group
$SU(3) \times SU(2) \times U(1)$.

A symmetry of the action is {\em not} 
automatically a symmetry of the groundstate of a physical system.
If the action is invariant under some symmetry group $G$
and the groundstate only under a subgroup $H$ of $G$, the 
symmetry group $G$ is said to be spontaneously broken down.
The symmetry is not completely lost though,
for the broken generators of $G$ transform one groundstate into another.

The physics of a broken global symmetry is quite different from a broken 
local gauge symmetry.
The signature of a broken continuous {\em global} symmetry group $G$ 
in a physical system is  the occurrence 
of massless scalar degrees of freedom, the so-called Goldstone bosons. 
Specifically, each broken generator of $G$
gives rise to a massless Goldstone boson field. 
Well-known realizations of Goldstone bosons 
are the long range spin waves in a ferromagnet,
in which  the rotational symmetry is broken below the Curie temperature
through the appearance of spontaneous magnetization.  
A beautiful example in particle physics is the low energy physics of 
the strong interactions, where the spontaneous
breakdown of (approximate) chiral symmetry leads to 
(approximately) massless pseudoscalar particles such as the pions.

In the case of a broken local gauge symmetry, on the other hand, 
the would be massless Goldstone bosons conspire with the massless 
gauge fields to form a massive vector field. 
This celebrated phenomenon is known as the Higgs mechanism. 
The canonical example in condensed matter physics 
is the ordinary superconductor. 
In the phase transition from the normal to the superconducting phase,
the $U(1)$ gauge symmetry is spontaneously broken 
by a condensate of Cooper pairs.
This leads to a mass $M_A$ for the photon field in the 
superconducting medium as witnessed by the Meissner effect:
magnetic fields are expelled from a superconducting region and 
have a characteristic penetration depth which in proper units 
is just the inverse of the photon mass $M_A$. 
The Higgs  mechanism also plays a key role in the
unified theory of weak and electromagnetic interactions, that is, the
Glashow-\-Weinberg-\-Salam model where the product gauge group
$SU(2)\times U(1)$ is broken to the $U(1)$ subgroup of
electromagnetism. Here, the massive vector particles 
correspond to the $W$ and $Z$ bosons mediating the short range weak 
interactions. 
More speculative applications of the Higgs mechanism 
are those where the
standard model of the strong, weak and electromagnetic interactions is
embedded in a grand unified model with a large simple gauge group. The
most ambitious attempts invoke supersymmetry as well.

In addition to the  characteristics in the spectrum of
fundamental excitations described above, there are in general 
other fingerprints of a broken symmetry in a physical system.
These are usually called {\em topological excitations} or 
just {\em defects}, see for 
example the references~\cite{cola, mermin, presbook, raja} for reviews.
Defects are collective degrees of freedom 
carrying `charges' or quantum numbers which are conserved for 
topological reasons, not related to a manifest symmetry of the action.  
It is exactly the appearance of these
topological charges which renders the corresponding collective
excitations stable.  
Topological excitations may manifest themselves
as particle-like, string-like or planar-like objects (solitons), 
or have to be interpreted as quantum mechanical tunneling processes
(instantons).  Depending on the model in which they occur,
these excitations carry evocative names like kinks, domain walls, vortices,
cosmic strings, Alice strings, monopoles, skyrmions, texture,
sphalerons and so on.
Defects are crucial for a full understanding of
the physics of systems with a broken symmetry and lead to a host
of rather unexpected and exotic phenomena, which are 
in general of a nonperturbative nature.

The prototypical example of a topological defect is the 
Abrikosov-Nielsen-Olesen flux tube 
in the type~II superconductor with broken $U(1)$ gauge 
symmetry~\cite{abri, niels}.
The topological quantum number characterizing these defects is the magnetic 
flux, which indeed can only take discrete values.  
A  beautiful example in particle physics is the 't Hooft-Polyakov 
monopole~\cite{thooftmon, polyamon} occurring 
in any  grand unified model in which  a simple gauge group $G$ is
broken to a subgroup $H$ which contains the electromagnetic $U(1)$
factor. Here, it is the quantized magnetic charge which is conserved for
topological reasons. 
In fact, the presence of 
magnetic monopoles in these models reconciles the
two well-known arguments for the quantization of electric
charge, namely Dirac's argument based on the existence of a
magnetic monopole~\cite{dirac} and 
the obvious fact that the $U(1)$ generator should
be compact as it belongs to a larger compact gauge group.

An example of a model with a broken global symmetry supporting 
topological excitations is  the effective sigma model describing the
low energy strong interactions for the mesons.  That is, the
phase with broken chiral symmetry alluded to before.  One may add a
topological term and a stabilizing term to the action and obtain a
theory that features topological particle-like objects called
skyrmions, which have exactly the properties of the baryons. 
See reference~\cite{skyrme} and also~\cite{fink, witsk}.
So, upon extending the effective model for the Goldstone bosons, 
we recover the
complete spectrum of the underlying strong interaction model
(quantum chromodynamics) and its low energy dynamics. Indeed,
this picture leads to an attractive phenomenological model for
baryons.

Another area of physics where defects may play a fundamental
role is cosmology.
See reference~\cite{brand} for a recent review. 
According to the standard cosmological hot big bang scenario, 
the universe cooled down through a sequence of symmetry breaking 
phase transitions in a very early stage. 
The question of the actual formation of defects in these phase 
transitions is of prime importance. It has been argued, for instance,
that magnetic monopoles might have been produced copiously.
As they tend to dominate the mass in the universe, however, 
magnetic monopoles are notoriously hard to accommodate and if indeed formed,
they have to be `inflated away'. 
In fact, phase transitions that see the production of cosmic strings
are much more interesting.
In contrast with magnetic monopoles,
cosmic strings do not lead to cosmological disasters
and according to an attractive but still speculative theory may even have 
acted as seeds for the formation 
of galaxies and other large scale structures in the present day universe.

Similar symmetry breaking 
phase transitions are extensively studied in condensed matter physics. 
We have already mentioned the transition 
from the normal to the superconducting phase in superconducting 
materials of type~II, which may give rise to
the formation of magnetic flux tubes. 
In the field of low temperature physics, there also exists a great body of 
both theoretical and experimental work 
on the transitions from the normal to the 
many superfluid phases of helium-3 in which  
line and point defects arise in a great variety, 
e.g.~\cite{volovik}. 
Furthermore, in uniaxial 
nematic liquid crystals,  point defects, 
line defects and texture arise
in the transition from the disordered to the ordered phase
in which the rotational global symmetry group $SO(3)$ is broken down 
to the semi-direct product group $U(1) \rtimes \Z_2$.
Bi-axial nematic crystals, in turn, exhibit a phase 
transition in which the global symmetry group is broken 
to the product group $\Z_2 \times \Z_2$ yielding 
line defects labeled by the elements of the (nonabelian)
quaternion group $\bar{D}_2$, e.g.~\cite{mermin}. 
Nematic crystals are cheap materials and as compared to superfluid helium-3,
for instance, relatively easy to work with in the laboratory. 
The symmetry breaking phase transitions typically appear at temperatures
that can be reached by a standard  kitchen oven, 
whereas the size of the occurring defects 
is such that these can be seen  by means of a simple microscope. 
Hence, these materials form an easily accessible experimental playground 
for the investigation of defect producing phase transitions
and as such may partly mimic the physics of the early universe in the 
laboratory.
For some recent ingenious experimental 
studies on the formation and the dynamics of topological 
defects in nematic crystals making use of high speed film 
cameras, the interested reader is referred to~\cite{bowick, turok}.

From a theoretical point of view, 
many aspects of topological defects have been studied and understood.  
At the classical level, one may roughly sketch
the following programme. One first uses simple topological
arguments, usually of the homotopy type, to see whether a given model
does exhibit topological charges. Subsequently, one may try to prove
the existence of corresponding classical solutions by functional
analysis methods or just by explicit construction of particular
solutions. On the other hand, one may in many cases determine the
dimension of the solution or moduli space and its dependence on the
topological charge using index theory.  
Finally, one may attempt to determine the
general solution space more or less explicitly.  In this respect,
one has been successful in varying degree. 
In particular, the self-dual
instanton solutions to the Yang-Mills theory (on $S^4$) have been
obtained completely.

The physical properties of topological defects can be probed by
their interactions with the ordinary particles or excitations
in the model.  This amounts to investigating
(quantum) processes in the background of the defect. In particular, one
may calculate the one-loop corrections to the various quantities
characterizing the defect, which involves studying 
the fluctuation operator.  
Here, one has to distinguish the modes with zero eigenvalue
from those with nonzero eigenvalues. The nonzero modes generically
give rise to the usual renormalization effects, such as mass and coupling
constant renormalization. The zero modes, which often arise as a
consequence of the global symmetries in the theory, lead to
collective coordinates.  Their quantization yields a semiclassical
description of the spectrum of the theory in a given topological
sector,  including the external quantum numbers of the soliton 
such as its energy and momentum and its internal quantum numbers such
as its electric charge.

In situations where the residual gauge group $H$ is nonabelian, 
this analysis is rather subtle. For instance, 
the naive expectation that a soliton can
carry internal electric charges which form representations of the
complete unbroken group $H$ is wrong. As only
the subgroup of $H$ which commutes with the topological charge can be
globally implemented, these internal charges form representations 
of this so-called centralizer subgroup. See~\cite{balglob, nelson, nelsonc} 
for the case of magnetic monopoles and~\cite{spm, balglob2} 
for magnetic vortices. 
This makes the full spectrum of topological and
ordinary quantum numbers in such a  broken phase rather intricate.

Also, an important effect on the spectrum and the interactions 
of a theory with a broken gauge group 
is caused by the introduction of additional
topological terms in the action, such as a nonvanishing
$\theta$ angle in 3+1 dimensional space time 
and the Chern-Simons term in 2+1 dimensions.  
It has been shown by Witten that 
in case of a nonvanishing $\theta$ angle, for example,
magnetic monopoles carry electric charges
which are shifted by an amount proportional to $\theta/2\pi$ and their
magnetic charge~\cite{thetaw}.

Other results are even more surprising. A broken gauge theory only
containing bosonic fields
may support topological excitations (dyons),
which on the quantum level carry half-integral spin and are fermions,
thereby realizing the counterintuitive possibility to make
fermions out of bosons~\cite{thora, jare}. 
It has subsequently been argued by Wilczek that in 2+1 dimensional space time
one can even have topological excitations, 
namely flux/charge composites, which behave as anyons, i.e.\
particles with fractional spin and quantum statistics 
interpolating between bosons and fermions~\cite{wilcchfl}. 
The possibility of anyons in two
spatial dimensions is not merely of academic interest, as many systems 
in condensed matter physics, for example, are effectively described
by 2+1 dimensional theories.
In fact, anyons are known to be realized as quasiparticles 
in fractional quantum Hall systems~\cite{hal, laugh}.
Furthermore, it has been been shown that an  anyon gas is  
superconducting~\cite{chen, fetter, laughsup}, see 
also~\cite{anyonbook}.
This new and rather exotic 
type of superconductivity still awaits an application
in nature.

To continue, remarkable calculations by 't Hooft
revealed a nonperturbative mechanism for baryon
decay in the standard model through instantons 
and sphalerons~\cite{thooftinst}.
Afterwards, Rubakov and Callan  discovered
the phenomenon of baryon decay catalysis induced by 
grand unified monopoles~\cite{callan, ruba}. Baryon number violating 
processes also occur in the vicinity of grand unified cosmic strings as 
has been established by Alford, March-Russell and 
Wilczek~\cite{alfbar}.

So far, we have enumerated properties and processes that only involve
the interactions between topological and ordinary excitations. 
However, the interactions between defects themselves 
can also be highly nontrivial. Here, one should not only
think of ordinary interactions corresponding to 
the exchange of field quanta. Consider, for instance, the case
of Alice electrodynamics which occurs if some nonabelian gauge group 
(e.g.\ $SO(3)$) is broken to the nonabelian subgroup $U(1) \rtimes \Z_2$,
that is, the semi-direct product of the electromagnetic group $U(1)$ and 
the additional cyclic group $\Z_2$ whose nontrivial element 
reverses the sign of the electromagnetic fields~\cite{schwarz}.
This model features magnetic monopoles and a magnetic $\Z_2$ 
string (the so-called Alice string) 
with the miraculous property that if a monopole 
(or an electric charge for that matter) 
is transported around the string, 
its charge will change sign.
In other words, a particle is converted into its own anti-particle.
This drastic effect is an example of a topological interaction, that is,
it only depends on the number of times the particle 
winds around the string and is independent of the distance 
between the particle and the string.
As alluded to before, 
this kind of interaction is not mediated by the exchange of  
field quanta, but should be seen as a nonabelian generalization 
of the celebrated Aharonov-Bohm effect~\cite{ahabo}.

Similar phenomena occur in models in which a continuous gauge group
is broken down to a {\em finite} subgroup $H$.
The topological defects arising in this case are string-like in three 
spatial dimensions and carry a magnetic flux corresponding to an element
$h$ of the residual gauge  group $H$. As these string-like objects trivialize
one spatial dimension, we may just as well descend to the plane, for 
convenience.
In this arena, these defects become magnetic vortices, i.e.\
particle-like objects of characteristic size $1/M_H$ with $M_H$ the 
symmetry breaking scale. 
Besides these topological particles, 
the broken phase features matter 
charges labeled by the unitary irreducible representations $\Gamma$ 
of the residual gauge group $H$. 
Since all gauge fields are massive, there are no ordinary 
long range interactions among these  particles.
The remaining long range interactions are the aforementioned 
topological Aharonov-Bohm interactions. 
If the residual gauge group $H$ is 
nonabelian, for instance, the nonabelian fluxes $h \in H$ carried by the 
vortices exhibit flux metamorphosis~\cite{bais}. In the process 
of circumnavigating one vortex with another vortex their fluxes may change.
Moreover, if a charge corresponding to some representation $\Gamma$ of 
$H$ is transported around a vortex carrying the magnetic flux $h \in H$, 
it returns transformed by the matrix $\Gamma(h)$ assigned to the 
element $h$ in the representation $\Gamma$.

The 2+1 dimensional spontaneously broken models briefly 
touched upon in the previous paragraph will be the subject of this thesis.
The organization is as follows. 
In chapter~\ref{chap2}, we present a self-contained discussion  
of planar gauge theories in which a continuous gauge group $G$ 
is spontaneously broken down to a finite subgroup $H$. 
The main focus will be 
on the discrete $H$ gauge theory that describes the long 
range physics of such a model. We establish the complete 
spectrum, which besides the aforementioned magnetic vortices 
and matter charges also consists of dyonic composites of the two,
and argue that as a result of the Aharonov-Bohm effect
these particles acquire braid statistics in the first quantized description.
The dyons appearing in the abelian case $H \simeq \Z_N$, for instance,
behave as anyons: upon interchanging two identical dyons, the wave 
function picks up a quantum statistical phase factor 
$\exp(\im \Theta) \neq 1,-1$. The particles featuring in nonabelian 
discrete $H$ gauge theories in general constitute nonabelian generalizations
of anyons: upon interchanging two identical particles 
the multi-component wave function transforms by means of a matrix.
Among other things, we will 
also address the issue of the spin-statistics connection for these particles, 
the cross sections for low energy scattering experiments 
involving these particles, 
and elaborate on the intriguing phenomenon of Cheshire charge. 
In fact, the key result of this chapter will be 
the identification of the quantum group or Hopf algebra 
underlying a discrete $H$ gauge theory. 
This is the so-called quantum double $D(H)$, which is completely 
determined in terms of the data of the residual finite gauge group $H$. 
The different particles in the spectrum of a  discrete $H$
gauge theory correspond to the inequivalent irreducible representations 
of the quantum double $D(H)$. Moreover, the quantum double $D(H)$ provides an 
unified description of the spin, braid and fusion properties of 
the particles.

Chapters~\ref{chap3} and~\ref{chap4} deal with the implications 
of adding a topological 
Chern-Simons term to these broken planar gauge theories.  
The abelian case  is treated in chapter~\ref{chap3}. Here, we consider 
Chern-Simons theories in which a gauge group $G$, 
being a direct product of various compact $U(1)$ gauge groups, is 
spontaneously broken down to a finite subgroup $H$.
Several issues will be addressed of which we only mention the main ones. 
To start with, a Chern-Simons term for the continuous gauge group 
$G$ affects the topological interactions in the broken phase.
It gives rise to additional Aharonov-Bohm interactions for the 
magnetic vortices. In fact, these additional topological interactions   
are governed by a 3-cocycle $\omega$ for 
the residual finite gauge group $H$ which is the remnant 
of the original Chern-Simons term for the broken gauge group $G$. 
Accordingly, the quantum double $D(H)$ underlying the discrete $H$ 
gauge theory in the absence of a Chern-Simons term is deformed into 
the quasi-quantum double $D^\omega (H)$.
To proceed, it turns out that not all conceivable 3-cocycles for 
finite abelian groups $H$ can be obtained from the spontaneous 
breakdown of a continuous abelian Chern-Simons theory.
The 3-cocycles that do not occur are actually the most interesting.
They render an abelian discrete $H$ gauge theory nonabelian.
We will show that, for example, a 
$\Z_2 \times \Z_2 \times \Z_2$ Chern-Simons theory defined by such a 
3-cocycle is dual to an ordinary $D_4$ gauge theory, 
with $D_4$ the nonabelian dihedral group of order $8$. The duality 
transformation relates the magnetic fluxes of one 
theory with the electric charges of the other.
Finally, in chapter~\ref{chap4} we study the 
nonabelian discrete $H$ Chern-Simons theories describing the long distance 
physics of Chern-Simons theories in which a continuous nonabelian 
gauge group $G$ is spontaneously broken down to a nonabelian finite 
subgroup $H$.

To conclude, throughout this thesis units in 
which $\hbar = c =1$ are employed.
Latin indices take the values $1,2$. Greek indices 
run from $0$ to $2$. Further, 
$x^1$ and $x^2$ denote spatial coordinates and $x^0=t$
the time coordinate. 
The signature of the three dimensional metric 
is taken as $(+,-,-)$.  Unless stated otherwise, we adopt 
Einstein's summation convention.

\chapter{Discrete  gauge theories}
\label{chap2}

\sectiona{Introduction}

In this chapter, we will study planar gauge theories
in which a gauge group $G$ is broken down to a finite subgroup $H$ 
via the Higgs mechanism. 
Such a model is governed by an action of the form
\bea                               \label{algz}
S &=& S_{\mbox{\scriptsize YMH} } + S_{\mbox{\scriptsize matter}},
\eea  
where the Yang-Mills Higgs part $S_{\mbox{\scriptsize YMH} }$ features 
a Higgs field whose nonvanishing vacuum expectation values are only 
invariant under the action of $H$
and where the matter part $S_{\mbox{\scriptsize matter}}$ 
describes matter fields minimally coupled to the gauge fields.
The incorporation of a Chern-Simons term for the broken gauge group $G$
will be dealt with in the next chapters.

As all gauge fields are massive, it seems that the 
low energy or equivalently the long distance 
physics of these spontaneously broken gauge theories
is completely trivial. This is not the case, 
however. It is the occurrence of topological defects and the persistence of 
the Aharonov-Bohm effect that renders 
the long distance physics nontrivial.
To be specific, the defects that occur in this model 
are (particle-like) vortices of characteristic size $1/M_H$, 
with $M_H$ the symmetry breaking scale. These vortices
carry magnetic fluxes labeled by the elements 
$h$ of the residual gauge group $H$. In other words, the vortices introduce 
nontrivial holonomies in the locally flat gauge fields.
Consequently, if the residual gauge group ${H}$ is nonabelian,
these fluxes exhibit nontrivial topological interactions.
In the process in which one vortex circumnavigates another, the 
associated magnetic fluxes feel 
each others holonomies and affect each other through conjugation.  
This is in a nutshell the long distance 
physics described by the Yang-Mills Higgs part 
$S_{\mbox{\scriptsize YMH}}$ of the action~(\ref{algz}).  As mentioned before,
in the matter part $S_{\mbox{\scriptsize matter}}$, we have 
matter fields minimally coupled to the gauge fields.
These charged matter fields form multiplets which transform irreducibly 
under the broken gauge group ${ G}$. In the 
broken phase, these branch to irreducible representations 
of the residual gauge group ${ H}$.
So, the matter fields introduce charges in the Higgs phase labeled by the 
unitary irreducible representations $\Gamma$ of $H$.
When such a charge encircles a magnetic 
flux $h \in H$, it exhibits an Aharonov-Bohm effect.
That is, it  returns transformed by the matrix $\Gamma(h)$ assigned to the 
group element $h$ in the representation $\Gamma$ of $H$.
Besides these matter charges and magnetic fluxes, the complete spectrum
of the discrete $H$ gauge theory describing  the long distance 
physics of the broken model~(\ref{algz}), 
consists of dyons obtained by composing  the charges and the fluxes. 

In this chapter, we set out to give a complete description of these discrete
gauge $H$ theories. 
The outline is as follows. 
In section~\ref{braidgroups}, we start by briefly recalling
the notion of braid groups which organize the interchanges of particles in 
the plane. 
Section~\ref{abznz} then contains a discussion of  
the planar abelian Higgs model in which the $U(1)$ gauge group 
is spontaneously broken to the cyclic subgroup $\Z_N$.
The main focus will be on the $\Z_N$ gauge theory that describes the 
long distance physics of this model. 
We show that the spectrum indeed consists of  
$\Z_N$ fluxes, $\Z_N$ charges and dyonic 
combinations of the two and establish the quantum mechanical 
Aharonov-Bohm interactions among these particles.
The subtleties involved 
in the generalization to models in which a nonabelian gauge group $G$ 
is broken to a nonabelian finite group $H$ are
dealt with in section~\ref{nonabz}.
In section~\ref{qdH}, we subsequently 
identify the algebraic structure underlying 
a discrete ${ H}$ gauge theory as the quantum double $D({ H})$.
Here, we also note that the wave functions of the 
multi-particle systems occurring in  discrete $H$ gauge theories,
in fact, realize unitary irreducible representations
of truncated braid groups, that is, factor groups of the 
ordinary braid groups, which simplifies matters considerably. 
In section~\ref{exampled2b}, we illustrate the previous 
general considerations with an explicit example of a nonabelian discrete
gauge theory, namely a $\bar{D}_2$ gauge theory, with $\bar{D}_2$ the 
double dihedral group.
We give the fusion rules for the particles, 
elaborate on the intriguing notions of Alice fluxes and Cheshire charges, 
calculate the cross section for an Aharonov-Bohm 
scattering experiment involving an Alice flux and a doublet charge,
and finally address the (nonabelian) braid 
statistical properties of the multi-particle systems that may emerge  
in this model. 
We have also included two appendices. Appendix~\ref{ahboverl} 
contains a short discussion of the cross sections  
for the Aharonov-Bohm scattering experiments 
involving the particles in (non)abelian discrete $H$ gauge theories.
Finally, in appendix~\ref{trubra} we give the group structure of two
particular truncated braid groups which enter
the treatment of the $\bar{D}_2$ gauge theory 
in section~\ref{exampled2b}.

\sectiona{Braid groups}    \label{braidgroups}

Consider a system of $n$ indistinguishable particles moving on a 
manifold $M$, which is assumed to be 
connected and path connected for convenience. 
The classical configuration space of this system 
is given by
\bea                                  \label{configi}
{\cal C}_n ({M}) &=& (M^{n} - D)/S_n,
\eea 
where the action of the permutation group $S_n$ on the particle 
positions is divided out to account for the indistinguishability 
of the particles. Moreover, the singular configurations $D$ 
in which two or more particles coincide are excluded.  
The configuration space~(\ref{configi}) is in 
general multiply-connected. This means that there are different 
kinematical options to  quantize this multi-particle system. 
To be precise, there is  a  consistent quantization
associated to each  unitary irreducible representation (UIR) of the 
fundamental group 
$\pi_1 ({\cal C}_n ({M}))$~\cite{laid, schul, schul2, imma}.

It is easily verified that for manifolds $M$ with dimension larger then 2, 
we have $\pi_1({\cal C}_n ({M})) \simeq S_n$. Hence, 
the inequivalent quantizations of multi-particle systems moving on 
such manifolds are labeled by the UIR's of the permutation group $S_n$. 
There are two  1-dimensional UIR's of $S_n$. The trivial representation
naturally corresponds with Bose statistics. In this case, 
the system is quantized by a (scalar) wave function, which 
is symmetric under all permutations of the particles.
The anti-symmetric representation, on the other hand,
corresponds with Fermi statistics, i.e.\
we are dealing with a wave function which acquires a minus sign 
under odd permutations of the particles.
Finally, parastatistics is also conceivable.
In this case, the system is quantized by a multi-component
wave function which transforms as a 
higher dimensional UIR of $S_n$.

\begin{figure}[tbh]    \epsfxsize=10cm
\centerline{\epsffile{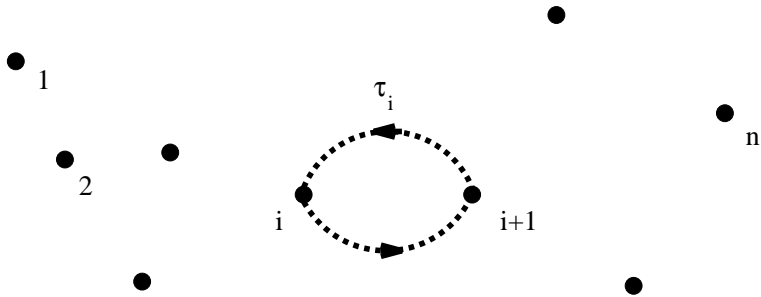}}
\caption{\sl The  braid operator $\tau_i$ establishes a counterclockwise 
interchange of the particles  $i$ and $i+1$ in a set of $n$
numbered indistinguishable particles in the plane.}
\label{touw}
\end{figure}

\begin{figure}[htb]    \epsfxsize=9.5cm
\centerline{\epsffile{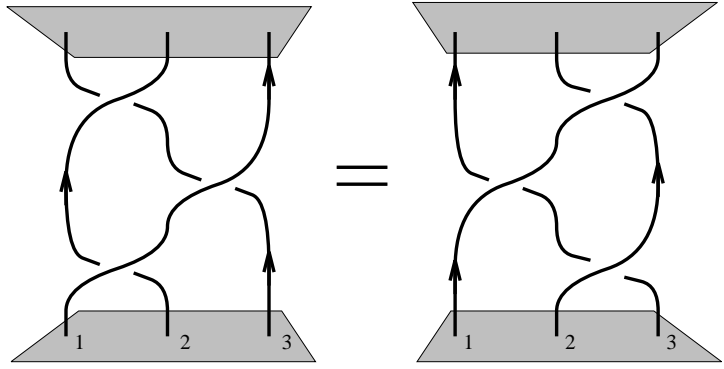}}
\caption{\sl 
Pictorial presentation of the braid relation 
$\tau_1 \tau_2 \tau_1= \tau_2 \tau_1 \tau_2$.
The particle trajectories corresponding to the composition of exchanges 
$\tau_1 \tau_2 \tau_1$ (diagram at the l.h.s.)
can be continuously deformed into the trajectories associated with 
the composition of exchanges 
$\tau_2 \tau_1 \tau_2$ (r.h.s.\ diagram).}
\label{ybes}
\end{figure}

\begin{figure}[htb]    \epsfxsize=11.5cm
\centerline{\epsffile{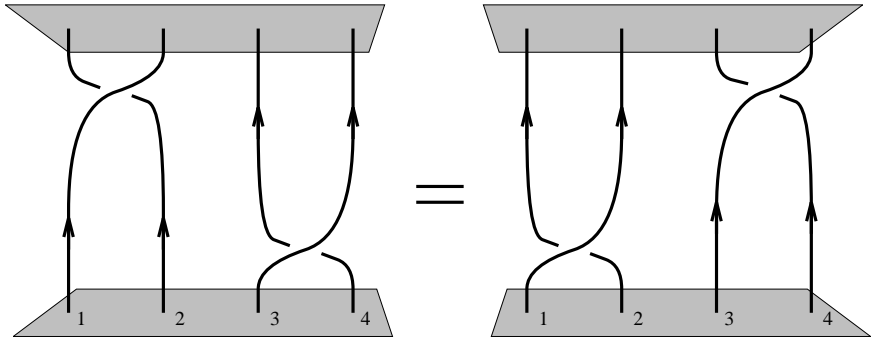}}
\caption{\sl
The braid relation $\tau_1 \tau_3 = \tau_3 \tau_1$
expresses the fact that the particle trajectories 
displayed in the l.h.s.\ diagram  can be continuously 
deformed into the trajectories in the r.h.s.\ diagram.}
\label{ybes2}
\end{figure}

It has been known for some time that quantum statistics for identical 
particles moving in the plane ($M={\mbox{\bf R}}^2$) 
can be much more exotic then 
in three or more dimensions~\cite{leinaas,wilcan}. 
The point is that the fundamental group of the associated 
configuration space  ${\cal C}_n ({\mbox{\bf R}}^2)$ is not given by 
the permutation group, but rather by the so-called braid 
group $B_n ({\mbox{\bf R}}^2)$~\cite{wu}.
In contrast with the permutation group $S_n$, 
the braid group $B_n ({\mbox{\bf R}}^2)$ is a nonabelian group of 
{\em infinite} order. It is 
generated by $n-1$ elements  $\tau_1, \ldots, \tau_{n-1}$, where $\tau_i$
establishes  a counterclockwise interchange 
of the particles $i$ and $i+1$ as depicted 
in figure~\ref{touw}. These generators are subject to the relations
\bea
\label{yangbax}
\ba{rcll} 
\tau_i\tau_{i+1}\tau_i &=& \tau_{i+1}\tau_i\tau_{i+1} & \qquad
 i=1,\ldots,n-2  \\
\tau_i\tau_j & = & \tau_j\tau_i & \qquad |i-j|\geq 2,
\ea
\eea  
which can be presented graphically as 
in figure~\ref{ybes} and~\ref{ybes2} respectively.
In fact, the permutation group $S_n$ ruling 
the particle exchanges in three or more dimensions,
is given by the same set of generators
with relations~(\ref{yangbax}) {\em and} the additional relations $\tau_i^2=e$
for all $i \in 1, \ldots, n-1$. 
These last relations are absent for 
$\pi_1({\cal C}_n ({\mbox{\bf R}}^2)) 
\simeq B_n ({\mbox{\bf R}}^2)$, since in the plane
a counterclockwise particle interchange $\tau_i$ ceases to 
be homotopic to the clockwise interchange $\tau_i^{-1}$.

The one dimensional UIR's of the braid group $B_n ({\mbox{\bf R}}^2)$ 
are labeled by an angular parameter $\Theta \in [0, 2\pi)$ 
and are defined  by assigning the 
same phase factor to all generators. That is, 
\bea                                \label{qstat} 
\tau_i  
&\mapsto& \exp (\im \Theta),
\eea
for all $i \in 1, \ldots, n-1$. 
The quantization of a system of $n$ identical particles in the plane
corresponding to an arbitrary but fixed $\Theta \in [0, 2\pi)$
is then given by a multi-valued (scalar) wave function that generates
the  quantum statistical phase $\exp(\im \Theta)$ upon a counterclockwise 
interchange of two adjacent particles.  
For $\Theta =0$ and $\Theta = \pi$,  we are 
dealing with bosons and fermions respectively.
The particle species related to other values 
of $\Theta$ have been called  anyons~\cite{wilcan}.
Quantum statistics deviating from conventional 
permutation statistics is known under various names in the literature,
e.g.\ fractional statistics, anyon statistics and exotic statistics.
We adopt the following nomenclature. 
An identical particle system  described by a (multi-valued) 
wave function that transforms as an one dimensional (abelian) UIR of 
the braid group $B_n ({\mbox{\bf R}}^2)$ ($\Theta \neq 0, \pi$)
is said to realize abelian braid statistics.
If an identical particle system is described by a multi-component 
wave function carrying an higher dimensional UIR of the braid group, 
then the particles are said to obey nonabelian braid statistics.

\begin{figure}[htb]    \epsfxsize=10cm
\centerline{\epsffile{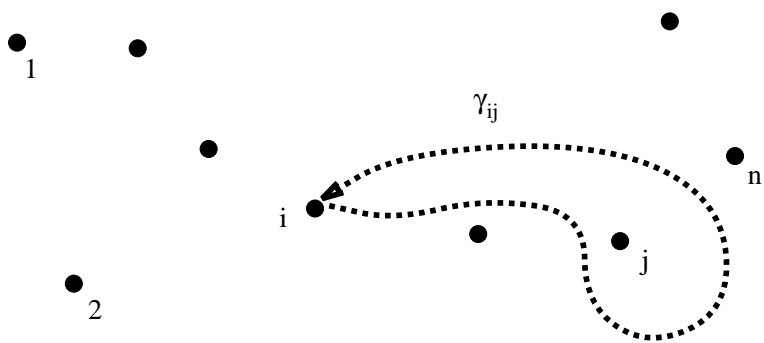}}
\caption{\sl The monodromy operator
$\gamma_{ij}$ takes particle $i$ counterclockwise
around particle $j$.}
\label{gammaije} 
\end{figure}

A system of $n$ distinguishable particles moving in the 
plane, in turn, is described by the 
non-simply connected configuration space
\bea
{\cal Q}_n ({\mbox{\bf R}}^{2}) &=& ({\mbox{\bf R}}^{2})^n - D.
\eea
The fundamental group of this configuration space is 
the so-called colored braid group $P_n({\mbox{\bf R}}^2)$, also known as the  
pure braid group. The colored braid group $P_n({\mbox{\bf R}}^2)$
is the subgroup of the ordinary braid group $B_n({\mbox{\bf R}}^2)$ 
generated by the monodromy operators
\bea                         \label{pbge}
\gamma_{ij} &:=& \tau_i \cdots \tau_{j-2} \tau_{j-1}^2 \tau_{j-2}^{-1}\cdots
             \tau_i^{-1}   \qquad \qquad \mbox{with} \; 1 \leq i<j \leq n.
\eea
Here, the $\tau_i$'s are the generators of $B_n({\mbox{\bf R}}^2)$ acting 
on the set of $n$ numbered 
distinguishable particles as displayed in figure~\ref{touw}.
It then follows from the definition~(\ref{pbge}) that 
the monodromy operator $\gamma_{ij}$ takes 
particle $i$ counterclockwise around particle $j$ as 
depicted in figure~\ref{gammaije}. The different 
UIR's of $P_n({\mbox{\bf R}}^2)$ now  label  
the inequivalent ways to quantize a system of $n$ 
distinguishable particles in the plane. 
Finally, a planar system that consists of a subsystem of identical particles 
of one type, a subsystem of identical particles of another type and so on, 
is of course also conceivable. 
The fundamental group of the configuration space of such a system 
is known as a partially colored braid group. 
Let the total number of particles of this system again be $n$, 
then the associated partially colored braid group 
is the subgroup of the ordinary braid group $B_n({\mbox{\bf R}}^2)$ 
generated by the braid operators that interchange identical 
particles and the monodromy operators acting on distinguishable 
particles. See for example~\cite{brekfa,brekke}.

To conclude, the fundamental excitations 
in planar discrete gauge theories, namely 
magnetic vortices and matter charges, are in principle bosons.
As will be argued in the next sections, 
in the first quantized description, these particles
acquire braid statistics through the Aharonov-Bohm effect.
Hence, depending on whether we are dealing with a system of 
identical particles, a system of distinguishable particles or 
a mixture, the associated multi-particle wave function
transforms as an representation 
of the ordinary braid group, colored braid group
or partially colored braid group respectively.

\sectiona{$\Z_N$ gauge theory} \label{abznz}

The simplest example of a broken gauge theory
is an $U(1)$ gauge theory broken down to the cyclic 
subgroup $\Z_N$.
This symmetry breaking scheme occurs in  an
abelian Higgs model in which the field that condenses carries
charge $Ne$, with $e$ the fundamental charge unit~\cite{krawil}.
The case $N=2$ is in fact realized in the ordinary BCS superconductor, as 
the field  that condenses in the BCS superconductor 
is that associated with the Cooper pair carrying charge $2e$.

This section is devoted to a 
discussion of such an abelian Higgs model focussing  
on the $\Z_N$ gauge theory describing the long range physics.
The outline is  as follows. 
In section~\ref{ahm}, we will start with a brief review of the 
screening mechanism for the electromagnetic fields 
of  external matter charges $q$ in the Higgs phase. 
We will argue that the external matter charges, which are multiples 
of the fundamental charge $e$ rather then multiples of the Higgs charge
$Ne$, are surrounded by screening charges provided by the Higgs condensate. 
These screening charges screen the electromagnetic fields around 
the external charges.  Thus 
no long range Coulomb interactions persist among  the external charges.
The main point of section~\ref{mavoab} will be, however,
that the screening charges
do {\em not} screen the Aharonov-Bohm interactions between the external 
charges and the magnetic vortices, which also occur in these 
models. {\em As a consequence, 
long range Aharonov-Bohm interactions persist between the 
vortices and the external matter charges in the Higgs phase.} 
Upon circumnavigating a magnetic
vortex (carrying a flux $\phi$ which is a multiple of the 
fundamental flux unit $\frac{2\pi}{Ne}$ in this case) with 
an external charge $q$ (being a multiple of the fundamental charge unit $e$) 
the wave function of the system picks up the Aharonov-Bohm 
phase $\exp(\im q\phi)$.  
These Aharonov-Bohm phases lead to observable low energy scattering 
effects from which we conclude that the physically distinct 
superselection sectors in the Higgs phase can be labeled 
as $(a,n)$, where $a$ stands for the number of fundamental 
flux units $\frac{2\pi}{Ne}$ and $n$ for the number of fundamental 
charge units $e$. In other words, 
the spectrum of the $\Z_N$ gauge theory in the 
Higgs phase consists of pure charges $n$, pure fluxes $a$ and dyonic
combinations. Given the remaining long range Aharonov-Bohm interactions, 
these charge and flux quantum numbers are defined modulo $N$.
Having identified the spectrum and the long range interactions
as the topological Aharonov-Bohm effect, we proceed with a
closer examination of this $\Z_N$ gauge theory in section~\ref{abdyons}.
It will be argued that multi-particle systems in 
general satisfy abelian braid statistics, 
that is, the wave functions realize one dimensional 
representations of the associated braid group. In particular,
identical dyons behave as anyons.
We will also discuss the composition rules 
for the charge/flux quantum numbers
when two particles are brought together.
A key result of this section is a topological proof of the spin-statistics
connection for the particles in the spectrum. 
This proof is of a general nature and applies to all the theories
that will be discussed in this thesis.

\subsection{Coulomb screening}  \label{ahm}

Let us start by emphasizing that we will work in 2+1 dimensional 
Minkowski space.  
The abelian Higgs model in which we are interested is given by 
\bea        \label{abhigg}
S &=& \int d \, ^3 x \;
({\cal L}_{\mbox{\scriptsize YMH}} + {\cal L}_{\mbox{\scriptsize matter}}) \\
{\cal L}_{\mbox{\scriptsize YMH}} &=& 
-\frac{1}{4}F^{\kappa\nu} F_{\kappa\nu} 
  +({\cal D}^\kappa \Phi)^*{\cal D}_\kappa \Phi - V(|\Phi|) \label{hiks} \\
{\cal L}_{\mbox{\scriptsize matter}} &=& -j^{\kappa}A_{\kappa}, \label{matcoup} 
\eea
where the Higgs field $\Phi$ is assumed to carry the charge $Ne$ 
w.r.t.\ the compact $U(1)$ gauge symmetry. In the conventions 
we will adopt, this means that the covariant derivative reads 
${\cal D}_{\rho}\Phi=(\partial_{\rho}+\im Ne A_{\rho})\Phi$. 
Furthermore, the potential  
\bea                         \label{poto}
V(|\Phi|) &=& \frac{\lambda}{4}(|\Phi|^2-v^2)^2  \qquad\qquad
 \lambda, v > 0,
\eea 
endows the Higgs field with a nonvanishing vacuum  expectation value
$|\langle \Phi \rangle|=v$, which implies that the 
the global continuous $U(1)$ symmetry is spontaneously broken. 
In this particular model the symmetry is 
not completely broken, however. Under global symmetry transformations 
$\Lambda(\alpha)$, with $\alpha \in [0,2\pi)$ being the $U(1)$ parameter,
the ground states transform as
\bea                     \label{reszn}
\Lambda(\alpha) \langle \Phi \rangle
&=& e^{\im N \alpha} \langle \Phi \rangle,
\eea 
since the Higgs field was assumed to carry  the charge $Ne$. Clearly,
the residual symmetry group of the ground states is the finite 
cyclic group $\Z_N$ corresponding to the 
elements  $\alpha= 2\pi k/N$ with $k \in 0,1,\ldots,N-1$.    

To proceed, the field equations following  from 
variation of the action~(\ref{abhigg}) 
w.r.t.\ the vector potential $A_{\kappa}$ and the Higgs field $\Phi$
are simply inferred as
\bea                            
\partial_{\nu} F^{\nu\kappa}
&=& j^\kappa+j^\kappa_H      \label{maxequ}     \\
{\cal D}_\kappa {\cal D}^\kappa \Phi^* &=& -\frac{\partial V}{\partial \Phi},
\label{hequ}
\eea   
where 
\bea       \label{Higgscur}
j^{\kappa}_H &=& \im Ne(\Phi^*{\cal D}^{\kappa}\Phi - 
({\cal D}^{\kappa}\Phi)^*\Phi),
\eea 
denotes the Higgs current.

In this section, we will only be concerned with the Higgs screening
mechanism for the electromagnetic fields of the matter charges, which are 
provided by the conserved matter current $j^\kappa$ in~(\ref{matcoup}). 
For convenience, we discard the dynamics of the fields that
are associated with this current and simply treat $j^\kappa$ 
as being external.
In fact, for our purposes the only important feature of 
the current $j^\kappa$ is that it allows us to introduce global 
$U(1)$ charges $q$ in the Higgs phase, which are multiples of the fundamental
charge $e$ rather then multiples of the 
Higgs charge $Ne$, so that all conceivable charge sectors can be discussed.

Let us start by recalling 
some of the basic dynamical features of  this model.
First of all, the complex Higgs field 
\bea
\Phi(x) &=& \rho(x)\exp (\im \sigma(x)) ,
\eea 
describes two physical degrees of freedom: the charged Goldstone boson field
$\sigma(x)$ and the physical field $\rho(x)-v$ with mass 
$M_H=v \sqrt{2\lambda}$ corresponding to the charged neutral Higgs particles.
The  Higgs mass $M_H$ sets the characteristic energy scale of this model.
At energies larger then $M_H$, the massive Higgs particles can 
be excited. At energies smaller then $M_H$ on the other hand,
the massive Higgs particles can not be excited. 
For simplicity we will restrict ourselves to 
the latter low energy regime. In this case,
the Higgs field is completely condensed, i.e.\ it acquires ground state values
everywhere
\bea                                     \label{simple}
\Phi(x) & \longmapsto & \langle \Phi(x) \rangle = v \exp (\im \sigma(x)).
\eea
The condensation of the Higgs field implies 
that the Higgs model in the low energy regime is 
governed by the effective action obtained from
the action~(\ref{abhigg}) by the following simplification 
\bea                               \label{efhig}
{\cal L}_{\mbox{\scriptsize YMH}} & \longmapsto & -\frac{1}{4}F^{\kappa\nu} F_{\kappa\nu} 
+\frac{M_A^2}{2} \tilde A^{\kappa}\tilde A_{\kappa} \\
\tilde{A}_{\kappa} & := & A_{\kappa} + \frac{1}{Ne}\partial_{\kappa}
\sigma
\label{Atild}  \\
 M_A & := & Ne v\sqrt{2}.      \label{mal}
\eea
In other words, the  dynamics of the Higgs medium arising here 
is described by the 
effective field equations inferred from varying the effective
action w.r.t.\ the gauge field $A_\kappa$ and the Goldstone boson $\sigma$
respectively    
\bea   \label{fieldhp}                            
\partial_\nu F^{\nu\kappa} &=& 
j^\kappa+j^\kappa_{\mbox{\scriptsize scr}}                         \\
\partial_\kappa  j^\kappa_{\mbox{\scriptsize scr}} &=& 0,       \label{fieldhp2}
\eea
with
\bea                                \label{scrcurr}
j^\kappa_{\mbox{\scriptsize scr}} &=& - M_A^2 \tilde{A}^\kappa,
\eea
the simple form the Higgs current~(\ref{Higgscur}) takes in the 
low energy regime.

It is easily verified that the field equations~(\ref{fieldhp}) 
and~(\ref{fieldhp2}) can be cast
in the following  form
\bea                     \label{kleingord}
(\partial_{\nu} \partial^{\nu} + M_A^2) \tilde{A}^\kappa &=& j^\kappa \\
\partial_{\kappa} \tilde{A}^\kappa &=& 0,
\eea
which clearly indicates that the gauge invariant 
vector field $\tilde{A}_\kappa$  has become massive.
More specifically, in this 2+1 dimensional setting 
it describes two physical degrees of freedom both
carrying the same mass $M_A$ defined in~(\ref{mal}). 
Consequently, the electromagnetic fields around sources in the Higgs medium
decay exponentially with mass $M_A$.
Of course, the number of degrees of freedom is conserved.
We started with an unbroken 
theory with two physical degrees of freedom $\rho-v$ and $\sigma$ 
for the Higgs field and one for the massless gauge field $A_{\kappa}$.
After spontaneous symmetry breaking the Goldstone boson $\sigma$
conspires with the gauge field $A_{\kappa}$ to form a massive 
vector field $\tilde{A}_{\kappa}$ with two  degrees of freedom,
while the real scalar field $\rho$ decouples in the low energy regime.

Let us finally turn to the response of  the Higgs medium to 
the external point charges $q=ne$ introduced by the matter current $j^\kappa$
in~(\ref{matcoup}). From~(\ref{kleingord}), we infer that 
the gauge invariant combined field 
$\tilde{A}_\kappa$  around this current 
drops off exponentially with mass $M_A$. Thus 
the gauge field $A_\kappa$ necessarily becomes pure gauge at 
distances much larger then $1/M_A$ from these point charges,
and the electromagnetic fields generated by this current
vanish accordingly.
In other words, the electromagnetic fields generated by 
the external charges $q$ are completely screened by the Higgs medium.
From the field equations~(\ref{fieldhp}) and~(\ref{fieldhp2}) 
it is clear how the Higgs screening mechanism works.
The external matter current $j^{\kappa}$ induces  
a screening current~(\ref{scrcurr}) in the Higgs medium proportional to 
the vector field $\tilde{A}_\kappa$. 
This becomes most transparent upon considering
Gauss' law  in this case 
\bea                    \label{higgsgaus}
Q \; = \; \int d \, ^2 x \; \nabla \cdot {\mbox{\bf E}} 
\; = \; q +  q_{\mbox{\scriptsize scr}} \; = \; 0,
\eea
which shows that the external point charge $q$ is surrounded
by a cloud of  screening charge density $j^0_{\mbox{\scriptsize scr}}$ with
support of characteristic size $1/M_A$.
The contribution of the screening charge 
$q_{\mbox{\scriptsize scr}} = \int d \, ^2x \, j^0_{\mbox{\scriptsize scr}}=-q$
to the long range Coulomb fields completely cancels 
the contribution of the external charge $q$. 
Thus we arrive at the well-known result that long range Coulomb 
interactions between  external charges vanish in the Higgs phase.

It has long been believed that with the vanishing of the Coulomb interactions,
there are no long range interactions left for the external charges in the 
Higgs phase.  However, it was indicated by Krauss, Wilczek 
and Preskill~\cite{krawil,preskra} that this is not the case.
They noted that when the $U(1)$ gauge
group is not completely broken, but instead we are left with
a finite cyclic manifest gauge group $\Z_N$ in the Higgs phase,
the external charges may still have long range Aharonov-Bohm 
interactions with the magnetic vortices  also appearing  in this model.
These interactions are of a purely quantum mechanical nature with no classical 
analogue.
The physical mechanism behind  the survival of  Aharonov-Bohm 
interactions was subsequently uncovered in~\cite{sam}: 
the screening charges $q_{\mbox{\scriptsize scr}}$ only couple to the Coulomb interactions 
and not to the Aharonov-Bohm interactions. As a result, the screening charges 
only screen the long range Coulomb interactions among the external charges,
but not the aforementioned long range Aharonov-Bohm 
interactions with the magnetic 
fluxes. We will discuss this phenomenon in further detail in the next section.

\subsection{Survival of the Aharonov-Bohm effect}
\label{mavoab}

A distinguishing feature of the abelian Higgs model~(\ref{hiks}) 
is that it supports stable vortices carrying 
magnetic flux~\cite{abri, niels}. 
These are static classical solutions of the 
field equations  with finite energy and correspond
to topological defects in the Higgs condensate, which are pointlike in 
this 2+1 dimensional setting. 
Here, we will briefly review the basic properties of these 
magnetic vortices and subsequently elaborate on their long range 
Aharonov-Bohm interactions with the screened external charges.

The energy density following from the action~(\ref{hiks}) for time 
independent field configurations reads
\bea                     \label{endensz}
{\cal E} &=& \frac{1}{2} (E^i E^i +B^2) +  (Ne A_0)^2 |\Phi|^2 +
{\cal D}_i \Phi ({\cal D}_i \Phi)^*
+V(|\Phi|).
\eea
All the terms occurring 
here are obviously positive definite. 
For field configurations of finite energy these 
terms should therefore vanish separately at spatial infinity. 
The potential~(\ref{poto}) vanishes for ground states only.
Thus the Higgs field is necessarily 
condensed~(\ref{simple}) at spatial infinity.
Of course, the Higgs condensate can still make a nontrivial 
winding in  the manifold of ground states. 
Such a winding at spatial infinity 
corresponds to a nontrivial holonomy  in the Goldstone boson field 
\bea                                      \label{goldhol}
\sigma(\theta+ 2\pi) - \sigma(\theta) &=& 2\pi a,
\eea
where $a$ is required to be an integer in order 
to leave the Higgs condensate~(\ref{simple}) 
itself single valued,
while $\theta$ denotes the polar angle. 
Requiring the fourth term in~(\ref{endensz}) to be integrable translates
into the condition
\bea        \label{atweg}
{\cal D}_i \Phi(r\rightarrow \infty) \sim 
\tilde{A}_i(r\rightarrow \infty)=0,
\eea  
with $\tilde{A}_i$ the gauge invariant combination of the Goldstone boson
and the gauge field defined in~(\ref{Atild}).
Consequently, the nontrivial holonomy in the Goldstone boson field
has to be compensated by an holonomy in the gauge fields and the vortices
carry  magnetic flux $\phi$  quantized as 
\bea         \label{quaflux}
\phi \;= \; \oint dl^i A^i  \; = \; 
\frac{1}{Ne} \oint dl^i \partial_i \sigma  \; = \; 
\frac{2\pi a}{Ne} \qquad \mbox{with $a \in \Z$.}
\eea
To proceed, the third term 
in the energy density~(\ref{endensz}) disappears at spatial infinity 
if and only if  $A_0 (r \rightarrow \infty)=0$,
and all in all we see that the gauge field $A_\kappa$ is pure gauge 
at spatial infinity, so the first two terms vanish automatically.
To end up with a regular field configuration corresponding to a  
nontrivial winding~(\ref{goldhol}) of the Higgs condensate at 
spatial infinity,
the Higgs field $\Phi$ should obviously become zero somewhere in the plane. 
Thus the Higgs phase is necessarily destroyed in some finite region in the 
plane. A closer evaluation of the energy density~(\ref{endensz}) shows 
that the Higgs field grows monotonically from its zero value to its 
asymptotic ground state value~(\ref{simple}) at the distance  $1/M_H$, 
the so-called core size~\cite{abri, niels}.
Outside the core we are in the Higgs phase, and the physics is described
by the effective Lagrangian~({\ref{efhig}), while inside the core
the $U(1)$ symmetry is restored.
The magnetic field associated with the flux~(\ref{quaflux}) of the vortex
reaches its maximum inside the core where the gauge fields are massless.
Outside the core the gauge fields become massive and the magnetic field
drops off exponentially with the mass $M_A$. 
The core size $1/M_H$ and the penetration depth $1/M_A$ of the magnetic field
are the two length scales characterizing the magnetic vortex.
The formation of magnetic vortices depends on the ratio of these 
two scales.  An evaluation of the free energy 
(see for instance~\cite{gennes}) yields that 
magnetic vortices can be formed iff $M_H/M_A =\sqrt{\lambda}/Ne \geq 1$.
We will always assume that this inequality is satisfied, 
so that magnetic vortices may indeed appear in the Higgs medium.
In other words, we assume that we are dealing with a 
superconductor of type~II.

We now have two dually charged 
types of sources in the Higgs medium.
On the one hand there are the vortices $\phi$ being sources for  
screened magnetic fields, and  on the other hand
the external charges $q$ being sources for screened  electric fields.
The magnetic fields of the vortices are localized within regions 
of length scale $1/M_H$ dropping off with mass $M_A$ at larger distances.
The external charges are point particles with Coulomb fields completely 
screened at distances $> 1/M_A$.
Henceforth, we will restrict our considerations to the low energy regime
(or alternatively send the Higgs mass $M_H$ and the mass $M_A$ of 
the gauge field to infinity by sending the symmetry breaking 
scale to infinity).  
This means that the distances between the sources remain much larger
then the Higgs length scale $1/M_H$. In other words, 
the electromagnetic fields
associated with the magnetic- and electric sources never overlap 
and the Coulomb interactions between
these sources vanish in the low energy regime.
Thus from a classical point of view 
there are no long range interactions left between the sources.
From a quantum mechanical perspective, however, it is known  that 
in ordinary electromagnetism 
shielded localized magnetic fluxes can affect electric charges 
even though their mutual electromagnetic fields do not interfere.
When an electric charge $q$ encircles  a localized magnetic
flux $\phi$, 
it notices the nontrivial holonomy in the locally
flat gauge fields around the flux and in this process the wave function
picks up  a quantum phase $\exp (\im q\phi)$ in the first quantized 
description.
This is the celebrated Aharonov-Bohm effect~\cite{ahabo}, which is  
a purely quantum mechanical effect with no classical analogue.
These long range Aharonov-Bohm interactions are of a topological 
nature, i.e.\ as long as the charge never enters the region where the 
flux is localized,
the Aharonov-Bohm interactions only depend on the number of windings of 
the charge around the flux and not on the distance between the charge and the 
flux.
Due to a remarkable cancellation in the effective action~(\ref{efhig}),
the screening charges $q_{\mbox{\scriptsize scr}}$ accompanying the external charges do not
exhibit the Aharonov-Bohm effect. As a result 
the long range Aharonov-Bohm effect persists between the external charges $q$
and the magnetic vortices $\phi$ in the Higgs phase. We will argue this
in further detail.

\begin{figure}[tbh]    \epsfxsize=9cm
\centerline{\epsffile{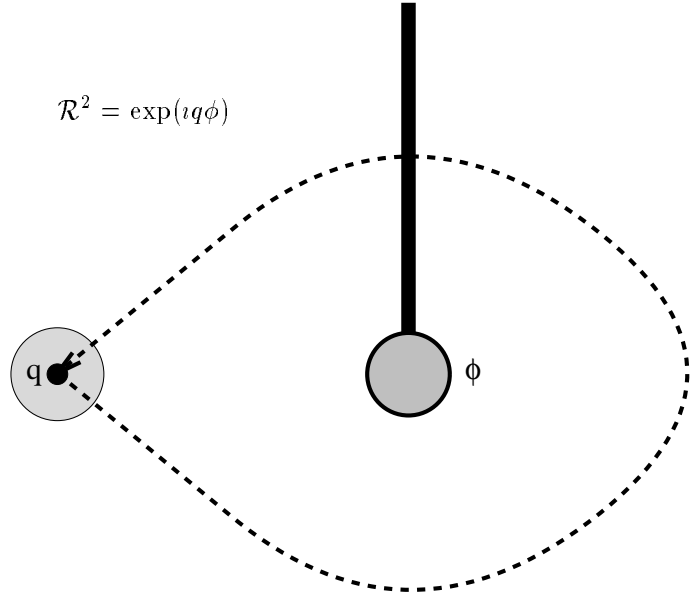}}
\caption{\sl Taking a screened
external charge $q$ around a magnetic vortex $\phi$ 
in the Higgs medium generates
the Aharonov-Bohm phase $\exp (\im q \phi)$. 
We have emphasized the extended structure of these 
sources, although this structure will not be probed 
in the low energy regime to which we confine ourselves here.
The shaded region around the external 
point charge $q$ represents the cloud of screening charge 
of characteristic size $1/M_A$.
The flux of the vortex is confined to the shaded circle bounded by the core 
at the distance $1/M_H$ from its centre. 
The string attached to the core represents the 
Dirac string of the flux, i.e.\ the  
strip in which the nontrivial parallel transport in the gauge fields 
takes place.}
\label{chargefluxbra}
\end{figure}

Consider the system depicted in figure~\ref{chargefluxbra} 
consisting of an external 
charge $q$ and a magnetic vortex $\phi$ in the Higgs phase
well separated from each other.
We have depicted
these sources as extended objects, but in the low energy regime their
extended structure will never be probed and it is legitimate
to describe these sources as point particles moving in the plane.
The magnetic vortex introduces a nontrivial holonomy~(\ref{quaflux})
in the gauge fields to which the external charge couples through  
the matter coupling~(\ref{matcoup})
\bea 
-\int \!\! d\, ^{2} x \; j^\kappa A_\kappa &=& \frac {q \phi}{2 \pi} 
\dot{\chi}_\phi({\mbox{\bf y}}(t)-{\mbox{\bf z}}(t)),
\label{int}
\eea 
where ${\mbox{\bf y}}(t)$ and ${\mbox{\bf z}}(t)$ respectively denote the 
worldlines of the external charge $q$ and magnetic vortex $\phi$ in the plane.
In the conventions we will use throughout this thesis, 
the nontrivial parallel transport in the gauge fields
around the magnetic vortices 
takes place in a thin strip (simply called Dirac string from now)
 attached  to the core of the vortex going
off to spatial infinity in the direction of the positive vertical axis.
This situation can always be reached by a smooth gauge transformation, and
simplifies the bookkeeping for the braid processes involving more than two 
particles. The 
multi-valued function $\chi_\phi({\mbox{\bf x}})$ with support 
in the aforementioned
strip of parallel transport is a direct translation of this convention.
It increases from $0$ to $2\pi$ if the strip is passed from right to left.
Thus when the external charge $q$ 
moves through this strip once in the counterclockwise
fashion indicated in figure~\ref{chargefluxbra},
the topological interaction Lagrangian~(\ref{int}) 
generates the action $q\phi$.
In the same process the screening charge $q_{\mbox{\scriptsize scr}}= -q$ 
accompanying the external charge $q$ also moves 
through this strip of parallel transport. Since the screening charge 
has a sign opposite to the sign of the external charge,
it seems, at first sight, that the total topological action associated 
with encircling a flux by a screened external charge vanishes. 
This is not the case though.
The screening charge $q_{\mbox{\scriptsize scr}}$ not only couples to the holonomy 
in the gauge field $A_\kappa$ around the vortex but also to the holonomy
in the Goldstone boson field $\sigma$. This follows directly from 
the effective low energy Lagrangian~(\ref{efhig}).
Let $j_{\mbox{\scriptsize scr}}^\kappa$ be the screening current~(\ref{scrcurr}) 
associated with the screening charge $q_{\mbox{\scriptsize scr}}$.
The interaction term in~(\ref{efhig}) couples this current to the 
massive gauge invariant  field $\tilde A_\kappa$ around the vortex:
$-j_{\mbox{\scriptsize scr}}^{\kappa}\tilde{A}_{\kappa}$.
As we have seen in~(\ref{atweg}), the holonomies in the gauge field
and the Goldstone boson field are related 
at large distances from the core of the vortex, such
that $\tilde A_\kappa$ strictly vanishes.  As a consequence, 
the interaction term $-j_{\mbox{\scriptsize scr}}^{\kappa}\tilde{A}_{\kappa}$ 
vanishes and indeed the matter coupling~(\ref{int}) summarizes 
all the remaining long range 
interactions in the low energy regime~\cite{sam}.

Being a  total time derivative, the topological interaction term~(\ref{int})
does not appear in the equations of motion and has no effect at
the classical level.   
In the first quantized description however, the appearance of this term 
has far reaching consequences. This is most easily seen using 
the path integral method for quantization. In the path integral formalism,
the transition amplitude or propagator
from one point in the configuration space at some
time to another point at some later time, is given by a weighed sum 
over all the paths connecting the two points. In this sum, the paths 
are weighed by their action $\exp (\im S)$. 
If we apply this prescription to our charge/flux system, we see that 
the Lagrangian~(\ref{int}) assigns amplitudes differing by $\exp (\im q\phi)$
to paths differing by an encircling 
of the external charge $q$ around the flux $\phi$.
Thus  nontrivial interference takes place between  paths 
associated with different winding numbers of the charge around the flux.
This is the Aharonov-Bohm effect which becomes observable 
in quantum interference experiments~\cite{ahabo},
such as low energy scattering 
experiments of external charges from the magnetic vortices. 
The cross sections measured in  these
Aharonov-Bohm scattering experiments can be found in 
appendix~\ref{ahboverl}.

There are two equivalent ways to present the appearance of
the Aharonov-Bohm interactions.
In the above discussion of the path integral formalism we kept 
the topological Aharonov-Bohm interactions in the Lagrangian for this 
otherwise free charge/flux system. In this description we work
with single valued wave functions on the 
configuration space for a given  time slice
\bea
\Psi_{q \phi}({\mbox{\bf y}}, {\mbox{\bf z}}, t) &=& \Psi_q ({\mbox{\bf y}}, t) 
\Psi_{\phi} ({\mbox{\bf z}}, t) \qquad\qquad \mbox{with ${\mbox{\bf y}} \neq {\mbox{\bf z}}$}.
\label{wave}
\eea
The factorization of the wave functions follows because there 
are no interactions between the external charge and the magnetic flux
other then the topological one~(\ref{int}).
The time evolution of these wave functions is given by the 
propagator associated with the two particle  Lagrangian
\bea
L &=& \frac{1}{2} m_q \dot{{\mbox{\bf y}}}^2 + \frac{1}{2} m_\phi \dot{{\mbox{\bf z}}}^2
+\frac {q \phi}{2 \pi}
\dot{\chi}_\phi({\mbox{\bf y}}(t)-{\mbox{\bf z}}(t)).
\eea
Equivalently, we may absorb the  topological interaction~(\ref{int}) 
in the boundary condition of the wave functions and 
work with multi-valued wave functions
\bea
\tilde{\Psi}_{q \phi}({\mbox{\bf y}}, {\mbox{\bf z}}, t) &:=&
e^{\im   \frac {q \phi}{2 \pi} \chi_\phi
({\mbox{\scriptsize \bf y}}-{\mbox{\scriptsize \bf z}}) } \;
\Psi_q ({\mbox{\bf y}}, t) \Psi_\phi ({\mbox{\bf z}}, t),
\label{wave2}
\eea                   
which propagate with the completely free two particle Lagrangian~\cite{wu}
(see also~\cite{forte})
\bea
\tilde{L} &=& 
\frac{1}{2} m_q \dot{{\mbox{\bf y}}}^2 + 
\frac{1}{2} m_\phi \dot{{\mbox{\bf z}}}^2.
\eea
We cling to the latter description from now on, that is,
we will always absorb the topological interaction terms in 
the boundary condition of the wave functions.
For later use and convenience we set some more conventions.
We will adopt a compact Dirac notation emphasizing the internal charge/flux
quantum numbers of the particles.
In this notation, the quantum state describing a  charge or flux 
localized at some position ${\mbox{\bf x}}$ in the plane is presented as
\bea
|{\mbox{charge/flux}} \rangle := |{\mbox{charge/flux}}, {\mbox{\bf x}}\rangle
= |{\mbox{charge/flux}}\rangle |{\mbox{\bf x}} \rangle.
\eea  
To proceed, the charges $q=ne$ will be abbreviated by the number $n$
of fundamental charge units $e$ and the fluxes $\phi$ by the number $a$  of 
fundamental flux units $\frac{2\pi}{Ne}$.
With the two particle quantum state  $|n \rangle |a\rangle$ 
we then indicate the multi-valued wave function
\bea
 | n \rangle |a \rangle &:=&  e^{\im \frac{n a}{N} 
\chi_{a} ({\mbox{\scriptsize \bf x}}- {\mbox{\scriptsize \bf y}})}  \;
|n, {\mbox{\bf x}} \rangle  |a, {\mbox{\bf y}} \rangle,
\label{Diracnota}
\eea
where by convention the particle that is located  most left in the plane 
(in this case the external charge $q=ne$), appears most 
left in the tensor product.
The process of transporting the charge adiabatically around the flux 
in a counterclockwise fashion as depicted in figure~\ref{chargefluxbra}
is now summarized by the action of the monodromy
operator on this two particle state
\bea         \label{monodro}
{\cal R}^2 \; | n \rangle |a \rangle 
&=& e^{\frac{2\pi \im}{N} na} \; | n \rangle |a \rangle,
\eea
which boils down to a residual global $\Z_N$  transformation 
by the flux $a$ of the vortex on the charge $n$.

Given the residual long range Aharonov-Bohm interactions~(\ref{monodro}) 
in the Higgs phase, the labeling of the charges 
and the fluxes by  integers is of course highly redundant.
Charges $n$ differing by a multiple of $N$ 
can not be distinguished. The  same holds for the fluxes $a$.
Thus the charge and flux quantum numbers are defined modulo $N$
in the residual manifest $\Z_N$ gauge theory arising in the Higgs phase.
Besides these pure $\Z_N$ charges and fluxes the full spectrum consists of 
charge/flux composites or dyons produced by fusing the charges and fluxes.
We return to a detailed discussion of this spectrum and the topological
interactions it exhibits in the next section.

Let us recapitulate our results from 
a more conceptual point of view (see also~\cite{alfrev, preskra, kli} in this 
connection). In unbroken (compact) 
quantum electrodynamics the quantized matter charges $q=ne$ (with $n \in \Z$),
corresponding to the different unitary irreducible representations
(UIR's) of the global symmetry group  $U(1)$,  
carry long range Coulomb fields.
In other words, the Hilbert space of this theory decomposes into a direct
sum of orthogonal charge superselection sectors that can be distinguished
by measuring the associated Coulomb fields at spatial infinity.   
Local observables preserve this decomposition, since they can not affect 
these long range properties of the charges. 
The charge sectors can alternatively be distinguished by their 
response to global $U(1)$  transformations, since these are  
related to physical measurements of the Coulomb fields at spatial infinity 
through Gauss' law. Let us emphasize that the states in the 
Hilbert space are of course invariant under local gauge transformations, 
i.e.\ gauge transformations with finite support, which 
become trivial at spatial infinity.

Here we touch upon the important distinction between global 
symmetry transformations and local gauge transformations. 
Although both leave the action of the model invariant, their physical 
meaning is rather different. A global symmetry (independent of the coordinates)  
is a true symmetry of the theory and in particular leads to a conserved 
Noether current.
Local gauge transformations, on the other hand, correspond 
to a redundancy in the variables describing this model and should therefore 
be modded out in the construction of the physical Hilbert space.
In the $U(1)$ gauge theory under consideration the fields that transform 
nontrivially under the global $U(1)$ symmetry are the matter fields. 
The associated Noether current $j^\kappa$ shows up in the Maxwell equations.
More specifically, the  conserved Noether charge $q=\int d\,^2 x \, j^0$, 
being the generator of the global symmetry, is identified with  
the Coulomb charge $Q=\int d\,^2 x \, \nabla \cdot {\mbox{\bf E}}$ 
through Gauss' law. This 
is the aforementioned relation between the global symmetry transformations
and physical Coulomb charge measurements at spatial infinity.

Although the long range Coulomb fields vanish
when this $U(1)$ gauge theory is spontaneously broken down to 
a finite cyclic group $\Z_N$, we are still able to detect
$\Z_N$ charge at arbitrary long distances through the Aharonov-Bohm effect.
In other words, there remains a relation between residual 
global symmetry transformations and physical charge measurements at 
spatial infinity.
The point is that we are left with a {\em gauged} $\Z_N$ 
symmetry in the Higgs phase, as witnessed by the appearance of 
stable magnetic fluxes in the spectrum. The magnetic fluxes introduce 
holonomies in the (locally flat) gauge fields, which take 
values in the residual manifest gauge group $\Z_N$ to 
leave the Higgs condensate single valued.  To be specific, 
the holonomy of a given flux
is classified by the group element picked up by the Wilson loop operator
\bea                \label{wilsonab}
W({\cal C}, {\mbox{\bf x}}_0)   &=& P \exp \, ( \im e \oint A^i dl^i) \; \in \Z_N,
\eea 
where ${\cal C}$ denotes a loop enclosing the flux starting  and 
ending at some fixed base point ${\mbox{\bf x}}_0$ at spatial infinity. 
The path ordering indicated by $P$ is trivial in this abelian case. 
These fluxes can be used for charge measurements in the Higgs phase
by means of the Aharonov-Bohm effect~(\ref{monodro}).  
This purely quantum mechanical effect,
boiling down to a global $\Z_N$ gauge transformation on the charge 
by the group element~(\ref{wilsonab}), is topological. 
It persists at arbitrary long ranges and therefore distinguishes 
the nontrivial $\Z_N$ charge  sectors in the Higgs phase. 
Thus the result of the Higgs mechanism 
for the charge sectors can  be summarized as follows:  
the charge superselection sectors of the original $U(1)$ gauge theory, 
which were in one-to-one correspondence with 
the UIR's of the global symmetry group $U(1)$, branch 
to UIR's of the residual  (gauged) symmetry group $\Z_N$ 
in the Higgs phase.

An important conclusion from this discussion is that a spontaneously broken 
$U(1)$ gauge theory in general can have distinct Higgs phases 
corresponding to different manifest gauge groups $\Z_N$. 
The simplest example is a  $U(1)$ gauge  theory 
with two Higgs fields; one carrying a charge $Ne$ and the 
other a charge $e$.  There are in principle
two possible Higgs phases  in this particular theory,
depending on whether the $\Z_N$ gauge symmetry
remains manifest or not. In the first case only  
the Higgs field with charge $Ne$ is condensed and we are left with 
nontrivial $\Z_N$ charge sectors.
In the second case the Higgs field carrying the fundamental 
charge $e$ is condensed. No charge sectors survive 
in this completely broken phase.
These two  Higgs phases, separated by a phase transition, can clearly 
be distinguished by probing the existence of $\Z_N$ charge sectors.
This is exactly the content of the nonlocal order parameter constructed 
by Preskill and Krauss~\cite{preskra} 
(see also~\cite{alfrev, alfmarc, alflee, lo1, lo2} in this context).  
In contrast with the Wilson loop operator 
and the 't Hooft loop operator  distinguishing the Higgs
and confining phase of a given gauge theory through the dynamics 
of electric and magnetic flux tubes~\cite{thooft, wilson},
this order parameter is of a topological nature.
To be specific, in this 2+1 dimensional setting
it amounts to evaluating the expectation value 
of  a closed electric flux tube linked with a closed magnetic flux 
loop corresponding to the worldlines of 
a minimal $\Z_N$ charge/anti-charge
pair linked with  the worldlines of a minimal $\Z_N$ magnetic 
flux/anti-flux pair. 
If the $\Z_N$ gauge symmetry is manifest, this order parameter
gives rise to the Aharonov-Bohm phase~(\ref{monodro}), whereas
it becomes trivial in the completely broken phase  with  minimal
stable flux  $\frac{2\pi}{e}$.

\subsection{Braid and fusion properties of the spectrum} \label{abdyons}

We proceed with a more thorough discussion of the topological interactions
described by the residual $\Z_N$ gauge theory in the Higgs phase 
of the model~(\ref{abhigg}). 
As we have argued in the previous section, the
complete spectrum of this discrete gauge theory consists
of pure $\Z_N$ charges labeled by $n$, pure $\Z_N$ fluxes labeled by $a$
and dyons produced by fusing these charges and fluxes 
\bea                              \label{compspectr}
|a \rangle \times |n \rangle  &=& 
|a, n \rangle \qquad  \qquad \mbox{with} 
\qquad a,n \in 0,1, \ldots, N-1.
\eea   
We have depicted this spectrum for a $\Z_4$ gauge theory 
in figure~\ref{z4z}.

\begin{figure}[htb] 
\begin{center}
\begin{picture}(85,80)(-15,-15)
\put(-5,-5){\dashbox(40,40)[t]{}}
\put(-15,0){\line(1,0){60}}
\put(0,-15){\line(0,1){60}}
\thinlines
\multiput(-10,-10)(0,10){6}{\multiput(0,0)(10,0){5}{\circle*{2.0}}}
\multiput(0,-10)(0,10){6}{\multiput(0,0)(20,0){3}{\circle*{2.0}}}
\put(0,40){\circle{3.6}}
%\put(0,40){\vector(0,-1){40}} 
\put(-3,42){\vector(0,1){5}}
\put(43,-3){\vector(1,0){5}}
\put(-10,50){\small$\phi[\frac{\pi}{2e}]$}
\put(50,-6){\small$q[e]$}
\end{picture}
\vspace{0.5cm}
\caption{\sl The spectrum of a Higgs phase with residual manifest
gauge group $\Z_4$  compactifies to the particles inside the dashed box.
The particles outside the box are identified with the ones inside by means 
of modulo $4$ calculus along the charge and flux axes.
The modulo $4$ calculus for the fluxes corresponds to Dirac 
monopoles/instantons, if these are present. The minimal monopole
tunnels the encircled flux into the vacuum.}
\label{z4z}
\end{center}
\end{figure}
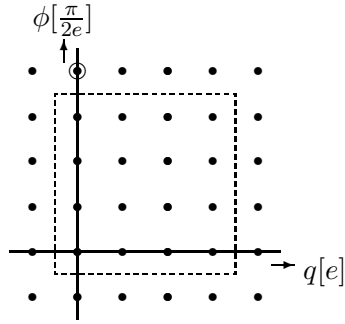

The topological interactions in these models are 
completely governed by the Aharonov-Bohm effect~(\ref{monodro}) 
and can simply be summarized as follows
\bea      
{\cal R}^2\;|a,n\rangle |a',n'\rangle &=&
e^{\frac{2 \pi \im}{N} (n a' +n'a)}
\;|a,n\rangle |a',n'\rangle      \label{monozoco}   \\
{\cal R} \; |a,n\rangle |a,n\rangle &=&
e^{\frac{2 \pi \im}{N} n a } \;|a,n\rangle |a,n\rangle
\label{brazoco}  \\
|a,n\rangle \times |a',n'\rangle 
       &=& |[a+a'],[n+n']\rangle  \label{fusionzoco} \\
{\cal C} \, |a,n \rangle &=& |[-a], [-n ]\rangle  \label{CC}  \\
T \, |a,n\rangle &=& e^{\frac{2 \pi \im}{N} na}  \;  |a,n \rangle.
\label{modulT}      
\eea 
The expressions~(\ref{monozoco}) and~(\ref{brazoco}) sum up 
the braid properties of the particles in the spectrum~(\ref{compspectr}). 
These realize abelian representations of the braid groups 
discussed in section~\ref{braidgroups}. For distinguishable 
particles only the monodromies, as contained in the pure braid 
groups~(\ref{pbge}), are relevant. In the present context, 
particles carrying different charge and magnetic flux are distinguishable. 
When a  particular particle $|a,n\rangle$ located at some position in the plane 
is adiabatically transported around another remote particle $|a',n'\rangle$
in the counterclockwise fashion depicted in figure~\ref{gammaije}, the 
total multi-valued wave function picks up the Aharonov-Bohm phase
displayed in~(\ref{monozoco}). In this process, the charge $n$
of the first particle moves through the Dirac string attached to the 
flux $a'$ 
of the second particle, while the charge $n'$ of the second particle moves through
the Dirac string of the flux $a$ of the first particle.
In short, the total 
Aharonov-Bohm effect for this monodromy 
is the composition of a global $\Z_N$ symmetry transformation on
the charge $n$ by the flux $a'$ and a global transformation on 
the charge $n'$ by the flux $a$.  
We confined ourselves to the case of two particles so far. 
The generalization  to systems containing more then two particles is 
straightforward. The quantum states describing these systems are 
tensor products of localized single particle states 
$|a,n, {\mbox{\bf x}}\rangle$, where we cling to the convention that the particle 
that appears most left in the plane appears most left in the tensor product.
These multi-valued wave functions carry abelian representations of the 
colored braid group: the action of the monodromy generators~(\ref{pbge}) 
on these wave functions boils down to the quantum phase     
in expression~(\ref{monozoco}).

For identical particles, i.e.\ particles carrying 
the same charge and flux, the braid operation depicted
in figure~\ref{touw} becomes meaningful.  In this braid process, in which
two adjacent identical particles $|a,n\rangle$ 
located at different positions in the 
plane are exchanged in a counterclockwise way, the charge of the particle
that moves `behind' the other dyon encounters the Dirac string attached to the 
flux of the latter. The result of this exchange in the multi-valued wave 
function is the quantum statistical phase factor (see expression~(\ref{qstat}))
presented in~(\ref{brazoco}).
In other words, the dyons in the spectrum of this $\Z_N$ theory 
are anyons.
In fact, these charge/flux composites are very close to 
Wilczek's original proposal for anyons~\cite{wilcchfl}. 

An important aspect of this theory is that the particles in the  
spectrum~(\ref{compspectr}) satisfy the canonical 
spin-statistics connection.
The proof of this connection is of a topological nature and applies 
in general to all the models that will be considered in this thesis.
The fusion rules play a role in this proof and we will discuss these first.

Fusion and braiding are intimately related. 
Bringing two particles together  is essentially a local process. 
As such, it can never affect global properties.
Thus the single particle state that arises after fusion should exhibit
the same global properties as the two particle state we started with.
In this topological theory, the global properties of a given 
configuration are determined by its braid properties with the 
different particles in the spectrum~(\ref{compspectr}). 
We have already established that the charges and fluxes 
become $\Z_N$ quantum numbers under these braid properties. 
Therefore the complete set of fusion rules, determining
the way the charges and fluxes of a two particle state 
compose into the charge and flux of a single particle state when the pair is 
brought together, can be summarized as~(\ref{fusionzoco}). 
The rectangular brackets denote modulo $N$ calculus such that the sum 
always lies in the range $0,1,\ldots, N-1$. 

It is worthwhile  to digress  a little
on the dynamical mechanism underlying the  modulo $N$ calculus 
compactifying the flux part of the spectrum. 
This modulo calculus is induced by magnetic monopoles, when these 
are present.
This observation will become important 
in chapter~\ref{abelsCS} where we will study the  incorporation 
of  Chern-Simons actions in this theory. 
The presence of magnetic monopoles can be accounted for by 
assuming that the compact $U(1)$ gauge theory~(\ref{abhigg}) 
arises from a spontaneously broken  $SO(3)$ gauge theory. The 
 monopoles we obtain in this particular model are 
the regular 't~Hooft-Polyakov monopoles~\cite{thooftmon,polyamon}.
Let us, alternatively, assume that we have singular 
Dirac monopoles~\cite{dirac} in this compact $U(1)$ gauge theory.  
In three spatial dimensions, these are  point particles carrying
magnetic charges $g$ quantized as  $\frac{2\pi}{e}$.
In the present  2+1 dimensional Minkowski setting,
they become instantons describing flux tunneling events 
$|\Delta \phi|= \frac{2\pi}{e}$.   As has been shown by 
Polyakov~\cite{polyakov}, the presence of these instantons 
in unbroken  $U(1)$ gauge theory has a striking dynamical effect.
It leads to linear confinement of electric charge. 
In the broken version of these theories, in which we are 
interested,  electric charge is screened and  the presence of 
instantons in the Higgs phase merely implies that the magnetic 
flux~(\ref{quaflux}) of the vortices is conserved modulo $N$ 
\bea                            \label{instans}
\mbox{instanton:}  & &  a \; \mapsto \; a -N.
\eea 
In other words, a flux $N$ moving in the plane (or $N$ minimal fluxes
for that matter) can disappear by ending on an instanton.
The fact that the instantons tunnel between
states that can not be distinguished 
by the braidings in this theory is nothing but 
the 2+1 dimensional translation of the unobservability of the  
Dirac string in three spatial dimensions.

We  turn to the connection between spin and statistics.
There are in principle two approaches to prove this deep relation, both 
having their own merits. 
One approach, originally due to Wightman~\cite{streater},
involves the axioms of local relativistic quantum field theory, and 
leads to the observation that integral spin fields commute,
while half integral spin fields anticommute.
The topological approach that we will take here was first
proposed by Finkelstein and Rubinstein~\cite{fink}. 
It does not rely upon the heavy framework
of local relativistic quantum field theory and among other things 
applies to the topological defects considered in this thesis.
The original formulation of Finkelstein and Rubinstein
was in the 3+1 dimensional context, but 
it naturally extends to 2+1 dimensional space time as we will 
discuss now~\cite{balach, sen}. See also~\cite{frohma, frogama}
for an algebraic approach.

The crucial ingredient in the topological proof of the spin-statistics 
connection for a given model is the existence of an anti-particle 
for every particle in the spectrum, such that the pair can annihilate 
into the vacuum after fusion. 
Consider the process depicted at the l.h.s. of the equality
sign in figure~\ref{spinstafig}. It describes the creation of 
two separate identical particle/anti-particle pairs from the vacuum,
a subsequent exchange of the particles of the two pairs and finally
annihilation of the pairs.
To keep track of the  writhing of the particle trajectories we 
depict them as ribbons with a white- and a dark side.
It is easily verified now that the closed ribbon associated with the 
process just explained  can be continuously deformed into the ribbon at the 
r.h.s., which corresponds to a rotation of the particle over an angle of
$2\pi$ around its own centre. In other words, the effect of interchanging
two identical particles in a consistent 
quantum description should be the same as the effect 
of rotating one particle over an angle of $2\pi$ around its centre. 
The effect of this rotation in the wave function is the spin factor
$\exp (2\pi \im s)$ with $s$ the spin of the particle, which in contrast with
three spatial dimensions may be any real number in two spatial dimensions.
Therefore the result of exchanging  the two identical particles
necessarily boils down to a quantum statistical phase 
factor $\exp (\im \Theta)$ in the wave function being
the same as the spin factor
\bea                          \label{spistath}
\exp (\im \Theta) &=& \exp (2\pi \im s).
\eea
This is the canonical spin-statistics connection. Actually, a further
consistency condition  can be inferred from this ribbon argument. The writhing 
in the particle trajectory can be continuously deformed to a writhing 
with the same orientation in the anti-particle trajectory. Therefore the 
anti-particle necessarily carries the same spin and statistics 
as the particle.

\begin{figure}[htb]    \epsfxsize=13cm
\centerline{\epsffile{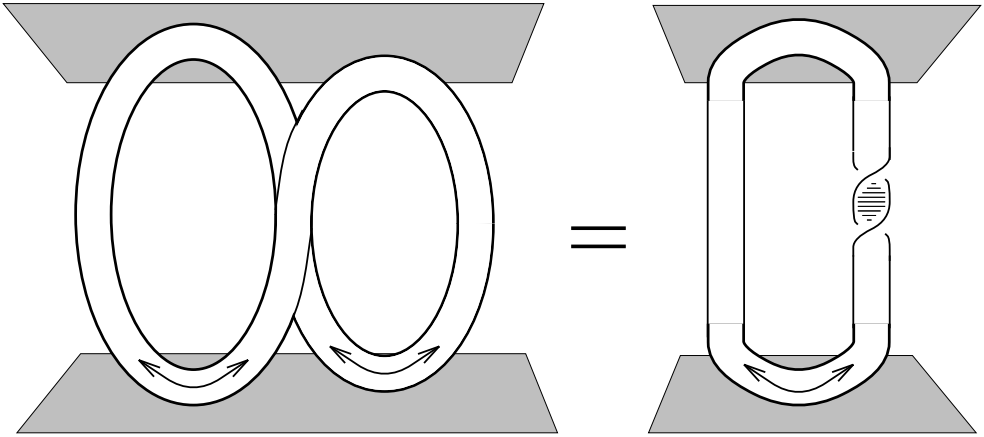}}
\caption{\sl Spin-statistics connection. The trajectories 
describing an exchange of two particles in
separate particle/anti-particle pairs (the 8 laying on its back)
can be continuously 
deformed into a single pair in which the particle undergoes a 
counterclockwise rotation
over an angle of $2\pi$ around its own centre (the 0 with a twisted leg).}
\label{spinstafig}
\end{figure}

Sure enough the topological proof of the spin-statistics theorem
applies to  the $\Z_N$ gauge theory at hand.
First of all, we can naturally assign an anti-particle to every 
particle in the spectrum~(\ref{compspectr}) through  the
charge conjugation operator~(\ref{CC}). Under charge conjugation
the charge and flux of the particles in the spectrum reverse sign and
amalgamating a particle with its charge conjugated partner yields
the quantum numbers of the vacuum as follows from the fusion 
rules~(\ref{fusionzoco}). Thus the basic assertion for the 
above ribbon argument is satisfied. 
From the quantum statistical phase factor~(\ref{brazoco}) assigned
to the particles and~(\ref{spistath}), we then conclude that the particles
carry spin. Specifically, under rotation over $2\pi$ the single particle
states should give rise to the spin factors displayed in~(\ref{modulT}). 
These spin factors can be interpreted as the Aharonov-Bohm phase 
generated when the charge of a given dyon rotates around its own flux.
Of course, a small separation between the charge and the flux of the dyon 
is required for this interpretation.
Also note that the particles and their anti-particles indeed 
carry the same spin and statistics, as follows immediately from the 
invariance of the Aharonov-Bohm effect under charge conjugation.

Having established a complete classification of the topological 
interactions in these abelian discrete gauge theories, we conclude with 
some remarks on the Aharonov-Bohm scattering  experiments 
by which these interactions can be probed.
(A concise discussion of these purely quantum mechanical  
experiments can be found in appendix~\ref{ahboverl}).
It is the monodromy effect~(\ref{monozoco}) that 
is measured in these two particle elastic scattering experiments.
To be explicit, the symmetric cross section for scattering a
particle~$|a,n\rangle$ from a particle~$|a',n'\rangle$ is given by
\bea                        \label{ZNab}
\frac{{\rm d} \sigma}{{\rm d} \theta} &=&
\frac{\sin^2 \left(\frac{\pi}{N}(na' +n'a )\right)}{2\pi p \sin^2 (\theta/2)},
\eea    
with  $p$ the relative momentum of the two particles and $\theta$ 
the scattering angle. 
A subtlety arises in scattering experiments involving two 
identical particles, however. Quantum statistics enters the scene:
exchange processes between the scatterer and the 
projectile have to be taken into account~\cite{anyonbook, preslo}.
This leads to the following cross section for Aharonov-Bohm 
scattering of two identical particles $|a,n\rangle$
\bea                        \label{ZNid}
\frac{{\rm d} \sigma}{{\rm d} \theta} &=&
\frac{\sin^2 (\frac{2\pi na}{N})}{2\pi p \sin^2 (\theta/2)} \; + \;
\frac{\sin^2 (\frac{2\pi na}{N})}{2\pi p \cos^2 (\theta/2)},
\eea     
where the second term summarizes the effect of the 
extra exchange contribution to the direct scattering amplitude.

\sectiona{Nonabelian discrete gauge theories} \label{nonabz}

The generalization of the foregoing analysis to spontaneously broken models
in which we are left with a {\em nonabelian} finite gauge group ${ H}$ 
involves essentially new features. 
In this introductory section, we 
will establish the complete flux/charge spectrum of such a  nonabelian 
discrete ${ H}$ gauge theory and discuss 
the basic topological interactions among the different flux/charge composites.
The outline is as follows. Section~\ref{topclas} contains
a general discussion on the topological classification of stable magnetic 
vortices and the subtle role magnetic monopoles play in this classification. 
In section~\ref{fluxmetamorphosis}, we  subsequently
review the properties of the nonabelian magnetic vortices that occur
when the residual symmetry group $H$ is nonabelian. 
The most important one being that these  vortices 
exhibit a nonabelian Aharonov-Bohm effect.
To be specific, the fluxes of the vortices, which are labeled by 
the group elements of ${ H}$, affect each other through conjugation 
when they move around each other~\cite{bais}.
Under the residual global symmetry group ${ H}$ the magnetic fluxes 
transform by conjugation as well, and the conclusion is that the vortices 
are organized in degenerate multiplets, corresponding to
the different conjugacy classes of ${ H}$.  
These classical properties will then be elevated
into the first quantized description in which the magnetic 
vortices are treated  
as point particles moving in the plane. 
In section~\ref{inclumatter}, we finally 
turn to the matter charges that may occur 
in these Higgs phases and their
Aharonov-Bohm interactions with the magnetic vortices.
As has been pointed out in~\cite{almawil,preskra}, these matter charges
are labeled by the different UIR's $\Gamma$ of the residual global symmetry 
group ${ H}$ and when such a charge encircles a nonabelian
vortex it picks up a global symmetry transformation by the matrix
$\Gamma(h)$ associated with the 
flux $h$ of the vortex in the representation $\Gamma$.
To conclude, we elaborate on the subtleties involved in the description
of dyonic combinations of the nonabelian magnetic fluxes and the matter
charges $\Gamma$.

\subsection{Classification of stable magnetic vortices}
\label{topclas}

Let us start by specifying the spontaneously broken gauge 
theories  in which we are left with
a nonabelian discrete  gauge theory.
In this case, we are dealing with a Higgs field $\Phi$ 
transforming according to some higher dimensional representation 
of a continuous nonabelian gauge group ${ G}$
\bea
S_{\mbox{\scriptsize YMH}} &=& \int d \, ^3 x \;  
(-\frac{1}{4}F^{a \; \kappa\nu} F_{\kappa\nu}^a
  +({\cal D}^\kappa \Phi)^\dagger \cdot {\cal D}_\kappa \Phi - 
V(\Phi) )     \label{nonhiks},
\eea
and a potential $V(\Phi)$ 
giving rise to a degenerate set of ground states 
$\langle \Phi \rangle \neq 0 $, which are only invariant under 
the action of a finite nonabelian subgroup ${ H}$ of ${ G}$.  
For simplicity, we make two assumptions. 
First of all, we assume that this Higgs potential is normalized such
that $V(\Phi) \geq 0$ and equals zero for the ground states 
$\langle \Phi \rangle $. More importantly, we assume 
that all ground states can be reached from any given one by global 
${ G}$ transformations. This last assumption  
implies that the ground state manifold 
becomes isomorphic to the coset ${ G}/{ H}$.
(Renormalizable examples of potentials doing the job for 
${ G}\simeq SO(3)$ and ${ H}$ some of its point groups 
can be found in~\cite{ovrut}). In the following, we will only be concerned
with the low energy regime of this theory,  so that the massive 
gauge bosons can be ignored.

The  stable vortices that can be formed in 
this spontaneously broken 
gauge theory  correspond to noncontractible maps
from the circle at spatial infinity (starting and ending at 
a fixed base point ${\mbox{\bf x}}_0$) 
into the ground state  manifold ${ G}/{ H}$. 
Different vortices are related to noncontractible maps 
that can not be continuously deformed into each other.
In short, the different vortices are labeled by the elements 
of the fundamental group $\pi_1$ of ${ G}/{ H}$ based at the 
particular ground state $\langle \Phi_0 \rangle$ the Higgs 
field takes at the base point ${\mbox{\bf x}}_0$ in the plane.
(Standard references on the use of homotopy groups 
in the classification of topological defects 
are~\cite{cola, mermin,presbook,trebin}. See also~\cite{poen} for an early 
discussion on the occurrence of nonabelian fundamental groups 
in models with a spontaneously broken global symmetry).

The content of the fundamental group $\pi_1 ({ G}/{ H})$  of the ground
state manifold for a specific spontaneously broken model~(\ref{nonhiks})
can be inferred from the exact sequence
\bea              \label{exactseq}
0 \simeq \pi_1 (H) \rightarrow \pi_1 (G) \rightarrow \pi_1 (G/H) \rightarrow
\pi_0 (H) \rightarrow \pi_0 (G) \simeq 0,
\eea
where the first isomorphism  follows from the fact that $H$ is discrete.
For convenience, we restrict our considerations to  continuous
Lie groups $G$ that are path connected, which 
accounts for the last isomorphism.
If  $G$ is simply connected as well, 
i.e.\ $\pi_1 ({ G}) \simeq 0$, 
then the exact sequence~(\ref{exactseq}) yields the isomorphism 
\bea   \label{flulaord}
\pi_1({ G}/{ H}) &\simeq& { H},
\eea 
where we used the result $\pi_0 (H) \simeq H$, which holds 
for finite $H$.
Thus the different magnetic vortices in this case
are in one-to-one correspondence
with the group elements $h$ of the residual symmetry group ${ H}$. 
When ${ G}$ is not simply connected, however,  
this is not a complete classification. This can be seen 
by the following simple argument. Let $\bar{ G}$ denote the universal covering
group of ${ G}$ and $\bar{ H}$ the corresponding lift 
of ${ H}$ into $\bar{ G}$.
We then have $G/H = \bar{ G}/\bar{ H}$ and in particular
$\pi_1({ G}/{ H}) \simeq \pi_1(\bar{ G}/\bar{ H})$.
Since the  universal covering group of $G$ is by definition simply connected,
that is, $\pi_1(\bar{G}) \simeq 0$, we obtain the following isomorphism
from the exact sequence~(\ref{exactseq}) for the lifted groups  
$\bar{G}$ and $\bar{H}$
\bea                                  \label{barh}
\pi_1({ G}/{ H}) \; \simeq \; \pi_1(\bar{ G}/\bar{ H}) \;\simeq \;
 \bar{ H}.
\eea   
Hence, for a non-simply connected broken gauge 
group $G$, the different stable
magnetic vortices are labeled by the elements of $\bar{H}$
rather then $H$ itself. 

It should be emphasized that the extension~(\ref{barh}) 
of the magnetic vortex spectrum is  
based on the tacit assumption that there are no 
Dirac monopoles featuring in this model. In any theory with a   
non-simply connected gauge group $G$, however, we have the freedom 
to introduce singular Dirac monopoles `by hand'~\cite{cola,later}. 
The  magnetic charges of these 
monopoles are characterized by the elements of the fundamental 
group $\pi_1(G)$, which is abelian for continuous Lie groups $G$.
The exact sequence~(\ref{exactseq}) for the present spontaneously 
broken model now implies the identification
\bea
\pi_1(G) &\simeq& \mbox{Ker} (\pi_1(G/H) \rightarrow \pi_0(H)) \\
         &\simeq& \mbox{Ker} (\bar{H} \rightarrow H). \nn 
\eea
In other words, the magnetic charges of the Dirac monopoles 
are in one-to-one correspondence with the
nontrivial elements of $\pi_1(G/H)\simeq \bar{H}$ 
associated with the trivial element  in $\pi_0(H) \simeq H$.
The physical interpretation of this formula is as follows. In the 
2+1 dimensional Minkowsky setting, in which we are interested,
the Dirac monopoles become instantons describing
tunneling events between magnetic vortices $\bar{h} \in \bar{H}$
differing by the elements of $\pi_1(G)$. 
Here, the decay or tunneling time will naturally depend exponentially
on the actual mass of the monopoles.
The important conclusion is  that in the presence of 
these Dirac monopoles the magnetic
fluxes $\bar{h} \in \bar{H}$ are conserved
modulo the elements of $\pi_1(G)$ and the proper labeling of the stable 
magnetic vortices boils down to the elements of the residual symmetry 
group~$H$ itself
\bea         \label{fluxspcompi}
\bar{H}/\pi_1(G) &\simeq& H.
\eea   
To proceed, the introduction of Dirac monopoles has a bearing on the matter 
content of the model as well. The only matter fields allowed in the theory
with monopoles are those that transform according to an ordinary 
representation of $G$. Matter fields carrying a faithful 
representation of the universal covering group 
$\bar{G}$ are excluded. This means that the 
matter charges appearing in the broken phase 
correspond to ordinary representations of $H$, while faithful 
representations of the lift $\bar{H}$ do not occur. As a result, 
the fluxes $\bar{h} \in \bar{H}$ related by tunneling events induced
by  the Dirac monopoles can not be distinguished through long 
range Aharonov-Bohm experiments with the available matter charges, 
which is consistent with the fact that the
stable magnetic fluxes are labeled by elements of $H$ 
rather then $\bar{H}$ in this case.

The whole discussion can now be summarized as follows. First of all,
if a simply connected  gauge group $G$ is spontaneously broken down 
to a finite subgroup $H$, we are left with a discrete $H$ gauge theory
in the low energy regime. 
The magnetic fluxes are labeled by the elements  of $H$, whereas the 
different electric charges correspond to the full set of UIR's of $H$.
When we are dealing with a non-simply connected gauge group $G$
broken down to a finite subgroup $H$, 
there are two possibilities depending on whether we allow 
for Dirac monopoles/instantons in the theory or not.
In case Dirac monopoles are ruled out, 
we obtain a discrete $\bar{H}$ gauge theory.
The stable fluxes are labeled by the elements of $\bar{H}$ and 
the different charges by the UIR's of $\bar{H}$.
If the model features singular Dirac monopoles, on the other hand,
then the stable fluxes simply correspond to the elements of the group $H$  
itself, while the allowed matter charges constitute UIR's of $H$.
In other words, we are left with a discrete $H$ gauge theory under these 
circumstances.

Let us illustrate these general considerations by some explicit examples.
First we return to the model discussed in the previous section, in which
the non-simply connected gauge group $G \simeq U(1)$ is spontaneously 
broken down to the finite cyclic group $H \simeq \Z_N$.
The topological classification~(\ref{barh}) for this particular model gives
\beas
\pi_1(U(1)/\Z_N) \; \simeq \; \pi_1({\mbox{\bf R}}/\Z_N \times \Z) \; \simeq 
\; \Z_N \times \Z \; \simeq \; \Z.
\eeas
Thus in the absence of Dirac monopoles 
the different stable vortices are labeled by the
integers in accordance with~(\ref{quaflux}), where we found  
that the  magnetic fluxes associated with these vortices are quantized as
$\phi = \frac{2\pi a}{Ne}$ with $a\in \Z$. In principle, we are dealing with
a discrete $\Z$ gauge theory now
and the complete magnetic flux spectrum could be distinguished 
by means of long range Aharonov-Bohm experiments with 
electric charges $q$ being fractions of the fundamental unit $e$,
which correspond to the UIR's of $\Z$.
Of course, this observation is rather academic in this context,
since free charges carrying fractions of the fundamental charge unit $e$
have never been observed. With matter charges $q$ being multiples of $e$, the 
low energy theory then boils down to a $\Z_N$ gauge theory, although
the topologically stable magnetic vortices in the broken phase are 
labeled by the integers $a$.
The Dirac monopoles/instantons 
that can be introduced in this theory correspond to the elements 
of $\pi_1(U(1)) \simeq \Z$. The presence of these monopoles, which
carry magnetic charge $g=\frac{2\pi m}{e}$ with $m \in \Z$, imply that 
the magnetic flux $a$ of the vortices is  conserved modulo $N$, 
as we have seen explicitly in~(\ref{instans}). 
In other words, the proper 
labeling of the stable magnetic fluxes is by the elements
of $\Z_N \times \Z /\Z \simeq \Z_N$, as indicated by~(\ref{fluxspcompi}).
Moreover, electric charge is necessarily quantized in 
multiples of the fundamental charge unit $e$ now, 
so that the tunneling events induced by the instantons 
are unobservable at long distances.
The unavoidable conclusion then becomes that 
in the presence of Dirac monopoles,
we are left with a $\Z_N$ gauge theory 
in the low energy regime of this spontaneously broken model.

When a gauge theory at some intermediate stage of symmetry breaking exhibits 
regular 't~Hooft-Polyakov monopoles, their effect on the stable magnetic 
vortex 
classification is automatically taken care of, as it should because the 
monopoles can not be left out in such a  theory.
Consider, for example, a model in which the non-simply connected
gauge group $G \simeq SO(3)$ is initially 
broken down to $H_1 \simeq U(1)$ and subsequently to $H_2 \simeq \Z_N$
\bea      \label{sbh}
     SO(3) \; \longrightarrow \; U(1) \; \longrightarrow \; \Z_N.
\eea
The first stage of symmetry breaking is accompanied by the appearance 
of regular 't~Hooft-Polyakov monopoles~\cite{thooftmon,polyamon} carrying
magnetic charges characterized by the  
elements of the second homotopy group 
$\pi_2  (SO(3)/U(1)) \simeq \Z$. A simple exact sequence
argument shows 
\bea
\pi_2  (SO(3)/U(1)) & \simeq & \mbox{Ker} (\pi_1(U(1)) \rightarrow 
\pi_1(SO(3)) \\
 & \simeq &  \mbox{Ker} (\Z \rightarrow \Z_2).   \nn 
\eea
Hence, the magnetic charges of the regular monopoles correspond to 
the elements of $\pi_1 (U(1))$
associated with the trivial element 
of $\pi_1(SO(3))$, that is, the even elements of $\pi_1(U(1))$. 
In short, the regular monopoles carry 
magnetic charge $g=\frac{4\pi m}{e}$ with $ m \in \Z$.
To proceed, the residual topologically stable magnetic vortices emerging
after the second symmetry breaking are labeled by the elements of
$\bar{H}_2 \simeq \Z_{2N}$, which follows
from~(\ref{barh}) 
\beas
\pi_1 (SO(3)/\Z_N) \; \simeq \; \pi_1 (SU(2)/\Z_{2N}) \; \simeq \; \Z_{2N}.
\eeas
As in the previous example, the magnetic fluxes carried by these vortices are 
quantized as $\phi=\frac{2\pi a}{Ne}$, while the presence of the regular
't Hooft-Polyakov monopoles now causes the fluxes $a$ to be conserved modulo
$2N$. 
The tunneling or decay time will depend on the 
mass of the regular monopoles, that is, 
the energy scale associated with the first symmetry breaking 
in the hierarchy~(\ref{sbh}).
Here it is assumed that the original $SO(3)$ gauge theory
does not feature Dirac monopoles ($g=\frac{2\pi m}{e}$, with $m=0,1$) 
corresponding to the elements of $\pi_1 (SO(3))\simeq \Z_2$.
This means that additional  matter fields carrying  
faithful (half integral spin) representations of the universal 
covering group $SU(2)$ are allowed in this model, which  leads to 
half integral charges $q=\frac{ne}{2}$ with $n\in \Z$ in the $U(1)$ phase.
In the final Higgs phase, the half integral charges 
$q$ and the quantized magnetic
fluxes $\phi$ then span the complete spectrum of the associated 
discrete $\Z_{2N}$ gauge theory. 

Let us now, instead, suppose that the original $SO(3)$ gauge 
theory contains Dirac monopoles.  
The complete monopole spectrum arising after the first symmetry breaking
in~(\ref{sbh}) 
then consists of the  magnetic charges $g=\frac{2 \pi m}{e}$ 
with $m \in \Z$, which implies that magnetic flux $a$ 
is conserved modulo $N$ in the final Higgs phase.
This observation is in complete agreement 
with~(\ref{fluxspcompi}), 
which states that the proper magnetic flux labeling is by the elements
of $\Z_{2N}/\Z_2 \simeq \Z_N$ under these circumstances.
In addition, the incorporation of Dirac monopoles 
rules out  matter fields which carry 
faithful representations of the universal covering group 
$SU(2)$. Hence, only integral electric charges are conceivable
($q=ne$ with $n\in \Z$) and all in all we end up with a discrete 
$\Z_N$ gauge theory in the Higgs phase.
This last situation can alternatively be 
implemented by embedding this spontaneously broken $SO(3)$ 
gauge theory in a $SU(3)$ gauge theory.
In other words,  the symmetry breaking hierarchy is extended to 
\bea    \label{sbh2}
SU(3) \; \longrightarrow \; SO(3) \; \longrightarrow \; U(1) 
\; \longrightarrow \; \Z_N.
\eea
The singular Dirac monopoles in the $SO(3)$ phase then turn into
regular 't Hooft-Polyakov monopoles 
\beas
\pi_2(SU(3)/SO(3)) \; \simeq \; \pi_1(SO(3)) \; \simeq \; \Z_2.
\eeas
The unavoidable presence of these monopoles automatically imply that
the magnetic flux $a$ of the vortices in the final Higgs phase 
is conserved modulo $N$. To be specific,  a magnetic flux $a=N$ 
can decay by ending on a regular monopole in this model, where the 
decay time will depend on the mass of the monopole or equivalently
on the energy scale associated with the first symmetry breaking 
in~(\ref{sbh2}). 
The existence of such a dynamical decay process is 
implicitly taken care of in the 
classification~(\ref{flulaord}), which indicates
that the stable magnetic fluxes are indeed labeled by the elements of
$\pi_1 (SU(3)/\Z_N)\simeq \Z_N$.

To conclude, in the above examples we restricted ourselves to the case 
where we are left with an abelian finite gauge group in the Higgs phase. 
Of course, the discussion extends to nonabelian finite groups as well. 
The more general picture then becomes as follows.
If the non-simply connected gauge 
group $G \simeq SO(3)$ is spontaneously broken to some (possibly nonabelian)
finite  subgroup $H \subset SO(3)$,
then the topologically stable magnetic fluxes correspond to the elements of 
the lift $\bar{H} \subset SU(2) \simeq \bar{G}$.  In the Higgs phase, we are 
then left with a discrete $\bar{H}$ gauge theory.
If we have embedded $SO(3)$
in $SU(3)$ (or alternatively introduced the conceivable $\Z_2$
Dirac monopoles), on the other hand, 
then the topologically stable magnetic fluxes 
correspond to the elements of $H$ itself and we end up with a discrete $H$
gauge theory.

\subsection{Flux metamorphosis}    \label{fluxmetamorphosis}

In the following, we assume for convenience that the spontaneously broken 
gauge group $G$ in our model~(\ref{nonhiks})
is simply connected. Hence, the stable magnetic vortices are  
labeled by the elements of the nonabelian residual symmetry
group $H$, as indicated by~(\ref{flulaord}). 

We start with a discussion of the classical field configuration 
associated with a static nonabelian  vortex in the plane.
In principle, this vortex is an extended object with a finite core 
size proportional to the inverse of the 
symmetry breaking scale $M_H$. In the low energy regime, however,
we can neglect this finite core size and we will idealize the 
vortex as a point singularity in the plane. 
For finite energy, the associated static classical field 
configuration then satisfies the equations
$V(\Phi)=0$, $F^{\kappa \nu}=0$, 
${\cal D}_i \Phi=0$ and $A_0=0$ outside the core.
These equations imply that  the Higgs field takes ground state values 
$\langle \Phi \rangle$ and the Lie algebra valued vector 
potential $A_\kappa$ is pure gauge so that all nontrivial 
curvature $F^{\kappa \nu}$ is localized inside the core. 
To be explicit, a path (and gauge) dependent 
solution w.r.t.\  an arbitrary but fixed 
ground state $\langle \Phi_0 \rangle$ 
at an arbitrary but fixed base point ${\mbox{\bf x}}_0$ can be presented as
\bea
\langle \Phi(\mbox{\bf x}) \rangle &=& W({\mbox{\bf x}}, {\mbox{\bf x}}_0, \gamma) 
\langle \Phi_0 \rangle ,
\eea 
where 
\bea
W({\mbox{\bf x}}, {\mbox{\bf x}}_0, \gamma) &=& 
P \exp ( \im e \int_{{\mbox{\scriptsize \bf x}}_0}^{\mbox{\scriptsize \bf x}} 
A^i dl^i),
\eea
is the  untraced path ordered 
Wilson line integral $W({\mbox{\bf x}}, {\mbox{\bf x}}_0, \gamma)$, which 
is evaluated along an oriented path $\gamma$ (avoiding the 
singularity) from the  base point to some other 
point ${\mbox{\bf x}}$ in the plane. 
Here we merely used the fact that the relation
${\cal D}_i \langle \Phi \rangle=0$ identifies the parallel transport
in the Goldstone boson fields  with that in the gauge fields,
as we have argued in full detail for the abelian case 
in section~\ref{mavoab}.
Now in order to keep the Higgs field
single valued, the magnetic flux of the vortex, picked up by 
the  Wilson line integral along a counterclockwise 
closed loop ${\cal C}$, which starts and ends at the base point 
and encloses the core, 
necessarily takes values in the subgroup  ${ H}_0$ of 
${ G}$ that leaves  the ground state  
$\langle \Phi_0 \rangle$ at the base point invariant
\bea                \label{paraflux}
W({\cal C}, {\mbox{\bf x}}_0 )   
&=& P \exp ( \im e \oint A^i dl^i) \; = \;  h \in { H}_0.
\eea   
This untraced Wilson loop operator~(\ref{paraflux}) completely classifies
the long range properties of the vortex solution.
It is invariant under a continuous deformation of the loop ${\cal C}$ that 
keeps the base point fixed and avoids the core of the vortex.
Moreover, it is invariant under continuous gauge transformations 
that leave the ground state $\langle \Phi_0 \rangle$
at the base point invariant. 
As in the abelian case, we fix this residual 
gauge freedom by sending all  nontrivial parallel transport 
into a narrow wedge or Dirac string from the core of the vortex to 
spatial infinity  as depicted in figure~\ref{sinvo}. 
It should be emphasized that our gauge fixing procedure for 
these vortex solutions involves two physically irrelevant choices.
First of all, we have chosen a fixed ground state $\langle \Phi_0 \rangle$
at the base point ${\mbox{\bf x}}_0$. This choice merely determines the embedding 
of the  residual symmetry group in ${ G}$ to be the stability group
${ H}_0$ of $\langle \Phi_0 \rangle$.
A different choice for this ground state gives rise to a 
different embedding of the residual symmetry group, 
but will eventually lead to an  unitarily equivalent 
quantum description of the discrete ${ H}$ gauge theory in the
Higgs phase.
For convenience, we subsequently fix the remaining gauge freedom by sending 
all nontrivial transport around the vortices to a small wedge.
Of course, physical phenomena will not depend on this choice.
In fact, an equivalent formulation of the low energy theory,
without fixing this residual gauge freedom for the vortices, 
can also be given~\cite{bucher}.

\begin{figure}[tbh]    \epsfxsize=3cm
\centerline{\epsffile{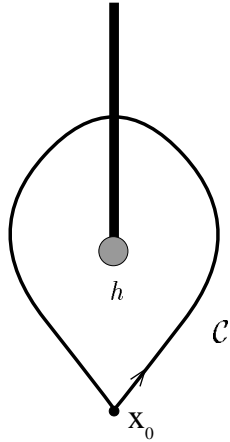}}
\caption{\sl  Single vortex solution. We have fixed the gauge freedom by 
sending all nontrivial parallel transport around the core in the Dirac 
string attached to the core. Thus outside the core,
the Higgs field takes the same 
ground state value $\langle \Phi_0 \rangle$ everywhere
except for the region where the Dirac string is localized. Here it makes a 
noncontractible winding in the ground state manifold. 
This winding corresponds to a holonomy in the gauge field classified 
by the result of the untraced Wilson loop operator
$W({\cal C}, {\mbox{\bf x}}_0) = h \in { H}_0$, 
which picks up the nonabelian  magnetic flux located inside the core.}
\label{sinvo}
\end{figure}

In this gauge fixed prescription, we are still able to perform 
global symmetry transformations  $g \in { H}_0$  
on these vortex solutions that leave  
the ground state $\langle \Phi_0 \rangle $ invariant.
These transformations affect the 
field configuration of the vortex in the following way 
\bea
\Phi ({\mbox{\bf x}})      &\longmapsto&  g \; \Phi ({\mbox{\bf x}}) \\
A_\kappa  ({\mbox{\bf x}}) &\longmapsto&  g \; A_\kappa  ({\mbox{\bf x}}) \; g^{-1},
\eea
as an immediate  consequence we then obtain 
\bea 
W({\cal C}, {\mbox{\bf x}}_0 ) &\longmapsto& g \; W({\cal C}, {\mbox{\bf x}}_0 ) \; g^{-1},
\eea 
which shows that the flux of the vortex becomes conjugated
$h \mapsto ghg^{-1}$ under such a  transformation. 
The conclusion is that  these nonabelian vortex solutions 
are in fact organized in degenerate multiplets under 
the residual global symmetry transformations ${ H}_0$, 
namely the different conjugacy classes of ${ H}_0$ denoted as  $^A C$,
where $A$ labels a particular conjugacy class. 
For convenience, we will refer to the stability group
of $\langle \Phi_0 \rangle $ as ${ H}$ from now on.

The different vortex solutions in a given conjugacy class $^A C$ of ${ H}$, 
being related by internal global symmetry transformations 
that leave the action~(\ref{nonhiks}) invariant, clearly 
carry the same external quantum numbers, that is, the  total energy of the 
configuration, the coresize etc. 
These solutions only differ by their internal magnetic flux quantum number.
This internal degeneracy becomes relevant in  
adiabatic interchange processes of  remote vortices in the plane.
Consider, for instance, the configuration of 
two remote vortices as presented in figure~\ref{metamo}. 
In the depicted adiabatic counterclockwise interchange of these vortices,
the  vortex initially carrying  the magnetic flux $h_2$
moves through the Dirac string attached to
the other vortex. As a result, its flux picks up a global symmetry 
transformation by the flux $h_1$ of the latter, i.e.\ 
$h_2 \mapsto h_1 h_2 h^{-1}_1$, such that the total flux 
of the configuration is conserved. 
This classical nonabelian Aharonov-Bohm effect 
appearing for noncommuting fluxes, which
has been called flux metamorphosis~\cite{bais}, 
leads to physical observable phenomena. Suppose, for example,
that  the magnetic flux $h_2$ was a member of a flux/anti-flux pair 
$(h_2,h_2^{-1})$ created from the vacuum. When $h_2$ encircles 
$h_1$, it returns as the flux $h_1 h_2 h^{-1}_1$ and will not
be  able to annihilate the flux $h_2^{-1}$ anymore.
Upon rejoining the pair we now obtain the stable flux
$h_1 h_2 h^{-1}_1 h_2$.  
Moreover, at the quantum level, 
flux metamorphosis leads to nontrivial Aharonov-Bohm scattering 
between nonabelian vortices as we will argue in more detail later on.

Residual global symmetry transformations naturally 
leave this observable Aharonov-Bohm effect for nonabelian 
vortices invariant. This simply follows from the fact that these 
transformations commute with this nonabelian Aharonov-Bohm effect.
To be precise,
a residual global symmetry transformation $g \in { H}$ on the two vortex 
configuration in figure~\ref{metamo}, for example,
affects the flux of both vortices through conjugation by the group 
element $g$, and it is easily verified that it makes no difference whether
such a transformation is performed before the interchange is started or 
after the interchange is completed. 
The extension of these classical 
considerations to configurations of more 
then two vortices in the plane is straightforward. Braid processes,
in which the fluxes of the vortices affect each other by conjugation, 
conserve the total flux of the configuration. The residual global symmetry
transformations $g \in { H}$ of the low energy regime,
which act by an overall conjugation of the fluxes 
of the vortices in the configuration by $g$, 
commute with these  braid processes.

\begin{figure}[tbh]    \epsfxsize=\textwidth
\centerline{\epsffile{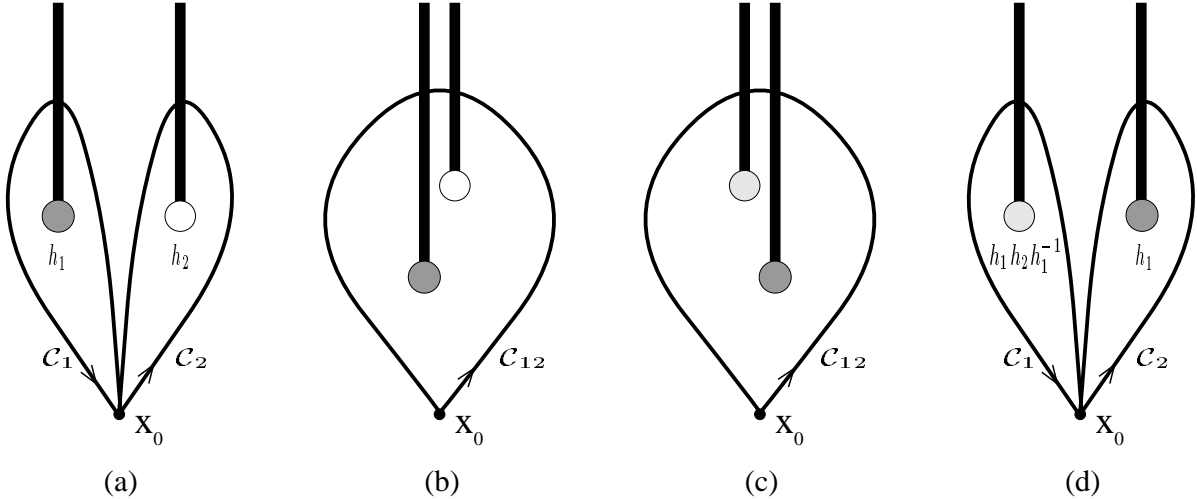}}
\caption{\sl  Flux metamorphosis. We start off with a classical configuration
of  two patched vortex solutions, as visualized in figure~(a). The vortices 
are initially assumed to carry the fluxes 
$W({\cal C}_{1}, {\mbox{\bf x}}_0 ) = h_1$ and 
$W({\cal C}_{2}, {\mbox{\bf x}}_0 ) = h_2$. The total flux of this 
configuration is picked up by the Wilson line integral along the loop
${\cal C}_{12}$ encircling both vortices as depicted in figure~(b): 
$W({\cal C}_{12}, {\mbox{\bf x}}_0 )
= W({\cal C}_{1} \circ {\cal C}_{2}, {\mbox{\bf x}}_0 )= 
W({\cal C}_{1}, {\mbox{\bf x}}_0 ) \cdot W({\cal C}_{2}, {\mbox{\bf x}}_0 )= h_1 h_2$.
Now suppose that the two vortices are interchanged in the counterclockwise 
fashion depicted in figures~(b)-(d). In this process  vortex~2 moves through
the Dirac string attached to vortex~1 and as a result its flux 
will be affected $h_2 \mapsto h_2'$. Vortex~1, on the other hand,
never meets any nontrivial parallel transport in the gauge 
fields and its flux remains the same. 
Since this local braid process should not be able to change the 
global properties of this system, i.e.\ the total flux, we have
$h_1 h_2= W({\cal C}_{12}, {\mbox{\bf x}}_0 )= W({\cal C}_{1}, {\mbox{\bf x}}_0 ) \cdot 
W({\cal C}_{2}, {\mbox{\bf x}}_0 )= h_2'h_1$. 
Thus the flux of vortex 2 becomes conjugated
$h_2'=h_1 h_2 h^{-1}_1$ by the flux of vortex~1 in this braid process.}
\label{metamo}
\end{figure}

As in the abelian case discussed in the previous sections, we wish 
to treat these nonabelian vortices as point particles in the first quantized 
description.  The degeneracy of these vortices under the residual 
global symmetry group ${ H}$ then indicates that 
we have to assign a finite dimensional  internal 
Hilbert space $V^A$ to these particles, which is 
spanned by the different fluxes in a given conjugacy 
class $^A C$ of ${ H}$ and endowed with the standard 
inner product~\cite{spm}
\bea
\langle h'| h \rangle &=& \delta_{h',h} \qquad \qquad \forall
\,  h,h'\in\, ^A C.
\eea    
Under the residual global symmetry transformations 
the flux eigenstates in this internal Hilbert space 
$V^A$ are affected through conjugation
\bea                         \label{fluxconjump}
g \in { H} : \qquad   |h \rangle &\longmapsto &
|g h g^{-1} \rangle.
\eea
In general, the particle can be in a 
normalized linear combination of the different 
flux eigenstates in the internal Hilbert space $V^A$. 
The residual global symmetry transformations~(\ref{fluxconjump}) 
act linearly on such states. 
Of course, the conjugated action of the residual symmetry group is in general 
reducible and, at first sight, it seems that we have to decompose 
this internal Hilbert space into the different irreducible components. 
This is not the case as we will see in more detail later on (see the 
discussion concerning relation~(\ref{scatamplo})). 
The point is that we can independently perform physical 
flux measurements by means of quantum interference experiments with 
electric charges. These measurements project 
out a particular flux eigenstate.
Clearly, these flux measurements do not commute with the residual global 
symmetry transformations and under their combined action the internal 
Hilbert spaces $V^A$ associated with the different conjugacy classes 
$^A C$ form irreducible representations.

The complete quantum state of these particles consists of an internal flux
part and an external part. The quantum state describing a single particle
in the flux eigenstate $|h_1 \rangle \in V^{A_1}$ at a fixed 
position ${\mbox{\bf y}}$ in the plane, for instance, is the formal tensor product 
$|h_1, {\mbox{\bf y}} \rangle = |h_1 \rangle |{\mbox{\bf y}} \rangle$. 
To proceed, the initial configuration depicted in 
figure~\ref{metamo} is described by 
the multi-valued two particle quantum state 
$|h_1, {\mbox{\bf y}} \rangle|h_2, {\mbox{\bf z}} \rangle$,
where again by convention the particle located most left in the plane 
appears most left in the tensor product.
The result of an adiabatic counterclockwise interchange 
of the two particles can now be summarized by the action of the 
braid operator
\bea                   \label{fluxmetam}
{\cal R} \; |h_1, {\mbox{\bf y}} \rangle|h_2, {\mbox{\bf z}} \rangle 
&=& |h_1 h_2 h^{-1}_1, {\mbox{\bf y}} \rangle|h_1, {\mbox{\bf z}}\rangle,
\eea
which acts linearly on linear combinations of these flux eigenstates.
What we usually measure in quantum interference experiments, however, 
is the effect in the internal wave function of a monodromy of 
the two particles 
\bea               \label{flumonodr}
{\cal R}^2  \; |h_1, {\mbox{\bf y}} \rangle|h_2, {\mbox{\bf z}} \rangle
&=& |(h_1 h_2) h_1 (h_1 h_2)^{-1}, {\mbox{\bf y}} \rangle|
h_1 h_2 h^{-1}_1, {\mbox{\bf z}}\rangle.
\eea 
This nonabelian Aharonov-Bohm effect can be probed either through 
a double slit experiment~\cite{colem,preslo} or through an Aharonov-Bohm 
scattering experiment as discussed in appendix~\ref{ahboverl}.
In the first case, we keep one particle fixed between the two slits, whereas
the other particle comes in as a plane wave. The geometry of the Aharonov-Bohm 
scattering experiment, depicted in figure~\ref{abverpres} 
is more or less similar. The interference pattern in both experiments
is determined by the internal transition amplitude
\bea             \label{transamp}
\langle u_2 |\langle u_1 | \; {\cal R}^2 \; | u_1 \rangle |u_2 \rangle,
\eea 
where $| u_1 \rangle$ and  $|u_2 \rangle$ respectively 
denote the properly normalized internal flux states of the two particles, 
which are generally linear combinations of the flux eigenstates in 
the corresponding internal Hilbert spaces $V^{A_1}$ and $V^{A_2}$.
The topological interference amplitudes~(\ref{transamp}) 
summarize all the physical 
obervables for vortex configurations in the low energy regime to which we 
confine ourselves here.
As we have argued before, the  residual global symmetry transformations
affect internal multi-vortex states  through an overall conjugation  
\bea                         \label{fluxconjump2}
g \in { H} : \qquad   
|h_1 \rangle|h_2 \rangle & \longmapsto &
|g h_1 g^{-1} \rangle|g h_2 g^{-1} \rangle,
\eea
which commutes with the braid operator and therefore
leave the interference amplitudes~(\ref{transamp}) invariant.

\subsection{Including matter}  \label{inclumatter}

Let us now suppose that the total model is 
of the actual form
\bea                          \label{totmod}
S &=& S_{\mbox{\scriptsize YMH}} + S_{\mbox{\scriptsize matter}},
\eea
where  $S_{\mbox{\scriptsize YMH}}$ denotes the action for the nonabelian 
Higgs model given in~(\ref{nonhiks}) and the action 
$ S_{\mbox{\scriptsize matter}}$ 
describes additional matter fields minimally coupled  
to the gauge fields. 
In principle, these matter fields correspond 
to multiplets which transform 
irreducibly under the spontaneously broken symmetry group ${ G}$. 
Under the residual symmetry group 
${ H}$ in the Higgs phase, however, these representations
will become reducible  and  branch to  
UIR's $\Gamma$ of ${ H}$. Henceforth, it is assumed that 
the matter content of the model is such that all UIR's 
$\Gamma$ of ${ H}$ are indeed realized. We will 
treat the different charges $\Gamma$, appearing 
in the Higgs phase in this way~\cite{almawil,preskra}, as point particles. 
In the first quantized description, these point charges 
then carry an internal Hilbert space, namely  
the representation space associated with $\Gamma$.
Now suppose we have a configuration of a nonabelian vortex in 
a flux eigenstate $|h \rangle$ at some fixed position in the plane 
and  a remote charge $\Gamma$ in a normalized internal charge 
state $|v\rangle$ fixed at another position. 
When the charge encircles the vortex in a counterclockwise fashion,
it meets the Dirac string and picks up a global symmetry 
transformation by the flux of the vortex
\bea                   \label{charmetam}
{\cal R}^2 \; |h, {\mbox{\bf y}} \rangle|v, {\mbox{\bf z}} \rangle 
&=& |h, {\mbox{\bf y}} \rangle|\Gamma(h) \, v, {\mbox{\bf z}} \rangle,
\eea 
where $\Gamma(h)$ is the matrix assigned to the group element $h$ in the 
representation $\Gamma$.
The residual global symmetry transformations on this two particle 
configuration
\bea
g \in { H} : \qquad |h, {\mbox{\bf y}} \rangle|v, {\mbox{\bf z}} \rangle
&\longmapsto&
|ghg^{-1}, {\mbox{\bf y}} \rangle |\Gamma (g) \, v, {\mbox{\bf z}} \rangle,
\eea 
again commutes with this monodromy operation. Thus the 
interference amplitudes
\bea               \label{scatamplo}
\langle v| \langle h| \; {\cal R}^2 \; |h \rangle | v \rangle &=&
\langle h|h \rangle \langle v | \Gamma (h) \, v  \rangle \; = \; 
\langle v | \Gamma (h) \, v  \rangle,
\eea                          
measured in double slit or Aharonov-Bohm scattering experiments 
involving these particles are invariant under the residual 
global symmetry transformations. As alluded to before,
these interference experiments can be used to measure the flux of 
a given vortex~\cite{colem,alfrev,preslo,vohgs}. To that end, we  
place the vortex between the two slits
(or alternatively use it as the scatterer in an Aharonov-Bohm 
scattering experiment)
and evaluate the interference pattern for  an incident beam of charges 
$\Gamma$ in the same internal state $| v \rangle$. In this way, we determine 
the interference amplitude~(\ref{scatamplo}). Upon repeating this experiment 
a couple of times with different internal states 
for the incident charge $\Gamma$, we can determine all matrix elements
of $\Gamma (h)$ and hence, iff $\Gamma$ corresponds to a faithful UIR of 
${ H}$, the group element $h$ itself. In a similar fashion, we may
determine the charge $\Gamma$ of a given particle 
and, moreover, its internal quantum state $|v \rangle$. In this case,
we put the unknown charge between the double slit
(or use it as the scatterer in an Aharonov-Bohm scattering experiment), 
measure the interference pattern for an incident beam of vortices
in the same flux eigenstate $|h\rangle$ and again repeat this experiment 
for all $h \in { H}$.

At this point, we have established the purely magnetic flux and the purely 
electric charge superselection sectors of the discrete gauge theory 
in the Higgs phase of the model~(\ref{totmod}).
The different magnetic sectors are labeled by the conjugacy classes  $^A C$
of  ${ H}$, whereas the different electric charge sectors 
correspond to the different UIR's $\Gamma$  of the residual symmetry 
group ${ H}$.
The complete spectrum of this discrete gauge theory 
also contains dyonic combinations of these sectors.  The relevant remark
in this context is that we have not yet completely exhausted the action 
of the residual global symmetry transformations on the internal magnetic
flux quantum numbers.  As we have seen in~(\ref{fluxconjump}),
the residual global ${ H}$  transformations 
affect the magnetic fluxes through conjugation. 
The transformations that slip through this conjugation may 
in principle be implemented on an additional internal charge degree of freedom 
assigned to these fluxes~\cite{spm}.
More specifically, the global symmetry transformations that leave 
a given flux $|h\rangle$ invariant are those that 
commute with this flux, i.e.\ the group elements in the centralizer 
$^h N \subset { H}$. The internal charges that we can assign to this flux
correspond to the different UIR's $\alpha$ of the group $^h N$. Hence, 
the inequivalent dyons that can be formed in the composition of 
a global ${ H}$ charge $\Gamma$ with a magnetic flux 
$|h\rangle$ correspond to the different irreducible
components of the subgroup $^h N$ of ${ H}$ contained in the representation
$\Gamma$.
Two remarks are pertinent now.
First of all, the centralizers of different fluxes in a given 
conjugacy class $^A C$ are isomorphic. Secondly, the full set of the residual 
global ${ H}$ symmetry transformations relate the fluxes in a given 
conjugacy class carrying unitary equivalent centralizer charge 
representations. In other words, the different dyonic sectors are labeled
by $(\, ^A \! C, \alpha \, )$, where $^A \! C$ runs over the different 
conjugacy classes of ${ H}$ and $\alpha$ over the different nontrivial 
UIR's of the associated centralizer. The explicit transformation properties
of these dyons under the full global group ${ H}$ involve some conventions, 
which will be  discussed in the algebraic approach to discrete ${ H}$ 
gauge theories we take in the following section.

The  physical observation behind this formal construction of 
the dyonic sectors is that we can in fact only measure the transformation 
properties of the charge of a given 
flux/charge composite under the centralizer of 
the flux of this composite~\cite{preslo}. 
A similar phenomenon occurs in the 3+1 dimensional
setting for monopoles carrying a nonabelian magnetic charge 
where it is known as the global color problem~\cite{nelson, balglob, nelsonc}. 
To illustrate this phenomenon,
we suppose that we have a composite of a pure flux $|h\rangle$
and a pure global ${ H}$ charge $\Gamma$ in some internal 
state $|v\rangle$. Thus the complete internal state of the composite becomes
$|h, v\rangle$.
As we have argued before, the charge of a given object can be determined 
through double slit or Aharonov-Bohm scattering experiments involving 
beams of vortices in the same internal flux state $|h'\rangle$ and repeating 
these experiments for all $h' \in { H}$.
The interference amplitudes measured in 
this particular case are of the form 
\bea                \label{topobstr}
\langle h,v| \langle h'| \; {\cal R}^2 \; | h' \rangle  | h,v \rangle 
&=&   \langle h,v| h' h h'^{-1}, \Gamma (h') \, v \rangle
\langle h'| (h'h) h'(h'h)^{-1} \rangle \\
&=& \langle v |\Gamma (h') \, v \rangle \: \delta_{h, h' h h'^{-1}} \; , \nn 
\eea 
where we used~(\ref{flumonodr}) and~(\ref{charmetam}). As a result of  the 
flux metamorphosis~(\ref{flumonodr}), the interference term is only 
nonzero for experiments involving fluxes $h'$ that commute with the 
flux of the composite, i.e.\ $h' \in$$\, ^h N$. Thus we are only 
able to detect the response of the charge $\Gamma$ of the composite 
to global symmetry transformations in $^h N$. 
This topological obstruction
is usually summarized with the 
statement~\cite{balglob2, schwarz, alice, preskra} 
that in the background of a single vortex $h$, the only `realizable' 
global symmetry transformations are those taking values 
in the centralizer $^h N$.

Let us close this section with a summary of the main conclusions. 
First of all, the complete spectrum  of the nonabelian 
discrete ${ H}$ gauge theory appearing 
in the Higgs phase of the model~(\ref{totmod}) can be 
presented as   
\bea                                            \label{conencen}
(\, ^A \! C, \alpha \, ) ,
\eea
where $^A \! C$ runs over the conjugacy classes of ${ H}$ and $\alpha$
denotes the different UIR's of the centralizer associated to a specific
conjugacy class $^A \! C$.  The purely magnetic sectors correspond to trivial
centralizer representations and are labeled by the different  nontrivial 
conjugacy classes. The pure charge sectors, on the other hand,
correspond to the trivial conjugacy class 
(with centralizer the full group ${ H}$) and are labeled by the different 
nontrivial UIR's of the residual symmetry group ${ H}$. 
The other sectors describe the dyons in this theory. 
Note that the sectors~(\ref{conencen}) boil down to the sectors of the 
spectrum~(\ref{compspectr}) in case ${ H} \simeq \Z_N$.

The remaining long range interactions
between the particles~(\ref{conencen}) are 
topological Aharonov-Bohm interactions.
In a counterclockwise braid process involving two given particles, the 
internal quantum state of the particle that moves through
the Dirac string attached to the flux of the other particle picks 
up a global symmetry transformation by this flux. 
This (in general nonabelian) Aharonov-Bohm effect conserves the total flux 
of the system and moreover commutes with the residual 
global ${ H}$ transformations, which act simultaneously 
on the internal quantum states 
of all the particles in the system. The last property ensures that the 
physical observables for a given system, which are all 
related to this Aharonov-Bohm effect, are invariant
under global ${H}$ transformations.

An exhaustive discussion of the braid and {\em fusion} properties 
of the particles in the spectrum~(\ref{conencen}) involves the 
algebraic structure underlying a discrete ${ H}$ gauge theory, 
which will be revealed in the next section. For notational
simplicity, we will 
omit explicit mentioning of the external degrees of freedom of the particles 
in the following. In our considerations, we usually work with  
position eigenstates for the particles unless we are discussing double slit-
or Aharonov-Bohm scattering experiments in which the incoming projectiles 
are in momentum eigenstates.

\sectiona{Quantum doubles}      \label{qdH}

It is by now well-established that 
there are deep connections between two dimensional rational conformal field 
theory, three dimensional topological field theory and quantum groups or
Hopf algebras (see for instance~\cite{alvarez1,alvarez2,witten} and references
therein).  
Discrete $H$ gauge theories, being examples of three dimensional 
topological field theories, naturally fit in this 
general scheme. The algebraic 
structure underlying a discrete $H$ gauge theory 
is the Hopf algebra $D(H)$~\cite{spm,spm1,sm}. This is the quasitriangular
Hopf algebra obtained from Drinfeld's
quantum double construction~\cite{drin,drinfeld} as applied to 
the abelian algebra ${\cal F}(H)$ of functions on the finite group ${ H}$.
(For a thorough treatment of Hopf algebras in general 
and related issues, the interested reader is referred to 
the excellent book by Shnider and Sternberg~\cite{shnider}).
Considered as a vector space, we then have 
$D(H) = {\cal F}(H) \ot {\bf C}[H]$, where ${\bf C}[H]$ denotes 
the group algebra  over the complex numbers ${\bf C}$.
Roughly speaking, the elements of  $D(H)$
signal the flux of the particles~(\ref{conencen}) and implement
the residual global symmetry transformations.
Under this action the particles form irreducible representations.
Moreover, the algebra  $D({ H})$  provides an unified description of 
the braiding and fusion properties of the particles.
Henceforth, we will simply refer to the  algebra $D({ H})$ 
as the quantum double. 
This name,  inspired by its mathematical construction, also 
summarizes nicely the physical content of a 
Higgs phase with a residual finite gauge group $H$.
The topological interactions between the particles 
are of a quantum mechanical nature, whereas the 
spectrum~(\ref{conencen}) exhibits an electric/magnetic self-dual 
(or double) structure.

In fact, the quantum double  $D({ H})$ was first proposed
by Dijkgraaf, Pasquier and Roche~\cite{dpr}. They identified it 
as the Hopf algebra associated with certain holomorphic orbifolds 
of rational conformal field theories~\cite{dvvv} 
and the related three 
dimensional topological field theories with finite gauge 
group $H$ as introduced by Dijkgraaf and Witten~\cite{diwi}. 
The new insight that emerged in~\cite{spm,spm1,sm}
was that such a topological field theory finds 
a natural realization as the residual discrete $H$ gauge theory 
describing the long range physics
of gauge theories in which a continuous gauge group $G$ is spontaneously 
broken down to a finite group $H$.

\subsection{$D({ H})$}   \label{thequdo}

As we have seen in the previous sections, we are basically left 
with two physical operations on the particles~(\ref{conencen}) in 
the spectrum of a discrete ${ H}$ gauge theory. 
We can independently measure their flux and their charge through quantum 
interference experiments.
The fluxes are the group elements $h \in { H}$, 
while the dyon charges are the representations of the centralizer of this 
particular flux. Flux measurements correspond to operators ${\mbox{P}}_h$
projecting out a particular flux $h$,
while the charge of a particle can be detected through 
its transformation properties under the residual global  
symmetry transformations $g \in { H}$ that commute with the 
flux of the particle.
The operators ${\mbox{P}}_h$ projecting out the 
flux $h \in { H}$ of a given quantum state   
naturally realize the projector  algebra
\bea           \label{multi}
{\mbox{P}}_h {\mbox{P}}_{h'} &=& \delta_{h,h'} \; {\mbox{P}}_h, 
\eea 
with $\delta_{h,h'}$ the kronecker delta function for the group elements
$h,h' \in { H}$.   As we have seen in~(\ref{fluxconjump}),
global symmetry transformations $g \in { H}$ 
affect the fluxes through  conjugation, which implies that 
the flux projection operators and global symmetry  transformations 
do not commute
\bea               \label{multip}
g \, {\mbox{P}}_h  &=& {\mbox{P}}_{ghg^{-1}} \, g. 
\eea
The combination 
of global symmetry transformations followed by flux measurements
\bea                          \label{lusien}
\{ {\mbox{P}}_h \, g \}_{h,g\in { H}},  
\eea
generate  the quantum double $D({ H})= {\cal F}(H) \ot {\bf C}[H]$ 
and the multiplication~(\ref{multi}) and~(\ref{multip}) 
of these elements can be recapitulated 
as~\footnote{In \cite{dpr,altsc1,spm,spm1,sm} 
the elements of the quantum double were presented as $\hook{h}{g}$.
For notational simplicity, we use the presentation ${\mbox{P}}_h \, g$  here.} 
\bea
{\mbox{P}}_h \, g \cdot {\mbox{P}}_{h'} \, g' &=& 
\delta_{h,g h'g^{-1}} \; {\mbox{P}}_h \, gg' .
\label{algeb}
\eea

The different particles~(\ref{conencen}) in the spectrum of the associated 
discrete ${ H}$ gauge theory constitute the complete set of inequivalent 
irreducible representations of the quantum double  $D({ H})$. 
To make explicit the irreducible action of the quantum double on these particles, 
we have to develop some further notation. To start with,
we will label the group elements in the different conjugacy classes of 
${ H}$ as
\bea
^A\!C &=& \{^A\!h_1,\;^A\!h_2, \ldots,\,^A\!h_k\}.
\eea 
Let $^A\!N \subset { H}$ be the centralizer of the group element 
$^A\!h_1$ and 
$\{^A\!x_1,\,^A\!x_2,\ldots,\,^A\!x_k\}$ a set of representatives for the 
equivalence classes of ${ H}/^A\!N$, 
such that $^A\!h_i=\,^A\!x_i\,^A\!h_1\,^A\!x_i^{-1}$. 
For convenience, we will always take $^A\!x_1=e$, with $e$ the unit
element in ${ H}$. 
To proceed, the  basis vectors  of the 
unitary irreducible representation ${\alpha}$  of the centralizer 
$^A\!N$ will be denoted by $^{\alpha}\!v_j$.  With these conventions the 
internal Hilbert space $V^A_{\alpha}$ is spanned by the quantum states
\bea                 \label{quantum states}
\{|\,^A\!h_i,\,^{\alpha}\!v_j\rangle\}_{i=1,\ldots,k}^{j=1,\ldots, 
\mbox{\scriptsize dim}{ \, \alpha}}. 
\eea
The combined action of a global symmetry  transformation $g \in { H}$ 
followed by a flux projection operator ${\mbox{P}}_{h}$
on these internal flux/charge eigenstates spanning the  
Hilbert space $V^A_{\alpha}$ can then be presented as~\cite{dpr} 
\bea \label{13zo}                                              
\Pi^A_{\alpha}(\, {\mbox{P}}_{h} \, {g}\, ) 
\; |\,^A\!h_i,\,^{\alpha}\!v_j \rangle &=&
\delta_{h,g\,^A\!h_i \, g^{-1}}\;\; |\, g\,^A\!h_i \, g^{-1},
\,{\alpha}(\tilde{g})_{mj}\,^{\alpha}v_m \rangle,
\eea
with
\bea   \label{centdef}
\tilde{g} &:=& \,^A\!x_k^{-1}\, g \,\, ^A\!x_i,
\eea
and  $\,^A\!x_k$  defined through $\,^A\!h_k := g\,^A\!h_i \, g^{-1}$.
It is easily verified that this element $\tilde{g}$ constructed from $g$ 
and the flux $^A\!h_i$ indeed commutes with $^A\!h_1$ and therefore can 
be implemented on the centralizer charge.  Two remarks are pertinent now.
First of all, there is of course arbitrariness involved in the 
ordering of the elements in the conjugacy classes and the choice of 
the representatives $\,^A\!x_k$ for the equivalence classes of the coset
${ H}/^A\!N$. However, different choices lead to unitarily equivalent 
representations of the quantum double. Secondly,
note that~(\ref{13zo}) is exactly the action anticipated in 
section~\ref{nonabz}.  
The flux $^A\!h_i$ of the associated particle is conjugated by the 
global symmetry transformation $g \in { H}$, 
while the part of $g$ that slips
through this conjugation is implemented on the 
centralizer charge of the particle.
The operator ${\mbox{P}}_{h}$ subsequently projects out the flux $h$.

We will now argue that the  
flux/charge eigenstates~(\ref{quantum states}) 
spanning the internal Hilbert space $V^A_{\alpha}$  carry
the same spin, i.e.\ a counterclockwise
rotation over an angle of $2\pi$ gives rise to 
the same spin factor for all quantum states in $V^A_{\alpha}$.
As in our discussion of abelian dyons in section~\ref{abdyons}, we assume 
a small seperation between the centralizer charge and the flux of the dyons.
In the aforementioned rotation, the centralizer charge of the dyon 
then moves through the Dirac string attached to the flux of the dyon
and as a result picks up a transformation by this flux. 
The element in the quantum double that implements  this effect 
on the internal quantum states~(\ref{quantum states})
is the central element 
\bea   \label{centraal}
\sum_h \; {\mbox{P}}_h \,  h.
\eea 
It signals the flux of the internal quantum state and implements this flux
on the  centralizer charge
\bea
\Pi^A_{\alpha}(\, \sum_h \; {\mbox{P}}_h \, h \,) \; 
|\,^A\!h_i,\,^{\alpha}\!v_j\rangle &=& 
|\,^A\!h_i,\,{\alpha}(^A\!h_1)_{mj}\,^{\alpha}v_m \rangle,
\eea
which boils down to the same matrix ${\alpha}(^A\!h_1)$
for all fluxes $^A\!h_i$  in $^A \! C$.
Here we used~(\ref{13zo}) and~(\ref{centdef}).
Since $^A\!h_1$ by definition commutes with all the elements in the 
centralizer $^A\!N$, it follows from  Schur's lemma  that it 
is proportional to the unit matrix in the irreducible representation $\alpha$
\bea                 \label{spin!}
{\alpha}(^A\!h_1) &=& e^{2\pi \im s_{(A,\alpha)}} \,  {\mbox{\bf 1}}_\alpha.
\eea
This proves our claim. The conclusion is that there is an overall spin value 
$s_{(A,\alpha)}$ assigned to the sector $(\, ^A \! C, \alpha \, )$.
Note that the only sectors carrying a nontrivial spin are the dyonic sectors
corresponding to nontrivial conjugacy classes paired with 
nontrivial centralizer charges.

The  internal Hilbert space  describing a system of two particles 
$(\, ^A C,\alpha \,)$ and  $(\,^B C,\beta\,)$
is the tensor product $V_{\alpha}^A \ot V_{\beta}^B$.
The extension of the action of the quantum double
$D({ H})$ on the single particle states~(\ref{13zo})
to the two particle states in $V_{\alpha}^A \ot 
V_{\beta}^B$ is given  by the comultiplication
\bea
 \Delta(\,{\mbox{P}}_h \, g \,) &=& 
\sum_{h' \cdot h''=h} {\mbox{P}}_{h'} \, g \ot {\mbox{P}}_{h''} \, {g},
\label{coalgeb}
\eea
which is an algebra morphism  from $D({ H})$ to 
$D({ H}) \ot D({ H})$. To be concrete, 
the tensor product representation of $D({ H})$ carried by 
the two particle internal Hilbert space $V_{\alpha}^A \ot V_{\beta}^B$  
is defined as  
$\Pi^A_{\alpha} \ot \Pi^B_{\beta}(\Delta (\, {\mbox{P}}_h \, g \,)  )$.
The action~(\ref{coalgeb}) of the quantum double on the internal
two particle quantum states in $V_{\alpha}^A \ot V_{\beta}^B$
can be summarized as follows. In accordance with our observations 
in the previous section, the residual 
global symmetry transformations $g \in { H}$ 
affect the internal quantum states of the two particles separately. 
The projection operator ${\mbox{P}}_h$ subsequently projects out the 
total flux of the two particle quantum state. 
Hence the action~(\ref{coalgeb}) of the quantum 
double determines the global properties of a given two particle
quantum state, which are conserved under the local process of fusing 
the two particles. 
It should be mentioned now that
the tensor product representation 
$(\Pi^A_{\alpha} \ot \Pi^B_{\beta}, V_{\alpha}^A \ot V_{\beta}^B)$ 
of $D({ H})$ is in general reducible, 
and can be decomposed into a direct sum of irreducible representations
$(\Pi^C_{\gamma}, V^C_\gamma)$.
The different single particle states that can be obtained by 
the aforementioned fusion process are the states in the different internal 
Hilbert spaces $V^C_\gamma$ that occur in this decomposition.
We will return to an  elaborate discussion of the fusion rules 
in section~\ref{amalgz}.

An important  property of the 
comultiplication~(\ref{coalgeb}) is that it is coassociative 
\bea \label{coas}
({\mbox{id}} \ot \Delta)\, \Delta(\, {\mbox{P}}_h\, g\,) = 
(\Delta \ot {\mbox{id}}) \, \Delta(\, {\mbox{P}}_h \, g \,) = 
\sum_{h' \cdot h''\cdot h'''=h} {\mbox{P}}_{h'} \, g \ot {\mbox{P}}_{h''} \, {g}
\ot {\mbox{P}}_{h'''} \, {g}.
\eea     
This means that the representation of the quantum double on 
the internal Hilbert space $V_\alpha^A \ot V_\beta^B \ot V_\gamma^C$
(describing a system of three particles) either through 
$({\mbox{id}} \ot \Delta)\, \Delta$ or through 
$(\Delta \ot {\mbox{id}}) \, \Delta$ is completely equivalent. 
Extending the action of the quantum double to systems containing  an 
arbitrary number of particles is now straightforward.
The global symmetry transformations
$g \in { H}$ are implemented on all the particles separately, 
while the operator ${\mbox{P}}_h$ projects out the total flux of the system.

\begin{figure}[htb]    \epsfxsize=11cm
\centerline{\epsffile{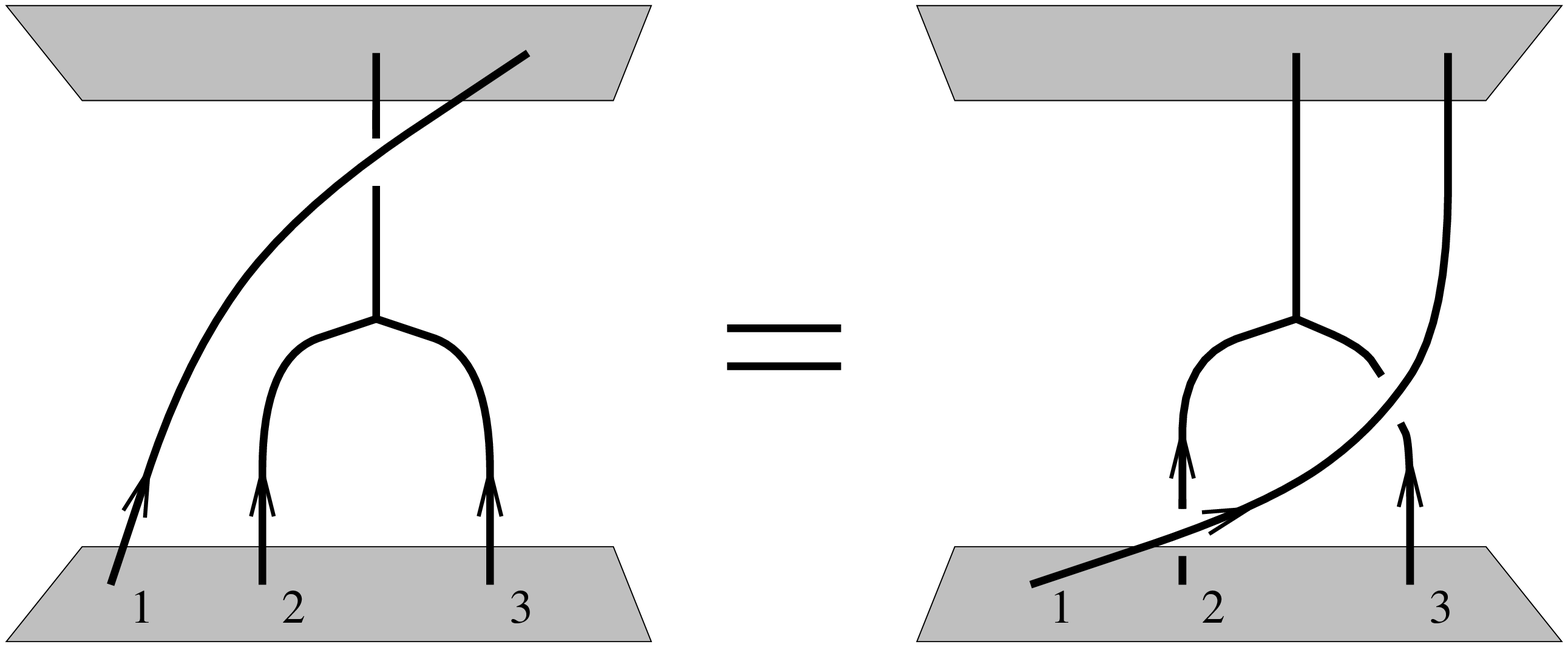}}
\vspace{2cm}
\epsfxsize=12cm
\centerline{\epsffile{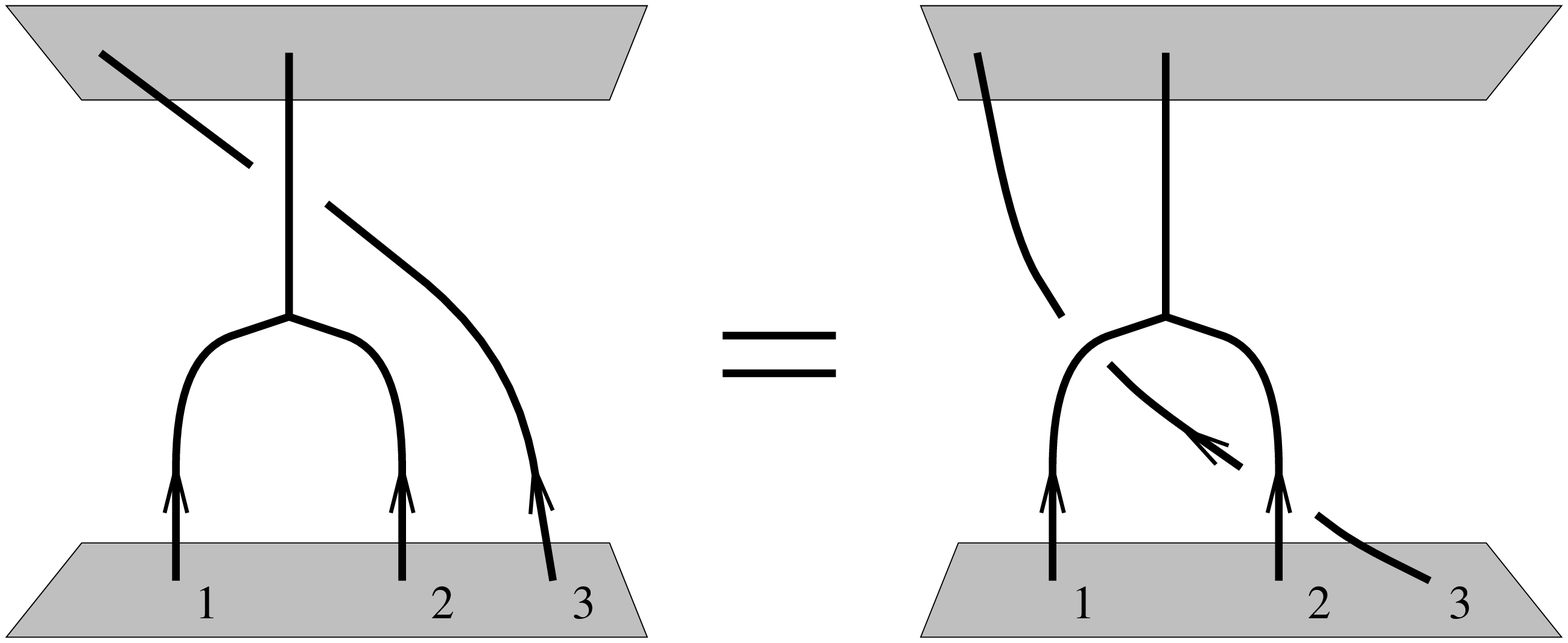}}
\caption{\sl  Compatibility of fusion
and braiding as expressed by the quasitriangularity conditions.
It  makes no difference whether a third particle  braids with two 
particles separately or with the composite that arises after fusing these two 
particles. We have depicted the trajectories of the particles as lines instead
of ribbons. This is what we will usually do when there  
is no writhing involved in the argument.}
\label{qu1zon}
\end{figure}

The braid operation is formally implemented by the universal $R$-matrix, 
which is an element of $D({ H}) \ot D({ H})$
\bea
R &=&\sum_{h, g}\, {\mbox{P}}_g \ot {\mbox{P}}_h \, {g} .    \label{cruijff}
\eea
The $R$ matrix acts on a two particle state as a global symmetry 
transformation on the  second particle by the flux of the first particle.
The physical braid operator ${\cal R}$ that 
effectuates a counterclockwise interchange of  
the two particles  is  defined as the action of this $R$ matrix 
followed by a permutation $\sigma$  of the two particles
\bea \label{jordi}
{\cal R}_{\alpha\beta}^{AB} &:=& 
\sigma\circ(\Pi_{\alpha}^A\otimes\Pi_{\beta}^B)(\, R \, ),
\eea
To be explicit, on the two particle state
$|\,^A\!h_i,\,^{\alpha}\!v_j \rangle |\,^B\!h_m,\,^{\beta}\!v_n \rangle
\in  V^A_{\alpha}\ot V^B_{\beta}$ we have
\bea                 \label{braidact}
{\cal R} \; |\,^A\!h_i,\,^{\alpha}\!v_j \rangle 
|\,^B\!h_m,\,^{\beta}\!v_n \rangle
&=& 
|\,^A\!h_i \, ^B\!h_m \,^A\!h_i^{-1},\,\beta(\,^A\!\tilde{h}_i\,)_{ln} \,
^{\beta}\!v_l \rangle
|\,^A\!h_i,\,^{\alpha}\!v_j \rangle,
\eea  
where the element  $ ^A\!\tilde{h}_i$ is defined as in~(\ref{centdef}). 
Note that the expression~(\ref{braidact}), which
summarizes the braid operation on all conceivable two particle states 
in this theory, contains the braid  effects
established
in the previous section, namely flux metamorphosis
for two pure magnetic fluxes~(\ref{fluxmetam}) and the Aharonov-Bohm
effect for a pure magnetic flux with a pure charge~(\ref{charmetam}).

It is easily verified  that the
braid operator~(\ref{braidact}) and the comultiplication~(\ref{coalgeb}) 
satisfy the quasitriangularity conditions 
\bea                              \label{grcommo}
{\cal R}\: \Delta(\, {\mbox{P}}_h\, g\,) 
&=& \Delta(\, {\mbox{P}}_h\, g\,) \:{\cal R}  \\
({\mbox{id}} \ot \Delta)({\cal R}) &=& 
{\cal R}_2 \; {\cal R}_1
\label{triazo2}       \\
(\Delta \ot {\mbox{id}})({\cal R}) &=& {\cal R}_1 \;  {\cal R}_{2}.
\label{triazo1}
\eea    
where the  braid operators ${\cal R}_1$ and ${\cal R}_2$ respectively 
act as ${\cal R} \ot {\mbox{\bf 1}}$ and ${\mbox{\bf 1}} \ot {\cal R}$ 
on three particle states in $V_\alpha^A \ot V_\beta^B \ot V_\gamma^C$.
The relation~(\ref{grcommo}) expresses the fact that the braid operator
commutes with the global symmetry transformations $g \in { H}$
and conserves the total magnetic flux of the configuration as
measured by ${\mbox{P}}_h$.
In addition, the quasitriangularity 
conditions~(\ref{triazo2}) and~(\ref{triazo1}), which can 
be presented graphically as in figure~\ref{qu1zon},  imply  
consistency between braiding and fusing.
From this set of quasitriangularity conditions, it follows that the 
braid operator satisfies the  Yang-Baxter equation
\bea                    \label{yobo}
{\cal R}_1 \;  {\cal R}_{2}\; {\cal R}_1 &=&
{\cal R}_2 \; {\cal R}_1\;  {\cal R}_{2}.
\eea
Thus the braid operators~(\ref{braidact}) define representations 
of the braid groups discussed in section~\ref{braidgroups}. 
These unitary representations are in general reducible.  
So the internal Hilbert space describing a  multi-particle  system 
in general splits up into a direct sum of irreducible subspaces 
under the action of the braid group.   
The braid properties of the system
depend on the particular irreducible subspace.
If the dimension of the irreducible
representation is one, we are dealing with abelian braid 
statistics or ordinary anyons. If the dimension is larger
then one, we are dealing with nonabelian braid statistics, i.e.\
the nonabelian generalization of anyons.
Note that these higher dimensional irreducible representations only 
occur for systems consisting  of more than two particles, 
because the braid group for two particles is abelian. 

To conclude, the internal Hilbert space describing 
a multi-particle system carries a representation of the internal symmetry
algebra $D({ H})$ and a braid group representation. 
Both representations are in general reducible. The quasitriangularity 
condition~(\ref{grcommo}) implies (see for instance~\cite{alvarez1,alvarez2})
that the action of the associated braid 
operators commutes with the action of the elements of $D({ H})$.
Thus the multi-particle internal Hilbert space can in fact be 
decomposed into a direct sum of irreducible subspaces under 
the direct product action of $D({ H})$ and the braid group. 
We discuss this in further detail in the next two sections. 
We first introduce the notion of truncated braid groups.

\subsection{Truncated braid groups}  \label{trunckbr}

We turn to a closer examination of the braid group representations 
that occur in discrete ${ H}$ gauge theories. 
An important observation  in this respect is that 
the braid operator~(\ref{braidact}) is of finite order
\bea                        \label{finR}
{\cal R}^m &=& {\mbox{\bf 1}} \ot {\mbox{\bf 1}},
\eea
with ${\mbox{\bf 1}}$ the identity operator and $m$ some integer depending on the 
specific particles on which the braid operator acts. 
In other words, we can assign a finite number  $m$ to any two particle 
internal Hilbert space $V_\alpha^A \ot V_\beta^B$, such that the effect
of $m$ braidings is trivial for all states in this internal Hilbert space.
This result, which can be traced back directly to the 
finite order of ${ H}$, 
implies that the multi-particle configurations appearing in a discrete ${ H}$ 
gauge theory actually realize representations of factor groups
of the braid groups discussed in section~\ref{braidgroups}.
Consider, for instance, a system consisting of $n$ 
indistinguishable particles. 
Thus all particles carry the same internal 
Hilbert space  $V^A_{\alpha}$ and  the $n$ particle   
internal Hilbert space describing this system is  the 
tensor product space $(V^A_{\alpha})^{\ot n}$. 
The abstract generator $\tau_i$, which establishes a counterclockwise 
interchange of the two adjacent particles $i$ and $i+1$, acts on
this internal Hilbert space by means of  the operator
\bea                                 \label{brare}
\tau_i &\longmapsto& {\cal R}_i  \, ,
\eea
with
\bea                          \label{ridef}
{\cal R}_i & := & {\mbox{\bf 1}}^{\ot (i-1)} \ot {\cal R} \ot {\mbox{\bf 1}}^{\ot (n-i-1)}.
\eea
Hence, the generator 
$\tau_i$ acts as~(\ref{braidact}) on the $i^{\rm th}$ and 
$(i+1)^{\rm th}$ entry in the tensor product space $(V^A_{\alpha})^{\ot n}$.
As follows from~(\ref{yobo}) and~(\ref{finR}),   
the homomorphism~(\ref{brare}) furnishes a representation of the 
braid group 
\bea
\label{eqy}
\ba{rcll}
\tau_i\tau_{i+1}\tau_i &=& \tau_{i+1}\tau_i\tau_{i+1} &
\qquad i=1,\ldots,n-2  \\
\tau_i\tau_j &=& \tau_j\tau_i & \qquad |i-j|\geq 2,
\ea
\eea
 with  the {\em extra} relation
\bea
\tau_i^m &=& e  \;\;\;\;\;\;\;\;\;\; i=1, \ldots , n-1.
\label{truncate}
\eea
where $e$ denotes the unit element or trivial braid.
For obvious reasons, we will call the factor groups
with defining relations~(\ref{eqy}) and the additional 
relation~(\ref{truncate}) {\em truncated} 
braid groups  $B(n,m)$, where $n$ stands 
for the number of particles and $m$ for the order of 
the generators $\tau_i$.

This picture naturally extends to a system containing 
$n$ distinguishable particles, i.e.\ 
the particles carry different internal Hilbert spaces or `colors' now.
The group that governs the monodromy properties of such a system 
is the truncated version $P(n,m)$ of the colored braid group $P_n({\mbox{\bf R}}^2)$
defined in~(\ref{pbge}). To be specific, the truncated colored braid 
group $P(n,m)$ is the subgroup of $B(n,m)$ generated by 
\bea                         
\gamma_{ij} &=& \tau_i \cdots \tau_{j-2}\; \tau_{j-1}^2 \; 
\tau_{j-2}^{-1}\cdots
\tau_i^{-1}   \qquad \qquad   1 \leq i<j \leq n,
\eea
with the extra relation~(\ref{truncate}) incorporated. 
Thus the generators of the pure braid group satisfy 
\bea     \label{puretru}           
\gamma_{ij}^{m/2} &=& e,  
\eea                                              
from which it is clear that the colored braid group $P(n,m)$ can only be 
defined for even $m$.
The representation of the colored braid group $P(n,m)$ realized by a system 
of $n$ different particles then  becomes  
\bea                                 \label{purarepo}
\gamma_{ij} &\longmapsto& {\cal R}_i \cdots {\cal R}_{j-1} \; {\cal R}_j^2 \;
{\cal R}_{j-1}^{-1} \cdots {\cal R}_i^{-1},
\eea
where the operators ${\cal R}_i$ defined by expression~(\ref{ridef}) 
now act on the tensor product space $ V^{A_{1}}_{\alpha_{1}} \ot \cdots \ot 
V^{A_{n}}_{\alpha_{n}}$ of $n$ different internal Hilbert spaces
$V^{A_{l}}_{\alpha_{l}}$.

Finally, a mixture of the above systems is of course also possible, that is,
a system containing a subsystem consisting of $n_1$ particles with `color' 
$V^{A_{1}}_{\alpha_{1}}$, a subsystem of $n_2$ particles carrying the 
different   
`color' $V^{A_{2}}_{\alpha_{2}}$ and so on.
Such a system  realizes a
representation of a truncated partially colored braid 
group (see for instance~\cite{brekfa,brekke} 
for the definition of ordinary partially colored braid groups).
Let $n$ again be the total number of particles in the system.
The truncated partially colored braid group associated with this system
then becomes the subgroup of $B(n,m)$, generated by 
the braid operations on  particles with the same `color' and the monodromy
operations on particles carrying different `color'.

The appearance of truncated rather than ordinary braid groups facilitates
the decomposition of a given multi-particle internal Hilbert space into 
irreducible subspaces under the braid/monodromy  operations. 
The point is that the representation theory of  
ordinary braid groups is rather complicated due to their infinite order.
The extra relation~(\ref{truncate}) for
truncated braid groups $B(n,m)$, however, causes these to become
finite for various values of the labels $n$ and $m$,  which 
leads to identifications  with well-known groups of 
finite order~\cite{pema}. It is instructive to consider
some of these cases explicitly.
The truncated braid group $B(2,m)$ for two indistinguishable 
particles, for instance, has 
only one generator $\tau$, which satisfies  $\tau^m = e$. 
Thus we obtain the isomorphism
\bea
B(2,m) &\simeq& \Z_m.
\label{eq:rela}
\eea
For $m=2$, the relations~(\ref{eqy}) and~(\ref{truncate}) are the 
defining relations of the permutation group $S_n$ on $n$ strands
\bea
B(n,2) &\simeq& S_n.
\label{eq:rela1}
\eea
A less trivial example is the nonabelian truncated braid group $B(3,3)$ for 3
indistinguishable particles.
By explicit construction  from the defining relations~(\ref{eqy}) 
and~(\ref{truncate}), we arrive at the identification   
\bea     
B(3,3) &\simeq& \bar{T},
\eea
with $\bar{T}$ the lift of the tetrahedral group into $SU(2)$.
The structure of the truncated braid group $B(3,4)$ and its subgroup 
$P(3,4)$, which for example occur in a $\bar{D}_2$ gauge theory 
(see section~\ref{d2bqst}), can be found in appendix~\ref{trubra}.

To our knowledge,
truncated braid groups have not been studied in the literature so far 
and a complete classification is not available. Although 
discrete ${ H}$ gauge theories just realize finite 
dimensional representations, 
it remains an interesting group theoretical question 
whether the truncated braid groups are of finite order for all values of the 
labels $n$ and $m$.

\subsection{Fusion, spin, braid statistics and all that $\dots$}   
\label{amalgz}

Let $(\Pi^A_{\alpha}, V_\alpha^A)$ and
$(\Pi^B_{\beta}, V_\beta^B)$  be  two irreducible 
representations of the quantum double $D({ H})$ as defined in~(\ref{13zo}).
The tensor product representation 
$(\Pi^A_{\alpha} \ot \Pi^B_{\beta},V_\alpha^A \ot V_\beta^B)$, constructed
by means of the comultiplication~(\ref{coalgeb}), 
need not be irreducible. In general, it  gives rise to a
decomposition
\bea               \label{pietk}
\Pi^A_{\al}\otimes\Pi^B_{\beta}& = & \bigoplus_{C,\gamma}
N^{AB\gamma}_{\alpha\beta C} \; \Pi^C_{\gamma}, 
\eea
where $N^{AB\gamma}_{\alpha\beta C}$  stands for  the 
multiplicity of the irreducible representation 
$(\Pi^C_{\gamma}, V^C_{\gamma})$.
From  the orthogonality relation for the characters of the irreducible 
representations of $D({ H})$, we infer~\cite{dpr}
\bea          \label{Ncoe}
N^{AB\gamma}_{\alpha\beta C} &=& \frac{1}{|{ H}|} \sum_{h,g}  \;
            \mbox{tr}  \left( \Pi^A_{\alpha} \ot \Pi^B_{\beta}
                      (\Delta (\, {\mbox{P}}_h \, g \,)  ) \right)   \;
            \mbox{tr}  \left( \Pi^C_{\gamma} (\, {\mbox{P}}_h \, g \,) \right)^*, 
\eea
where $|{ H}|$ denotes the order of the group ${ H}$ and $*$ indicates
complex conjugation.
The fusion rule~(\ref{pietk}) 
now determines which particles $(\,^C C,\gamma)$ can be formed in the composition 
of the two particles $(\, ^A C,\alpha \,)$ and  $(\,^B C,\beta\,)$,
or if read backwards, gives the decay channels of the particle $(\,^C C,\gamma\,)$.

The fusion algebra, spanned by the elements
$\Pi^A_{\alpha}$ with multiplication rule~(\ref{pietk}), is 
commutative and associative and can therefore be diagonalized.
The matrix  implementing this diagonalization is the 
modular $S$ matrix~\cite{ver0} 
\bea                                \label{fusionz}   
S^{AB}_{\alpha\beta} &:=& \frac{1}{|{ H}|} \, \mbox{tr} \; {\cal 
R}^{-2 \; AB}_{\; \; \; \; \; \alpha\beta}       \\
&=&
\frac{1}{|{ H}|} \, \sum_{\stackrel{\,^A\!h_i\in\,^A\!C\,,^B\!h_j\in\,
^B\!C}{[\,^A\!h_i,\,^B\!h_j]=e}} 
\mbox{tr} \left(\alpha (\,^A\!x_i^{-1}\,^B\!h_j\,^A\!x_i) \right)^*
\; \mbox{tr} \left( \beta(\,^B\!x_j^{-1}\,^A\!h_i\, ^B\!x_j) \right)^*, \nn
\eea
which contains all information concerning 
the fusion algebra~(\ref{pietk}).
In particular, the multiplicities~(\ref{Ncoe}) can be expressed in 
terms of the modular $S$ matrix by means of Verlinde's formula~\cite{ver0}
\bea      \label{verlindez}
N^{AB\gamma}_{\alpha\beta C}&=&\sum_{D,\delta}\frac{
S^{AD}_{\alpha\delta}S^{BD}_{\beta\delta}
(S^{*})^{CD}_{\gamma\delta}}{S^{eD}_{0\delta}}.
\eea 
Whereas the modular $S$ matrix is determined through the monodromy operator
following from~(\ref{braidact}),
the modular matrix $T$ contains the spin 
factors~(\ref{spin!}) assigned to the particles 
\bea                                         \label{modutz}
T^{AB}_{\alpha\beta} &:=& 
\delta_{\alpha,\beta} \, \delta^{A,B} \; \exp(2\pi \im s_{(A,\alpha)})
\; = \;
\delta_{\alpha,\beta} \, \delta^{A,B} 
\frac{1}{d_\alpha}   \, \mbox{tr} \left(  \alpha(^A\!h_1) \right),
\eea
with $d_\alpha$ the dimension of the centralizer charge 
representation $\alpha$ of the particle $(\, ^A C, \alpha \,)$. 
The  matrices~(\ref{fusionz}) and~(\ref{modutz}) now 
realize an unitary representation of the modular group $SL(2,\Z)$
with the following  relations~\cite{dvvv}
\begin{eqnfourarray}
{\cal C} &=&(ST)^3 \; = \; S^2,      &         \label{charconj} \\
S^* &=& {\cal C} S \; = \;S^{-1}, & $\qquad S^t \;=\; S, $        \label{sun}      \\
T^* &=&T^{-1}, & $\qquad T^t \;=\; T.$                   \label{tun}
\end{eqnfourarray}  
The  relations~(\ref{sun}) and~(\ref{tun}) express the fact that 
the matrices~(\ref{fusionz}) and~(\ref{modutz}) 
are symmetric and unitary. To proceed, the matrix ${\cal C}$ defined 
in~(\ref{charconj}) represents the charge conjugation operator, which
assigns an unique anti-partner 
${\cal C} \, (\, ^A C,\alpha \,)=(\, ^{\bar{A}} C,\bar{\alpha} \,)$
to every particle  $(\, ^A C,\alpha \,)$ in the spectrum,
such that the vacuum channel occurs in the fusion rule~(\ref{pietk}) 
for the particle/anti-particle pairs.
Also note that the complete set of relations imply that 
the charge conjugation matrix ${\cal C}$ commutes with 
the modular matrix $T$, from which we conclude
that a given particle carries the same spin as its anti-partner.

Having determined the fusion rules and the associated modular algebra,
we turn to the issue of braid statistics and the fate of the spin
statistics connection in this nonabelian context.
Let us emphasize from the outset that much of what follows has 
been established elsewhere in a more general setting.
See~\cite{alvarez1,alvarez2} and the references therein 
for the conformal field theory point 
of view and~\cite{frohma,frogama,witten} for the related 
2+1 dimensional space time perspective.

We first discuss a system consisting of two distinguishable 
particles $(\, ^A C,\alpha \,)$ and  $(\,^B C,\beta\,)$. The associated 
two particle internal Hilbert space $V_\alpha^A \ot V_\beta^B$ carries 
a representation of the abelian truncated colored braid group $P(2,m)$ with 
$m/2 \in \Z$ the order of the monodromy matrix ${\cal R}^2$ for this 
particular two-particle system.
This representation decomposes into a direct sum 
of one dimensional irreducible subspaces, each being labeled by the 
associated eigenvalue of the monodromy matrix ${\cal R}^2$. 
Recall from section~\ref{thequdo}, that the monodromy operation
commutes with the action of the quantum double. 
This implies that the decomposition~(\ref{pietk}) 
simultaneously diagonalizes
the monodromy matrix. To be specific, the two particle 
total flux/charge eigenstates spanning a given 
fusion channel $V_\gamma^C$ all carry the same monodromy eigenvalue,
which in addition can be shown to satisfy 
the generalized spin-statistics connection~\cite{dpr}
\bea             \label{gekspist}
K^{ABC}_{\alpha\beta\gamma} \;
{\cal R}^2 &=& 
e^{2\pi \im(s_{(C,\gamma)}-s_{(A,\alpha)}-s_{(B,\beta)})} \;\;
K^{ABC}_{\alpha\beta\gamma},
\eea
where  $K^{ABC}_{\alpha\beta\gamma}$ stands for the projection on  
the irreducible component $V_\gamma^C$  of  $V_\alpha^A \ot V_\beta^B$.  
In other words, the monodromy operation on a two particle state in a given
fusion channel is the same as a rotation over an angle of $-2\pi$ of 
the two particles separately accompanied by a rotation over an angle 
of $2\pi$ of the single particle state emerging after fusion. 
This is consistent with the observation that these two processes can be 
continuously deformed into each other, as can be seen from the associated
ribbon diagrams depicted in figure~\ref{kanaal}.
The discussion can now be summarized by the statement that 
the total internal Hilbert space $V_\alpha^A \ot V_\beta^B$ decomposes 
into the following  direct sum of irreducible representations of the direct 
product  $D({ H}) \times P(2,m)$ 
\bea
 \bigoplus_{C,\gamma} N^{AB\gamma}_{\alpha\beta C} \; 
(\Pi^C_{\gamma}, \Lambda_{C-A-B} ),
\eea 
where $\Lambda_{C-A-B}$ denotes the one dimensional 
irreducible representation of $P(2,m)$ in which the monodromy 
generator $\gamma_{12}$ acts as~(\ref{gekspist}).

\begin{figure}[htb]    \epsfxsize=11cm
\centerline{\epsffile{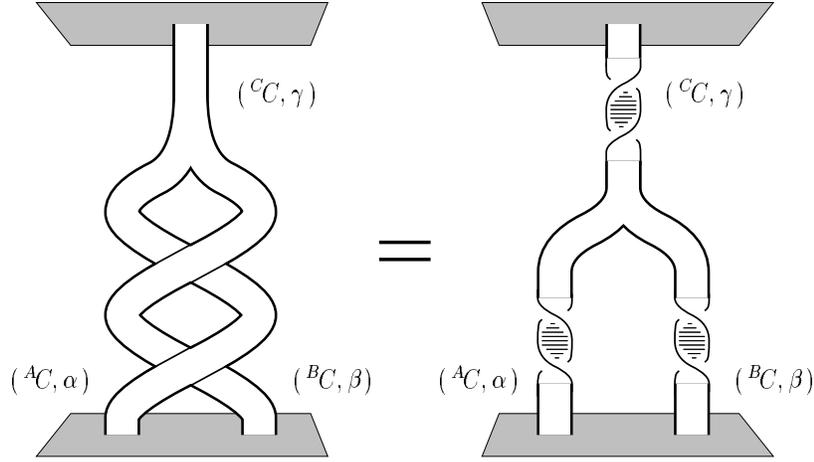}}
\caption{\sl Generalized spin-statistics connection. 
The displayed ribbon diagrams are homotopic as can 
be checked with the pair of pants you are presently wearing. 
This means that
a monodromy of two particles in a given fusion channel followed  
by fusion of the pair  can be continuously deformed into the process
describing a rotation over an angle of $-2\pi$ of 
the two particles seperately followed by fusion of the pair and 
a final rotation over an angle of $2\pi$ of the composite. }
\label{kanaal}
\end{figure}

The analysis for a configuration 
of two indistinguishable particles $(\, ^A C,\alpha \,)$
is analogous. The total internal Hilbert space 
$V_\alpha^A \ot V_\alpha^A$ decomposes
into one dimensional irreducible subspaces under the action 
of the truncated braid group $B(2,m)$ with $m$ the order 
of the braid operator ${\cal R}$, which depends on the system
under consideration. 
By the same argument 
as before, the two particle total flux/charge eigenstates spanning 
a given fusion channel $V_\gamma^C$ all 
carry  the same one dimensional representation
of $B(2,m)$.  The quantum statistical parameter assigned to this channel
now satisfies the square root version of the generalized spin-statistics 
connection~(\ref{gekspist})
\bea             \label{gespietst}
K^{AAC}_{\alpha\alpha\gamma} \;
{\cal R} &=& \epsilon \;
e^{ \pi \im(s_{(C,\gamma)}-2 s_{(A,\alpha)})} \; \;
K^{AAC}_{\alpha\alpha\gamma},
\eea
with $\epsilon$ a sign depending on whether the 
fusion channel $V_\gamma^C$ appears in a symmetric or an
anti-symmetric fashion~\cite{alvarez1}. 
In other words, the internal space Hilbert space for a system of two 
indistinguishable particles $(\, ^A C,\alpha \,)$ breaks up into the following
irreducible representations of the direct product $D(H) \times B(2,m)$
\bea                           \label{nil!}
 \bigoplus_{C,\gamma} N^{AA\gamma}_{\alpha\alpha C} \; 
(\Pi^C_{\gamma}, \Lambda_{C-2A} ),
\eea 
with $\Lambda_{C-2A}$ the one dimensional representation 
of the truncated braid group $B(2,m)$ defined in~(\ref{gespietst}).

The result~(\ref{gespietst}) is actually rather surprising.
It states that  indistinguishable particle systems
in a nonabelian discrete $H$ gauge theory quite
generally violate the canonical spin-statistics connection~(\ref{spistath}).
More accurately, in a nonabelian discrete gauge theory
we are dealing with the generalized connection~(\ref{gespietst}), which 
incorporates the canonical one. In fact, the canonical spin-statistics
connection is retrieved in some particular channels occurring in~(\ref{nil!}),
as we will argue now.
Let us first emphasize that the basic assertions for the ribbon proof 
depicted in figure~\ref{spinstafig} are naturally  
satisfied in the nonabelian setting as well. For every particle 
$(\, ^A C,\alpha \,)$ in the spectrum there exists an 
anti-particle $(\, ^{\bar{A}} C,\bar{\alpha} \,)$ such that 
under the proper composition the pair acquires the quantum numbers of the 
vacuum and may decay. Moreover, every particle carries the same spin as its
anti-partner, as indicated by the fact that the 
charge conjugation operator ${\cal C}$ commutes with the modular matrix $T$. 
It should be noted now that the ribbon proof in figure~\ref{spinstafig}
actually {\em only} applies to states in which
the particles that propagate along the exchanged ribbons are 
in strictly  identical internal states. Otherwise the ribbons can not be
closed. Indeed, we find that the action~(\ref{braidact}) of the braid operator 
on two particles in identical internal flux/charge eigenstates
\bea                 \label{braidactid}
{\cal R} \; |\,^A\!h_i,\,^{\alpha}\!v_j \rangle 
|\,^A\!h_i,\,^{\alpha}\!v_j \rangle
&=& 
|\,^A\!h_i,\,\alpha\,(\,^A\! h_1\,)_{mj} \,
^{\alpha}\!v_j \rangle
|\,^A\!h_i,\,^{\alpha}\!v_j \rangle ,
\eea  
boils down to the diagonal matrix~(\ref{spin!}) and therefore to the 
spin factor~(\ref{spist}) for all  $i,j$
\bea                             \label{spist}
\exp(\im \Theta_{(A,\alpha)}) &=& \exp(2\pi \im s_{(A,\alpha)}).
\eea
The conclusion is that the canonical spin-statistics connection is 
restored in the fusion channels spanned by linear combinations 
of the states~(\ref{braidactid}) in which the particles are in strictly 
identical internal flux/charge eigenstates. 
The quantum statistical parameter~(\ref{gespietst}) assigned to 
these channels reduces to the spin factor~(\ref{spist}), thus the 
effect of a counterclockwise interchange 
of the two particles in the states in these channels is the same 
as the effect of rotating one of the particles over an angle of $2\pi$.
To conclude, the closed ribbon proof does not apply to the other channels
and we are left with the more involved connection~(\ref{gespietst}) following 
from the open ribbon argument displayed in figure~\ref{kanaal}.

Finally, higher dimensional irreducible braid group representations 
are conceivable for a system of more than two particles. The occurrence of such 
representations simply means that the generators of the braid group can 
not be diagonalized simultaneously. What happens in this situation is that
under the full set of braid operations, the system jumps between isotypical 
fusion channels, i.e.\ fusion channels of the same type or `color'.
Let us make this
statement more precise. To keep the discussion general, we
do not specify the nature of the particles in the system.
Depending  on whether the system consists of 
distinguishable particles, indistinguishable particles or some mixture, we are dealing
with a truncated braid group, a colored braid group or a 
partially colored braid group respectively.
The internal Hilbert for such a system again decomposes  
into a direct sum of irreducible subspaces (or fusion channels) under 
the action of the quantum double.  
Given the fact that the action of the associated  
braid group commutes with that of the quantum double, we are left 
with two possibilities.
First of all, there will in general be some 
fusion channels separately  being  invariant under
the action of the full  braid group.
As in the two particle systems discussed before, 
the total flux/charge eigenstates spanning such a 
fusion channel, say $V_\gamma^C$, carry the same one 
dimensional irreducible representation $\Lambda_{ab}$ of the braid group, 
that is, these states realize abelian braid statistics with the same
quantum statistical parameter.
The fusion channel $V_\gamma^C$ then carries the irreducible representation 
$(\Pi_\gamma^C, \Lambda_{ab})$ of the direct product of the 
quantum double and the braid group.
In addition, it is also feasible 
that states carrying the {\em same} total flux and charge 
in {\em different} (isotypical) fusion channels 
are mixed under the action of 
the full  braid group.  In that case, we are dealing with 
a higher dimensional irreducible representation of the 
truncated braid group or nonabelian braid statistics. 
Note that nonabelian braid statistics is conceivable, 
if and only if some fusion channel, say $V_\delta^D$, occurs more 
then once  in the decomposition of the Hilbert space
under the action of the quantum double. 
Only then there are some orthogonal states with 
the same total flux and charge available to span an higher dimensional 
irreducible representation of the braid group.
The number $n$ of fusion channels $V_\delta^D$ 
related by the action of the braid operators
now constitutes the dimension of the irreducible representation 
$\Lambda_{nonab}$ of the braid group and the multiplicity of
this representation 
is the dimension $d$ of the fusion channel $V_\delta^D$. 
To conclude, the direct sum of these $n$ fusion 
channels $V_\delta^D$ carries an $n \cdot d$ dimensional 
irredicible representation $(\Pi_\delta^D, \Lambda_{nonab})$
of the direct product of the quantum double and the braid group.

\sectiona{$\bar{D}_2$  gauge theory}   \label{exampled2b}

Here, we will illustrate the general considerations of the foregoing sections 
with one of the simplest nonabelian discrete ${ H}$ gauge theories,
namely that with  gauge group the double dihedral group
${ H} \simeq \bar{D}_2$, see also~\cite{spm,spm1,sm}. 

A $\bar{D}_2$  gauge theory may, for instance, 
appear as `the long distance remnant' of a Higgs model of the 
form~(\ref{totmod}) in which 
the gauge group  $G \simeq SU(2)$ is spontaneously broken 
down to the double dihedral 
group $H \simeq \bar{D}_2 \subset SU(2)$.
Since $SU(2)$ is simply connected, the fundamental group 
$\pi_1(SU(2)/\bar{D}_2)$ coincides with the residual symmetry group
$\bar{D}_2$. Hence, the stable magnetic fluxes in this broken 
theory are indeed labeled by the group elements of $\bar{D}_2$.
See the discussion concerning 
the isomorphism~(\ref{flulaord}) in section~\ref{topclas}.
In the following, we will not dwell any further on the explicit
details of this or other possible
embeddings in broken gauge theories and simply focus on the
features of the $\bar{D}_2$ gauge theory itself. 
We start with a discussion of the spectrum.

\begin{table}[htb] 
\begin{center}
\begin{tabular}[t]{lcl} \hline
$ \mbox{Conjugacy class}                      $ & & $ \qquad 
\mbox{Centralizer}       $  \\       \hl \\[-4mm] 
$ e=\{ e \}                                      $& &$ \qquad \bar{D}_2$\\ 
$ \bar{e}=\{\bar{e}\}                           $& &$ \qquad \bar{D}_2$\\ 
$ X_1=\{X_1,\bar{X}_1\}\ $& &$ 
\qquad \Z_4\simeq \{e,X_1,\bar{e},\bar{X}_1\} $\\ 
$ X_2=\{X_2,\bar{X}_2\} $& &$ 
\qquad \Z_4\simeq \{e,X_2,\bar{e},\bar{X}_2\} $\\ 
$ X_3=\{X_3,\bar{X}_3\} $& &$ 
\qquad \Z_4 \simeq \{e,X_3,\bar{e},\bar{X}_3\}$\\[1mm]
\hl
\end{tabular}
\end{center} 
\caption{\sl Conjugacy classes of $\bar{D}_2$ together with their 
centralizers.}  
\label{tabcond2}  
\end{table}
\begin{table}[htb] 
\begin{center}
\begin{tabular}[t]{lrrrrr}     \hline   \\[-4mm]
$\str \bar{D}_2\qquad$&$ e $&$ \bar{e}$&$ X_1  $&$ X_2  $&$ X_3  $\\ 
\hl \\[-4mm]
$ 1   \str$&$ 1 $&$ 1      $&$ 1  $&$ 1  $&$ 1  $\\ 
$ J_1 \str$&$ 1 $&$ 1      $&$ 1  $&$-1  $&$-1  $\\ 
$ J_2 \str$&$ 1 $&$ 1      $&$-1  $&$ 1  $&$-1  $\\ 
$ J_3 \str$&$ 1 $&$ 1      $&$-1  $&$-1  $&$ 1  $\\ 
$ \chi \str$&$ 2 $&$ -2     $&$ 0  $&$ 0  $&$ 0  $\\[1mm]   \hl
\end{tabular} \qquad \qquad
\begin{tabular}[t]{lrrrrr} \hline \\[-4mm]
$\str \Z_4\qquad$&$ e $&$ X_a      $&$ \bar{e} $&$ \bar{X}_a $\\ 
\hl  \\[-4mm]
$   {\Gamma^0}\str$&$ 1 $&$ 1       $&$ 1  $&$ 1     $\\ 
$   {\Gamma^1}\str$&$ 1 $&$ \imath  $&$-1  $&$-\imath$\\ 
$   {\Gamma^2}\str$&$ 1 $&$ -1      $&$ 1  $&$-1     $\\ 
$   {\Gamma^3} \str$&$ 1 $&$ -\imath $&$-1  $&$\imath $\\[1mm]   \hl
\end{tabular}
\end{center}
\caption{\sl Character tables of $\bar{D}_2$ and $\Z_4$.}
\label{chartabd2}
\end{table}

The double dihedral group $\bar{D}_2$ is a group of order~$8$ with a nontrivial
centre of order~$2$. The fluxes 
associated with its group elements are organized in the 
conjugacy classes exhibited in  table~\ref{tabcond2}.
There are five conjugacy classes denoted by $e,\bar{e},X_1,X_2$ and $X_3$.   
The conjugacy class $e$ corresponds to the trivial flux sector, while
$\bar{e}$ contains the nontrivial centre element. 
The conjugacy classes  $X_1,X_2$ and $X_3$ consist of two commuting 
elements of order~$4$. Thus there are four nontrivial purely magnetic 
flux sectors: one singlet flux $\bar{e}$ 
and three different doublet fluxes $X_1,X_2$ and $X_3$. 
The purely electric charge sectors, on the other hand, correspond to 
the UIR's of $\bar{D}_2$.
From the character table displayed in table~\ref{chartabd2}, 
we infer that there are four nontrivial pure charges in the spectrum:
three singlet charges $J_1,J_2,J_3$ and one doublet charge $\chi$.  
The fluxes $X_a/\bar{X}_a$ act on the doublet charge $\chi$ 
as $\pm \im \sigma_a$, with $\sigma_a$ being the Pauli matrices.
Let us now turn to the dyonic sectors. These are constructed by 
assigning nontrivial centralizer representation to the nontrivial
fluxes. The centralizers associated with the different flux sectors 
can be found in table~\ref{tabcond2}. The flux $\bar{e}$ 
obviously commutes with the full group $\bar{D}_2$, while the centralizer
of the other flux sectors is the cyclic group $\Z_4$.  
Thus  we arrive at thirteen different dyons:
three singlet dyons and one doublet dyon
associated with the flux $\bar{e}$ and nine doublets dyons 
associated with the fluxes $X_1,X_2$ and $X_3$ paired with 
nontrivial $\Z_4$ representations.
All in all, the spectrum of this theory features 22 particles, which 
will be labeled as
\bea   \label{spectred2}       \ba{rclrcl}
1        &:=& (e,\, 1 )
& \qquad \qquad \bar{1}        &:=& (\bar{e},\, 1 )       \\
J_a      &:=& (e,\, J_a )  
& \qquad \qquad \bar{J}_a      &:=& (\bar{e},\, J_a)       \\
\chi     &:=& (e,\, \chi )     
& \qquad \qquad \bar{\chi}     &:=& (\bar{e},\, \chi)     \\
\sigma_a^+&:=& (X_a,\, {\Gamma^0})
& \qquad \qquad \sigma_a^-&:=& (X_a,\, {\Gamma^2})  \\ 
\tau_a^+&:=& (X_a,\, {\Gamma^1})
& \qquad \qquad \tau_a^-&:=& (X_a,\,{\Gamma^3}),
\ea
\eea 
for convenience.
Note that the square of the dimensions of the internal Hilbert spaces
carried by these particles indeed 
add up to the order of the quantum double $D(\bar{D}_2)$: 
$8 \cdot 1^2 +14 \cdot 2^2 = 8^2$.

We proceed with a detailed analysis of  the topological interactions 
between the particles in the spectrum~(\ref{spectred2}). The discussion
is organized as follows.
In section~\ref{chesd2b}, we will establish 
the fusion rules. This is the natural setting to 
discuss a feature special for nonabelian discrete ${ H}$ 
gauge theories: a pair of nonabelian fluxes can carry charges that are not 
localized on any of the two fluxes nor anywhere else. 
We will show that these so-called Cheshire charges can be excited 
by monodromy  processes with for instance the doublet charges $\chi$.
To proceed, section~\ref{abd2b} contains a discussion of the cross sections
associated with  Aharonov-Bohm scattering experiments 
with the particles in this theory.
Finally, the issue of nonabelian braid statistics will be dealt with in 
section~\ref{d2bqst}.

\subsection{Alice in physics}     \label{chesd2b}

As we have seen in                
section~\ref{amalgz}, the topological interactions for a particular discrete 
${ H}$ gauge theory are classified by the content of the associated 
modular matrices $S$ and $T$. The modular $T$ matrix~(\ref{modutz})
contains the spin factors of the particles. For the particles in the 
spectrum of  this $\bar{D}_2$ gauge theory we easily infer
\bea   \label{spind2}  \ba{cc} 
\mbox{particle}       & \qquad   \exp (2\pi \im  s)    \\
1,J_a             & \qquad    1   \\
\bar{1},\bar{J}_a & \qquad    1  \\
\chi / \bar{\chi} & \qquad    \pm 1 \\
\sigma_a^{\pm}    & \qquad    \pm 1  \\
\tau_a^{\pm}      & \qquad    \pm \im.
\ea
\eea  
The modular $S$ matrix~(\ref{fusionz}), on the other hand, 
is determined  by the monodromy matrix following from~(\ref{braidact}).
It can be verified that the modular $S$ matrix
for this model is real and therefore orthogonal.   
For future reference, we have displayed it in table~\ref{modsd2}. 
In this section, we will focus on the fusion rules 
obtained from Verlinde's formula~(\ref{verlindez}) and the key role 
they play as overall selection rules for the flux/charge exchanges 
occurring when the particles encircle each other.

\begin{table}[h]
\begin{center}
\begin{tabular}{crrrrrrrrrrr}      \hline \\[-4mm]
$S$& &$1 $& $\bar{1} $&$J_a $&$\bar{J}_a $&$\chi $&$\bar{\chi} 
$&$\sigma^+_a$ &$\sigma^-_a $&$\tau^+_a $&$\tau^-_a $\\ \hl
\\[-4mm]  
$1$& &$1 $&$1 $&$1 $&$1 $&$2 $&$2 $&$2 $&$2 $&$2 $&$2 $\\ 
$\bar{1}$& &$1 $&$1 $&$1 $&$1 $&$-2 $&$-2 $&$2 $&$2 $&$-2 $
&$-2 $\\ 
$J_b$& &$1 $&$1 $&$1 $&$1 $&$2 $&$2 $&$2\ep_{ab} $
&$2\ep_{ab} $&$2\ep_{ab} 
$&$2\ep_{ab} $\\ 
$\bar{J}_b$& &$1 $&$1 $&$1 $&$1 $
&$-2 $&$-2 $&$2\ep_{ab} $&$2\ep_{ab} $&$-2\ep_{ab} 
$&$-2\ep_{ab} $\\ 
$\chi$& &$2$&$-2$&$2$&$-2$&$4 $&$-4 $&$0 $&$0 $&$0 $&$0 $\\ 
$\bar{\chi}$& &$2$&$-2$&$2$&$-2$&$-4$&$4  $&$0 $&$0 $&$0 
$&$0 $\\ 
$\sigma^+_b$& &$2 $&$2 $&$2\ep_{ab} $&$2\ep_{ab} $&$0 $&$0  
$&$4\delta_{ab} $&$-4\delta_{ab} $&$0 $&$0 $\\ 
$\sigma^-_b$& &$2 $&$2 $&$2\ep_{ab} $&$2\ep_{ab} $&$0 $&$0  
$&$-4\delta_{ab}     $&$4\delta_{ab} $&$0 $&$0 $\\ 
$\tau^+_b$& &$2 $&$-2$&$2\ep_{ab} $&$-2\ep_{ab} $&$0 $&$0 $&$0 
$&$0 $&$-4\delta_{ab} $&$4\delta_{ab} $\\ 
$\tau^-_b$& &$2 $&$-2$&$2\ep_{ab} $&$-2\ep_{ab} $&$0 $&$0 $&$0 
$&$0 $&$4\delta_{ab} $&$-4\delta_{ab} $\\[1mm]   \hline
\end{tabular}
\end{center}
\caption{\sl Modular S-matrix of the quantum double 
$D(\bar{D}_2)$ up to an overall factor
$\frac{1}{8}$. We defined $\epsilon_{ab}= 1$ if $a=b$ 
and $-1$ otherwise.}
\label{modsd2}
\end{table}

We start with the fusion rules for the purely electric charges.
These are dictated by the representation ring of $\bar{D}_2$
\bea                \label{puch}
J_a \times J_a=1, \qquad J_a \times J_b = J_c, \qquad J_a \times \chi=\chi,
\qquad \chi \times \chi = 1+ \sum_a J_a.
\eea
The dyons associated with the flux $\bar{1}$ are obtained by simply
composing this flux with the purely electric charges
\bea                           \label{zendy}
J_a \times \bar{1}= \bar{J}_a, \qquad \chi \times \bar{1} = \bar{\chi}.
\eea 
In a similar fashion, we produce the other dyons 
\bea                                \label{ooknog}
J_a \times \sigma^+_a = \sigma^+_a, \qquad 
J_b \times \sigma^+_a= \sigma^-_a, \qquad 
\chi \times \sigma^+_a = \tau^+_a + \tau^-_a.
\eea
We now have all the constituents of the spectrum~(\ref{spectred2}). 
Recall that the fusion algebra is commutative and associative.
This implies that the complete set of fusion rules is 
actually determined by a minimal subset. 
Bearing this in mind, amalgamation involving the 
flux  $\bar{1}$ is unambiguously prescribed by~(\ref{zendy}) and
\bea                          \label{centflufu}
\bar{1} \times \bar{1} = 1, \qquad
\bar{1} \times \sigma^{\pm}_a = \sigma^{\pm}_a, \qquad 
\bar{1} \times \tau^{\pm}_a = \tau^{\mp}_a. 
\eea
The complete set of fusion rules is fixed by the previous ones 
together with
\bea
J_a \times \tau^{\pm}_a  = \tau^{\pm}_a, \qquad
J_b \times \tau^{\pm}_a  = \tau^{\mp}_a, \qquad
\chi \times \tau^{\pm}_a = \sigma^+_a + \sigma^-_a,
\eea
and 
\bea
\sigma^{\pm}_a \times \sigma^{\pm}_a &=&1+J_a+\bar{1}+\bar{J}_a 
\label{remark} \\
\sigma^{\pm}_a \times \sigma^{\pm}_b &=& \sigma^+_c+\sigma^-_c \\
\sigma^{\pm}_a \times \tau^{\pm}_a &=& \chi+\bar{\chi}    \\
\sigma^{\pm}_a \times \tau^{\pm}_b  &=& \tau^+_c + \tau^-_c \\
\tau^{\pm}_a \times \tau^{\pm}_a &=& 1+J_a+\bar{J}_b+\bar{J}_c \label{spreek}\\
\tau^{\pm}_a \times \tau^{\pm}_b &=& \sigma^+_c+\sigma^-_c.
\eea
A few remarks are pertinent at this stage. First of all,
the class algebra of $\bar{D}_2$ is respected as an 
overall selection rule. 
The class multiplication in the fusion rule~(\ref{remark}), for instance,
reads $X_a * X_a = 2e+2\bar{e}$. The appearance of the class algebra
expresses magnetic flux conservation.
In establishing the fusion rule,  all fluxes in the consecutive 
conjugacy classes are multiplied out.
To proceed, the modular $S$ matrix as given in table~\ref{modsd2}
is real and therefore equal to its inverse as follows from~(\ref{sun}).
As a consequence, the charge conjugation operator ${\cal C}$ acts on the  
spectrum~(\ref{spectred2}) as the unit matrix ${\cal C}=S^2={\mbox{\bf 1}}$,
thus the particles in this $\bar{D}_2$ gauge theory feature as 
their own anti-partner. Only two similar particles are able to annihilate, 
as witnessed  by the 
occurrence of the vacuum representation $1$ in the fusion rule for two 
similar particles.

At first sight, the message of the fusion rule~(\ref{remark}) is rather 
remarkable. It seems that the fusion of two pure fluxes $\sigma_a^+$ 
may give rise to electric charge creation. 
One could start wondering about electric charge conservation at this point. 
Electric charge is conserved though. Before fusion 
this charge was present in the form of so-called nonlocalizable Cheshire 
charge~\cite{preskra,alice,spm,sm}, i.e.\ the nontrivial representation 
of the global symmetry group $\bar{D}_2$ carried by the pair.
This becomes clear upon writing the fusion rule~(\ref{remark})
in terms of the two particle flux states for the different channels
\bea
\frac{1}{\sqrt{2}}
\{ |\bar{X}_a \rangle |X_a \rangle 
\; + \; |X_a \rangle  |\bar{X}_a \rangle \}
&\longmapsto& 1 \label{corvac}   \\
\frac{1}{\sqrt{2}}
\{ |\bar{X}_a \rangle |X_a \rangle
\; - \; |X_a \rangle  |\bar{X}_a \rangle \}
&\longmapsto& J_a  \label{corstat} \\
\frac{1}{\sqrt{2}}
\{ |X_a \rangle |X_a \rangle
\; + \; |\bar{X}_a \rangle  |\bar{X}_a \rangle \}
&\longmapsto& \bar{1} \label{corfluch} \\
\frac{1}{\sqrt{2}}
\{ |X_a \rangle |X_a \rangle
\; - \; |\bar{X}_a \rangle  |\bar{X}_a \rangle \}
&\longmapsto& \bar{J}_a.
\eea
The identification of the two particle flux states with the single
particle states is established by  the action~(\ref{coalgeb}) 
of the quantum double $D(\bar{D}_2)$ on these two particle states. 
On the one hand, we can perform global $\bar{D}_2$
symmetry transformations from which we learn the 
charge carried by the flux pair.
As indicated by the 
comultiplication~(\ref{coalgeb}), these act as an overall conjugation. 
The total flux of the pair, on the other hand, is obtained
by applying the flux projection operators~(\ref{multi}).
Note that  the above quantum states describing the flux pairs are nonseparable.
The two fluxes are correlated: by measuring the flux of 
one particle of the pair we 
instantaneously fix the flux of the other.  This is 
the famous  Einstein-Podolsky-Rosen (EPR) paradox~\cite{epr}.
It is no longer possible to make a flux measurement on one particle 
without affecting the other instantaneously, just as in the notorious
experiment with two spin $1/2$ particles in the singlet state. 
The Cheshire charge carried by the flux pair depends on the symmetry 
properties of these nonseparable quantum states.
The symmetric quantum states correspond to the trivial charge $1$, whereas 
the anti-symmetric quantum states carry the nontrivial charge $J_a$.
It is clear that the charge  $J_a$ 
can not be localized on any of the fluxes  
nor anywhere else. It is a property of the pair 
and only becomes localized when the fluxes are 
brought together in a fusion process.  It is this elusive 
nature, reminiscent of the 
smile of the Cheshire cat in Alice's adventures in 
wonderland~\cite{carroll}, 
that was the motivation to call such a charge Cheshire charge.

\begin{figure}[htb] 
   \epsfxsize=6cm
\centerline{\epsffile{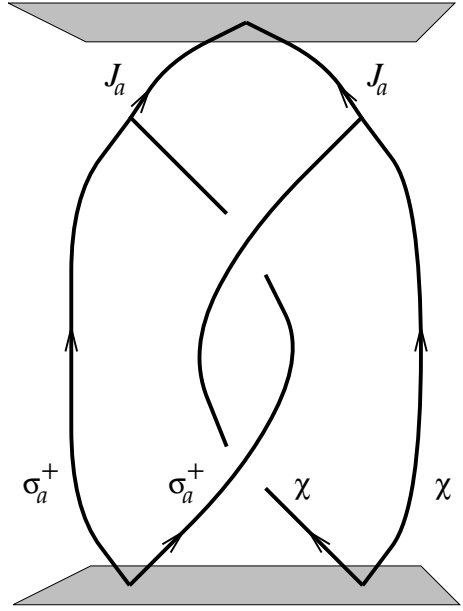}} 
\caption{\sl  A  charge/anti-charge pair $\chi$ and a 
flux/anti-flux pair $\sigma_a^+$ are created from the vacuum
at a certain time slice. 
The lines denote the worldlines of the particles.
After the charge $\chi$  has encircled the flux $\sigma_a^+$, 
both particle/anti-particle pairs carry Cheshire charge 
$J_a$. These Cheshire charges become  localized upon bringing the members
of the pairs together again. Subsequently, the two charges 
$J_a$ annihilate.}
\label{ccatd2} 
\end{figure}

The Cheshire charge $J_a$ of the flux pair can be excited 
by encircling one flux in the pair by the 
doublet charge $\chi$~\cite{preskra,sm,colem}.
Here we draw on a further analogy with 
Alice's adventures.
The magnetic fluxes 
$\{X_a,\bar{X}_a\}$ act
by means of the Pauli matrices $\pm \im \sigma_a$ on 
the doublet charge $\chi$.
This means that  when a charge $\chi$ with its orientation down 
is adiabatically transported around,  for example, the flux $X_2$, 
it returns with its orientation up
\bea                              \label{vlaflip}
{\cal R}^2 \; |X_2 \rangle | \left( \ba{c} 0 \\ 1  \ea \right) \rangle
&=&
|X_2 \rangle | \left( \ba{c} 1 \\ 0  \ea \right) \rangle ,
\eea
as follows from~(\ref{braidact}). In terms of Alice's adventures:
the charge has gone through the looking-glass. 
For this reason the flux $X_2$ is called 
an Alice flux~\cite{schwarz,alice,preskra}.  
The other fluxes $X_a,\bar{X}_a$ affect the doublet charge $\chi$ 
in a similar way.
Now consider the process depicted in 
figure~\ref{ccatd2}.   We start with the creation of a 
charge/anti-charge pair $\chi$ and a flux/anti-flux pair $\sigma_a^+$ 
from the vacuum. Thus both pairs do not carry 
Cheshire charge at this stage. They are in the vacuum channel of the 
corresponding fusion rules~(\ref{puch}) and~(\ref{remark}).
Next, one member of the charge pair encircles a flux in the flux pair.
The flip of the charge orientation~(\ref{vlaflip})  
leads to an exchange of the internal  quantum numbers of the pairs: 
both pairs carry Cheshire charge $J_a$  after this process, i.e.\
both pairs are in the $J_a$ channel of the associated fusion rules.
The global charge of the configuration is conserved. Both charges   
$J_a$  can be annihilated  by bringing them together
as follows from the fusion rules~(\ref{puch}).
These phenomena can be made explicit  by  writing this process 
in terms of the corresponding correlated  internal quantum states 
\bea   \label{uitzd2}
1 & \longmapsto & \frac{1}{2} 
\{ |\bar{X}_2 \rangle |X_2 \rangle 
\; + \; |X_2 \rangle  |\bar{X}_2 \rangle \} 
\{ | \left( \ba{c} 1 \\ 0  \ea \right) \rangle
| \left( \ba{c} 0 \\ 1  \ea \right) \rangle  \; - \;
| \left( \ba{c} 0 \\ 1  \ea \right) \rangle
| \left( \ba{c} 1 \\ 0  \ea \right) \rangle  \}  \nn \\
  & \stackrel{{\mbox{\scriptsize \bf 1}} \ot {\cal R}^2 \ot 
{\mbox{\scriptsize \bf 1}}}{\longmapsto} &
\frac{1}{2}      
\{ |\bar{X}_2 \rangle |X_2 \rangle
\; - \; |X_2 \rangle  |\bar{X}_2 \rangle \} 
\{ | \left( \ba{c} 0 \\ 1  \ea \right) \rangle
| \left( \ba{c} 0 \\ 1  \ea \right) \rangle  \; + \;
| \left( \ba{c} 1 \\ 0  \ea \right) \rangle
| \left( \ba{c} 1 \\ 0  \ea \right) \rangle  \}  \nn \\
& \longmapsto & |J_2 \rangle  |J_2 \rangle \nn \\
  & \longmapsto & 1.            
\eea 
Here we used~(\ref{braidact}) and the fact that the fluxes 
$\{X_2,\bar{X}_2\}$ act
by means of the Pauli matrices $\pm \im \sigma_2$  
on the charge $\chi$.
After the charge has encircled the flux, the flux pair is in the
anti-symmetric quantum state~(\ref{corstat}) with Cheshire charge $J_2$,
while the same observation holds for the  quantum state of  
the charge pair. Before fusion the charge pair was in the anti-symmetric
vacuum representation $1$, while the state that emerges after the monodromy 
carries the Cheshire charge $J_2$.
For convenience, we restricted ourselves to  the flux pair $\sigma_2^+$ here.
The argument for the other flux pairs is completely similar.

This discussion naturally extends to the exchange of
magnetic quantum numbers in monodromy processes involving 
noncommuting fluxes~(\ref{flumonodr}). 
If we replace the doublet charge pair
by a flux pair $\sigma_b^+$ starting off in the 
vacuum channel~(\ref{corvac}), both flux pairs end up in the 
nontrivial flux channel~(\ref{corfluch}) after the monodromy
and both pairs now carry the total flux $\bar{1}$. These fluxes 
become localized upon fusing the members of the pairs, and subsequently 
annihilate each other according to their fusion rule~(\ref{centflufu}).

We close this section by emphasizing the profound role that the fusion rules
play as overall selection rules in the flux/charge exchange processes 
among the particles. It is natural
to confine our considerations to multi-particle systems which are
in the vacuum sector $1$, 
i.e.\ the overall flux and charge of this system vanishes.
Thus the particles necessarily appear in pairs,
 as we have seen in the example of 
such a system in figure~\ref{ccatd2}.
The fusion rules classify the different 
total fluxes and Cheshire charges these
pairs can carry and determine the flux/charge exchanges that may occur
in monodromy processes involving particles in different pairs.

\subsection{Scattering doublet charges off Alice fluxes}   \label{abd2b}

The Aharonov-Bohm interactions among the particles in the 
spectrum~(\ref{spectred2}) roughly fall into two classes. 
First of all there are the interactions in which no internal 
flux/charge quantum numbers are exchanged between the particles. 
In this case, the monodromy  matrix following from~(\ref{braidact})
is diagonal in the two particle flux/charge eigenbasis, 
with possibly different 
Aharonov-Bohm phases as diagonal elements.  The cross sections 
measured in Aharonov-Bohm scattering experiments with the associated 
particles simply follow from the well-known cross section~(\ref{ababcs}) 
derived by Aharonov and Bohm~\cite{ahabo}. The more interesting
Aharonov-Bohm interactions are those in which internal 
flux/charge quantum numbers are exchanged between the particles.
In this case, the monodromy matrix is off diagonal in the
flux/charge  eigenbasis.
The cross sections appearing in Aharonov-Bohm scattering experiments
involving such particles are discussed in appendix~\ref{ahboverl}. 
In this section, we will focus on a nontrivial example, namely 
an Aharonov-Bohm experiment in which a doublet charge $\chi$ 
scatters from an Alice flux $\sigma_2^+$.

The total 
internal Hilbert space associated with the two particle system consisting 
of a pure doublet charge $\chi$ together with
a pure doublet flux $\sigma_2^+$ is four dimensional. 
We define the following natural flux/charge eigenbasis  in this internal 
Hilbert space
\bea        \label{cfbasis}     \ba{rclrcl}
e_{\uparrow \uparrow} 
&=& 
|X_2 \rangle| \left( \ba{c} 1 \\ 0  \ea \right) \rangle 
             \\
e_{ \uparrow\downarrow} 
&=& |X_2 \rangle | \left( \ba{c} 0 \\ 1  \ea \right) \rangle 
           \\
e_{\downarrow\uparrow } 
&=& |\bar{X}_2 \rangle| \left( \ba{c} 1 \\ 0  \ea \right) \rangle 
         \\
e_{\downarrow\downarrow} 
&=& |\bar{X}_2 \rangle| \left( \ba{c} 0 \\ 1  \ea \right) \rangle. 
\ea    
\eea
The fluxes ${X}_2/\bar{X}_2$ are represented by the Pauli matrices 
$\pm \im\sigma_2$ in the doublet charge representation $\chi$.
From~(\ref{braidact}), we then infer that the monodromy matrix 
takes the following block diagonal 
in this basis
\bea                                   \label{allesd2}
{\cal R}^2 &=&  \left( \ba{rrrr} 0 & 1 & &  \\ -1 & 0 &   &    \\
                                   &   & 0 & -1 \\ 
                                   &   & 1 & 0 \ea \right),
\eea  
which summarizes the phenomenon discussed in the previous section:
the orientation of the charge $\chi$ is flipped, when
it is transported around the Alice fluxes ${X}_2$ or $\bar{X}_2$.

Let us now consider the Aharonov-Bohm scattering experiment in 
which the doublet charge 
$\chi$ scatters from the Alice flux $\sigma_2^+$.
We  assume that we are measuring with  a detector that only gives a signal 
when a scattered charge  $\chi$ enters the device with a specific orientation 
(either $\uparrow$ or $\downarrow$). Here we may, for instance,
think of an apparatus in which we have captured the associated 
anti-particle. This is the charge with opposite orientation, as we 
have seen in~(\ref{uitzd2}).
If the orientation of the scattered charge entering the device matches
that of the anti-particle, the pair annihilates 
and we assume that the apparatus somehow gives a signal when such an
annihilation process occurs.
The cross section measured with such a detector 
involves the matrix elements
of  the scattering matrix 
\beas                     \label{scatmd2}
{\cal R}^{-\theta/\pi}({\mbox{\bf 1}} - {\cal R}^{2}) 
&=&     
\sqrt{2} e^{-\im \theta/2} 
\left( \ba{rrrr} \cos{\frac{\pi - \theta}{4}} & \sin{\frac{\pi - \theta}{4}} 
&   &  \\ 
                -\sin{\frac{\pi - \theta}{4}}  & \cos{\frac{\pi - \theta}{4}} 
&  &   \\
      &   & \cos{\frac{\pi - \theta}{4}} & -\sin{\frac{\pi - \theta}{4}} \\
      &   & \sin{\frac{\pi - \theta}{4}} & \cos{\frac{\pi - \theta}{4}}
        \ea \right).
\eeas
for the flux/charge eigenstates~(\ref{cfbasis}). 
This scattering matrix is determined using the prescription~(\ref{exclur})
in the monodromy eigenbasis in which the above monodromy 
matrix~(\ref{allesd2})  is diagonal,
and subsequently transforming back to the flux/charge 
eigenbasis~(\ref{cfbasis}).
Now suppose that the scatterer is 
in a particular flux eigenstate, while 
the projectile that  comes in is a charge with a specific orientation and
the detector is only sensitive for scattered charges with this 
specific orientation.                              
Under these circumstances, the two particle in and out state are the same, 
$|\, \mbox{in}\rangle = |\, \mbox{out}\rangle $, and equal to one 
of the flux/charge eigenstates in~(\ref{cfbasis}). 
In other words, we are measuring 
the scattering amplitudes on the diagonal of the 
scattering matrix~(\ref{scatmd2}). 
Note that the formal sum of the out state $|\mbox{out} \rangle$ 
over a complete basis of flux eigenstates for the scatterer, 
as indicated in appendix~\ref{ahboverl},
boils down to one term here, 
namely the flux eigenstate of the scatterer in the 
in state $|\mbox{in} \rangle$. The other flux eigenstate
does {\em not} contribute. The corresponding matrix element vanishes,
because the flux of the scatterer is not
affected when it is encircled by  the charge $\chi$.
From~(\ref{exclus}) we obtain the following  
exclusive cross section for this scattering experiment
\bea              \label{c+d2}
\frac{{\rm d} \sigma_+}{{\rm d} \theta}   &=&
\frac{1+\sin{(\theta/2)}}{8 \pi p \sin^2{(\theta/2)}}   \; .
\eea
The charge flip cross section, in turn, is measured by a detector
which only signals scattered charges with an orientation opposite 
to the orientation of the charge of the projectile.
In this case, the state $|\mbox{in} \rangle$ is again
one of the flux/charge eigenstates in~(\ref{cfbasis}), while
the $|\mbox{out} \rangle$ state we measure is the same as the 
in state, but with the orientation of the charge flipped. 
Thus we are now measuring the off diagonal matrix elements of the scattering
matrix~(\ref{scatmd2}). 
In a similar fashion as before, we find the following form for the 
charge flip cross section
\bea              \label{c-d2}
\frac{{\rm d} \sigma_-}{{\rm d} \theta} &=& 
\frac{1-\sin{(\theta/2)}}{8 \pi p \sin^2{(\theta/2)}}  \;.
\eea
The exclusive cross sections~(\ref{c+d2}) and~(\ref{c-d2}), 
which are the same as 
derived for scattering of electric charges from Alice fluxes in Alice 
electrodynamics by Lo and Preskill~\cite{preslo}, are clearly multi-valued
\bea 
\frac{{\rm d} \sigma_{\pm}}{{\rm d} \theta} \; (\theta + 2\pi) &=& 
\frac{{\rm d} \sigma_{\mp}}{{\rm d} \theta} \;  (\theta).
\eea
This merely reflects the fact that a detector only signalling  
charges $\chi$ with their orientation up, becomes a detector only
signalling charges with orientation down (and vice versa), when
it is transported over an angle $2\pi$ around the scatterer.
Specifically, in this parallel transport the anti-particle in our 
detector feels the holonomy in the gauge fields associated with
the flux of the scatterer and returns with its orientation flipped.
As a consequence, the device becomes sensitive for the opposite
charge orientation after this parallel transport.

Verlinde's detector does not suffer from this multi-valuedness.
It does not discriminate  between the orientations 
of the scattered charge, and gives a signal whenever a charge $\chi$ 
enters the device. This detector measures the total or 
inclusive cross section, i.e.\ both branches of the multi-valued 
cross section~(\ref{c+d2}) (or~(\ref{c-d2}) for that matter). 
To be specific, the exclusive cross sections~(\ref{c+d2})
and~(\ref{c-d2})
combine in the following fashion 
\bea
\frac{{\rm d} \sigma}{{\rm d}\theta} \; = \;
\frac{{\rm d}\sigma_-}{{\rm d}\theta}  + 
\frac{{\rm d}\sigma_+}{{\rm d}\theta} \; = \; 
 \frac{1}{4 \pi p \sin^2{(\theta/2)}} \; .
\eea
into Verlinde's single valued inclusive cross section~(\ref{Aharonov}) 
for this scattering experiment.

The above analysis is easily extended to Aharonov-Bohm scattering
experiments involving other particles in the spectrum~(\ref{spectred2}) 
of this $\bar{D}_2$ gauge theory. It should be stressed however, that 
a crucial ingredient in the derivation of the {\em multi-valued} exclusive
cross sections~(\ref{c+d2}) and~(\ref{c-d2}) 
is that the monodromy matrix~(\ref{allesd2}) is off diagonal
and has imaginary eigenvalues  $\pm \im$. In the other cases, where
the monodromy matrices are diagonal or off diagonal 
with eigenvalues $\pm 1$, as it appears for scattering noncommuting
fluxes $\sigma_a^+$ and $\sigma_b^+$ from each other,
we arrive at {\em single valued} exclusive cross sections.

\subsection{Nonabelian braid statistics}  \label{d2bqst}

We finally turn to the issue of nonabelian braid statistics. As we have argued 
in section~\ref{trunckbr}, the braidings and monodromies 
for multi-particle configurations appearing in discrete
${ H}$ gauge theories are governed by truncated braid groups. 
To be precise, the total internal Hilbert space for a given multi-particle
system carries a representation of some truncated braid group,
which in general decomposes into a direct sum of irreducible representations.
In this section, we identify the truncated braid groups ruling in this 
particular $\bar{D}_2$ gauge theory and elaborate on the aforementioned
decomposition. We first consider the indistinguishable particle configurations 
in this model.

It can easily be verified that the  braid operators  acting on a 
configuration, which  only  contains  singlet charges $J_a$,
are of order one. 
The same holds for the singlet dyons $\bar{1}$ and $\bar{J}_a$.
In other words, 
these particles behave as ordinary bosons, in accordance with the trivial
spin factors~(\ref{spind2}) assigned to them.
To proceed, the braid operators acting on a system of $n$
doublet charges $\chi$ are of order two and therefore realize 
a (higher dimensional) representation of the permutation group $S_n$. 
The same observation appears for the doublet dyons $\bar{\chi}$ 
and $\sigma^{\pm}_a$. The total internal Hilbert spaces for these 
indistinguishable particle systems can then be decomposed into a direct sum 
of  subspaces, each carrying an irreducible representation
of the permutation group. The one dimensional representations that  
appear in this decomposition correspond to Bose or Fermi statistics, while
the higher dimensional representations describe parastatistics.
Finally, braid statistics occurs for a system consisting of $n$ 
dyons $\tau_a^{\pm}$. The braid operators that act on such a system are 
of order four, thus the associated internal Hilbert space splits up 
into a direct sum of 
irreducible representations  of the truncated braid group $B(n,4)$. 
The one dimensional representations
that occur in this decomposition realize abelian anyon statistics, 
whereas the higher dimensional representations  correspond to nonabelian 
braid statistics or nonabelian anyons. 
We  will illustrate these features with two 
representative examples. We first examine a system 
containing two dyons $\tau_1^{+}$. The irreducible 
braid group representations available for this system are one dimensional,
since the truncated braid group  $B(2,4)$ for two particles is abelian. 
We then turn to the more interesting system consisting of 
three dyons $\tau_1^{+}$. In this case, we are dealing with nonabelian braid
statistics. The associated total internal Hilbert space
breaks up into four 1-dimensional irreducible subspaces and two 
2-dimensional irreducible subspaces 
under the action of the nonabelian truncated braid group
$B(3,4)$.

We start by setting some conventions. First of all, the 
two fluxes in the conjugacy class associated with the 
dyon $\tau_1^{+}$ are ordered as indicated in table~\ref{tabcond2}
\beas
^1 h_1 &=& X_1   \\
^1 h_2 &=& \bar{X}_1,
\eeas 
while we take the following coset representatives appearing 
in the definition~(\ref{13zo}) of the centralizer charge
\beas
^1 x_1 &=& e   \\
^1 x_2 &=& X_2.
\eeas
To lighten the  notation a bit,  we furthermore 
use the following  abbreviation for  the internal flux/charge 
eigenstates of the dyon $\tau_1^{+}$
\beas
|\uparrow \,\rangle &:=& |X_1, ^{1} v\rangle \\
|\downarrow \,\rangle &:=& |\bar{X}_1, ^{1} v\rangle.
\eeas 

Let us now consider a system consisting of two dyons $\tau^{+}_1$.
Under the action of the quantum  double  $D(\bar{D}_2)$,
the internal Hilbert space $V_{\tau^{+}_1} \ot V_{\tau^{+}_1}$ 
associated with this system decomposes according to the fusion 
rule~(\ref{spreek}), which we repeat for convenience
\bea                     \label{repeat}
\tau^{+}_1 \times \tau^{+}_1 &=& 1+J_1+\bar{J}_2+\bar{J}_3.
\eea                                   
The two particle states corresponding to the different fusion channels
carry an one dimensional (irreducible) 
representation of the abelian truncated braid group
$B(2,4)=\Z_4$. We first establish the different irreducible pieces
contained in the $B(2,4)$ representation carried by the total internal 
Hilbert space $V_{\tau^{+}_1} \ot V_{\tau^{+}_1}$.
This can be done by calculating the traces of the elements 
$\{e, \tau, \tau^2, \tau^3 \}$ of $B(2,4)$ in this representation
using the standard diagrammatic techniques (see for 
instance~\cite{kaufman,adams}).
From the character vector obtained in this way, we learn that this 
representation breaks up as
\bea                    \label{brad24}
\Lambda_{B(2,4)} &=& 3\;  \Gamma^1 +   \Gamma^3,
\eea 
with $\Gamma^1$ and $\Gamma^3$ the irreducible $\Z_4$ representations 
displayed in the character table~\ref{tabcond2}. 
After some algebra,  we then arrive at the following  basis of 
mutual eigenstates under the combined action of the quantum double 
and the truncated braid group 
\begin{eqnfourarray} 
 V_{\tau^{+}_1} \ot V_{\tau^{+}_1}  \qquad 
& \qquad &  D(\bar{D}_2)   & $\qquad B(2,4) $   \nn \\
\frac{1}{\sqrt{2}} \{ 
|\uparrow \, \rangle |\downarrow \, \rangle-|\downarrow \, \rangle
|\uparrow \, \rangle \} 
&\qquad & \;\; 1 & $\qquad  \Gamma^{1} \;\;\; \,$  \label{vac11} \\
\frac{1}{\sqrt{2}} \{ 
|\uparrow \, \rangle |\downarrow \, \rangle + |\downarrow \, \rangle
|\uparrow \,  \rangle \} 
&\qquad & \;\;  J_1& $\qquad  \Gamma^{3} \;\;\;\,$  \label{j13}   \\
\frac{1}{\sqrt{2}} \{ 
|\uparrow \, \rangle |\uparrow \, \rangle+|\downarrow \, \rangle
|\downarrow \, \rangle \} 
&\qquad & \;\;  \bar{J}_2 & $\qquad   \Gamma^{1} \;\;\;\, $  \label{j21}\\
\frac{1}{\sqrt{2}} \{ 
|\uparrow \, \rangle |\uparrow \, \rangle- |\downarrow \,\rangle
|\downarrow \, \rangle  \} 
&\qquad & \;\; \bar{J}_3 & $\qquad   \Gamma^{1} , \;\;\; $   \label{j31}
\end{eqnfourarray}   
from which we conclude that 
the two particle internal Hilbert space 
$V_{\tau^{+}_1} \ot V_{\tau^{+}_1}$ decomposes
into the following  direct sum of one dimensional 
irreducible representations of the direct product  
$D(\bar{D}_2) \times B(2,4)$
\bea
(1,\Gamma^{1}) \: + \: (J_1,\Gamma^{3}) \: + \:
(\bar{J}_2,\Gamma^{1}) \: + \: (\bar{J}_3, \Gamma^{1}).
\eea
The two particle states contained in~(\ref{j21}) and~(\ref{j31})
satisfy the canonical spin-statistics connection~(\ref{spist}), that is,   
$\exp (\im \Theta) = \exp(2\pi \im s_{\tau^{+}_1})= \im$.
In other words, these states realize semion statistics.
Accidentally, the same observation appears for the state~(\ref{vac11}). 
Finally, the two particle state displayed in~(\ref{j13})
satisfies the generalized spin-statistics connection~(\ref{gespietst}) and 
describes semion statistics with quantum statistical parameter
$\exp (\im \Theta)=-\im$.

We now extend our analysis to a system containing 
three dyons $\tau^{+}_1$. From~(\ref{repeat}) and 
the fusion rules~(\ref{zendy}), (\ref{ooknog}) 
and (\ref{centflufu}), we infer that the decomposition of 
the total internal Hilbert space under the action of the quantum double 
becomes
\bea                  \label{dd2bardeco}
\tau^{+}_1 \times \tau^{+}_1 \times \tau^{+}_1  &=& 4 \; \tau^{+}_1.
\eea
The occurrence of four equivalent fusion channels indicates that 
nonabelian braid statistics is conceivable and it turns out that  
higher dimensional irreducible representations of the 
truncated braid group $B(3,4)$ indeed appear. The
structure of this group and its irreducible representations 
are discussed in appendix~\ref{trubra}.  A lengthy but straightforward 
diagrammatic calculation of the character vector associated with
the $B(3,4)$ representation carried by the three particle internal 
Hilbert space $V_{\tau^{+}_1} \ot V_{\tau^{+}_1} \ot V_{\tau^{+}_1}$
reveals the following irreducible pieces
\bea    \label{b34dec}
\Lambda_{B(3,4)}  &=&  4 \; \Lambda_1 + 2 \; \Lambda_5,
\eea 
with  $\Lambda_1$ and $\Lambda_5$ the irreducible representations of 
$B(3,4)$ exhibited in the character table~\ref{tab:tab4}.
The one dimensional representation $\Lambda_1$ describes 
abelian semion statistics,
while the two dimensional representation $\Lambda_5$ corresponds to 
nonabelian braid statistics.
From~(\ref{dd2bardeco}) and~(\ref{b34dec}), we can immediately conclude that 
this three particle internal Hilbert space breaks up into  
the following direct sum of irreducible subspaces under the action of the  
direct product $D(\bar{D}_2) \times B(3,4)$  
\bea
2 \; (\tau^{+}_1, \Lambda_1) \: + \: ( \tau^{+}_1, \Lambda_5 ),
\eea 
where $(\tau^{+}_1, \Lambda_1)$ labels a two dimensional  
and $( \tau^{+}_1, \Lambda_5 )$ a four dimensional representation.
A basis adapted to this decomposition can be cast in the following form
\begin{eqnfourarray}
 V_{\tau^{+}_1} \ot V_{\tau^{+}_1} \ot V_{\tau^{+}_1} \qquad \qquad  \qquad
& \qquad &  D(\bar{D}_2)   & $\qquad B(3,4) $   \nn \\
|\downarrow  \, \rangle |\downarrow  \, \rangle |\downarrow  \, \rangle 
& \qquad &
\; |\uparrow \, \rangle_1 & $\Lambda_1 \;\;\;\, $  \\
|\uparrow  \, \rangle|\uparrow  \, \rangle|\uparrow  \, \rangle 
& \qquad & 
\; |\downarrow \, \rangle_1 & $\Lambda_1 \;\;\;\,$  \\   
\frac{1}{\sqrt{3}}\{ |\uparrow  \, \rangle|\uparrow  \, \rangle
                     |\downarrow  \, \rangle
                    -|\uparrow  \, \rangle |\downarrow  \, \rangle
                     |\uparrow  \, \rangle
                    +|\downarrow  \, \rangle |\uparrow  \, \rangle
                     |\uparrow  \, \rangle \} 
& \qquad & 
\; |\uparrow \, \rangle_2 & $\Lambda_1 \;\;\;\,$ \\
\frac{1}{\sqrt{3}}\{ |\downarrow  \, \rangle|\downarrow  \, \rangle 
                     |\uparrow  \, \rangle
                    -|\downarrow  \, \rangle|\uparrow  \, \rangle
                     |\downarrow  \, \rangle
                    +|\uparrow  \, \rangle|\downarrow  \, \rangle
                     |\downarrow  \, \rangle\} 
&\qquad& 
\; |\downarrow \, \rangle_2 & $\Lambda_1 \;\;\;\,$ \\
\frac{1}{2} \{2|\uparrow  \, \rangle|\uparrow  \, \rangle
               |\downarrow  \, \rangle
              +|\uparrow  \, \rangle |\downarrow  \, \rangle
               |\uparrow  \, \rangle
              -|\downarrow  \, \rangle |\uparrow  \, \rangle
               |\uparrow  \, \rangle \} 
& \qquad &
\; |\uparrow \, \rangle_3 & $\Lambda_5 \;\;\;\,$     \label{la35} \\
\frac{1}{2}\{2|\downarrow  \, \rangle|\downarrow  \, \rangle 
              |\uparrow  \, \rangle
             +|\downarrow  \, \rangle|\uparrow  \, \rangle
              |\downarrow  \, \rangle
             -|\uparrow  \, \rangle|\downarrow  \, \rangle
              |\downarrow  \, \rangle\} 
& \qquad &
\; |\downarrow \, \rangle_3 & $\Lambda_5 ' \;\;\;\,$ \label{la35p}  \\
\frac{1}{\sqrt{2}} \{|\uparrow  \, \rangle |\downarrow  \, \rangle
                     |\uparrow  \, \rangle
                    +|\downarrow  \, \rangle |\uparrow  \, \rangle
                     |\uparrow  \, \rangle \} 
& \qquad &
\; |\uparrow \, \rangle_4 & $\Lambda_5 \;\;\;\, $     \label{la45}\\
\frac{1}{\sqrt{2}}\{ |\downarrow  \, \rangle|\uparrow  \, \rangle
                     |\downarrow  \, \rangle
                    +|\uparrow  \, \rangle|\downarrow  \, \rangle
                     |\downarrow  \, \rangle\}
& \qquad &
\; |\downarrow \, \rangle_4 & $\Lambda_5 ' ,\;\;\; $     \label{la45p}
\end{eqnfourarray}   
The subscript attached to the single particle states in the second column 
label the four fusion channels showing up in~(\ref{dd2bardeco}). 
In other words, these states summarize the global properties of 
the three particle states in the first column, that is, the total flux 
and charge, which are conserved under braiding. 
Each of the three particle states in the first four rows  
carry the one dimensional representation $\Lambda_1$  
of the truncated braid group $B(3,4)$.
The particles in these states obey semion statistics
with quantum statistical parameter $\exp (\im \Theta)= \im$, and 
satisfy the canonical spin-statistics connection.
Finally, the states in the last four rows constitute a basis for the 
representation  $( \tau^{+}_1, \Lambda_5 )$. To be specific,
the states~(\ref{la35}) and~(\ref{la45}),
carrying the same total flux and charge, form a basis for 
a two dimensional irreducible representation $\Lambda_5$ of the 
truncated braid group. The same remark holds for the states~(\ref{la35p}) 
and~(\ref{la45p}).
For convenience, we have distinguished these 
two irreducible representations by a prime. 
Note that we have chosen a basis which 
diagonalizes the braid operator ${\cal R}_1$  acting on the   
first two particles with eigenvalues either $\im$ or $-\im$, 
whereas the braid operator ${\cal R}_2$ for the last 
two particles mixes the states in the different fusion channels.
Of course, this choice is quite arbitrary.
By another basis choice, we could have reversed this situation.  

Let us also comment briefly on the distinguishable particle systems
that can occur in this theory.
The maximal order of the monodromy operator for distinguishable particles
in this model is four. Thus the distinguishable particle systems 
in this theory
are governed by the truncated colored braid 
groups $P(n,8)$ and their subgroups. 
A system consisting of the three different particles 
$\sigma_1^+$, $\sigma_2^+$ and $\tau_3^{+}$, for instance, realizes a 
representation of the colored braid group $P(3,4) \subset P(3,8)$.
(The group structure of $P(3,4)$ and a classification of its irreducible 
representations are given in appendix~\ref{trubra}).  
The internal Hilbert space for this system breaks up into the following 
two 4-dimensional irreducible representations of  
$D(\bar{D}_2) \times P(3,4)$
\beas
(\chi, \Omega_8) \: + \: (\bar{\chi}, \Omega_9).
\eeas  
This result summarizes 
\beas
\sigma_1^+ \times \sigma_2^+ \times \tau_3^{+} &=& 2 \; \chi + 
2 \;\bar{\chi} \\
\Lambda_{P(3,4)} &=& 2 \; \Omega_8 + 2 \; \Omega_9,
\eeas
with $\Omega_8$ and $\Omega_9$ the two dimensional irreducible 
representations displayed in the character table~\ref{char}. 
The conclusion is that this system obeys nonabelian `monodromy statistics',
that is, the three monodromy operators displayed in~(\ref{p34mongen}) 
can not be diagonalized simultaneously.

As a last blow, we return to the process described by~(\ref{uitzd2}).
After the double pair creation, we are dealing with a four particle system
consisting of a subsystem of two indistinguishable particles $\sigma_2^+$ 
and a subsystem of two indistinguishable particles $\chi$. 
Initially, the two particle state for the fluxes $\sigma_2^+$ is bosonic,
whereas  the two particle state for the charges $\chi$ is fermionic.
After the monodromy has taken place, the situation is reversed. 
The two particle state for the fluxes $\sigma_2^+$ has become fermionic 
and the 
two particle state for the charges $\chi$ bosonic. In other words,
the exchange of Cheshire charge is accompanied by an exchange of quantum 
statistics~\cite{brekke}. 
The total  four particle system now realizes a two dimensional
irreducible representation of the associated 
truncated partially colored braid group. 
The two braid operators ${\cal R}_1$ and ${\cal R}_3$ 
for the particle exchanges in the two subsystems 
act diagonally with eigenvalues $\pm 1$ and $\mp 1$ respectively.
Furthermore, under the repeated action of the monodromy 
operator ${\cal R}_2^2$,
the subsystems simultaneously jump back and forth between the 
fusion channels $1$ and $J_2$ with their associated 
Cheshire charge and quantum statistics.

\aanhangsel

\sectiona{Aharonov-Bohm scattering}  \label{ahboverl}

The only experiments in which 
the particles in a discrete ${ H}$ gauge theory 
leave `long range fingerprints' 
are of a quantum mechanical nature, namely quantum interference experiments, 
such as the double slit experiment~\cite{colem,preslo}
and the Aharonov-Bohm scattering experiment~\cite{ahabo}.
What we are measuring in these experiments is the way the 
particles affect their mutual internal flux/charge quantum numbers when 
they encircle each other. In other words, we are probing the content 
of the monodromy matrix ${\cal R}^2$ following from~(\ref{braidact}).   
In this appendix, we will give a concise discussion of 
two particle Aharonov-Bohm scattering and provide the details entering
the calculation of the cross sections in section~\ref{abd2b}.
For a recent review of the experimental status of the Aharonov-Bohm effect,
the reader is referred to~\cite{peshkin}.

\begin{figure}[tbh]    \epsfxsize=14cm
\centerline{\epsffile{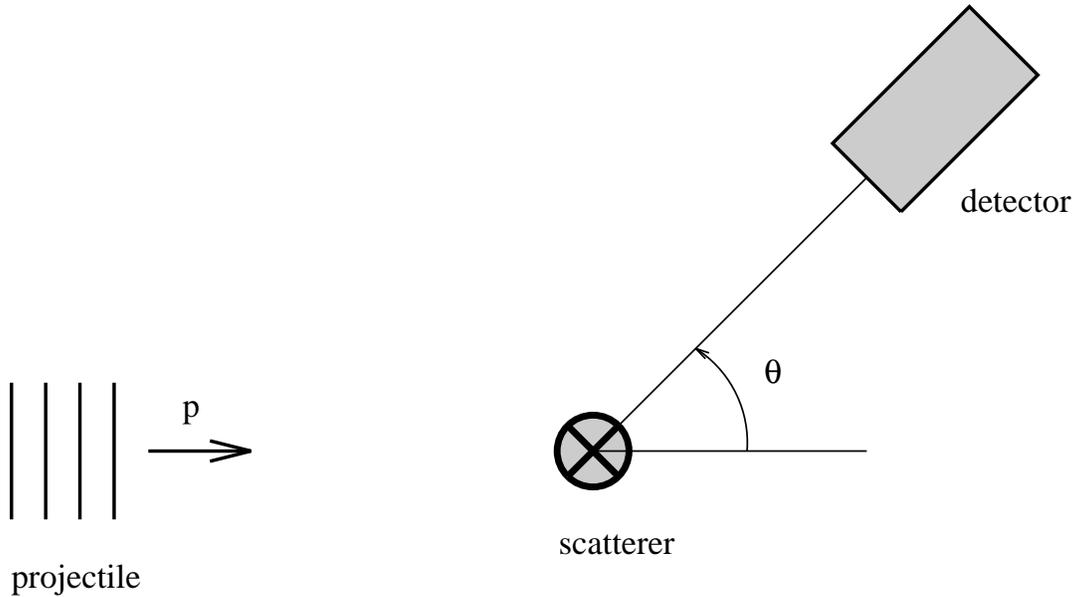}}
\caption{\sl The  geometry of the Aharonov-Bohm scattering experiment. 
The projectile comes 
in as a plane wave with momentum $p$ and scatters elastically 
from a scatterer fixed at the origin. It is assumed that the 
projectile never enters the region where
the scatterer is located. 
The cross section for the scattered projectile
is measured by a detector placed at the scattering angle $\theta$.}
\label{abverpres}
\end{figure}

The geometry of the Aharonov-Bohm scattering experiment is depicted 
in figure~\ref{abverpres}. 
It involves two particles, a projectile and a scatterer fixed at the origin.
The incoming external part of the total wave function is a plane wave 
for the projectile vanishing at the location of the scatterer.
Nontrivial scattering takes place if the 
monodromy matrix ${\cal R}^2$ acting on the internal part of the 
wave function is nontrivial. 

In the abelian discrete gauge theory discussed in section~\ref{abznz},
we only encountered the abelian version,
that is,  the effect of a monodromy of the two particles 
in the internal wave function is just a phase 
\bea
{\cal R}^2 &=& e^{ 2 \pi \im \alpha}.
\eea 
The differential cross section for the quantum mechanical 
scattering experiment  involving such particles 
has been derived by Aharonov and Bohm~\cite{ahabo} 
\bea                        \label{ababcs}
\frac{{\rm d} \sigma}{{\rm d} \theta} &=&
\frac{\sin^2 {(\pi \alpha})}{2\pi p \sin^2 (\theta/2)}    \; ,
\eea
with $\theta$ the scattering angle and  
$p$ the momentum of the incoming plane wave of the projectile.

The particles appearing in a  nonabelian discrete 
gauge theory, can exchange 
internal flux/charge quantum numbers when they encircle each other.
This effect is described by 
nondiagonal monodromy {\em matrices} ${\cal R}^2$ acting on multi-component 
internal wave functions. The cross section measured in 
Aharonov-Bohm scattering experiment involving these particles
is a  `nonabelian' generalization of 
the abelian one given in~(\ref{ababcs}).
An elegant closed formula for these nonabelian cross sections has been
derived by Erik Verlinde~\cite{ver1}.
The crucial insight was that the monodromy matrix ${\cal R}^2$ for 
two particles can always be diagonalized, since 
the braid group for two particles is abelian.
In the monodromy eigenbasis in which the monodromy matrix ${\cal R}^2$ 
is diagonal, the nonabelian problem then
reduces to the abelian one solved by Aharonov and Bohm.
The solution can subsequently be cast in the  basis independent form
\bea \label{Aharonov}
\frac{{\rm d} \sigma}{{\rm d} \theta}|_{\mbox{in} \rightarrow \mbox{all}} 
&=&
\frac{1}{4\pi p \sin^2(\theta/2)} \;\; 
[1-\mbox{Re} \langle \mbox{in}|{\cal R}^2|\mbox{in}\rangle],
\eea 
with $|\mbox{in} \rangle$ the normalized two particle 
incoming internal quantum state.   Note that this cross section 
boils down to~(\ref{ababcs}) for the abelian case.
We will always work in the natural two particle 
flux/charge eigenbasis being the tensor product of the single particle 
internal basis states~(\ref{quantum states}).
In fact,  in our applications the $|\mbox{in} \rangle$ state 
usually is a particular two particle flux/charge eigenstate. 
The detector  measuring the cross section~(\ref{Aharonov}) 
is a device which does not 
discriminate between the different internal `disguises' the scattered 
projectile can take. 
Specifically, in the scattering process 
discussed in section~\ref{abd2b}, the Verlinde
detector gives a signal, when the scattered pure doublet charge $\chi$ enters
the apparatus with its charge orientation either up or down.
In this sense, Verlinde's cross section~(\ref{Aharonov}) is inclusive.  

Inspired by this work, 
Lo and Preskill subsequently introduced a finer detector~\cite{preslo}.
Their device is able to distinguish between the different internal 
appearances of the projectile. 
In the scattering process studied in section~\ref{abd2b}, for example,
we can use a device, which  only gives a signal 
if the projectile enters the device with its internal charge orientation up.
The  exclusive cross section measured with such a detector can be expressed
as
\bea \label{exclus}
\frac{{\rm d} \sigma}{{\rm d} \theta}|_{\mbox{in} \rightarrow \mbox{out}} 
&=&  
\frac{1}{8 \pi p \sin^2 (\theta/2)} \; \; 
|\langle \mbox{out}|{\cal R}^{-\theta/\pi} 
({\mbox{\bf 1}} - {\cal R}^2) | \mbox{in} \rangle |^2,
\eea
where $|\mbox{in} \rangle$ and $|\mbox{out} \rangle$ denote
normalized two particle incoming-  and  outgoing internal quantum states.
The outgoing  state we observe depends on  
the detector we have installed, but since we only measure the projectile, 
so `half' of the out state,  
the state $| \mbox{out} \rangle$ in~(\ref{exclus})
should always be summed over a complete basis for the internal Hilbert space
of the scatterer.  
The new ingredient in the exclusive cross section~(\ref{exclus}) 
is the matrix ${\cal R}^{-\theta/\pi}$. This matrix is defined 
as the diagonal matrix in the monodromy eigenbasis, which acts as
\bea  \label{exclur}
{\cal R}^{-\theta/\pi} &:=&  e^{-\im \alpha\theta}   \qquad\qquad
\mbox{with $\alpha \in [0,1)$,}
\eea
on a monodromy eigenstate characterized by the
eigenvalue $\exp (2 \pi \im \alpha)$ under ${\cal R}^2$.
By a basis transformation, we then find the matrix elements of 
${\cal R}^{-\theta/\pi}$ in our favourite 
two particle flux/charge eigenbasis.

A peculiar property of the exclusive cross section~(\ref{exclus}) 
is that it is in general multi-valued. This is just 
a reflection of the fact that the detector can  generally 
change its nature, when it is parallel transported around
the scatterer. An apparatus that only detects projectiles with 
internal charge orientation up
in the scattering process studied in section~\ref{abd2b}, for example, 
becomes an apparatus,
which only detects projectiles with charge orientation down, after a rotation
over an angle of $ 2 \pi$ around the scatterer. 
Verlinde's detector, giving a signal independent of the internal 
`disguise' of the projectile entering the device, 
obviously does {\em not} suffer from this multi-valuedness. 
As a matter of fact, extending the aforementioned 
sum of  the $|\mbox{out} \rangle$ state  in~(\ref{exclus})
over a complete basis of the internal Hilbert space 
for the scatterer by a sum 
over a complete basis of the internal Hilbert space for the projectile
and subsequently using the partition of unity,  yields
the single valued inclusive cross section~(\ref{Aharonov}).

As a last remark, the cross sections 
for Aharonov-Bohm scattering experiments in which
the projectile and the scatterer are indistinguishable
particles contain an extra contribution 
due to conceivable exchange processes between the scatterer 
and the projectile~\cite{preslo,anyonbook}. The 
incorporation of this exchange contribution
amounts to diagonalizing the braid matrix ${\cal R}$ 
instead of the monodromy matrix ${\cal R}^2$.

\sectiona{$B(3,4)$ and $P(3,4)$}         \label{trubra}

In this appendix, we give the structure of the truncated braid group 
$B(3,4)$ and the truncated colored braid group $P(3,4)$, which enter
the discussion of the nonabelian braid properties of certain three particle 
configurations in a $\bar{D}_2$ gauge theory in section~\ref{d2bqst}.

According to the general definition~(\ref{eqy})-(\ref{truncate}),
the truncated braid group $B(3,4)$ for three indistinguishable particles 
is generated
by two elements $\tau_1$ and $\tau_2$ subject to the relations
\beas
\tau_1 \tau_2 \tau_1 &=& \tau_2   \tau_1 \tau_2   \\
\tau_1^4 &=& \tau_2^4 \; \, = \; \, e.
\eeas
By  explicit construction from these defining relations, which is a 
lengthy and not at all trivial job, it can be inferred
that $B(3,4)$ is a  group of order 96, which splits up into the 
following conjugacy classes 
\bea 
C_0^1 & = & \{ e \}    \\
C_0^2 & = & \{\tau_1\tau_2\tau_1\tau_2\tau_1\tau_2 \} \nn \\
C_0^3 & = & \{\tau_2^2\tau_1^2\tau_2^2\tau_1^2\} \nn \\
C_0^4 & = & \{\tau_2^2\tau_1^3\tau_2^2\tau_1^3\} \nn \\
C_1^1 & = & \{\tau_1\; , \; \tau_2\; , \; \tau_2\tau_1\tau_2^3\; , \; 
\tau_2^2\tau_1\tau_2^2\; , \; \tau_2^3\tau_1\tau_2\; , \; 
\tau_1^2\tau_2\tau_1^2\} \nn \\
C_1^2 & = & \{\tau_1^3\tau_2\tau_1^2\tau_2\; , \; 
\tau_2^3\tau_1\tau_2^2\tau_1\; , \; 
\tau_2\tau_1^3\tau_2\tau_1^2\; , \; 
\tau_2^2\tau_1^2\tau_2^2\tau_1\; , \; 
\tau_1\tau_2^3\tau_1\tau_2^2\; , \; 
\tau_1^2\tau_2^2\tau_1^2\tau_2\} \nn \\
C_1^3 & = & \{\tau_2\tau_1^3\tau_2\tau_1^3\tau_2\; , \; 
\tau_1^2\tau_2\tau_1^3\tau_2^2\tau_1\; , \; 
\tau_2^3\tau_1\tau_2^3\tau_1^2\; , \; 
\tau_1\tau_2^2\tau_1^3\tau_2^2\tau_1\; , \; 
\tau_2\tau_1\tau_2^3\tau_1^2\tau_2^2\; , \; 
\tau_2\tau_1^2\tau_2^3\tau_1^2\tau_2\} \nn \\
C_1^4 & = & \{\tau_2^2\tau_1^3\tau_2^2\; , \; 
\tau_1^2\tau_2^3\tau_1^2\; , \; 
\tau_2^3\tau_1^3\tau_2\; , \; 
\tau_1^3\; , \; \tau_2\tau_1^3\tau_2^3\; , \; 
\tau_2^3\}  \nn \\
C_2^1 & = & \{\tau_1\tau_2\; , \; \tau_2\tau_1\; , \; 
\tau_1^2\tau_2\tau_1^3\; , \; 
\tau_1^3\tau_2\tau_1^2\; , \; 
\tau_2\tau_1^2\tau_2^2\tau_1\; , \; 
\tau_2^2\tau_1\tau_2^3\; , \; 
\tau_2^3\tau_1\tau_2^2\; , \; 
\tau_1\tau_2^2\tau_1^2\tau_2\} \nn \\
C_2^2 & = & \{\tau_1^2\tau_2\tau_1^3\tau_2\tau_1\; , \; 
\tau_1\tau_2\tau_1^3\tau_2\tau_1^2\; , \; 
\tau_2\tau_1^2\tau_2^2\tau_1^3\; , \; 
\tau_1\tau_2\tau_1^2\tau_2^2\tau_1^2\; , \; \nn \\
      &   & \; \tau_2\tau_1^3\tau_2^2\tau_1^2\; , \; 
\tau_1\tau_2\tau_1^3\tau_2^2\tau_1\; , 
\; \tau_1^2\tau_2\tau_1^3\tau_2^2\; , 
\; \tau_1^2\tau_2^2\tau_1^3\tau_2\} \nn \\
C_2^3 & = & \{\tau_1^3\tau_2\tau_1^3\tau_2^2\tau_1\; , \; 
\tau_1\tau_2^2\tau_1^3\tau_2\tau_1^3\; , \; 
\tau_2\tau_1^3\tau_2^2\; , \; \tau_2^2\tau_1^3\tau_2\; , \; 
 \tau_2^3\tau_1^3\; , \; 
\tau_1\tau_2^3\tau_1^2\; , \; 
\tau_1^2\tau_2^3\tau_1\; , \; 
\tau_1^3\tau_2^3\}  \nn  \\
C_2^4 & = & \{\tau_1^3\tau_2^3\tau_1^2\; , \; 
\tau_1^2\tau_2^3\tau_1^3\; , \; 
\tau_2^3\tau_1\; , \; \tau_1\tau_2^3\; , \; 
\tau_1\tau_2\tau_1\tau_2\; , \; \tau_1^3\tau_2\; , \; 
\tau_2\tau_1^3\; , \; \tau_2\tau_1\tau_2\tau_1\} \nn \\
C_3^1 & = & \{\tau_1^2\; , \; \tau_2^2\; , \; 
\tau_1\tau_2^2\tau_1^3\; , \; 
\tau_2^2\tau_1^2\tau_2^2\; , \; 
\tau_1^2\tau_2^2\tau_1^2\; , \; 
\tau_1^3\tau_2^2\tau_1\} \nn \\
C_3^2 & = & \{\tau_2\tau_1^2\tau_2\; , \; 
\tau_1\tau_2^2\tau_1\; , \; \tau_1^2\tau_2^2\; , \; 
\tau_2^3\tau_1^2\tau_2^3\; , \; 
\tau_1^3\tau_2^2\tau_1^3\; , \; 
\tau_2^2\tau_1^2\} \nn \\
C_4^1 & = & \{\tau_1\tau_2\tau_1\; , \; \tau_1^2\tau_2\; , \; 
\tau_2^2\tau_1\; , \; \tau_2\tau_1^2\; , \; 
\tau_1\tau_2^2\; , \; \tau_1^3\tau_2\tau_1^3\; , \; 
\tau_1^3\tau_2\tau_1^3\tau_2^2\tau_1^2\; , \; \nn \\
      &   & \; \tau_2\tau_1^3\tau_2^2\tau_1\; , \; 
\tau_1^2\tau_2^2\tau_1^3\; , \; 
\tau_1\tau_2^2\tau_1^3\tau_2\; , \; 
\tau_1^3\tau_2^2\tau_1^2\; , \; 
\tau_1\tau_2\tau_1^3\tau_2^2\} \nn \\
C_4^2 & = & \{\tau_1\tau_2\tau_1\tau_2\tau_1\tau_2\tau_1\tau_2\tau_1\; , \; 
\tau_2\tau_1^2\tau_2^2\; , \; \tau_1\tau_2^2\tau_1^2\; , \; 
\tau_2^2\tau_1^2\tau_2\; , \; \tau_1^2\tau_2^2\tau_1\; , \; 
\tau_2\tau_1^3\tau_2\; , \; \nn \\
      &   & \; \tau_2^3\tau_1^3\tau_2^3\; , \; 
\tau_2^3\tau_1^2\; , \; \tau_1^3\tau_2^2\; , \; 
\tau_2^2\tau_1^3\; , \; \tau_1^2\tau_2^3\; , \; 
\tau_1\tau_2^3\tau_1\}.           \nn
\label{B34}
\eea
We organized the conjugacy classes such 
that $C_k^{i+1} = z C_k^i$, with 
$z = \tau_1\tau_2\tau_1\tau_2\tau_1\tau_2$ the 
generator of the centre of $B(3,4)$. 
The character table of the truncated braid group
$B(3,4)$ is displayed in table~\ref{tab:tab4}.

The truncated colored braid group $P(3,4)$, which  contains the monodromy  
operations on a configuration of three distinguishable particles,
is the subgroup of $B(3,4)$ generated by 
\bea
\gamma_{12} &=& \tau_1^2  \nn  \\
\gamma_{13} &=& \tau_1 \tau_2^2 \tau_1^{-1}= \tau_1 \tau_2^2\tau_1^3 
\label{p34mongen} \\
\gamma_{23} &=& \tau_2^2,    \nn 
\eea
which satisfy
\beas
\gamma_{12}^2 \; = \; \gamma_{13}^2 \;= \; \gamma_{23}^2 \;=\; e.
\eeas
It can be verified  that $P(3,4)$ is a group
of order 16 splitting up in the following 10 conjugacy classes
\bea          \label{p34}
\ba{lcl lcl}
C_0 &=& \{e \}  &   \qquad  C_1 &=& 
\{\tau_1 \tau_2 \tau_1 \tau_2 \tau_1 \tau_2\}  \\
C_2 &=& \{ \tau_2^2\tau_1^2\tau_2^2\tau_1^2 \} &   \qquad
C_3 &=& \{ \tau_2^2\tau_1^3\tau_2^2\tau_1^3 \} \\
C_4 &=& \{\tau_1^2\; , \; \tau_2^2\tau_1^2\tau_2^2 \} &   \qquad
C_5 &=& \{\tau_2^2\; , \; \tau_1^2\tau_2^2\tau_1^2 \}  \\
C_6 &=& \{\tau_1\tau_2^2\tau_1^3\; , \;
\tau_1^3\tau_2^2\tau_1 \}  &      \qquad
C_7 &=& \{\tau_1\tau_2^2\tau_1 \; , \;
\tau_1^3\tau_2^2\tau_1^3 \}   \\
C_8 &=& \{\tau_2\tau_1^2\tau_2 \; , \;
\tau_2^3\tau_1^2\tau_2^3 \}    &  \qquad
C_9 &=& \{\tau_1^2\tau_2^2 \; , \; \tau_2^2 \tau_1^2\} .
\ea
\eea   
For convenience, we expressed the elements of $P(3,4)$ in terms of the 
braid generators $\tau_1$ and $\tau_2$ rather than the monodromy generators
$\gamma_{12}$, $\gamma_{13}$ and $\gamma_{23}$.
It turns out that $P(3,4)$ is the coxeter group denoted 
as $16/8$ in~\cite{thomas}. Its centre
of order four contained in the first four 
conjugacy classes naturally coincides with that of $B(3,4)$. 
The character table of this group is exhibited in 
table~\ref{char}.

\newpage

\begin{table}[t]
\begin{tabular}{crrrrrrrrrrrrrrrr}
 \hline  \\[-4mm]
 & ${\ssc C_0^1}$ & ${\ssc C_0^2}$ & ${\ssc C_0^3}$ & 
            ${\ssc C_0^4}$ & ${\ssc C_1^1}$ & ${\ssc C_1^2}$ & 
            ${\ssc C_1^3}$ & ${\ssc C_1^4}$ & ${\ssc C_2^1}$ & 
            ${\ssc C^2_2}$ & ${\ssc C^3_2}$ & ${\ssc C^4_2}$ & 
            ${\ssc C_3^1}$ & ${\ssc C_3^2}$ & ${\ssc C_4^1}$ & 
            ${\ssc C_4^2}$  \\ \hline   \\[-4mm]
${\ssc \Lambda_0}$ & ${\ssc 1}$ & ${\ssc 1}$ & ${\ssc 1}$ & ${\ssc 1}$ & 
                     ${\ssc 1}$ & ${\ssc 1}$ & ${\ssc 1}$ & ${\ssc 1}$ &
                     ${\ssc 1}$ & ${\ssc 1}$ & ${\ssc 1}$ & ${\ssc 1}$ & 
                     ${\ssc 1}$ & ${\ssc 1}$ & ${\ssc 1}$ & ${\ssc 1}$   
                     \\ 
${\ssc \Lambda_1}$ & ${\ssc 1}$ & -${\ssc 1}$ & ${\ssc 1}$ & -${\ssc 1}$ & 
                     ${\ssc \im}$ & -${\ssc \im}$ & ${\ssc \im}$ & -${\ssc \im}$ & 
                    -${\ssc 1}$ &  ${\ssc 1}$ & -${\ssc 1}$ & ${\ssc 1}$ & 
                    -${\ssc 1}$ & ${\ssc 1}$ & -${\ssc \im}$ & ${\ssc \im}$ 
                     \\
${\ssc \Lambda_2}$ & ${\ssc 1}$ & ${\ssc 1}$ & ${\ssc 1}$ & ${\ssc 1}$ & 
                    -${\ssc 1}$ & -${\ssc 1}$ & -${\ssc 1}$ & -${\ssc 1}$ & 
                     ${\ssc 1}$ & ${\ssc 1}$ & ${\ssc 1}$ & ${\ssc 1}$ & 
                     ${\ssc 1}$ & ${\ssc 1}$ & -${\ssc 1}$ & -${\ssc 1}$  
                     \\
${\ssc \Lambda_3}$ & ${\ssc 1}$ & -${\ssc 1}$ & ${\ssc 1}$ & -${\ssc 1}$ & 
                    -${\ssc \im}$ & ${\ssc \im}$ & -${\ssc \im}$ & ${\ssc \im}$ & 
                    -${\ssc 1}$ & ${\ssc 1}$ & -${\ssc 1}$ & ${\ssc 1}$ & 
                    -${\ssc 1}$ & ${\ssc 1}$ & ${\ssc \im}$ & -${\ssc \im}$ 
                     \\ 
${\ssc \Lambda_4}$ & ${\ssc 2}$ & ${\ssc 2}$ & ${\ssc 2}$ & ${\ssc 2}$ & 
                     ${\ssc 0}$ & ${\ssc 0}$ & ${\ssc 0}$ & ${\ssc 0}$ & 
                    -${\ssc 1}$ & -${\ssc 1}$ & -${\ssc 1}$ & -${\ssc 1}$ & 
                     ${\ssc 2}$ & ${\ssc 2}$ & ${\ssc 0}$ & ${\ssc 0}$ \\ 
${\ssc \Lambda_5}$ & ${\ssc 2}$ & -${\ssc 2}$ & ${\ssc 2}$ & -${\ssc 2}$ & 
                     ${\ssc 0}$ & ${\ssc 0}$ & ${\ssc 0}$ & ${\ssc 0}$ & 
                     ${\ssc 1}$ & -${\ssc 1}$ & ${\ssc 1}$ & -${\ssc 1}$ & 
                    -${\ssc 2}$ & ${\ssc 2}$ & ${\ssc 0}$ & ${\ssc 0}$ \\ 
${\ssc \Lambda_6}$ & ${\ssc 2}$ & ${\ssc 2\im}$ & -${\ssc 2}$ & -${\ssc 2\im}$ & 
                   ${\ssc \oq}$ & -${\ssc \oq^*}$ & -${\ssc \oq}$ & 
                   ${\ssc \oq^*}$ & ${\ssc \im}$ & -${\ssc 1}$ & -${\ssc \im}$ & 
                   ${\ssc 1}$ & ${\ssc 0}$ & ${\ssc 0}$ & ${\ssc 0}$ & 
                   ${\ssc 0}$  \\ 
${\ssc \Lambda_7}$ & ${\ssc 2}$ & ${\ssc 2\im}$ & -${\ssc 2}$ & -${\ssc 2\im}$ & 
                    -${\ssc \oq}$ & ${\ssc \oq^*}$ & ${\ssc \oq}$ & 
                    -${\ssc \oq^*}$ & ${\ssc \im}$ & -${\ssc 1}$ & -${\ssc \im}$ & 
                     ${\ssc 1}$ & ${\ssc 0}$ & ${\ssc 0}$ & ${\ssc 0}$ & 
                     ${\ssc 0}$ \\ 
${\ssc \Lambda_8}$  & ${\ssc 2}$ & -${\ssc 2\im}$ & -${\ssc 2}$ & ${\ssc 2\im}$ & 
                      -${\ssc \oq^*}$ & ${\ssc \oq}$ & ${\ssc \oq^*}$ & 
                      -${\ssc \oq}$ & -${\ssc \im}$ & -${\ssc 1}$ & ${\ssc \im}$ & 
                      ${\ssc 1}$ & ${\ssc 0}$ & ${\ssc 0}$ & ${\ssc 0}$ & 
                      ${\ssc 0}$  \\ 
${\ssc \Lambda_9}$  & ${\ssc 2}$ & -${\ssc 2\im}$ & -${\ssc 2}$ & ${\ssc 2\im}$ & 
                      ${\ssc \oq^*}$ & -${\ssc \oq}$ & -${\ssc \oq^*}$ & 
                      ${\ssc \oq}$ & -${\ssc \im}$ & -${\ssc 1}$ & ${\ssc \im}$ & 
                      ${\ssc 1}$ & ${\ssc 0}$ & ${\ssc 0}$ & ${\ssc 0}$ & 
                      ${\ssc 0}$ \\ 
${\ssc \Lambda_{10}}$ & ${\ssc 3}$ & ${\ssc 3}$ & ${\ssc 3}$ & ${\ssc 3}$ & 
                        ${\ssc 1}$ & ${\ssc 1}$ & ${\ssc 1}$ & ${\ssc 1}$ & 
                        ${\ssc 0}$ & ${\ssc 0}$ & ${\ssc 0}$ & ${\ssc 0}$ & 
                       -${\ssc 1}$ & -${\ssc 1}$ & -${\ssc 1}$ & -${\ssc 1}$  
                        \\ 
${\ssc \Lambda_{11}}$ & ${\ssc 3}$ & -${\ssc 3}$ & ${\ssc 3}$ & -${\ssc 3}$ & 
                        ${\ssc \im}$ & -${\ssc \im}$ & ${\ssc \im}$ & -${\ssc \im}$ & 
                        ${\ssc 0}$ &  ${\ssc 0}$ & ${\ssc 0}$ & ${\ssc 0}$ & 
                        ${\ssc 1}$ & -${\ssc 1}$ & ${\ssc \im}$ & -${\ssc \im}$ 
                        \\ 
${\ssc \Lambda_{12}}$ & ${\ssc 3}$ & ${\ssc 3}$ & ${\ssc 3}$ & ${\ssc 3}$ & 
                       -${\ssc 1}$ & -${\ssc 1}$ & -${\ssc 1}$ & -${\ssc 1}$ & 
                        ${\ssc 0}$ & ${\ssc 0}$ & ${\ssc 0}$ & ${\ssc 0}$ & 
                       -${\ssc 1}$ & -${\ssc 1}$ & ${\ssc 1}$ & ${\ssc 1}$ 
                       \\ 
${\ssc \Lambda_{13}}$ & ${\ssc 3}$ & -${\ssc 3}$ & ${\ssc 3}$ & -${\ssc 3}$ & 
                        -${\ssc \im}$ & ${\ssc \im}$ & -${\ssc \im}$ & ${\ssc \im}$ & 
                         ${\ssc 0}$ & ${\ssc 0}$ & ${\ssc 0}$ & ${\ssc 0}$ & 
                         ${\ssc 1}$ & -${\ssc 1}$ & -${\ssc \im}$ & 
                         ${\ssc \im}$ \\ 
${\ssc \Lambda_{14}}$ & ${\ssc 4}$ & ${\ssc 4}$ & -${\ssc 4}$ & -${\ssc 4}$ & 
                          ${\ssc 0}$ & ${\ssc 0}$ & ${\ssc 0}$ & ${\ssc 0}$ & 
                          ${\ssc 1}$ & ${\ssc 1}$ & -${\ssc 1}$ & -${\ssc 1}$ &
                          ${\ssc 0}$ & ${\ssc 0}$ &  ${\ssc 0}$ & ${\ssc 0}$  
                          \\ 
${\ssc \Lambda_{15}}$ & ${\ssc 4}$ & -${\ssc 4}$ & -${\ssc 4}$ & ${\ssc 4}$ & 
                          ${\ssc 0}$ & ${\ssc 0}$ & ${\ssc 0}$ & ${\ssc 0}$ &
                         -${\ssc 1}$ & ${\ssc 1}$ & ${\ssc 1}$ & -${\ssc 1}$ & 
                          ${\ssc 0}$ & ${\ssc 0}$ & ${\ssc 0}$ & ${\ssc 0}$ 
                         \\[1mm] \hline
\end{tabular}  
\caption{\sl Character table of the truncated braid group 
$B(3,4)$. We used $\oq := \im+1$.}
\label{tab:tab4}
\end{table}

\begin{table}[tb]  
\begin{center}
\begin{tabular}{crrrrrrrrrrr} \hl
$P(3,4)$ & & $C_0$ & $C_1$ & $C_1$ & $C_3$ & $C_4$ & 
$C_5$ & $C_6$ & $C_7$ & $C_8$ & $C_9$    \\ \hl  \\[-4mm] 
$\Omega_0$ & & $1$ & $1$  & $1$ & $1$ & $1$ & $1$ & $1$ & $1$ & $1$ & $1$ \\ 
$\Omega_1$ & & $1$ & $1$ & $1$ & $1$ & $-1$ & $-1$ & $1$ & $-1$ & $-1$ & $1$ \\ 
$\Omega_2$ & & $1$ & $1$ & $1$ & $1$ & $-1$ & $1$ & $-1$ & $1$ & $-1$ & $-1$ \\ 
$\Omega_3$ & & $1$ & $1$ & $1$ & $1$ & $1$ & $-1$ & $-1$  & $-1$ & $1$ & $-1$ 
 \\   $\Omega_4$ & & $1$ & $-1$ & $1$ & $-1$ & $-1$ & $1$ & $1$ & $-1$ & 
$1$ & $-1$ \\ 
$\Omega_5$ & & $1$ & $-1$ & $1$ & $-1$ & $1$ & $-1$ &
$1$ & $1$ & $-1$ & $-1$ \\ 
$\Omega_6$ & & $1$ & $-1$ & $1$ & $-1$ & $1$ & $1$ & $-1$ & $-1$ &
$-1$ & $1$ \\ 
$\Omega_7$ & & $1$ & $-1$  & $1$  & $-1$  & $-1$ & $-1$ & $-1$ & $1$ & $1$ & 
             $1$ \\  
$\Omega_8$& & $2$ & $2\im$  & $-2$ & $-2\im$ & $0$ & $0$ & $0$ & $0$ & $0$
           & $0$ \\
$\Omega_{9}$& & $2$ & $-2\im$  & $-2$ & $2\im$ & $0$ & $0$ & $0$ & $0$ 
               & $0$ & $0$ \\[1mm]              \hl
\end{tabular}
\end{center}
\caption{\sl Character table of 
the truncated colored braid group $P(3,4)$.}
\label{char}
\end{table}

\sectie

\chapter{Abelian Chern-Simons theories}  \label{abelsCS}
\label{chap3}

\sectiona{Introduction}

A characteristic feature of three dimensional space time is the possibility
to endow a gauge theory with a so-called Chern-Simons 
term~\cite{des1,schonfeld}. It is well-known that the incorporation 
of such a term renders a gauge theory topological.
That is, the gauge fields acquire a topological mass, whereas  
the charges coupled to the gauge fields 
now induce magnetic fluxes and as a result exhibit 
nontrivial braid statistics (e.g.~\cite{goldmac,witten}). 
This is roughly speaking the effect 
of adding a Chern-Simons term to an unbroken continuous gauge theory.
Here, we will study the implications 
of adding a Chern-Simons term  to the
spontaneously broken planar gauge theories discussed in the previous 
chapter~\cite{spm1,sm,sam}. 
Hence, the models under consideration are governed by an
action of the form
\bea                               \label{alg}
S &=& S_{\mbox{\scriptsize YMH} } + S_{\mbox{\scriptsize matter}} 
+ S_{\mbox{\scriptsize CS}} \, ,
\eea
where the Yang-Mills Higgs action $S_{\rm{\mbox{\scriptsize YMH}}}$ again gives rise to
the spontaneous breakdown of the continuous compact 
gauge group $G$ to a finite 
subgroup ${ H}$ and $S_{\mbox{\scriptsize matter}}$ describes 
a conserved matter current  minimally coupled to the gauge fields. 
Finally, $S_{\rm{CS}}$ denotes the Chern-Simons action 
for the gauge fields.
For convenience, we restrict ourselves to abelian broken 
Chern-Simons gauge theories~(\ref{alg}) 
and return to the nonabelian case in chapter~\ref{chap4}. 
To be specific, in the present chapter we  
focus on symmetry breaking schemes 
\bea                                        \label{scheme}
{ G} \; \simeq \; U(1)^k   &\longrightarrow&  { H} \, ,
\eea 
with  $U(1)^k$ the direct product of $k$ compact $U(1)$ gauge 
groups and the finite subgroup $H$ 
a direct  product of $k$ cyclic groups  $\Z_{N^{(i)}}$ of order $N^{(i)}$
\bea       \label{genab}
{ H} &\simeq& \Z_{N^{(1)}} \times \Z_{N^{(2)}} \times \cdots \times
\Z_{N^{(k)}} \, .
\eea

The Chern-Simons terms for the gauge group $U(1)^k$ are known to 
fall into two  types, see for example~\cite{wesolo} and references therein. 
On the one hand, there are terms that describe self-couplings
of the various $U(1)$ gauge fields. These  will be called
Chern-Simons terms of type~I for convenience. On the other hand, 
there are terms (type~II) that establish couplings
between two different $U(1)$ gauge fields.
To be concrete, the most general Chern-Simons action 
for the gauge group $U(1) \times U(1)$, for instance, is of the form
\bea 
S_{\mbox{\scriptsize CS}} &=& \int d\,^3x \;(\, 
\frac{\mu^{(12)}}{2} \epsilon^{\kappa\sigma\rho}
                              A^{(1)}_{\kappa} \partial_{\sigma} 
                              A^{(2)}_{\rho}
\: + \: 
\sum_{i=1}^2 \frac{\mu^{(i)}}{2} \epsilon^{\kappa\sigma\rho}
                              A^{(i)}_{\kappa} \partial_{\sigma} 
                              A^{(i)}_{\rho} \,) ,
\eea 
with $A_{\kappa}^{(1)}$ and $A_{\kappa}^{(2)}$ 
the two $U(1)$ gauge fields. The parameters 
$\mu^{(1)}$, $\mu^{(2)}$ denote the topological masses 
characterizing the two Chern-Simons terms of type~I 
and  $\mu^{(12)}$ the topological mass 
characterizing the Chern-Simons term of type~II. 
In the unbroken phase, these Chern-Simons terms assign 
magnetic fluxes to the quantized matter charges $q^{(1)}$ and $q^{(2)}$
coupled to the two compact $U(1)$ gauge fields. 
Specifically, the type~I Chern-Simons term for the gauge 
field $A_{\kappa}^{(i)}$ attaches a magnetic 
flux $\phi^{(i)}=-q^{(i)}/\mu^{(i)}$ to a 
matter charge $q^{(i)}=n^{(i)} e^{(i)}$ with $n^{(i)} \in \Z$ and 
$e^{(i)}$ the fundamental charge for $A_{\kappa}^{(i)}$.
As a consequence, there are nontrivial topological 
interactions among these charges. 
When a charge $q^{(i)}$ encircles a remote  charge 
$q^{(i)'}$ in a counterclockwise fashion,
the wave function acquires the Aharonov-Bohm phase 
$\exp(-\im q^{(i)} q^{(i)'}/\mu^{(i)})$~\cite{goldmac}.  
The Chern-Simons term of type~II, in turn, attaches  
fluxes which belong to one $U(1)$ gauge group to the matter charges of the 
other. That is, a charge $q^{(1)}$ induces a flux 
$\phi^{(2)}=-2q^{(1)}/\mu^{(12)}$ and a charge
$q^{(2)}$ induces a flux $\phi^{(1)}=-2q^{(2)}/\mu^{(12)}$. 
Hence, the type~II Chern-Simons term gives rise to topological interactions
among matter charges of the two different $U(1)$ gauge groups.
A counterclockwise monodromy of a 
charge  $q^{(1)}$ and a charge  $q^{(2)}$, for example, yields
the Aharonov-Bohm phase 
$\exp(-2\im q^{(1)} q^{(2)}/\mu^{(12)})$, 
e.g.~\cite{ezawa,hagen,kimko,wesolo,wilccs}.

The presence of a Chern-Simons term 
for the continuous gauge group $U(1)^k$  naturally has a bearing on the 
topological interactions in  the broken phase. 
As we will argue, it gives rise to nontrivial Aharonov-Bohm phases 
among the vortices labeled by the elements 
of the residual gauge group~(\ref{genab}). 
To be specific, the $k$ different vortex species carry quantized 
flux $\phi^{(i)} = \frac{2\pi a^{(i)}}{N^{(i)} e^{(i)}}$ with  
$a^{(i)} \in \Z$ and $N^{(i)}$ the order of the $i^{\rm th}$ cyclic group
of the product group~(\ref{genab}). A type~I Chern-Simons term 
for the gauge field  $A_{\kappa}^{(i)}$  then implies the Aharonov-Bohm 
phase $\exp(\im \mu^{(i)} \phi^{(i)} \phi^{(i)'})$ for a 
counterclockwise monodromy of  a vortex $\phi^{(i)}$ and 
a vortex $\phi^{(i)'}$. 
A Chern-Simons term of type~II coupling the 
gauge fields $A_{\kappa}^{(i)}$ and $A_{\kappa}^{(j)}$, in turn,
gives rise to the Aharonov-Bohm
phase $\exp(\im \mu^{(ij)} \phi^{(i)} \phi^{(j)})$ for the process
in which a vortex $\phi^{(i)}$ circumnavigates
a vortex $\phi^{(j)}$ in a 
counterclockwise fashion.
In fact, these additional Aharonov-Bohm phases among the vortices  
form the only distinction with the abelian discrete 
$H$ gauge theory describing the long distance physics 
in the absence of a Chern-Simons action for the broken gauge group $U(1)^k$. 
That is, the Higgs mechanism removes the 
fluxes attached the matter charges $q^{(i)}$ 
in the unbroken Chern-Simons phase. 
Hence, contrary to the unbroken Chern-Simons phase,  
there are {\em no} Aharonov-Bohm interactions among the matter charges
in the Chern-Simons Higgs phase~\cite{spm1,sm,sam}.  
The canonical Aharonov-Bohm interactions 
$\exp(\im q^{(i)}\phi^{(i)})$ between the  
matter charges $q^{(i)}$ and 
the magnetic vortices $\phi^{(i)}$ persist though.

%We will  also addres the implications of introducing 
%Dirac monopoles in these compact $U(1)^k$ Chern-Simons theories.
%There are $k$ different species corresponding to 
%the $k$ different compact $U(1)$ gauge groups.
%It is known that a consistent incorporation of these monopoles, in fact, 
%requires the quantization of the topological masses 
%characterizing the type~I and~II Chern-Simons terms~\cite{hen,pisar}. 
%Moreover, it has been argued that the presence of 
%monopoles does {\em not} lead to 
%confinement of the charges $q^{(i)}$
%in the unbroken Chern-Simons phase~\cite{affleck,pisar,klee}. 
%Instead, they describe tunneling events leading to the creation
%or annihilation of charges $q^{(i)}$, 
%i.e.\ there is only a finite number
%of stable particles in the spectrum depending on the integral Chern-Simons 
%parameter~\cite{klee}.
%The matter charges are now conserved modulo the integral
%Chern-Simons parameter.
%In the broken phase, the presence of the Dirac monopoles
%implies that the magnetic fluxes $a^{(i)}$ carried by the vortices 
%are conserved modulo $N^{(i)}$,
%but the flux decay driven by the monopoles is accompanied by the creation 
%of matter charge with magnitude proportional to the integral Chern-Simons 
%parameter~\cite{spm1,sm}.

The organization of this chapter is as follows.
In section~\ref{bgt}, we very briefly recall 
that the Chern-Simons actions for a 
compact gauge group $G$ are classified by the 
cohomology group $H^4(BG, \Z)$ of the classifying 
space $BG$~\cite{diwi}. 
For finite groups $H$, this classification boils 
down to the cohomology group $H^3(H,U(1))$ of the group $H$ itself.
In other words, the different Chern-Simons theories for a finite 
gauge group $H$ correspond to the independent
3-cocycles $\omega \in H^3(H,U(1))$,
which describe additional Aharonov-Bohm interactions among the 
fluxes labeled by the elements of $H$.
We then note that the inclusion
$H\subset G$ induces a natural 
homomorphism $H^4(BG, \Z) \to H^3(H,U(1))$.
This homomorphism determines the discrete 
$H$ Chern-Simons theory $\omega \in H^3(H,U(1))$
describing  the long distance physics of the 
Chern-Simons theory $S_{\mbox{\scriptsize CS}} \in H^4(BG, \Z)$ in which the continuous 
gauge group $G$ is broken to the finite subgroup $H$.
Section~\ref{fincoh} subsequently 
contains a short introduction to the cohomology groups $H^n(H,U(1))$
of finite abelian groups $H$.
In particular, we give the explicit realization of the complete set 
of independent 3-cocycles $\omega \in H^3(H,U(1))$
for the abelian groups~(\ref{genab}).
It turns out that these split up into three different types,
namely 3-cocycles (type~I) 
which give rise to Aharonov-Bohm interactions 
among fluxes of the same cyclic gauge group in 
the direct product~(\ref{genab}), 
those (type~II) that describe interactions between 
fluxes corresponding to two different cyclic gauge groups 
and finally 3-cocycles (type~III) that
lead to additional Aharonov-Bohm interactions between 
fluxes associated to three different cyclic gauge groups.
In section~\ref{multiple}, we turn to the classification of 
Chern-Simons actions for the compact gauge group $U(1)^k$ and 
establish that the homomorphism 
$H^4(B(U(1)^k), \Z) \to H^3(H,U(1))$ induced by the spontaneous symmetry 
breakdown~(\ref{scheme}) is not onto.
That is, 
the only Chern-Simons theories with finite abelian 
gauge group~(\ref{genab}) that may 
arise from a spontaneously broken $U(1)^k$ Chern-Simons theory are those
corresponding to a 3-cocycle of type~I and/or type~II, while 3-cocycles of 
type~III do not occur.   Further, the introduction of a 3-cocycle 
$\omega \in H^3(H,U(1))$ in an abelian discrete $H$ gauge theory 
leads to a natural deformation of
the related quantum double $D(H)$ into the quasi-quantum double 
$\DW$.  This deformation is discussed in section~\ref{symalg}.

In the next sections, 
these general considerations are illustrated by some representative 
examples. Specifically, section~\ref{typeI} contains an analysis of 
the abelian Chern-Simons Higgs model in which the compact 
gauge group $G \simeq U(1)$ is broken down 
to the cyclic subgroup $H \simeq \Z_N$. 
We briefly review the unbroken phase of this model and
recall that a consistent 
implementation of Dirac monopoles requires 
the topological mass to be quantized
as $\mu=\frac{pe^2}{\pi}$ with $p\in\Z$~\cite{hen,pisar}.
This is  in accordance  with the fact that the 
different  Chern-Simons actions for a compact gauge group $U(1)$
are classified by the integers: $H^4(BU(1), \Z) \simeq \Z$. 
We then turn to the broken phase of the model and 
establish that the long distance physics is indeed described by a $\Z_N$ 
Chern-Simons theory with 3-cocycle $\omega \in H^3( \Z_N ,U(1))\simeq \Z_N$
fixed by the natural homomorphism $H^4(BU(1), \Z) \to H^3( \Z_N ,U(1))$.
That is, the integral Chern-Simons parameter $p$ becomes periodic 
in the broken phase with period $N$. 
Section~\ref{typeII} contains a similar treatment of a 
Chern-Simons theory of type~II with gauge group
$G \simeq U(1) \times U(1)$ spontaneously broken down 
to $H \simeq \Z_{N^{(1)}} \times \Z_{N^{(2)}}$. The long distance physics 
of this model is described by a $\Z_{N^{(1)}} \times \Z_{N^{(2)}}$
Chern-Simons theory defined by a 3-cocycle of type~II.
The abelian discrete $H$ Chern-Simons theories 
which do not occur in spontaneously broken $U(1)^k$ Chern-Simons 
theories are actually the most interesting. 
These are the theories defined by  the aforementioned 
3-cocycles of type~III.
The simplest example of such a theory, 
namely that with gauge group $H \simeq \Z_2 \times \Z_2 \times \Z_2$,
is discussed in section~\ref{typeIII}. We will show that 
the introduction of the corresponding 3-cocycle of type~III renders
this theory nonabelian. In fact, this 
theory turns out to be  dual  
to an ordinary $D_4$ gauge theory with $D_4$ the nonabelian 
dihedral group of order $8$. 

In section~\ref{dijwit}, we briefly evaluate 
the Dijkgraaf-Witten invariant for some lens spaces 
using the three different types of  3-cocycles for 
various finite abelian groups $H$.
Finally, in an appendix we have collected
some results in the theory of cohomology which will be used in this 
chapter. In particular, it contains a derivation   
of the cohomology group $H^3(H,U(1))$ 
of an arbitrary  abelian finite group~(\ref{genab}).

As a last remark, the treatment of the examples 
in sections~\ref{typeI}, \ref{typeII}
and~\ref{typeIII} is more or less self contained. 
The reader could well start with section~\ref{typeI} 
and occasionally go back to earlier sections to fill in some details.

\sectiona{Group cohomology and symmetry breaking}
\label{bgt}

As has been argued by Dijkgraaf and Witten~\cite{diwi}, 
the  Chern-Simons actions $S_{\mbox{\scriptsize CS}}$ for 
a compact gauge group $G$ are in one-to-one correspondence
with the elements of the cohomology 
group $H^4 (B{ G}, \Z)$ of the classifying space
$B{ G}$ with integer coefficients $\Z$. 
(Let $EG$ be a contractible space characterized 
by a free action of $G$. The classifying space $BG$ is then given 
by dividing out the action of $G$ on $EG$, that is, $BG=EG/G$.
See for instance~\cite{novikov}).
In particular, this classification includes 
the case of finite gauge groups $H$. 
The isomorphism~\cite{milnor}
\bea                            \label{miln}
H^n(B{ H},{\mbox{\bf Z}}) &\simeq& H^n({ H},{\mbox{\bf Z}}),
\eea 
which only holds for finite groups ${ H}$, shows that the cohomology 
of the classifying space $BH$ is the same as that of the group $H$ itself.
In addition, we have the isomorphism 
\bea                  \label{clasi}         
H^{n} (H, \Z)  & \simeq &   H^{n-1} ({ H}, U(1))    
\qquad \forall  \, n>1.
\eea
A derivation of this result, using the universal coefficients theorem, 
is contained in appendix~\ref{gc}.
Especially, we now arrive at the identification 
\bea          \label{clasor}                 
H^4 ({BH}, \Z) & \simeq & H^3 ({ H}, U(1)).
\eea 
This expresses the fact that the different Chern-Simons theories 
for a finite gauge group $H$ are defined by the elements 
$\omega \in H^3({ H}, U(1))$, i.e.\ algebraic 
3-cocycles $\omega$ taking values in $U(1)$. 
These 3-cocycles can then be interpreted
as $\omega = \exp(\im S_{\mbox{\scriptsize CS}})$, where $S_{\mbox{\scriptsize CS}}$ denotes a
Chern-Simons action for the finite gauge group $H$~\cite{diwi}.
With abuse of language, we will  usually call $\omega$ itself
a Chern-Simons action for $H$.

Let $K$ be a subgroup of a compact group $G$.
The inclusion $K \subset G$ induces a 
natural homomorphism 
\bea    \label{restric} 
H^4 (B{ G}, \Z)  &\longrightarrow&   H^4 (BK, \Z) \, ,
\eea 
called the restriction (e.g.~\cite{brown,cartan}). 
This homomorphism determines the fate 
of a given Chern-Simons action $S_{\mbox{\scriptsize CS}} \in H^4 (B{ G}, \Z)$ 
when the gauge group $G$ is spontaneously broken 
down to $K$.
That is, the mapping~(\ref{restric}) 
fixes the Chern-Simons action~$\in H^4 (BK, \Z)$ for the residual
gauge group $K$ to which $S_{\mbox{\scriptsize CS}}$ reduces in the broken phase.
In the following, we will only 
be concerned with Chern-Simons theories
in which a continuous (compact) gauge  group $G$ is broken down 
to a finite subgroup $H$.  
The long distance physics of such a model 
is described by a discrete $H$ Chern-Simons theory 
with 3-cocycle $\omega \in H^3 ({ H}, U(1))$ determined by 
the original Chern-Simons action $S_{\mbox{\scriptsize CS}}$ 
for the broken gauge group $G$.
The 3-cocycle $\omega$ now governs the additional 
Aharonov-Bohm phases among the magnetic fluxes $h \in H$ implied 
by the Chern-Simons action $S_{\mbox{\scriptsize CS}}$.
This is roughly speaking the physical background
to the natural homomorphism 
\bea                          
H^4 (B{ G}, \Z)  &\longrightarrow&   H^3 ({ H}, U(1)) \, , 
\label{homo}   
\eea 
which is the composition of the restriction 
$H^4 (B{ G}, \Z)  \to  H^4 (BH, \Z)$ induced by the inclusion 
$H \subset G$, and the isomorphism~(\ref{clasor}).

The restrictions~(\ref{restric}) and~(\ref{homo}) for continuous 
subgroups $K \subset G$ and finite subgroups $H \subset G$, respectively,
are not necessarily onto. Hence, it is not 
guaranteed that all Chern-Simons theories with gauge group $K$ (or $H$)
can be obtained from spontaneously  broken Chern-Simons theories
with gauge group $G$. 
Particularly, in the following  sections we will see that the natural 
homomorphism $H^4(B(U(1)^k), \Z) \to H^3(H,U(1))$ induced
by the symmetry breaking~(\ref{scheme}) is not onto.

\sectiona{Cohomology of finite abelian groups}
\label{fincoh}

Here, we give a brief introduction to the 
cohomology groups $H^n({ H},U(1))$ of a finite abelian group $H$.
The discussion is organized as follows. 
In section~\ref{reldef}, we begin by recalling  
the basic definitions and subsequently focus on the cocycle 
structure occurring in an abelian discrete $H$
Chern Simons theory. 
Finally, the explicit realization
of all independent 3-cocycles $\omega \in H^3({ H},U(1))$
for an arbitrary abelian group ${ H}$ is given in section~\ref{aa3c}.

\subsection{$H^n(H,U(1))$}
\label{reldef}

In the (multiplicative) 
algebraic description of the cohomology groups  $H^n(H,U(1))$, the 
$n$-cochains are represented as $U(1)$ valued functions 
\bea                \label{4}
c :  \; \underbrace{{ H} \times \cdots \times { H}}_{n\; 
\mbox{\scriptsize times}} 
&\longrightarrow& U(1).
\eea
The set of all $n$-cochains forms the 
abelian group $C^n({ H},U(1)) := C^n$ 
with pointwise multiplication 
\beas
(c \cdot d) \, (A_1,\ldots,A_n) &=&   
c \, (A_1,\ldots,A_n)
\; d \, (A_1,\ldots,A_n),
\eeas
where  the capitals $A_j$ (with $1 \leq j \leq n$) 
denote  elements of the finite group ${ H}$ and $c,d \in C^n$. 
The coboundary operator $\delta$ then establishes
a  mapping 
\beas \delta :   \;
 C^n & \longrightarrow & C^{n+1}  \\
           c & \longmapsto & \delta c, \nn
\eeas
given by 
\bea                     \label{coboundop}
\delta
c \, (A_1, \ldots , A_{n+1}) &=& c \, (A_2,\ldots,A_{n+1})
\; c \, (A_1,\ldots  ,  A_n)^{(-1)^{n+1}} \\           
&  &  \qquad \qquad \times \prod_{i=1}^{n} 
c \, (A_1,\dots,A_i \cdot A_{i+1},\ldots,A_{n+1})^{(-1)^i}, \nn
\eea
which acts as  a derivation $\delta(c \cdot d) = \delta c \cdot \delta d$.
It can be checked explicitly that $\delta$ is indeed 
nilpotent $\delta^2 =1$. 
The coboundary operator $\delta$ 
naturally defines two subgroups $Z^n$ and $B^n$
of $C^n$. Specifically, the subgroup $Z^n\subset C^n$
consists of $n$-cocycles  being the $n$-cochains $c$ in the kernel of $\delta$
\bea                                         
\label{ach}
\delta c &=& 1 \qquad \forall \,\, c \in Z^n,
\eea
whereas the subgroup $B^n \subset Z^n \subset C^n$ 
contains the $n$-coboundaries or exact $n$-cocycles 
\bea                            \label{cobou}
c &=& \delta b  \qquad \forall \,\,c \in B^n.
\eea 
with $b$ some cochain $\in C^{n-1}$.  The cohomology group
$H^n({ H},U(1))$ is now defined as 
\bea
H^n({ H},U(1)) &:=& Z^n/B^n.
\eea 
In other words, the elements of $H^n({ H},U(1))$ correspond 
to the $n$-cocycles~(\ref{ach}) with equivalence relation  
$c \sim c \delta b$.

The so-called slant product $i_A$ with $A \in H$ is  a 
mapping in the opposite direction 
to the coboundary operator (see for instance~\cite{spanier} and also 
appendix~\ref{laap} of chapter~\ref{chap4})
\beas
i_A : \; C^n & \longrightarrow & C^{n-1}\\
      c & \longmapsto & i_A c , \nn 
\eeas
defined as
\bea   
i_A c \, (A_1,\ldots,A_{n-1}) &:=& c \, (A,A_1,\ldots  ,  
A_{n-1})^{(-1)^{n-1}}  \label{ig}\\
 & &  \qquad  \times \prod_{i=1}^{n-1} 
c \, (A_1,\ldots,A_i, A, A_{i+1},\ldots,A_{n-1})^{(-1)^{n-1+i}}.     \nn  
\eea
It can be shown that the slant product satisfies the following  relation 
\bea
\delta ( i_A  c )&=& i_A \, \delta c.  \label{w1}
\eea
for all $n$-cochains $c$. Notably, if $c$ is a $n$-cocycle,
we immediately infer from~(\ref{w1}) that  $i_A  c$ becomes 
a $(n-1)$-cocycle: 
$\delta (i_A  c) = i_A \, \delta c =  1$.
Hence, the slant product establishes an homomorphism 
\bea
i_A: 
H^n({ H},U(1))
&\longrightarrow&  H^{n-1}({ H},U(1)),
\eea
for each $A \in H$.

Let us finally turn to the cocycle structure appearing  in 
an abelian discrete $H$ gauge theory with Chern-Simons 
action $\omega \in H^3({ H},U(1))$. 
First of all, as indicated by~(\ref{coboundop}) and~(\ref{ach}), 
the 3-cocycle $\omega$ satisfies the relation
\bea
\label{pentagon}
\omega(A,B,C)\;\omega(A,B \cdot C,D)\;\omega(B,C,D) &=& 
\omega(A \cdot B,C,D)\;\omega(A,B,C \cdot D),   \qquad
\eea
for all $A,B,C \in H$.  To continue,
the slant product~(\ref{ig}) as applied to $\omega$
gives rise to a set of 2-cocycles $c_A \in H^2({ H},U(1))$
\bea     \label{c}
c_A (B,C) \; := \; i_A \omega(B,C) \; = \;
\frac{\omega(A,B,C)\;\omega(B,C,A)}{\omega(B,A,C)},
\eea
which are labeled by the different elements $A$ of $H$.
As will become clear in section~\ref{symalg}, 
these 2-cocycles enter the definition  of the projective dyon charge 
representations associated to the magnetic fluxes in this
abelian discrete $H$ Chern-Simons gauge theory. 
To be specific, the different charges we can assign to 
the abelian magnetic flux $A \in H$ to form dyons
are labeled by the inequivalent unitary irreducible projective 
representations $\alpha$ of $H$ defined as
\bea                \label{project} 
{\alpha}(B)\, \cdot \, {\alpha}(C) &=& c_A(B,C) \;  {\alpha}(B \cdot C).
\eea
Here, the 2-cocycle relation satisfied by $c_A$ 
\bea
c_A (B,C) \; c_A (B \cdot C, D) &=& c_A (B,C\cdot D) \; c_A (C, D),
\label{tweeko} 
\eea
implies that the representations $\alpha$ are associative.
To conclude, as follows from~(\ref{coboundop}) and~(\ref{ach}),
the 1-cocycles obey the relation $c \, (B)\; c \, (C) = c \, (B \cdot C)$.
In other words, the different 1-cocycles being the elements of 
the cohomology group $H^1({ H},U(1))$ correspond to the inequivalent 
ordinary UIR's of the group ${ H}$. These label the conceivable
{\em free} charges in a Chern-Simons theory with finite abelian 
gauge group $H$.

\subsection{Chern-Simons actions for finite abelian groups}   
\label{aa3c}

In this section, we present the explicit realization of the different 
3-cocycles~(\ref{pentagon}) for the  finite abelian groups~(\ref{genab})
and subsequently evaluate the 2-cocycles obtained from these 3-cocycles
by means of the slant product~(\ref{c}).

For convenience, we start with the abelian groups of the 
particular form ${ H} \simeq \Z_N^k$, that is, 
$H$ is the direct product of $k$ cyclic groups of the 
{\em same} order $N$. An abstract group cohomological derivation 
(contained in appendix~\ref{gc}) reveals  
the following content of the relevant cohomology groups
\bea
H^1(\Z_N^k,U(1)) &\simeq& \Z_N^k   \label{conj1e}            \\
H^2(\Z_N^k,U(1)) &\simeq& \Z_N^{\frac{1}{2}k(k-1)} \label{conj2e}  \\
H^3(\Z_N^k,U(1)) 
&\simeq& \Z_N^{k+\frac{1}{2}k(k-1) +\frac{1}{3!}k(k-1)(k-2)}.   \label{conj3e}
\eea
As we have seen in the previous section, the 
first result labels the inequivalent  UIR's of $\Z_N^k$, 
the second labels the different 2-cocycles entering the  
projective representations of $\Z_N^k$,
whereas the last result gives the number of different 3-cocycles or 
Chern-Simons actions for  $\Z_N^k$.  
The derivation of the isomorphism~(\ref{conj3e}) in appendix~\ref{gc} 
pointed out that there are,
in fact, three dissimilar types of 3-cocycles.
The explicit realization of these 3-cocycles involves some 
notational conventions, which we establish first.
Let $A,B$ and $C$ denote elements of $\Z_N^k$, i.e.\
\bea 
A &:=& (a^{(1)} , a^{(2)}, 
\ldots, a^{(k)}) \qquad \mbox{with 
$a^{(i)} \in \Z_N$   
for $i=1,\ldots, k$}\, ,
\eea
and similar decompositions for  $B$ and $C$. We adopt the additive 
presentation for the abelian 
group $\Z_N^k$, that is, the elements $a^{(i)}$ of $\Z_N$
take values in the range $ 0,\ldots, N-1$, and group multiplication 
is defined as
\bea
A \cdot B \; = \; [A+B] \; := \; ([a^{(1)}+b^{(1)}],  \ldots ,
[a^{(k)}+b^{(k)}]).
\eea
Here, the rectangular brackets again denote modulo $N$ 
calculus, such that the sum always lies in the range $0, \ldots, N-1$.
With these conventions, the three  types of 3-cocycles
for the direct product group $\Z_N^k$ take the following form
\begin{eqnfourarray} 
\omega_{\rm I}^{(i)}(A,B,C) \!\!\!   &=& \!\!\!
\exp \left( \frac{2 \pi \im p^{(i)}_{\mbox{\scriptsize I}}}{N^2} 
a^{(i)}(b^{(i)} +c^{(i)} -[b^{(i)}+c^{(i)}]) \right)
 &  $\!\! 1 \leq i \leq  k$ \, \label{type1}    \\
\omega_{\mbox{\scriptsize II}}^{(ij)}(A,B,C) \!\!\!&=& \!\!\!
\exp \left( 
\frac{2 \pi \im p^{(ij)}_{\mbox{\scriptsize II}}}{N^2} a^{(i)}(b^{(j)} +c^{(j)} -
[b^{(j)}+c^{(j)}]) \right) &    $\!\! 1 \leq i < j \leq k$
\label{type2} \,  \\
\omega_{\mbox{\scriptsize III}}^{(ijl)}(A,B,C) \!\!\! &=& \!\!\!
\exp \left( \frac{2 \pi \im  
p^{(ijl)}_{\mbox{\scriptsize III}}}{N} 
a^{(i)}b^{(j)}c^{(l)} \right) &
 $\!\! 1 \leq i < j < l \leq k$ , \label{type3}
\end{eqnfourarray}
where the integral parameters $p_{\rm I}^{(i)}$, $p_{\mbox{\scriptsize II}}^{(ij)}$ and 
$p^{(ijl)}_{\mbox{\scriptsize III}}$  label the different elements of the 
cohomology group $H^3(\Z_N^k,U(1))$.
In accordance with~(\ref{conj3e}), the 3-cocycles are periodic functions
of these parameters with period $N$. 
For the 3-cocycles of type~III
this periodicity is obvious, while for the 3-cocycles 
of type~I and~II  it is immediate
after the observation that the factors  
$(b^{(i)} +c^{(i)} - [b^{(i)}+c^{(i)}])$, with $1\leq i \leq  k$,
either vanish or equal $N$.  Moreover, it is also readily checked that these 
3-cocycles indeed satisfy the relation~(\ref{pentagon}).

Let us proceed with a closer examination of these three types of 3-cocycles.
The $k$ different 3-cocycles of type~I  describe self-couplings, that is, 
couplings between  the magnetic fluxes 
($a^{(i)}$,$b^{(i)}$ and $c^{(i)}$) associated to the same gauge 
group $\Z_N$ in the direct product $\Z_N^k$. 
In this counting procedure, it is, of course, 
understood that every 3-cocycle actually stands for a set of 
$N-1$ nontrivial 3-cocycles labeled by the periodic 
parameter $p_{\mbox{\scriptsize I}}^{(i)}$. 
The 3-cocycles of type~II, in turn,  establish pairwise 
couplings between the magnetic fluxes corresponding to
different gauge groups $\Z_N$  
in the direct product $\Z_N^k$.
Note that the 3-cocycles 
$\omega_{\mbox{\scriptsize II}}^{(ij)}$ and $\omega_{\mbox{\scriptsize II}}^{(ji)}$ are 
equivalent, since they differ by a 3-coboundary~(\ref{cobou}).
In other words, there are only $\frac{1}{2}k(k-1)$ 
distinct 3-cocycles of type~II.
A similar argument holds for the 3-cocycles of type~III.
A permutation of the labels $i$, $j$ and $k$ in these 
3-cocycles yields an equivalent 3-cocycle. Hence, we end up with
$\frac{1}{3!}k(k-1)(k-2)$ different 3-cocycles of type~III, which
realize couplings between the fluxes associated to three 
distinct $\Z_N$ gauge groups.

We are now well prepared to discuss the 3-cocycle structure for
general abelian groups $H$ being direct 
products~(\ref{genab}) of cyclic groups possibly of 
different order. Let us assume that ${ H}$ consists of $k$ cyclic factors.
The abstract analysis in appendix~\ref{gc}  shows that depending 
on the divisibility of the orders of the different 
cyclic factors, there are again $k$ distinct 3-cocycles of type~I,
$\frac{1}{2}k(k-1)$ different 3-cocycles of type~II and 
$\frac{1}{3!}k(k-1)(k-2)$ different 3-cocycles of type~III.
It is easily verified that the associated generalization of the 3-cocycle 
realizations~(\ref{type1}),~(\ref{type2}) and~(\ref{type3})
becomes 
\bea
\omega_{\mbox{\scriptsize I}}^{(i)}(A,B,C)    &=& 
\exp \left( \frac{2 \pi \im p^{(i)}_{\mbox{\scriptsize I}}}{N^{(i)\;2}} \;
a^{(i)}(b^{(i)} +c^{(i)} -[b^{(i)}+c^{(i)}]) \right)  \label{type1do} \\
\omega_{\mbox{\scriptsize II}}^{(ij)}(A,B,C) &=&             
\exp \left( \frac{2 \pi \im p_{\mbox{\scriptsize II}}^{(ij)}}{N^{(i)}N^{(j)}}  \;
a^{(i)}(b^{(j)} +c^{(j)} - [b^{(j)}+c^{(j)}]) \right)  \label{type2do} \\
\omega_{\mbox{\scriptsize III}}^{(ijl)} (A,B,C) &=& \exp \left( \frac{2 \pi \im
p_{\mbox{\scriptsize III}}^{(ijl)}}{{\gcd}(N^{(i)}, N^{(j)},N^{(l)})} \;
a^{(i)}b^{(j)}c^{(l)} \right),            \label{type3do}
\eea
where $N^{(i)}$ (with $1\leq i \leq  k$) denotes the order 
of the $i^{\rm th}$
cyclic factor of the direct product group $H$. 
In accordance with~(\ref{conj1do}), 
the 3-cocycles of type~III are cyclic in the integral parameter
$p_{\mbox{\scriptsize III}}^{(ijl)}$ with period the greatest common
divisor ${\gcd}(N^{(i)}, N^{(j)},N^{(l)})$ of $N^{(i)}$, $N^{(j)}$ and 
$N^{(l)}$. The periodicity of the 3-cocycles of 
type~I coincides with the order $N^{(i)}$ of the associated cyclic factor of
$H$. Finally, the 3-cocycles of type~II  are  
periodic in the integral parameter $p_{\mbox{\scriptsize II}}^{(ij)}$ with
period the greatest common divisor  ${\gcd}(N^{(i)},N^{(j)})$ 
of $N^{(i)}$ and $N^{(j)}$.  This last periodicity  becomes clear upon
using the theorem 
\bea            \label{theorem}
\frac{\gcd(N^{(i)},N^{(j)})}{N^{(i)}N^{(j)}} &=& 
\frac{x}{N^{(i)}} + \frac{y}{N^{(j)}} \qquad\qquad \mbox{with $x,
y\in \Z$},
\eea
which indicates that~(\ref{type2do}) boils down to a 3-coboundary 
or trivial 3-cocycle, if we set 
$p_{\mbox{\scriptsize II}}^{(ij)} = {\gcd}(N^{(i)},N^{(j)})$.

We close this section with a brief examination
of the 2-cocycles related to the three different types 
of 3-cocycles through the slant product~(\ref{c}). 
Let us start with the 3-cocycles of type~I and type~II.
Upon substituting the expressions~(\ref{type1do}) 
and~(\ref{type2do}) respectively in~(\ref{c}), we simply infer 
that the associated 2-cocycles $c_A$ 
correspond to the trivial element 
of the second cohomology group $H^2({ H}, U(1))$. To be precise,
these 2-cocycles are 2-coboundaries 
\be      \label{repphase}
c_A(B,C) \; = \; \delta \varepsilon_A (B,C) \; = \;
\frac{\varepsilon_A(B) \; \varepsilon_A(C)}{\varepsilon_A(B \cdot C)} \, ,
\ee
where the 1-cochains $\varepsilon_A$ of type~I and type~II read
\bea                                  \label{epi}
\varepsilon^{\mbox{\scriptsize I}}_{A}(B) &=& \exp 
\left( \frac{2 \pi \im p^{(i)}_{\mbox{\scriptsize I}}}
{N^{(i)\,2}} \; a^{(i)} b^{(i)} \right) \\
\varepsilon^{\mbox{\scriptsize II}}_{A}(B) &=& 
\exp \left(
\frac{2 \pi \im p^{(ij)}_{\mbox{\scriptsize II}}}{N^{(i)}N^{(j)}} 
\; a^{(i)} b^{(j)} \right). \label{epii}
\eea
Hence, Chern-Simons actions of type~I and/or type~II  
for the abelian gauge group $H$
give rise to trivial projective representations~(\ref{project}) labeled
by the ordinary UIR's of $H$, that is, the elements of the cohomology 
group $H^1 ({ H}, U(1))$. The one dimensional 
dyon charges then take the form 
$\alpha = \varepsilon_A \Gamma$, where $\Gamma$ denotes an UIR of $H$.
In contrast, the 2-cocycles $c_A$ obtained from the 
3-cocycles~(\ref{type3do}) of type~III
correspond to nontrivial elements
of the  cohomology group $H^2({ H}, U(1))$. 
The conclusion is that the dyon charges featuring in
an abelian discrete $H$ gauge theory
with a Chern-Simons action of type~III are 
nontrivial (higher dimensional) 
projective representations of ${ H}$.

\sectiona{Chern-Simons actions for $U(1)^k$ gauge theories}      
\label{multiple}

Chern-Simons theory with gauge group $U(1)^k$ endowed with minimally 
coupled matter fields has received considerable attention recently
(see~\cite{ezawa,hagen,kimko,wesolo,wilccs} and references therein). 
The motivation to study such a  theory is that it may
possibly find an application in multi-layered Hall systems. 
Here, we confine ourselves to the classification of 
the conceivable Chern-Simons actions for the gauge group $U(1)^k$.
In addition, we establish which Chern-Simons theories with finite abelian
gauge group $H$ may result from a spontaneous breakdown
of these $U(1)^k$ Chern-Simons gauge theories.

The most general Chern-Simons action for a planar  
$U(1)^k$ gauge theory is of the form
\bea    \label{act}
S_{\mbox{\scriptsize CS}}&=& 
\int d\,^3x \;( {\cal L}_{\mbox{\scriptsize CSI}} + {\cal L}_{\mbox{\scriptsize CSII}}) 
\\
{\cal L}_{\mbox{\scriptsize CSI}} &=& 
\sum_{i=1}^k  \; \frac{\mu^{(i)}}{2} \epsilon^{\kappa\sigma\rho}
                              A^{(i)}_{\kappa} \partial_{\sigma} 
                              A^{(i)}_{\rho} \label{CSt1}  \\
{\cal L}_{\mbox{\scriptsize CSII}} &=&
\sum_{i<j=1}^k  \frac{\mu^{(ij)}}{2} \epsilon^{\kappa\sigma\rho}
                              A^{(i)}_{\kappa} \partial_{\sigma} 
                              A^{(j)}_{\rho},     \label{CSt2}
\eea
where $A^{(i)}_\kappa$ (with $i=1,\dots,k$)
denote the  various $U(1)$  gauge fields and $\mu^{(i)}$,
$\mu^{(ij)}$ the topological masses. Hence,
there are  $k$ distinct Chern-Simons terms~(\ref{CSt1}), which
describe self couplings of the  $U(1)$ gauge fields.
In analogy with the terminology developed in the previous section,
we call these terms Chern-Simons terms of type~I.
Moreover, there are  $\frac{1}{2}k(k-1)$ 
distinct Chern-Simons terms of type~II 
establishing  pairwise couplings between different $U(1)$ gauge fields. 
Note that by a partial integration a 
term labeled by $(ij)$ becomes a term $(ji)$. 
Therefore, these  terms are equivalent and should not be counted separately.
Also note that up to a total 
derivative the Chern-Simons terms of type~I and type~II are 
indeed invariant under $U(1)^k$ gauge transformations
\bea       \label{cher}
A^{(i)}_{\rho}  &\longrightarrow& A^{(i)}_{\rho} - 
\partial_{\rho} \Omega^{(i)} \qquad \qquad    i=1,\ldots,k,
\eea
while the requirement of abelian gauge invariance immediately 
rules out `Chern-Simons terms of type~III'
\be                  \label{CSt3}
\sum_{i<j<l=1}^k \frac{\mu^{(ijl)}}{2}
\epsilon^{\kappa\sigma\rho}   A^{(i)}_{\kappa} A^{(i)}_{\sigma}
                              A^{(l)}_{\rho},
\ee
which would establish a coupling  between three 
different $U(1)$ gauge fields.

We now make the assumption that this abelian 
gauge theory is compact and features Dirac monopoles/instantons.
As indicated by the general discussion in section~\ref{topclas}, 
the Dirac monopoles that can be introduced in this particular theory
are labeled by the elements of the fundamental group 
$\pi_1(U(1)^k) \simeq \Z^k$. This is nothing but the obvious statement
that there is a family of Dirac monopoles 
related to each compact $U(1)$ gauge group. Hence, the complete spectrum 
of Dirac monopoles consists of the magnetic charges 
$g^{(i)} = \frac{2\pi m^{(i)}}{e^{(i)}}$ with $m^{(i)} \in \Z$ and 
$1 \leq i \leq k$. Here, $e^{(i)}$ denotes 
the fundamental charge associated with the  compact
$U(1)$ gauge  group being the $i^{\rm th}$ factor 
in the direct product $U(1)^k$. 
A consistent implementation of these monopoles requires  
that the topological masses in~(\ref{CSt1}) and~(\ref{CSt2}) are 
quantized as
\bea                                  
\mu^{(i)} &=& \frac{p^{(i)}_{\mbox{\scriptsize I}} e^{(i)}e^{(i)}}{\pi} \;\qquad \qquad 
\mbox{with $p^{(i)}_{\mbox{\scriptsize I}} \in {\mbox{\bf Z}}$}       \label{quantmui}   \\
\mu^{(ij)} &=&\frac{p^{(ij)}_{\mbox{\scriptsize II}} e^{(i)}e^{(j)}}{\pi}
\qquad \qquad
\mbox{with $p^{(ij)}_{\mbox{\scriptsize II}} \in {\mbox{\bf Z}}$}.       \label{quantmuij}  
\eea
This will be shown  in sections~\ref{rev} and~\ref{revii},
where we will discuss these models in further detail.
The integral Chern-Simons parameters 
$p^{(i)}_{\mbox{\scriptsize I}}$ 
and $p^{(ij)}_{\mbox{\scriptsize II}}$ now  label the different 
elements of the cohomology group 
\bea                    \label{u1k}
H^4(B(U(1)^k), \Z) &\simeq& \Z^{ k + \frac{1}{2}k(k-1)},
\eea
where a derivation of the isomorphism~(\ref{u1k}) 
is contained in appendix~\ref{gc}.

To conclude,  we have all the ingredients to make explicit the 
homomorphism~(\ref{homo}) accompanying the spontaneous symmetry breakdown 
of the gauge group $U(1)^k$ to the finite abelian group 
$H \simeq \Z_{N^{(1)}} \times \Z_{N^{(2)}} \times \cdots \times
\Z_{N^{(k)}}$.
In terms of the integral Chern-Simons parameters 
in~(\ref{quantmui}) and~(\ref{quantmuij}), it takes the form
\bea
H^4(B(U(1)^k), \Z) &\longrightarrow & H^3({ H}, U(1))   \label{homo1en2} \\
p^{(i)}_{\mbox{\scriptsize I}} & \longmapsto & p^{(i)}_{\mbox{\scriptsize I}} \qquad \; \bmod  N^{(i)} 
\label{homoI} \\
p^{(ij)}_{\mbox{\scriptsize II}} & \longmapsto & p^{(ij)}_{\mbox{\scriptsize II}} \qquad \bmod  
\gcd(N^{(i)}, N^{(j)}).     \label{homoII}
\eea
Here, the periodic parameters being the images of this mapping label the 
different 3-cocycles~(\ref{type1do}) and~(\ref{type2do}) of type~I and 
type~II. The natural conclusion then becomes that
 the long distance physics of a spontaneously broken 
$U(1)^k$ Chern-Simons theory of type~I/II is described by 
a Chern-Simons theory of type~I/II with the residual finite 
abelian gauge group $H$.
We will illustrate this result with two representative examples 
in sections~\ref{typeI} and~\ref{typeII}. 
As a last obvious remark, from~(\ref{homo1en2}) we also learn that
abelian discrete $H$ gauge theories with 
a Chern-Simons action of type~III can not be obtained from a spontaneously 
broken $U(1)^k$ Chern-Simons theory.

\sectiona{Quasi-quantum doubles}
\label{symalg}

The introduction of a Chern-Simons action $\omega \in H^3({ H}, U(1))$ 
in a discrete $H$ gauge theory leads to a natural  deformation
of the associated quantum double $D(H)$ (discussed in section~\ref{qdH})
into the quasi-quantum double $D^\omega ({ H})$.
Here, we recall the basis features of the quasi-quantum double 
$D^\omega ({ H})$ for  abelian finite groups $H$~\cite{dpr}.
For a general study of quasi-Hopf algebras, the reader
is referred to the original papers by Drinfeld~\cite{drin,drinfeld} 
and  the book by Shnider and Sternberg~\cite{shnider}.

The quasi-quantum double $D^\omega ({ H})$ 
for an abelian finite group $H$ is spanned by the basis elements  
\bea
\{ {\mbox{P}}_A \, B \}_{A,B\in { H}},      \label{els}  
\eea
representing a global symmetry transformation $B \in H$ followed
by the operator ${\mbox{P}}_A$, which  projects 
out the particular magnetic flux $A\in H$.
The deformation of the quantum double $D({ H})$ into the 
{\em quasi}-quantum double $D^\omega ({ H})$ by means of 
the 3-cocycle $\omega$
amounts to relaxing the coassociativity condition~(\ref{coas}) for  the 
comultiplication into {\em quasi}-coassociativity~\cite{dpr}
\bea \label{quasicoas}
({\mbox{id}} \ot \Delta) \, \Delta( \, {\mbox{P}}_A \, B \, ) &=& 
\varphi\cdot (\Delta \ot {\mbox{id}}) \, \Delta( \, {\mbox{P}}_A B \, ) 
\cdot\varphi^{-1}.
\eea
Here, the  invertible associator  
$\varphi \in D^\omega ({ H})^{\ot 3}$ is defined 
in terms of the 3-cocycle $\omega$ as
\bea                  
\varphi &:=& \sum_{A,B,C}\,\omega^{-1}(A,B,C) \;
{\mbox{P}}_A \otimes {\mbox{P}}_B \otimes {\mbox{P}}_{C}.  \label{isom}  
\eea
The multiplication and  comultiplication are  deformed accordingly
\bea
{\mbox{P}}_A \, B \cdot {\mbox{P}}_D \, C &=& 
\delta_{A,D} \;\; {\mbox{P}}_A  \, B \cdot C 
\;\; c_A(B,C) \label{algebra}       \\
 \Delta(\,{\mbox{P}}_A \, B \,) &=& 
\sum_{C\cdot D=A} \; {\mbox{P}}_C \, B \ot {\mbox{P}}_D \, B       \;\;
c_B(C,D), \label{coalgebra}
\eea
where  $c$ denotes the 2-cocycle obtained from $\omega$  through  the 
slant product~(\ref{c}) and $\delta_{A,B}$ the kronecker delta function
for the group elements of $H$. 
The 2-cocycle relation~(\ref{tweeko}) now implies that  the 
multiplication~(\ref{algebra}) is associative and, in addition, 
that the comultiplication~(\ref{coalgebra}) is indeed
quasi-coassociative~(\ref{quasicoas}).
By repeated use of the 3-cocycle relation~(\ref{pentagon})
for $\omega$, one also easily verifies the relation
\bea
c_A (C,D) \; c_B (C,D) \; c_C (A,B) \; c_D (A,B)  &=& 
c_{A \cdot B} (C,D) \;  c_{C \cdot D} (A,B),     \label{alamos}
\eea
which, in turn, indicates that the comultiplication~(\ref{coalgebra}) defines 
an algebra morphism from  $D^{\omega}({ H})$ to 
$D^{\omega}({ H})^{\ot 2}$.

As alluded to before, the dyons in the associated 
discrete $H$ gauge theory with Chern-Simons action $\omega$
are labeled by a magnetic flux $A \in H$ paired with an 
unitary irreducible  projective representation of $H$ 
defined as~(\ref{project}).
Thus the complete spectrum can be presented as
\bea         \label{repo}
( \, A, {\alpha} \,),
\eea
where $A$ runs over the different elements of $H$ 
and $\alpha$ over the  related
range of inequivalent projective UIR's of $H$.
This spectrum
constitutes the complete set of inequivalent irreducible representations 
of the quasi-quantum double $D^{\omega}({ H})$.   Let
$^{\alpha}\!v_j$ denote a basis vector in the representation 
space associated with $\alpha$. 
A basis for the internal Hilbert space $V_{\alpha}^A$ assigned
to a particular  dyon $( \, A, {\alpha} \,)$  then becomes 
\bea                 \label{quantumstatescs}
\{|\, A,\,^{\alpha}\!v_j\rangle\}_{j=1,\ldots, 
\mbox{\scriptsize dim}{\, \alpha}} \,.
\eea
The irreducible representation of the quasi-quantum double carried by this 
internal Hilbert space is now given by the action~\cite{dpr} 
\bea \label{13}                                              
\Pi^A_{\alpha}(\, {\mbox{P}}_B \, C \,) \; |\,A ,\,^{\alpha}\!v_j \rangle &=&
\delta_{A,B}\;\; |\,A,\,\alpha(C)_{ij}\,^{\alpha}\!v_i \rangle.
\eea
In other words, the global symmetry transformations $C\in H$ affect
the projective dyon charge $\alpha$ and 
leave the abelian magnetic flux $A$ invariant. 
The projection operator ${\mbox{P}}_B$
subsequently projects out the flux $B \in H$.
Note that although the dyon charges $\alpha$ are projective 
representations of ${ H}$,
the action~(\ref{13}) defines an ordinary representation 
of the quasi-quantum double 
\bea                
\Pi^A_{\alpha}(\, {\mbox{P}}_B \, C \,) \cdot \Pi^A_{\alpha}(\, {\mbox{P}}_D \, E\,)
&=& \Pi^A_{\alpha}( \, {\mbox{P}}_B \, C \cdot {\mbox{P}}_D \, E \,).
\eea

As indicated by our discussion in section~\ref{aa3c},
we may now  distinguish two cases.
Depending on the actual 3-cocycle $\omega$ at hand, 
the 2-cocycle $c_A$ obtained from the slant product~(\ref{c})
is either trivial or nontrivial.  
When $c_A$ is trivial, it can be written as the coboundary~(\ref{repphase}) 
of a 1-cochain or phase factor $\varepsilon_A$. 
This situation occurs for the 2-cocycles $c_A$
related to the 3-cocycles of type~I and~II or products thereof.
From the relations~(\ref{project}) and~(\ref{repphase}),
we then obtain that the inequivalent (trivial) 
projective dyon charge representations are of the form 
\bea                   
{\alpha}(C) &=&             \varepsilon_A(C) \;  \;
 \Gamma^{n^{(1)} \cdots \;  n^{(k)}} (C),    \label{rei}      
\eea
where $\Gamma^{n^{(1)}\cdots \; n^{(k)}}$ denotes an ordinary 
UIR of  ${ H}$
\bea              \label{hrepz}
\Gamma^{n^{(1)} \cdots \; n^{(k)}} (C) 
&=&   \exp \left(\sum_{l=1}^k \frac{2 \pi \im}{N^{(l)}} \, n^{(l)} c^{(l)}
\right).
\eea 
Hence, the projective dyon charge 
representations remain one dimensional in this case.
To be specific, for a 3-cocycle of type~I, the epsilon factor appearing
in the  dyon charge representation~(\ref{rei}) 
is given by~(\ref{epi}), while a 3-cocycle of type~II leads to the 
factor~(\ref{epii}). If we are dealing with a 3-cocycle $\omega$ being 
a product of various 3-cocycles of type~I and~II, then the total
epsilon factor obviously becomes the product of the epsilon factors 
related to the 3-cocycles of type~I and~II constituting the total 
3-cocycle~$\omega$.
The 2-cocycles $c_A$ associated to the 3-cocycles of type~III, 
in contrast, are nontrivial. 
As a consequence, the dyon charges correspond to  
nontrivial  higher dimensional irreducible projective 
representations of ${ H}$, if the total 3-cocycle $\omega$ 
contains a factor of type~III.

The presence of a Chern-Simons action $\omega$ for the gauge group $H$
naturally affects the spins assigned to the dyons 
in the spectrum~(\ref{repo}).
The associated spin factor is determined by considering 
the action of the central
element on the internal quantum state describing a given dyon 
$( \, A, {\alpha} \,)$
\bea \label{spin13}                                              
\Pi^A_{\alpha}(\, \sum_B \; {\mbox{P}}_B \, B \,) \; 
|\,A ,\,^{\alpha}\!v_j \rangle &=&
 |\,A,\,\alpha(A)_{ij}\,^{\alpha}\!v_i \rangle.
\eea   
Upon using~(\ref{project}) and subsequently~(\ref{c}), we infer that 
the matrix $\alpha(A)$ commutes with all other matrices appearing in 
the projective UIR $\alpha$ of ${ H}$
\bea
\alpha(A) \, \cdot \, \alpha(B) 
\;= \; \frac{c_A(A,B)}{c_A(B,A)} \;  \alpha(B) \, \cdot \, \alpha(A)
\; = \; \alpha(B) \, \cdot \, \alpha(A)    \qquad \forall B \in { H}.
\eea 
From Schur's lemma, we then conclude that $\alpha (A)$ is 
proportional to the unit matrix in this 
irreducible projective  representation of $H$
\bea                           \label{anp}
\alpha(A) &=& e^{2 \pi \im s_{(A,\alpha)}} \; {\mbox{\bf 1}}_\alpha,
\eea
where $s_{(A,\alpha)}$ denotes the spin carried by the dyon 
$( \, A, {\alpha} \,)$. Relation~(\ref{anp}), in particular, 
reveals the physical relevance of the epsilon factors
entering the definition~(\ref{rei}) of the dyon charges 
in the presence of Chern-Simons actions of type~I and/or type~II.
Under a rotation over an angle of $2\pi$,
they give rise to an additional spin factor $\varepsilon_A(A)$ 
in the internal quantum state describing a 
dyon carrying the magnetic flux $A$.

The action~(\ref{13}) of the quasi-quantum double is extended to 
two particle states by means of the comultiplication~(\ref{coalgebra}).
In other words, the tensor product 
representation $(\Pi^A_{\alpha} \ot \Pi^B_{\beta}, V_{\alpha}^A \ot 
V_{\beta}^B)$ of $D^\omega(H)$ associated to
a system consisting of the two dyons $(\,A, \alpha \,)$ 
and $(\,B, \beta \,)$ is defined by the action
$\Pi^A_{\alpha} \ot \Pi^B_{\beta}( \Delta (\, {\mbox{P}}_A \, B \, ))$. 
The tensor product representation of the quasi-quantum double related to
a system of three dyons $(\,A, \alpha \,)$, $(\,B, \beta \,)$ and 
$(\,C, \gamma \,)$ may now be defined
either through $(\Delta \ot {\mbox{id}} ) \, \Delta$
or through $({\mbox{id}} \ot \Delta) \, \Delta$.
Let $(V_{\alpha}^A \ot V_{\beta}^B) \ot V_{\gamma}^C$ denote 
the representation space corresponding to $(\Delta \ot {\mbox{id}} ) \, \Delta$
and $V_{\alpha}^A \ot (V_{\beta}^B \ot V_{\gamma}^C)$ the one 
corresponding to $({\mbox{id}} \ot \Delta) \, \Delta$. 
The quasi-coassociativity condition~(\ref{quasicoas})  indicates that
these representations are equivalent. To be precise, their equivalence 
is established by the nontrivial isomorphism or intertwiner
\bea                              \label{fi}     
\Phi : \;  (V_{\alpha}^A \ot V_{\beta}^B) \ot V_{\gamma}^C 
&\longrightarrow&  V_{\alpha}^A \ot (V_{\beta}^B \ot V_{\gamma}^C),
\eea
with
\bea   
\Phi \;  := \; \Pi_{\alpha}^A \! \ot \! \Pi_{\beta}^B \! \ot 
\Pi_{\gamma}^C \,(\varphi) \; = \; \omega^{-1} (A,B,C). \nn
\eea 
Here,  we used~(\ref{isom}) in the last equality sign.
Finally, the 3-cocycle relation~(\ref{pentagon}) 
implies  consistency in rearranging the brackets,
that is, commutativity of the pentagonal diagram~\footnote{Here, 
we use the compact notation $V_A:= V_{\alpha}^A$.}
\[
\ba{ccccc}
\! ((V_A \ot V_B) \ot V_C)  \ot  V_D   \! \! \!
&\stackrel{\scriptscriptstyle{\Phi \ot {\mbox{\tiny \bf 1}}} }{\rightarrow} 
& \! \! \!
(V_A \ot (V_B \ot V_C)) \ot V_D
\!  \! \! &
\stackrel{\scriptscriptstyle{
({\mbox{\tiny id}} \ot \Delta \ot {\mbox{\tiny id}})(\Phi)}}{\longrightarrow} 
& \! \!  \! \!
V_A \ot ((V_B \ot V_C) \ot V_D)
\\
           & & & &            \\
\downarrow {\scriptstyle (\Delta \ot  {\mbox{\scriptsize id}} 
\ot {\mbox{\scriptsize id}})(\Phi)}
& & & & 
\downarrow {\scriptstyle {\mbox{\scriptsize \bf 1}} \ot \Phi}           \\
           & & & &            \\
\! (V_A \ot V_B) \ot (V_C \ot V_D) 
& &\stackrel{({\mbox{\scriptsize id}} \ot 
{\mbox{\scriptsize id}} \ot \Delta)(\Phi)}{\longrightarrow} & &
\! \! \!  V_A \ot (V_B \ot (V_C \ot V_D)).
\ea
\]

To proceed, the definition of the universal $R$-matrix remains the same 
\bea
R & = &  \sum_{C,D} \; {\mbox{P}}_C \ot {\mbox{P}}_D \, C \, ,
\eea
and the  action of the  braid operator
\bea \label{jordicr}
{\cal R}_{\alpha\beta}^{AB} &:=& 
\sigma\circ(\Pi_{\alpha}^A\otimes\Pi_{\beta}^B)( \, R \, ),
\eea  
on the two particle internal Hilbert space $V_{\alpha}^A \ot V_{\beta}^B$
can be summarized as
\bea                 \label{braidaction}
{\cal R} \;| \, A,\, ^{\alpha}\!v_j\rangle
|\, B,\,^{\beta}\!v_l\rangle &=& 
|\,B,\,{\beta}(A)_{ml} \, ^{\beta}\!v_m\rangle 
|\,A,\, ^{\alpha}\!v_j\rangle.
\eea   
It then follows from~(\ref{rei}) and~(\ref{braidaction}) that
the dyons in an abelian discrete $H$ gauge theory endowed with a 
Chern-Simons action of type~I and/or type~II  
obey abelian braid statistics, where the epsilon 
factors~(\ref{epi}) and~(\ref{epii}) represent  
additional Aharonov-Bohm phases generated between the magnetic fluxes.
This picture changes drastically in the presence of a Chern-Simons action 
of type~III.  In that case, the expression~(\ref{braidaction})
indicates that the higher dimensional internal charge 
of a dyon $( \, B,\beta \,)$ picks up an 
Aharonov-Bohm {\em matrix} $\beta(A)$ upon encircling another  remote
dyon $(\, A, \alpha \,)$.
Thus, the introduction of a Chern-Simons action
of type~III in an abelian discrete gauge theory
leads to {\em non}abelian phenomena.  In particular, 
the multi-dyon configurations in such a theory may 
realize nonabelian braid statistics.

The quasitriangularity conditions now involve 
the comultiplication~(\ref{coalgebra}),
the associator~(\ref{fi}) and the braid operator~(\ref{jordicr})
\bea
{\cal R}\: \Delta(\, {\mbox{P}}_A\, B\,)           \label{grcom}
&=& \Delta(\, {\mbox{P}}_A\, B\,) \:{\cal R}  \\
({\mbox{id}} \ot \Delta)({\cal R}) &=& 
\Phi^{-1} \; {\cal R}_2 \; \Phi \; {\cal R}_1 \;
\Phi^{-1}
\label{tria2}       \\
(\Delta \ot {\mbox{id}})({\cal R}) &=& \Phi \; {\cal R}_1 \; \Phi^{-1} \; 
{\cal R}_{2} \; \Phi.
\label{tria1}
\eea
Here, the braid operator ${\cal R}_1$ 
acts as ${\cal R} \ot {\mbox{\bf 1}}$ on the three particle
internal Hilbert space $(V_{\alpha}^A \ot V_{\beta}^B) \ot V_{\gamma}^C$ and 
${\cal R}_2$ as  ${\mbox{\bf 1}} \ot {\cal R}$ on 
$V_{\alpha}^A \ot (V_{\beta}^B \ot V_{\gamma}^C)$.
The relation~(\ref{c}) implies that these conditions 
are indeed satisfied.
The condition~(\ref{grcom}) obviously states that the action of the 
quasi-quantum double commutes with the braid operation, whereas 
the conditions~(\ref{tria2}) and~(\ref{tria1}), in turn, indicate that 
the following hexagonal diagrams commute
\beas
\ba{ccccc} 
V_A \ot (V_B \ot V_C)
& 
\stackrel{\Phi^{-1}}{\rightarrow}
& 
(V_A \ot V_B) \ot V_C   
& 
\stackrel{{\cal R}_1}{\rightarrow}
&
(V_B \ot V_A) \ot V_C  \\
\downarrow {\scriptstyle ({\mbox{\scriptsize id}} 
\ot \Delta) ({\cal R})} & & & & \downarrow 
{\scriptstyle \Phi } \\
(V_B \ot V_C) \ot V_A & 
\stackrel{\Phi^{-1}}{\leftarrow}
& 
V_B \ot (V_C \ot V_A) & 
\stackrel{{\cal R}_2}{\leftarrow}
& V_B \ot (V_A \ot V_C)       \\
& & & & \\
(V_A \ot V_B) \ot V_C       
& 
\stackrel{\Phi}{\rightarrow}   
& 
V_A \ot (V_B \ot V_C)            
& 
\stackrel{{\cal R}_2}{\rightarrow}    
&
V_A \ot (V_C \ot V_B)     \\
\downarrow {\scriptstyle (\Delta \ot {\mbox{\scriptsize id}})({\cal R})} 
& & & & 
\downarrow {\scriptstyle \Phi^{-1}}  \\
V_C \ot (V_A \ot V_B)         & 
\stackrel{\Phi}{\leftarrow}
& 
(V_C \ot V_A) \ot V_B      & 
\stackrel{{\cal R}_1}{\leftarrow}
&   (V_A \ot V_C) \ot V_B \, .
\ea
\eeas    
In other words, these conditions express 
the compatibility of braiding and fusion depicted in figure~\ref{qu1zon}. 
From the complete set of quasitriangularity
conditions, we then infer that instead of the ordinary Yang-Baxter 
equation~(\ref{yobo}), the braid operators now satisfy 
the quasi-Yang-Baxter equation
\bea       \label{qujaba}
{\cal R}_1 \; \Phi^{-1} \, {\cal R}_2 \, \Phi \; {\cal R}_1 &=&
\Phi^{-1} \, {\cal R}_2 \, \Phi \; {\cal R}_1 \; 
\Phi^{-1} \, {\cal R}_2 \, \Phi.
\eea  
Hence, the truncated braid group representations 
(see section~\ref{trunckbr} for the definition of truncated braid groups) 
realized by the multi-dyon
configurations in these discrete Chern-Simons gauge theories 
involve the associator~(\ref{isom}), which takes care of the rearrangement of 
brackets.
Let $(((V_{\alpha_1}^{A_1} \ot V_{\alpha_2}^{A_2})\ot \cdots 
\ot V_{\alpha_{n-1}}^{A_{n-1}}) \ot V_{\alpha_n}^{A_n})$ denote 
an internal Hilbert space for a system of $n$ dyons.
Thus, all left brackets occur at the beginning. 
Depending on the actual nature of the dyons,
this internal Hilbert space then carries a representation 
of an ordinary truncated braid group, a partially colored braid group
or a colored braid group on $n$ strands.
This representation is defined by the formal assignment~\cite{altsc1} 
\bea                                           \label{brareco}
\tau_i & \longmapsto& \Phi_i^{-1}  \; {\cal R}_i \;  \Phi_i   \, ,
\eea
with  $1 \leq i \leq n-1$ and 
\bea 
{\cal R}_i  &:= &
{\mbox{\bf 1}}^{\ot (i-1)} \ot {\cal R} \ot {\mbox{\bf 1}}^{\ot (n-i-1)}   \\
\Phi_i & :=& 
\left(\bigotimes_{i=1}^n \Pi_{\alpha_i}^{A_i}\right) 
(\Delta_L^{i-2} (\varphi) \ot 1^{\ot (n-i-1)}) .    \label{intermi}
\eea
Here, $\varphi$ is the associator~(\ref{isom}), whereas the object 
$\Delta_L$  stands for  the mapping 
\beas
\Delta_L (\, {\mbox{P}}_{C_1} \, D_1  \ot {\mbox{P}}_{C_2} \, D_2
\ot \cdots \ot {\mbox{P}}_{C_m} \, D_m\,) 
& :=& \Delta(\, {\mbox{P}}_{C_1} \, D_1 \, ) \ot {\mbox{P}}_{C_2} \, D_2 \ot 
\cdots \ot {\mbox{P}}_{C_m} \, D_m\, ,
\eeas 
from $D^{\omega}({ H})^{\ot m}$ to $D^{\omega}({ H})^{\ot (m+1)}$ and
$\Delta_L^k$ for the associated mapping from 
$D^{\omega}({ H})^{\ot m}$ to $D^{\omega}({ H})^{\ot (m+k)}$ being the
result of applying $\Delta_L$ $k$ times.
The isomorphism~(\ref{intermi}) 
now  parenthesizes the adjacent internal Hilbert spaces
$V_{\alpha_i}^{A_i}$  and $V_{\alpha_{i+1}}^{A_{i+1}}$ 
and  ${\cal R}_i$ acts as~(\ref{braidaction}) 
on this pair of internal Hilbert spaces.   
At this point, it is important to note that 
the 3-cocycles of type~I and type~II, 
displayed in~(\ref{type1do}) and~(\ref{type2do}), are 
symmetric in the two last entries, i.e.\ $\omega(A,B,C) = \omega(A,C,B)$. 
This implies that the isomorphism $\Phi_i$  
commutes with the braid operation ${\cal R}_i$ for these 3-cocycles. 
A similar observation appears for the 3-cocycles of type~III 
given in~(\ref{type3do}). To start with, 
$\Phi_i$ obviously commutes with ${\cal R}_i$, iff the exchanged dyons
carry the same fluxes, that is, $A_i=A_{i+1}$. Since the 3-cocycles 
of type~III are not symmetric in their last two entries, 
this no longer holds when the particles carry different fluxes
$A_i \neq A_{i+1}$. In this case, however, only the monodromy operation
${\cal R}_i^2$ is relevant, which clearly  commutes with the isomorphism
$\Phi_i$.
The conclusion is that the isomorphism $\Phi_i$ drops out 
of the formal definition~(\ref{brareco}) of the 
truncated braid group representations in 
Chern-Simons theories with an abelian finite gauge group $H$.
It should be stressed, though, 
that this simplification only occurs for abelian 
gauge groups $H$. In Chern-Simons theories with a nonabelian 
finite gauge group, in which the fluxes exhibit flux metamorphosis,
the isomorphism $\Phi_i$ has to be taken into account. 
Finally, the relation~(\ref{grcom}) again extends to internal Hilbert spaces 
for an arbitrary number of dyons and states that the action of the 
quasi-quantum double commutes with the action of the associated 
truncated braid groups.

Let us continue with some  brief remarks  on the fusion rules 
of the quasi-quantum double $D^{\omega}({ H})$
\bea               \label{piet}
\Pi^A_{\al}\otimes\Pi^B_{\beta}& = & \bigoplus_{C , \gamma}
N^{AB\gamma}_{\alpha\beta C} \; \Pi^C_{\gamma}.
\eea
The fusion coefficients are given by~\cite{dpr}
\bea          \label{Ncoef}
N^{AB\gamma}_{\alpha\beta C} &=& \frac{1}{|{ H}|} \, \sum_{D,E}
            \mbox{tr}  \left( \Pi^A_{\alpha} \ot \Pi^B_{\beta}
                      (\Delta (\, {\mbox{P}}_E \, D \, )) \right)  \;
            \mbox{tr} \left(  \Pi^C_{\gamma} (\,{\mbox{P}}_E \,D \,) \right)^* \\
    &=& \delta_{C,A \cdot B} \; \frac{1}{|{ H}|} \,
\sum_D \mbox{tr} \left( \alpha(D) \right) \; 
       \mbox{tr} \left( \beta(D)  \right) \; 
\mbox{tr} \left( \gamma(D) \right)^* \;   c_D(A,B),      \nn
\eea
where $|{ H}|$ denotes the order of 
the abelian group ${ H}$ and $*$ complex conjugation. The Kronecker delta 
appearing here expresses the fact that the various composites, 
which  may result  from fusing the 
dyons $(\, A, \alpha \, )$ and $(\, B, \beta \, )$, carry 
the flux $A \cdot B$, whereas the rest of the formula determines the 
composition rules for the dyon charges $\alpha $ and $\beta$.
Furthermore, the  modular matrices take the form
\bea                                
S^{AB}_{\alpha\beta} &=& \frac{1}{|{ H}|} \, \mbox{tr} \; {\cal 
R}^{-2 \; AB}_{\; \; \; \; \; \alpha\beta} \; = \;\frac{1}{|{ H}|} \,
\mbox{tr} \left(\alpha(B) \right)^* \; \mbox{tr} \left(\beta (A)\right)^*  
\label{fusion}  \\
T^{AB}_{\alpha\beta} &=& 
\delta_{\alpha,\beta} \, \delta^{A,B} \; 
\exp(2\pi \im s_{(A,\alpha)})
 \; = \;  \delta_{\alpha,\beta} \, \delta^{A,B} 
\frac{1}{d_\alpha} \, \mbox{tr} \left( \alpha(A) \right),     \label{modut}
\eea
with $d_\alpha$ the dimension of the projective dyon charge 
representation $\alpha$. These matrices naturally 
satisfy the relations~(\ref{charconj}),   
(\ref{sun}) and~(\ref{tun}), while the fusion rules~(\ref{Ncoef})
can be expressed in terms of the modular $S$
matrix~(\ref{fusion}) by means of Verlinde's 
formula~(\ref{verlindez}).

Finally, it should be noted~\cite{dpr}  
that the deformation of the quantum double 
$D(H)$ into the quasi-quantum double $D^{\omega}({ H})$ only depends on the 
cohomology class of  $\omega$ in $H^3(H,U(1)$, i.e.\ the quasi-quantum double 
$D^{\omega\delta \beta}({ H})$ with $\delta \beta$ a 3-coboundary is 
isomorphic to $D^{\omega}({ H})$.

\sectiona{$U(1)$ Chern-Simons theory}
\label{typeI}  

We turn to an explicit example of a spontaneously broken
Chern-Simons gauge theory, namely  
the planar abelian Higgs model treated in section~\ref{abznz}
equipped with a Chern-Simons term~(\ref{CSt1}) for 
the gauge fields~\cite{spm1,sm,sam}
\bea        \label{action}
S &=& \int d \, ^3x \;
({\cal L}_{\mbox{\scriptsize YMH}} + {\cal L}_{\mbox{\scriptsize matter}} + {\cal L}_{\mbox{\scriptsize CSI}}) \\
{\cal L}_{\mbox{\scriptsize YMH}} &=& 
-\frac{1}{4}F^{\kappa\nu} F_{\kappa\nu} 
  +({\cal D}^\kappa \Phi)^*{\cal D}_\kappa \Phi - V(|\Phi|)  
\label{higgspa} \\
{\cal L}_{\mbox{\scriptsize matter}} &=& -j^{\kappa}A_{\kappa} \label{maco} \\
{\cal L}_{\mbox{\scriptsize CSI}} &=&  
\frac{\mu}{2} \epsilon^{\kappa\nu\tau} A_{\kappa} \partial_{\nu}
A_{\tau},     \label{CStyp1}
\eea
with
\bea                         \label{pot}
V(|\Phi|) &=& \frac{\lambda}{4}(|\Phi|^2-v^2)^2  \qquad\qquad
 \lambda, v > 0.
\eea 
Recall from section~\ref{ahm} that  
the Higgs field $\Phi$ is assumed to carry the charge $Ne$, which 
gives rise to the spontaneous symmetry breakdown  
$U(1) \rightarrow \Z_N$ at the energy scale $M_H=v\sqrt{2\lambda}$. 
Moreover, the matter charges introduced by the current in~(\ref{maco})
are quantized as $q=ne$ with $n\in \Z$.

With the incorporation of the topological Chern-Simons term~(\ref{CStyp1}),
the complete phase diagram for a compact planar $U(1)$
gauge theory endowed with matter now exhibits the following structure.
Depending on the parameters in our model~(\ref{action}) 
and the presence of Dirac monopoles/instantons, 
we can distinguish the phases:
\begin{itemize}
\item $\mu=v=0$ $\Rightarrow$ Coulomb phase.
   The spectrum consists of the quantized matter charges $q=ne$ with Coulomb 
   interactions, where the Coulomb potential depends logarithmically
   on the distances between the charges in two spatial dimensions.
\item $\mu=v=0$ with Dirac monopoles $\Rightarrow$ confining phase.
   As has been shown by Polyakov, the contribution of monopoles to the 
   partition function leads to linear confinement 
   of the quantized charges $q$~\cite{polyakov}.
\item $v \neq 0, \,  \mu = 0$ $\Rightarrow$ $\Z_N$ Higgs phase. 
   The spectrum  consists of screened matter charges $q=ne$, magnetic 
   fluxes quantized as $\phi= \frac{2\pi a}{Ne}$ with $a \in \Z$
   and dyonic combinations. 
   The long range interactions are topological Aharonov-Bohm 
   interactions under which the  charges and fluxes become $\Z_N$ 
   quantum numbers. In the presence of Dirac monopoles, 
   magnetic flux $a$ is  conserved modulo $N$.  
   (See section~\ref{abznz} and section~\ref{topclas}).
\item $v=0, \, \mu \neq 0 $ $\Rightarrow$  Chern-Simons electrodynamics. 
   The gauge fields 
   carry the topological mass $|\mu|$. 
   The charges $q=ne$ constituting the spectrum are screened by induced 
   magnetic fluxes $\phi= -q/\mu$. 
   The long range interactions between the matter charges
   are Aharonov-Bohm interactions with coupling constant $\sim 1/\mu$. 
   It has been argued that the presence 
   of Dirac monopoles does {\em not} lead to confinement~\cite{pisar,affleck} 
   of the matter charges in this massive Chern-Simons phase. 
   Instead, it implies that the topological mass is quantized as
   $\frac{pe^2}{\pi}$ with $p \in \Z$. Moreover, the Dirac monopoles
   now describe tunneling events between particles with charge difference 
   $\Delta q = 2pe$, with $p$ the integral Chern-Simons parameter.
   Thus, the spectrum only contains a total number of $2p-1$ 
   distinct stable charges in this case.
\item $v \neq 0, \, \mu \neq 0$ $\Rightarrow$ $\Z_N$ Chern-Simons Higgs phase.
   Again, the spectrum features screened matter charges $q=ne$, magnetic
   fluxes quantized as $\phi= \frac{2\pi a}{Ne}$ with $a \in \Z$
   and dyonic combinations. In this phase, 
   we have the conventional long range
   Aharonov-Bohm 
   interaction $\exp (\im q \phi)$ between charges and fluxes, and,
   in addition, Aharonov-Bohm interactions $\exp (\im \mu \phi \phi')$
   between the fluxes themselves~\cite{spm1,sam}. 
   Under these interactions, the charges 
   then obviously remain  $\Z_N$ quantum numbers, whereas a compactification
   of the magnetic flux quantum numbers only occurs for fractional values
   of the topological mass $\mu$~\cite{spm1}. In particular, the aforementioned
   quantization of the topological mass required in the presence of Dirac 
   monopoles renders the magnetic fluxes to be $\Z_N$ quantum numbers.
   The flux tunneling $\Delta a = -N$ induced by the minimal 
   Dirac monopole is now accompanied by a charge jump $\Delta n = 2p$, with
   $p$ the integral Chern-Simons parameter. Finally, as implied by the 
   homomorphism~(\ref{homoI}) for this case, the Chern-Simons parameter
   becomes periodic in this broken phase, that is, there are just
   $N-1$ distinct $\Z_N$ Chern-Simons Higgs phases in which both charges
   and fluxes are $\Z_N$ quantum numbers~\cite{spm1,sm}.
\end{itemize}

Here,  we focus on the phases summarized in the last two items. 
The discussion is organized as follows.
Section~\ref{ub} contains a brief exposition of 
Chern-Simons electrodynamics featuring Dirac monopoles.
In section~\ref{bp}, we then turn to  the Chern-Simons Higgs 
screening mechanism for the electromagnetic fields generated by  
the matter charges and the magnetic vortices in the broken phase
and establish the above mentioned long range 
Aharonov-Bohm interactions between these particles.
To conclude, a detailed discussion of the 
discrete $\Z_N$ Chern-Simons gauge theory 
describing the long distance physics in the broken phase is presented
in section~\ref{rev}.

\subsection{Dirac monopoles and topological mass quantization}
\label{ub}

Let us begin by briefly recalling the basic features of Chern-Simons 
electrodynamics, i.e.\ we set the symmetry breaking scale 
in our model~(\ref{action}) to zero for the moment ($v=0$) and 
take $\mu \neq 0$. 
Varying the action~(\ref{action}) w.r.t.\ the vector 
potential $A_{\kappa}$ yields the field equations
\bea                      \label{fieldequation}
\partial_{\nu} F^{\nu\kappa} + 
\mu\epsilon^{\kappa\nu\tau} \partial_{\nu}A_{\tau} &=& j^\kappa+j^\kappa_H \, , 
\eea   
where $j^{\kappa}_H$ denotes the Higgs current~(\ref{Higgscur}) and 
$j^{\kappa}$ the minimally coupled matter current in~(\ref{maco}).
These field equations indicate that the gauge fields are massive.
To be precise, this model features a single component 
photon, which carries the topological mass $|\mu|$~\cite{des1}.
In other words, the electromagnetic fields generated by the currents 
in~(\ref{fieldequation}) are screened, that is, they fall 
off exponentially with mass $|\mu|$. Hence, at distances 
$\gg 1/|\mu|$ 
the Maxwell term in~(\ref{fieldequation}) can be neglected, which reveals 
how the screening mechanism operating in Chern-Simons 
electrodynamics works. 
The currents $j^{\kappa}$ and $j^{\kappa}_H$  induce 
magnetic flux currents 
$-\frac{1}{2}\epsilon^{\kappa\nu\tau} \partial_{\nu} A_{\tau}$
exactly screening the electromagnetic fields 
generated by $j^{\kappa}$ and $j^{\kappa}_H$. 
Specifically, from Gauss' law
\be                      \label{gauss}
Q=q +q_H+\mu \phi=0,
\ee  
with $Q=\int\! d\, ^2x\, \nabla \!\cdot\! {\mbox{\bf E}} =0$, 
$q=\int\! d\, ^2x \, j^0 $, $q_H=\int\! d\,^2x \, j^0_H $ 
and $\phi = \int \! d\,^2x \,\epsilon^{ij}\partial_i A^j$,
we learn  that the Chern-Simons  screening  mechanism 
attaches fluxes $\phi=-q/\mu$ and 
$\phi_H=-q_H/\mu$ of characteristic size $1/|\mu|$ to the
point charges $q$ and $q_H$ respectively~\cite{des1}.

The remaining long range interactions between these screened  
charge $q$ are the topological  Aharonov-Bohm  interactions 
implied by the matter coupling~(\ref{maco}) and the Chern-Simons 
coupling~(\ref{CStyp1})~\cite{goldmac}
\bea           \label{ons}
{\cal R}^2 \; |q\rangle|q'\rangle &=&
e^{-\im \frac{q q'}{\mu}}|q\rangle|q'\rangle \\
{\cal R} \; |q\rangle |q\rangle &=& e^{-\im \frac{q q}{2\mu}} \, 
|q\rangle|q\rangle.     \label{onsan}
\eea
Thus the particles in this theory realize abelian braid statistics.
In particular, identical particle configurations exhibit
anyon statistics  with quantum statistical 
parameter~(\ref{onsan}) 
depending on the specific charge of the particles and the inverse of the 
topological mass $\mu$.

\begin{figure}[htb] 
\begin{center}
\begin{picture}(125,125)(-60,-60)
\put(-60,0){\line(1,0){120}}
\put(0,-60){\line(0,1){120}}
\thinlines
\multiput(-60,0)(10,0){13}{\line(0,1){2}}
\multiput(0,-60)(0,10){13}{\line(1,0){2}}
\multiput(-30,60)(20,-40){4}{\circle*{2.0}}
\multiput(-20,40)(20,-40){3}{\circle*{2.0}}
\put(-20,40){\circle{3.6}}
\put(-20,40){\vector(1,-2){20}} 
\put(5,50){\vector(0,1){5}}
\put(43,-3){\vector(1,0){5}}
\put(8,55){\small$\phi[\frac{\pi}{2e}]$}
\put(50,-6){\small$q[e]$}
\end{picture}
\vspace{0.5cm}
\caption{\sl Spectrum of unbroken $U(1)$ Chern-Simons theory.
We depict the flux $\phi$ 
versus the global $U(1)$  charge $q$.
The Chern-Simons parameter $\mu$ is set to its minimal nontrivial value 
$\mu=\frac{e^2}{\pi}$, i.e.\ $p=1$. 
The arrow represents the effect of a {\em charged} Dirac 
monopole/instanton, which shows that there is just one stable particle
in this theory.}
\label{u1}
\end{center}  
\end{figure}
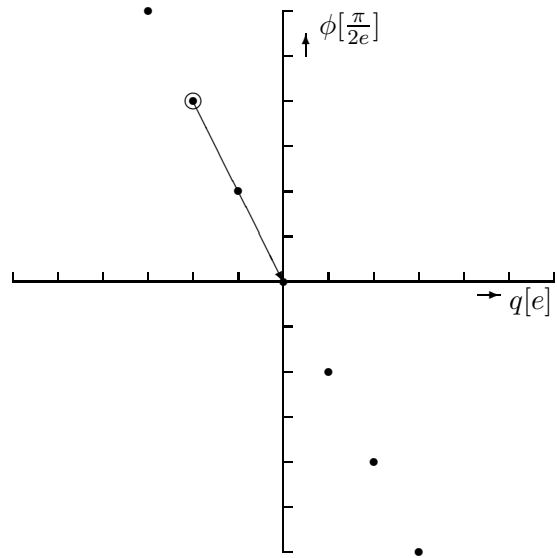

We now suppose that this compact $U(1)$ gauge  theory contains
Dirac monopoles carrying magnetic charges quantized
as $g=\frac{2\pi m}{e}$ with $m \in \Z$.
In this 2+1 dimensional model,
these monopoles become instantons, which correspond to tunneling events 
between states with flux difference, and, 
as result of  Gauss' law~(\ref{gauss}), also charge 
difference~\cite{hen,affleck,pisar,klee}. To be explicit, the minimal
Dirac monopole induces the tunneling 
\bea                     \label{dcharge}  
\mbox{instanton:} & &
\left\{ \ba{lcl}    
\Delta \phi &=& -\frac{2\pi}{e} \\
\Delta q &=& \mu \frac{2\pi}{e}.  
\ea
\right.
\eea
A consistent implementation of these 
Dirac monopoles requires the  quantization of 
the matter charges $q$ in multiples of $e$, 
and, as a direct consequence, quantization of the topological mass $\mu$.  
Dirac's argument~\cite{dirac} works
also in the presence of a Chern-Simons term. 
In this case, the argument goes as follows. 
The tunneling event~(\ref{dcharge}) 
corresponding to the  minimal Dirac monopole should 
be invisible to the monodromies~(\ref{ons}) 
with the charges $q$ present in our model.
In other words, the Aharonov-Bohm phase 
$\exp (-\im \frac{ q \Delta q}{\mu})=\exp ( -\im \frac{2\pi q}{e})$
should be trivial, 
which implies the charge quantization $q=ne$ with $n \in \Z$.
Furthermore, the tunneling event~(\ref{dcharge}) 
should respect this quantization rule for $q$,
that is, the charge jump has to be a multiple of $e$: 
$\Delta q = \mu \frac{2\pi}{e}=pe$ with $p \in \Z$, 
which leads to the quantization $\mu = \frac{pe^2}{2\pi}$.
There is, however, a further restriction on the values of 
the topological mass $\mu$.
So far, we have only considered the monodromies in this theory, 
but the particles connected by Dirac monopoles
should as a matter of course also have the same spin factor or equivalently 
the same quantum statistical 
parameter~(\ref{onsan}). In particular, the spin factor
for the charge $\Delta q$ connected to the vacuum $q=0$ should be trivial: 
$\exp (-\im \frac{(\Delta q)^2}{2\mu})
=\exp (-\im \frac{\mu}{2} (\frac{2\pi}{e})^2)=1$. The conclusion is that 
in the presence of Dirac monopoles the 
topological mass is necessarily quantized as
\bea                 \label{mu}
\mu &=& \frac{pe^2}{\pi}   \qquad \qquad \mbox{with $p \in \Z$},
\eea
which is the result alluded to in~(\ref{quantmui}).
The observation that the presence of Dirac monopoles
implies quantization of the topological mass $\mu$ 
was first made by Henneaux and Teitelboim~\cite{hen}. 
However, they only used the monodromy part of the above argument and 
did not implement the demand that the particles connected by 
Dirac monopoles should give rise the same
spin factor. As a consequence, they  
arrived at the erroneous finer
quantization $\mu=\frac{pe^2}{2\pi}$. Subsequently, Pisarski 
derived the correct quantization~(\ref{mu}) by considering 
gauge transformations in the background of a Dirac monopole~\cite{pisar}.

To conclude,
as indicated by~(\ref{dcharge})
and~(\ref{mu}), the Dirac monopoles  drive a modulo 
$2p$ calculus for the quantized charges $q=ne$, with 
$p$ being the integral Chern-Simons parameter. 
Thus, the spectrum of  this unbroken $U(1)$ 
Chern-Simons theory consists of $2p-1$ stable charges  
$q=ne$ screened by the induced magnetic 
fluxes $\phi=-q/\mu$ (see for example~\cite{sm,klee,moorseib}).
We have depicted this spectrum for $p=1$ 
in figure~\ref{u1}.

\subsection{Dynamics of the Chern-Simons Higgs medium}
\label{bp}

We continue with an analysis of  the Higgs 
phase of the model~(\ref{action}), i.e.\  we  set $v \neq 0$ and take the 
topological mass $\mu$ to be nonvanishing.
The discussion is kept general, which means that 
the topological mass $\mu$  may take any real value in this section. 
The incorporation of Dirac monopoles in this phase, which requires the 
quantization~(\ref{mu}) of the topological mass,
will be discussed in the next section.

As we have seen in section~\ref{ahm},
at energies well below the symmetry breaking scale 
$M_H = v \sqrt{2\lambda}$, 
the Higgs part~(\ref{higgspa}) of the action reduces to 
\bea                               \label{efhi}
{\cal L}_{\mbox{\scriptsize YMH}} & \longmapsto & -\frac{1}{4}F^{\kappa\nu} F_{\kappa\nu} 
+\frac{M_A^2}{2} \tilde A^{\kappa}\tilde A_{\kappa}    \label{mhed} \\
\tilde{A}_{\kappa} & := & A_{\kappa} + \frac{1}{Ne}\partial_{\kappa}
\sigma
\label{Atilde}  \\
 M_A & := & Ne v\sqrt{2},     \label{ma}
\eea
with $\sigma$ the charged Goldstone boson field.
Hence, in the low energy regime, 
to which we confine ourselves,
our model  is governed by the effective action 
obtained from substituting~(\ref{efhi}) in~(\ref{action}).
The field equations which  follow from varying the effective
action w.r.t.\  $A_\kappa$ and the Goldstone boson $\sigma$
respectively, then become    
\bea                      \label{fequus}
\partial_\nu F^{\nu\kappa} + 
\mu\epsilon^{\kappa\nu\tau} \partial_{\nu}A_{\tau} &=& 
j^\kappa+j^\kappa_{\mbox{\scriptsize scr}}                         \\
\partial_\kappa  j^\kappa_{\mbox{\scriptsize scr}} &=& 0,
\eea
where the Higgs current~(\ref{Higgscur}) again boils down to the 
screening current
\bea                                \label{scrcur}
j^\kappa_{\mbox{\scriptsize scr}} &=& - M_A^2 \tilde{A}^\kappa.
\eea
From these equations, it is readily inferred that the two 
polarizations $+$ and $-$ of the photon field $\tilde{A}_\kappa$
now carry the masses~\cite{pisa}
\bea
M_{\pm} &=& \sqrt{M_A^2 + \frac{1}{2}\mu^2 \pm \frac{1}{2}\mu^2\sqrt{
\frac{4 M_A^2}{\mu^2}+1}},
\label{mass}
\eea
which differ by the topological mass~$|\mu|$. Note that by setting 
$\mu = 0$ in~(\ref{mass}), we  restore the 
fact that in the  ordinary Higgs phase 
both polarizations of the photon carry the same mass 
$M_+ = M_-=M_A$ (see section~\ref{ahm}). 
Taking the limit $v \rightarrow 0$,
on the other hand, yields $M_+=|\mu|$ and $M_-=0$. The $-$ component then
ceases to be a physical degree of freedom~\cite{pisa} and we  
recover the fact that  unbroken Chern-Simons electrodynamics features
a single component photon with mass $|\mu|$.

There are  now two different types of sources for electromagnetic 
fields in this Chern-Simons Higgs medium: the 
quantized point charges $q=ne$ introduced 
by the matter current $j^\kappa$ and 
the vortices of characteristic size $1/M_H$ carrying 
quantized magnetic flux $\phi=\frac{2\pi a}{Ne}$ with $a\in \Z$.
The latter enter the 
field equations~(\ref{fequus}) by means of the 
flux current $-\frac{1}{2}\epsilon^{\kappa\nu\tau} \partial_{\nu}A_{\tau}$.
The field equations~(\ref{fequus}) then show  that 
both the matter current and the flux current generate electromagnetic 
fields, which are screened at large distances by an 
induced current $j^\kappa_{\mbox{\scriptsize scr}}$ in the 
Chern-Simons Higgs medium~\cite{sam}.
This becomes clear from Gauss' law  for this case
\bea                    \label{higgsgauss}
Q=q +  q_{\mbox{\scriptsize scr}}  +\mu \phi  =0,
\eea
with
\bea 
q_{\mbox{\scriptsize scr}}  =  \int d \, ^2x \, j^0_{\mbox{\scriptsize scr}} = - \int d \, ^2x \, 
M_A^2 \tilde{A}^0,
\eea 
which implies that both 
the  matter charges $q$ and the  magnetic vortices $\phi$
are surrounded by localized screening charge densities $j^0_{\mbox{\scriptsize scr}}$.
At large distances, the contribution to the long range Coulomb fields 
of the induced screening charges 
\bea               \label{scherm}     \ba{rcl}
q=ne        &\Rightarrow& q_{\mbox{\scriptsize scr}} =-q    \\
\phi=\frac{2\pi a}{Ne}   &\Rightarrow& q_{\mbox{\scriptsize scr}} = -\mu \phi,
\ea
\eea 
then completely cancel those of the matter charges $q$ and the fluxes 
$\phi$ respectively.
Here, it is of course understood that the screening charge density 
$j^0_{\mbox{\scriptsize scr}}$ accompanying a magnetic vortex 
is localized in a ring outside the core, 
since inside the core the Higgs field vanishes and the 
Chern-Simons Higgs medium is destroyed. 
Let us also stress that just as in the ordinary Higgs medium 
(see section~\ref{ahm}) the matter charges $q$ are screened by 
charges $q_{\mbox{\scriptsize scr}}=-q$ provided by the Higgs condensate  
in this Chern-Simons Higgs medium 
and {\em not} by attaching fluxes to them as in the case of 
unbroken Chern-Simons electrodynamics. This is already 
apparent from the fact that the irrational `screening' fluxes 
$\phi=-q/\mu$ would render the Higgs condensate multi-valued.

Recall from our discussion in section~\ref{mavoab} that the 
induced screening charges~(\ref{scherm}) do not couple to the 
long range Aharonov-Bohm interactions~\cite{sam}. 
Hence, taking a screened charge $q$ around a screened magnetic flux 
$\phi$ gives rise to the conventional Aharonov-Bohm phase 
\bea          \label{klam}
{\cal R}^2 \; |q \rangle |\phi  \rangle &=& e^{\im q \phi } \;
|q \rangle |\phi  \rangle,  
\eea 
as implied by the coupling~(\ref{maco}). 
This summarizes the remaining long range interactions for the matter charges.
That is, in contrast with unbroken Chern-Simons electrodynamics, 
there are no long range Aharonov-Bohm interactions between 
the matter charges themselves in this broken phase.
Instead, we now obtain nontrivial Aharonov-Bohm interactions among the 
screened magnetic fluxes
\bea        \label{fluxAB}
{\cal R}^2 \; |\phi \rangle |\phi ' \rangle &=& e^{\im \mu \phi \phi '} \;
|\phi \rangle |\phi ' \rangle   \\
{\cal R} \; |\phi \rangle |\phi  \rangle &=& 
e^{\im \frac{\mu}{2} \phi \phi} \;
|\phi \rangle |\phi \rangle,     \label{qs}
\eea
entirely due to the Chern-Simons coupling~(\ref{CStyp1}). 
From~(\ref{qs}), we conclude that depending on their flux 
and the topological mass, identical magnetic vortices 
realize anyon statistics.

In retrospect, the basic characteristics of the Chern-Simons Higgs 
screening mechanism uncovered in~\cite{sam} 
and briefly outlined above find their confirmation 
in results established in earlier  studies of the 
static magnetic vortex solutions of the abelian Chern-Simons Higgs model.
In fact, the analysis of the properties of these 
so-called Chern-Simons vortices was started 
by Paul and Khare~\cite{pakh}, who noted that they correspond to finite 
energy solutions carrying both magnetic flux and electric charge. 
Subsequently, various authors have obtained both analytical and numerical
results on these static vortex solutions. 
See for example~\cite{boya}, \cite{hong}-\cite{jacko}, 
\cite{vega} and for a review~\cite{boyarev}.
Here, we just collect the main results.
In general, one takes the following Ansatz for a static 
vortex solution of the field equations corresponding to~(\ref{action})
\bea
\Phi(r,\theta) &=& \rho(r) \exp \left(\im \sigma(\theta) \right)
 \label{higans} \\
A_0(r,\theta) &=& A_0(r)                             \\
A_i (r,\theta)&=& -A(r) \partial_i \sigma(\theta).    \label{jawel}
\eea
Here, $r$ and $\theta$ denote the polar coordinates 
and $\sigma$ the multi-valued Goldstone boson 
\bea
\sigma(\theta+ 2\pi) - \sigma(\theta) &=& 2\pi a,   \label{mvs}
\eea  
with $a \in \Z$ to render the Higgs field itself single valued.
Regularity of the solution imposes the following boundary conditions as 
$r \rightarrow 0$ 
\bea
\ba{lll}
\rho \rightarrow 0, \qquad  &  A_0 \rightarrow {\mbox{constant}}, \qquad &  
A \rightarrow 0,
\ea
\eea  
whereas, for finite energy, 
the asymptotical behavior for $r \rightarrow \infty$  becomes 
\bea                                         \label{asy}
\ba{lll}
\rho \rightarrow v, \qquad   &  A_0 \rightarrow 0, \qquad &  
A \rightarrow \frac{1}{Ne}.
\ea
\eea 
From~(\ref{asy}),~(\ref{jawel}) and~(\ref{mvs}) it then follows that this 
solution corresponds to  the quantized magnetic flux                      
\bea         \label{quflux}
\phi \; = \; 
\oint dl^i A^i(r=\infty) \;= \; 
\frac{1}{Ne} \oint  dl^i\partial_i \sigma  \; = \; 
\frac{2\pi a}{Ne} \qquad \mbox{with $a \in \Z$.}
\eea     
Since the two polarizations of the photon carry 
distinct masses~(\ref{mass}),
it seems, at first sight, 
that there are two different  vortex solutions 
corresponding to a long range exponential decay of 
the electromagnetic fields either with mass $M_-$ or with mass $M_+$. 
However, a careful analysis~\cite{ino} (see also~\cite{boya}) of the 
differential equations following from the field equations with this Ansatz
shows that the $M_+$ solution does not exist for 
finite $r$. Hence,  we are left with the $M_-$ solution.
To proceed, it turns out that 
the modulus $\rho$ of the Higgs field~(\ref{higans})
grows monotonically from zero (at $r=0$) to its 
asymptotic ground state value~(\ref{asy}) at $r=1/M_H$, where the profile 
of this growth does not change much in the full range of the 
parameters (see for example~\cite{boya}).

\begin{figure}[p]    \epsfxsize=12.5cm
\centerline{\epsffile{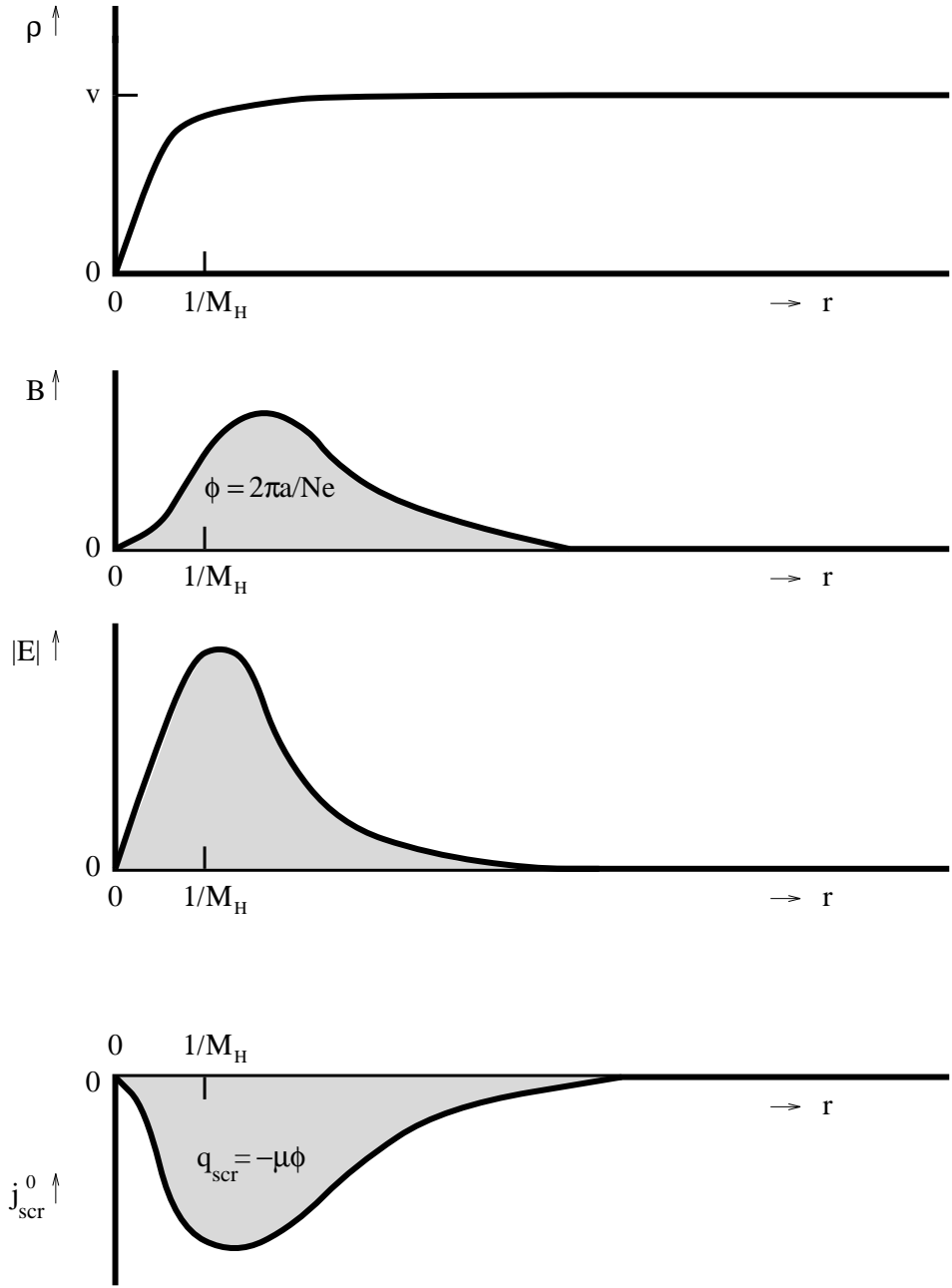}}
\caption{\sl Qualitative behavior of the vortex solution carrying 
the quantized magnetic flux $\phi=\frac{2\pi a}{Ne}$ 
in the Chern-Simons limit.  
We have respectively depicted the modulus of 
the Higgs field $\rho$, the magnetic
field $B$, the electric field $|{\mbox{\bf E}}|$ 
and the screening charge density $j^0_{\mbox{\scriptsize scr}}=
-2(Ne)^2 \rho^2 A^0$ 
versus the radius $r$. 
The electromagnetic fields and the screening charge density vanish at $r=0$,
reach there maximal value 
outside the core at $r=1/M_H$ and subsequently drop off exponentially 
with mass $M_-$ at larger distances.}
\label{qCShigg}
\end{figure}

An important issue is, of course, 
whether vortices will actually form or not, that is, 
whether the superconductor we are describing is type~II or I
respectively. In this context, the competition between  
the penetration depth $1/M_-$ of the electromagnetic fields
and the coresize $1/M_H$ becomes important.
In ordinary superconductors ($\mu=0$), 
an evaluation of the free energy yields that we are dealing 
with a type~II superconductor if $M_H/M_A=\sqrt{\lambda}/Ne \geq 1$, 
and a type~I superconductor otherwise~\cite{gennes}.
Since $M_-$ is smaller then $M_A$, it is expected that in the presence 
of a Chern-Simons term the type~II region is extended. A perturbative  
analysis for small topological mass $\mu$ shows that this is indeed the
case~\cite{jacobs}.
In the following, we will always 
assume that our parameters are adjusted  
such that we are in the type~II region.

Let us now briefly recall the structure of the electromagnetic 
fields of the vortex solution in the full range of parameters. 
To start with, 
the distribution of the magnetic field $B=\partial_1 A^2-\partial_2 A^1$ 
strongly depends on the topological mass $\mu$. 
For $\mu=0$,  we  are dealing with 
the Abrikosov-Nielsen-Olesen vortex discussed in section~\ref{mavoab}.
In that case, the magnetic field reaches its maximal value 
at the center ($r=0$) of the vortex and 
drops off exponentially with mass $M_A$ at distances $r > 1/M_H$.
For $\mu \neq 0$, the magnetic field 
then decays exponentially with mass $M_-$ at distances $r > 1/M_H$.
Moreover, as the topological mass $|\mu|$ increases from zero, 
the magnetic field at the origin $r=0$ diminishes 
until it completely vanishes in the 
so-called Chern-Simons limit: $e,|\mu| \rightarrow\infty$, 
with fixed ratio $e^2/\mu$~\cite{boya}. 
(Note that in case the topological mass $\mu$ is quantized as~(\ref{mu}), 
this limit simply means $e \rightarrow \infty$ leaving 
the Chern-Simons parameter $p$ fixed). 
Hence, in the Chern-Simons limit, which amounts to 
neglecting the Maxwell term in~(\ref{higgspa}), the magnetic 
field is localized in a ring-shaped region around 
the core at $r=1/M_H$, as depicted
in figure~\ref{qCShigg}, see~\cite{boya,hong,jacobs,jacko,jakh}. 
To proceed, as indicated by the zeroth component of the 
field equation~(\ref{fieldequation}), a magnetic field distribution 
$B$ generates an electric field distribution ${\mbox{\bf E}}$ 
iff $\mu \neq 0$.  
These electric fields are localized in a ring shaped region around 
the core at $1/M_H$ for all values of $\mu \neq 0$. 
To be specific, 
they vanish at $r=0$ and fall off with mass $M_-$ at distances 
$r>1/M_H$.
We have seen in~(\ref{scherm}) how these electric fields, 
induced by the magnetic field of the vortex, 
are screened by the Chern-Simons Higgs medium occurring at $r>1/M_H$. 
A screening charge density $j_{\mbox{\scriptsize scr}}^0$
develops in the neighborhood of the core of the vortex,
which falls off with mass $M_-$. In this static case, the screening 
charge density boils down to $j^0_{\mbox{\scriptsize scr}} = -  M_A^2 A^0$, 
that is, the Goldstone boson does not contribute. 
The analytical and numerical evaluations in for 
example~\cite{boya,hong,jacobs,jacko,jakh} 
show that the distribution of $A_0$ is indeed of 
the shape described above.

The spin that can be calculated for this classical 
Chern-Simons vortex solution takes the 
value (e.g.\ \cite{boya,hong,jacobs,jacko,jakh}) 
\bea           \label{spinajax}
s \; = \; \int d\,^2 x \, \epsilon^{ij} x^i \, T^{0j} \;  = 
\; \frac{\mu \phi^2}{4\pi},
\eea
where $T^{0j}$ denotes the energy momentum tensor. 
Note that this spin value is consistent with the quantum statistical 
parameter~(\ref{qs}). That is, these vortices satisfy the spin statistics
connection $\exp({\im \Theta}) = \exp({2\pi \im s})$.
This is actually a good point to resolve some inaccuracies in the 
literature. It  is often stated (see for example~\cite{boya,hong})
that it is the fact that the Chern-Simons vortices carry 
the charge~(\ref{scherm}) which leads to nontrivial
Aharonov-Bohm interactions between these vortices. 
As we have argued, however, the screening 
charges $q_{\mbox{\scriptsize scr}}$ do not couple to the Aharonov-Bohm 
interactions~\cite{sam} and the  phases in~(\ref{fluxAB}),~(\ref{qs}) 
are entirely due to the Chern-Simons term~(\ref{CStyp1}).
In fact, erroneously assuming that the screening charges 
accompanying the vortices do couple to the Aharonov-Bohm 
interactions leads to the quantum statistical parameter 
$\exp(-\im \mu \phi^2/2)$, which is inconsistent with 
the spin~(\ref{spinajax}) carried by these vortices. 
In this respect, we  remark  that 
the correct quantum statistical parameter~(\ref{qs}) for the vortices 
has also been derived in the dual formulation of this model~\cite{kleep}.

To our knowledge, the nature of the static point charge solutions 
$j=(q \delta({\mbox{\bf x}}), 0,0)$ of the field equations~(\ref{fequus}) 
have not been studied in the literature so far. 
An interesting question in this context is with which mass~(\ref{mass})
the electromagnetic fields fall off around these  matter charges. 
We conjecture that this exponential decay corresponds to the mass $M_+$. 
The overall picture then becomes that the magnetic vortices $\phi$ 
excite the $-$~polarization of the massive 
photon in the Chern-Simons Higgs medium, 
whereas the $+$~polarization is excited around the matter charges $q$.

\subsection{ $\Z_N$ Chern-Simons  theory}
\label{rev}

Here, we turn to the incorporation of Dirac monopoles in the 
$\Z_N$ Chern-Simons Higgs phase discussed in the previous section.
In other words, the topological mass is quantized as~(\ref{mu}) in the 
following. We will argue that with this particular quantization,
the $\Z_N$ Chern Simons theory describing 
the long distance physics in this Higgs phase corresponds to the 
3-cocycle $\omega_I$ determined by the homomorphism~(\ref{homoI}) 
for this case~\cite{spm1,sm}.

As we have seen in the previous section, the Higgs mechanism causes the 
identification of charge and flux occurring in unbroken Chern-Simons 
electrodynamics to disappear.
That is, the  spectrum of the $\Z_N$ Chern-Simons Higgs phase consists
of the quantized matter charges $q=ne$, the quantized magnetic fluxes 
$\phi=\frac{2\pi a}{Ne}$ and dyonic combinations of the two. 
We will label these particles as $(a,n)$ with  $a,n \in \Z$.
Upon implementing the quantization~(\ref{mass}) 
of the topological mass $\mu$, the Aharonov-Bohm 
interactions~(\ref{klam}), (\ref{fluxAB}) and~(\ref{qs}) 
can  be cast in the form
\bea                      \label{monoI}
{\cal R}^2\;|a,n\rangle |a',n'\rangle &=&
e^{\frac{2 \pi \im}{N} (n a' +n'a
+\frac{2p}{N}aa')}
\;|a,n\rangle |a',n'\rangle \\    \label{bradenI}
{\cal R} \; |a,n\rangle |a,n\rangle &=&
e^{\frac{2 \pi \im}{N} (n a +\frac{p}{N}aa)}
\;|a,n\rangle |a,n\rangle \\ \label{sfI}
T \;  |a,n\rangle &=& e^{\frac{2 \pi \im}{N}(na + \frac{p}{N}aa)} 
\; |a,n \rangle. 
\eea 
Here, $p$ denotes the integral Chern-Simons parameter, while
expression~(\ref{sfI}) contains the spins assigned 
to the particles. Under these remaining long range topological 
interactions, the charge label $n$ obviously becomes a $\Z_N$ quantum 
number, i.e.\ at large distances 
we are only able to distinguish the charges $n$ modulo $N$.
Furthermore, in the presence of the Dirac monopoles/instantons~(\ref{dcharge}) 
magnetic flux $a$ is conserved modulo $N$. However, the 
flux decay events are now accompanied by charge creation~\cite{spm,sm}.
To be specific, in terms of the integral 
flux and charge quantum numbers $a$ and $n$,
the tunneling event induced by the minimal Dirac monopole
can be recapitulated as 
\bea                            \label{instanz}
\mbox{instanton:}  & &      \left\{ \ba{lcl}    
a &\mapsto& a -N  \\
n &\mapsto& n+2p.  
\ea
\right.          
\eea
We have depicted this effect of a Dirac monopole in the spectrum 
of a $\Z_4$ Chern-Simons Higgs phase in figure~\ref{z41}.
Recall from section~\ref{ub} that  
the quantization~(\ref{mu}) of the topological mass  was such that 
the particles connected by monopoles were invisible to the 
monodromies~(\ref{ons}) and carried the same spin
in the unbroken phase.
This feature naturally  persists in this broken phase.
It is readily checked that the particles connected by 
the monopole~(\ref{instanz}) can not be distinguished 
by the Aharonov-Bohm interactions~(\ref{monoI}) and give rise to the same 
spin factor~(\ref{sfI}). 
As a result, the spectrum of this broken phase can be presented as
\bea                              \label{compsp}
(a,n)  \qquad \qquad \qquad \mbox{with} \qquad a,n \in 0,1, \ldots, N-1,
\eea 
where it is understood that the modulo $N$ calculus for the magnetic 
fluxes $a$ involves the charge jump~(\ref{instanz}).

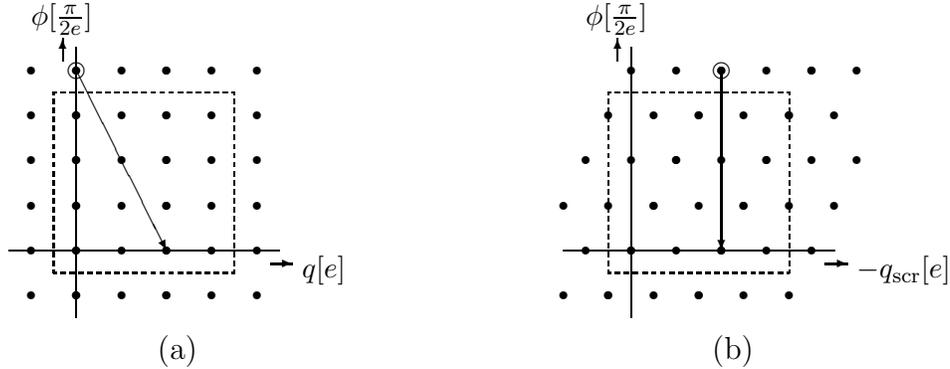
\begin{figure}[tbh] 
\begin{center}
\begin{picture}(85,80)(-15,-15)
\put(-5,-5){\dashbox(40,40)[t]{}}
\put(-15,0){\line(1,0){60}}
\put(0,-15){\line(0,1){60}}
\thinlines
\multiput(-10,-10)(0,10){6}{\multiput(0,0)(10,0){5}{\circle*{2.0}}}
\multiput(0,-10)(0,10){6}{\multiput(0,0)(20,0){3}{\circle*{2.0}}}
\put(0,40){\circle{3.6}}
\put(0,40){\vector(1,-2){20}} 
\put(-3,42){\vector(0,1){5}}
\put(43,-3){\vector(1,0){5}}
\put(-10,50){\small$\phi[\frac{\pi}{2e}]$}
\put(50,-6){\small$q[e]$}
\put(18,-24){(a)}
\end{picture}
\hspace{2cm}
\begin{picture}(85,80)(-15,-15)
\put(-5,-5){\dashbox(40,40)[t]{}}
\put(-15,0){\line(1,0){60}}
\put(0,-15){\line(0,1){60}}
\thinlines
\multiput(-15,-10)(5,10){6}{\multiput(0,0)(10,0){5}{\circle*{2.0}}}
\multiput(-5,-10)(5,10){6}{\multiput(0,0)(20,0){2}{\circle*{2.0}}}
\multiput(-15,10)(5,10){4}{\multiput(0,0)(20,0){1}{\circle*{2.0}}}
\multiput(35,-10)(5,10){4}{\multiput(0,0)(20,0){1}{\circle*{2.0}}}
\put(-3,42){\vector(0,1){5}}
\put(43,-3){\vector(1,0){5}}
\put(20,40){\circle{3.6}}
\put(20,40){\vector(0,-1){40}}
\put(-10,50){\small$\phi[\frac{\pi}{2e}]$}
\put(50,-6){\small$-q_{\mbox{\scriptsize scr}}[e]$}
\put(18,-24){(b)}
\end{picture}
\end{center} 
\vspace{0.5cm}
\caption{\sl The spectrum of a $\Z_4$ Chern-Simons Higgs phase compactifies
to the particles inside the dashed box.
We depict the flux $\phi$ versus the matter charge $q$
and the screening charge  $-q_{\mbox{\scriptsize scr}}=q+\mu\phi$ 
respectively. 
The Chern-Simons parameter $\mu$ is set to its minimal nontrivial value 
$\mu=\frac{e^2}{\pi}$, that is,  $p=1$. 
The arrows visualize the tunneling event induced by a minimal Dirac monopole.}
\label{z41}
\end{figure}

Let us now explicitly verify that we are indeed dealing with a 
$\Z_N$ gauge theory with Chern-Simons action~(\ref{type1}), i.e.
\bea     \label{seedorf}
\omega_{\mbox{\scriptsize I}} (a,b,c)    &=& 
\exp \left( \frac{2 \pi \im p}{N^2} \,
a(b +c -[b+c]) \right),
\eea 
where the rectangular brackets denote modulo $N$
calculus such that the sum always lies in the range $ 0,1, \ldots, N-1$.
First of all, the different particles~(\ref{compsp}) constitute
the compactified spectrum on which the quasi-quantum double 
$D^{\omega_I}(\Z_N)$ acts. The additional Aharonov-Bohm interactions 
among the fluxes are then absorbed in the definition of the 
dyon charges.~\footnote{In fact, the more accurate statement 
at this point~\cite{spm1} is that the fluxes $\phi$ enter 
the Noether charge $\tilde{Q}$ which generates the residual $\Z_N$ symmetry
in the presence of a Chern-Simons term.  That is, 
$\tilde{Q}=q +\frac{\mu}{2} \phi$, with $q$ the usual 
contribution of a matter charge.}
To be specific, the dyon charge~(\ref{rei}) corresponding  
to the flux $a$ is given by  
\bea                   
{\alpha}(b) &=&             \varepsilon_a(b) \;
 \Gamma^{n} (b)  ,
\eea
with $\varepsilon_a(b)$ defined in~(\ref{epi})
\bea        \label{vareen}
\varepsilon_a(b) &=& \exp \left( \frac{2 \pi \im p}{N^2} ab \right) ,
\eea 
and 
\bea  \label{znorrep}
\Gamma^{n} (b) &=& \exp \left( \frac{2 \pi \im }{N} nb \right),
\eea
an UIR of $\Z_N$.
The action of the braid operator~(\ref{braidaction}) now gives rise to the 
Aharonov-Bohm phases presented in~(\ref{monoI}) and~(\ref{bradenI}),
whereas the action of the central element~(\ref{spin13}) yields the 
spin factor~(\ref{sfI}).   Furthermore, the fusion rules for 
$D^{\omega_I}(\Z_N)$  following from~(\ref{Ncoef}) 
\bea   \label{CSfusion}
(a,n) \times (a',n') &=& 
\left( [a+a'],[n+n'+\frac{2p}{N}(a+a'-[a+a'])] \right),
\eea  
express the tunneling properties of the Dirac monopoles.    
Specifically, iff the sum of the fluxes $a+a'$ exceeds $N-1$, the composite 
carries unstable flux and tunnels back to the range~(\ref{compsp}) 
by means of the charged monopole~(\ref{instanz}).
Note that the charge jump induced by the monopole 
for Chern-Simons parameter $p\neq 0$  implies that the fusion 
algebra now equals $\Z_{kN} \times \Z_{N/k}$~\cite{dvvv}.
Here, we defined $k := N/{\gcd}(p,N)$  for odd $N$ and 
$k := N/{\gcd}(2p,N)$ for even $N$, where  $\gcd$ stands for the 
greatest common divisor.
In particular, for odd $N$ and Chern-Simons parameter $p=1$, 
the complete spectrum is generated by the single magnetic flux $a=1$.
Finally, the charge conjugation operator 
${\cal C}=S^2$ following from~(\ref{fusion})   takes the form
\bea
{\cal C} \; (a,n) &=& \left(
[-a], [-n + \frac{2p}{N}(-a-[-a])]\right).
\eea 
In other words, we have the usual action of the charge conjugation
operator. The fluxes $a$ and charges $n$ reverse sign. Subsequently,
the `twisted' modulo $N$ calculus for the fluxes~(\ref{instanz})  
and the ordinary modulo $N$ calculus for the charges are applied 
to return to the range~(\ref{compsp}). Also note 
that the particles and anti-particles in this theory
naturally carry the same spin, that is,  
the action~(\ref{sfI}) of the modular $T$ matrix indeed commutes with 
${\cal C}$.

Having established the fact that the $U(1)$ Chern-Simons 
term~(\ref{CStyp1}) gives rise to the 3-cocycle~(\ref{seedorf}) in the 
residual $\Z_N$ gauge theory in the Higgs phase, we now 
turn to  the periodicity $N$ of 
the Chern-Simons parameter $p$ as indicated by the 
homomorphism~(\ref{homoI}).
This periodicity can be made explicit as follows.
From the braid properties~(\ref{monoI}), the spin factors~(\ref{sfI}) and 
the fusion rules~(\ref{CSfusion}),
we infer that setting  the Chern-Simons parameter to $p=N$ 
amounts to an automorphism 
\bea
(a,n) &\longmapsto& (a,[ n+2a]),
\eea
of the spectrum~(\ref{compsp}) for $p=0$. In other words,
for $p=N$ the theory describes the same topological 
interactions between the particles as for $p=0$, 
we just have relabeled the dyons.

\begin{figure}[tbh]    \epsfysize=10cm
\centerline{\epsffile{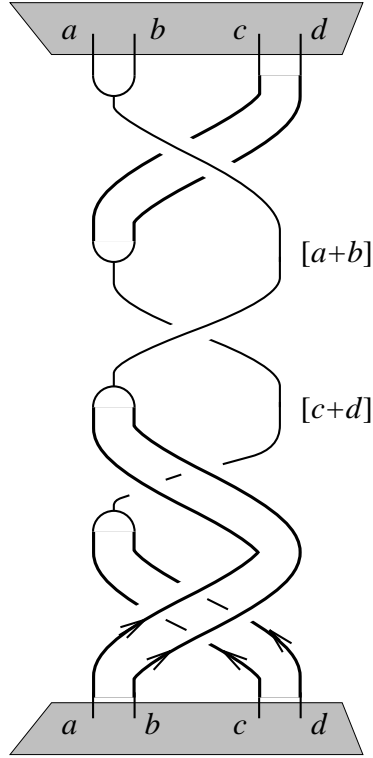}}
\caption{\sl  The 3-cocycle condition states that the topological
action $\exp(\im S_{\mbox{\scriptsize CSI}})$ for this process is trivial. The vertices 
in which the fluxes are fused correspond to a  minimal  
Dirac instanton iff the total flux of the composite is larger then $N-1$.}
\label{dreip}
\end{figure}    
 
Let us close this section by identifying  the process corresponding to the 
Chern-Simons action~(\ref{seedorf}). To start with, a comparison of the 
expressions~(\ref{seedorf}) and~(\ref{vareen}) yields
\bea                     \label{acmilan}
\omega_{\mbox{\scriptsize I}} ( a,b,c) &=& \varepsilon_b(a) \; \varepsilon_c(a) 
\; \varepsilon^{-1}_{[b+c]} (a),
\eea 
from which we immediately conclude 
\bea    \label{intermilan}
\omega_{\mbox{\scriptsize I}} ( a,b,c)
&=&
\exp (\im S_{\mbox{\scriptsize CSI}}) 
\left\{\vcenter{\epsfxsize=1.5cm\epsfbox{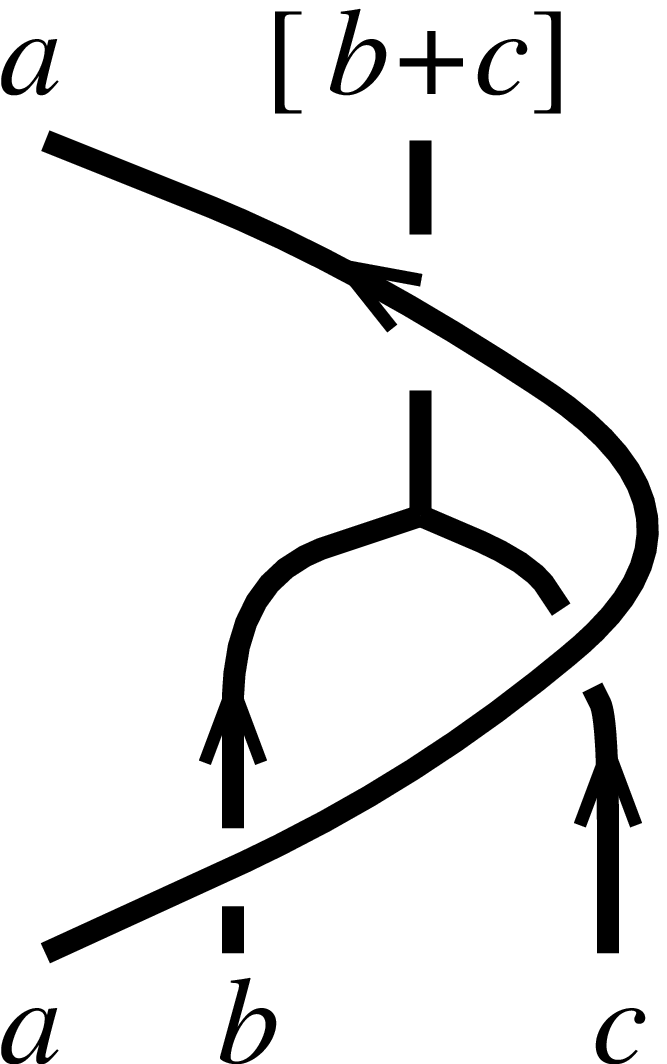}}\right\} \,.
\eea
Here, the fluxes $a$,$b$ and $c$ 
are again assumed to take values in the range $0,1,\ldots, N-1$. 
The vertex corresponding to fusion of  the fluxes $b$ and $c$
then describes the tunneling event~(\ref{instanz}) induced by the minimal 
Dirac monopole iff the total flux $b+c$ of the composite exceeds $N-1$.
Of course, the total Aharonov-Bohm phase for the process depicted 
in~(\ref{intermilan}), which also involves the matter 
coupling~(\ref{maco}), is trivial as witnessed by the 
fact that the quasitriangularity condition~(\ref{tria2}) is satisfied.
The contribution~(\ref{intermilan}) 
of the Chern-Simons term~(\ref{CStyp1}) to this total Aharonov-Bohm 
phase, however, is nontrivial iff the vertex corresponds to a monopole. 
It only generates Aharonov-Bohm phases between magnetic fluxes and therefore
only notices the flux tunneling at the vertex and not the charge creation. 
Specifically, in the first  braiding of the
process~(\ref{intermilan}), the Chern-Simons coupling 
generates the Aharonov-Bohm phase $\varepsilon_b(a)$, 
in the second  $\varepsilon_c(a)$ and 
in the last $\varepsilon^{-1}_{[b+c]}(a)$. 
Hence, the total Chern-Simons action for this process 
indeed becomes~(\ref{acmilan}).
With the prescription~(\ref{intermilan}), 
factorization of the topological action, 
the so-called skein relation
\bea     \label{skein}
\exp (\im S_{\mbox{\scriptsize CSI}})
\left\{\vcenter{\epsfxsize=1.5cm\epsfbox{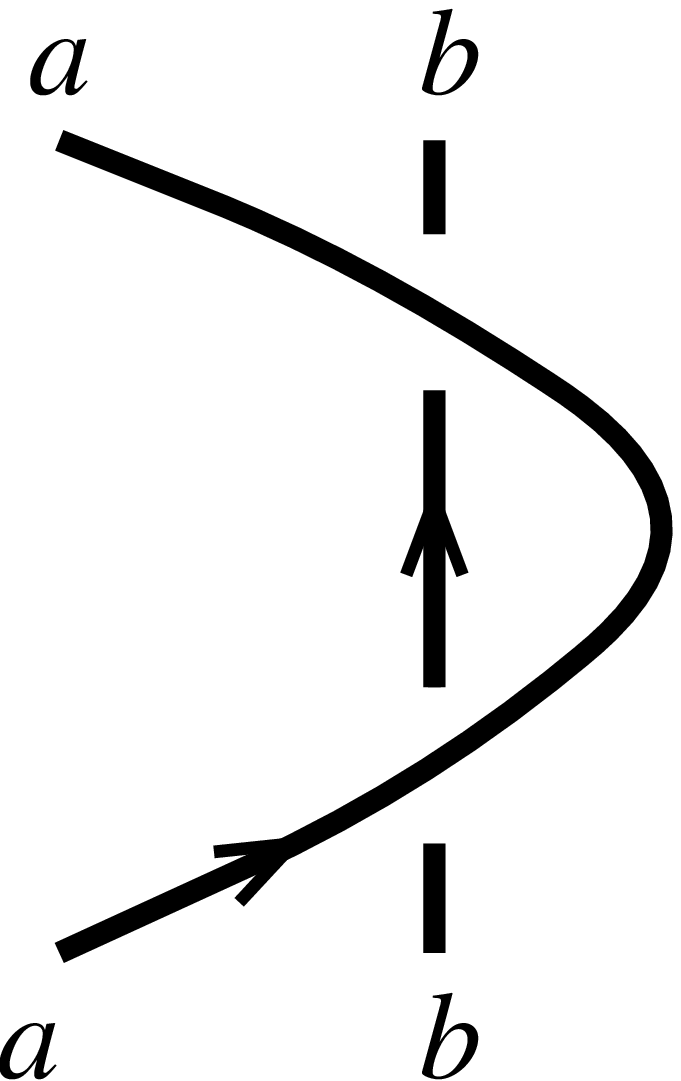}}\right\} \; = \;
\exp (\im S_{\mbox{\scriptsize CSI}})
\left\{\vcenter{\epsfxsize=1.1cm\epsfbox{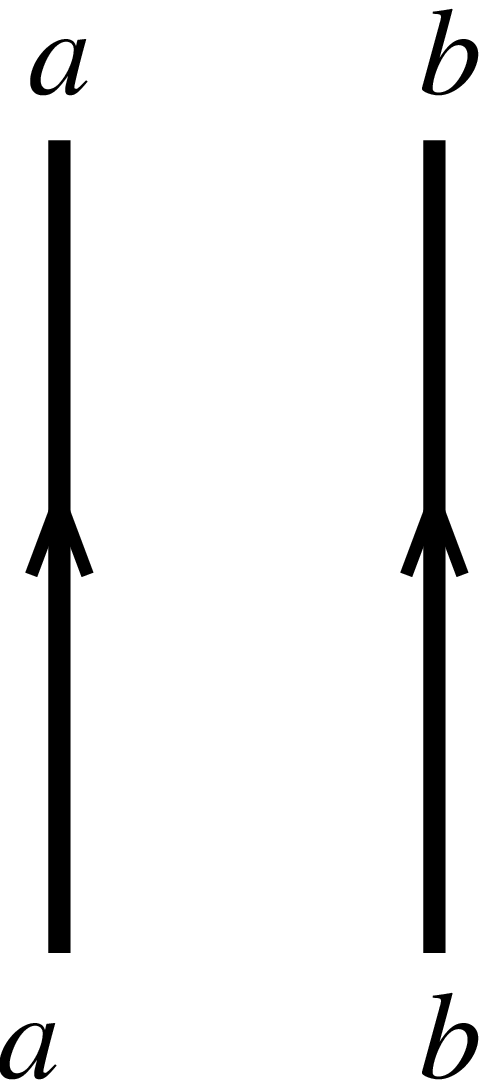}}\right\} \; = \;1,
\eea   
and the obvious relation
\bea
\omega_{\mbox{\scriptsize I}}^{-1} ( a,b,c)
&=&
\exp (\im S_{\mbox{\scriptsize CSI}}) 
\left\{\vcenter{\epsfxsize=1.5cm\epsfbox{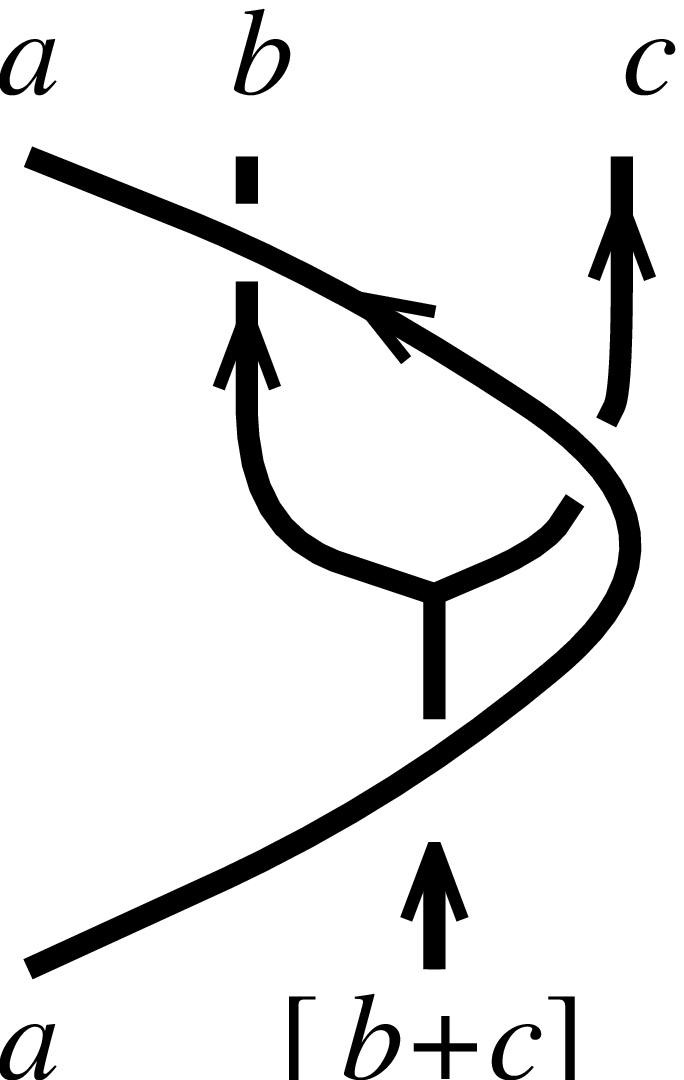}}\right\}\, ,
\eea
it is then  readily verified that  the 3-cocycle condition
\bea           \label{repetee}
\omega_{\mbox{\scriptsize I}} ( a,b,c) \;\,  
\omega_{\mbox{\scriptsize I}} ( a,[b+c], d) \;\,
\omega_{\mbox{\scriptsize I}}^{-1}  ( a,b,[c+d]) \;\,    
\omega_{\mbox{\scriptsize I}}^{-1}  ( [a+b],c,d ) \;\,    
\omega_{\mbox{\scriptsize I}} ( b,c,d ) \; = \; 1  \, , \qquad \; 
\eea
boils down to the statement that the topological 
action $\exp(\im S_{\mbox{\scriptsize CSI}})$ for the process depicted 
in figure~\ref{dreip} is trivial. In fact, this condition
can now be interpreted as the requirement that the particles connected
by Dirac monopoles should give rise to the same spin factor, 
which, in turn, imposes the quantization~(\ref{mu}) of the topological mass.
To that end, we first note that iff the total flux of either one of the 
particle pairs in figure~\ref{dreip} does not exceed $N-1$, i.e.\ 
$a+b <N-1$ and/or $c+d < N-1$,  
the 3-cocycle condition~(\ref{repetee}) is trivially satisfied,
as follows from the skein relation~(\ref{skein}).
When both pairs carry flux larger then $N-1$, all vertices 
in figure~\ref{dreip} correspond to Dirac monopoles~(\ref{instanz}), 
transferring fluxes $N$ into the charges $2p$ and vice versa.
The requirement that the action $\exp(\im S_{\mbox{\scriptsize CSI}})$ for this process 
is trivial now becomes nonempty. Let us, for example, consider the case
$a+b=N$ and $c+d = N$. Each pair may 
then be viewed as a single  particle carrying either unstable flux $N$ 
or charge $2p$ depending on the vertex it has crossed.
The total Chern-Simons  action $\exp(\im S_{\mbox{\scriptsize CSI}})$ for this case
then reduces to the quantum statistical parameter (or spin factor)
$\varepsilon_N(N)= \exp(2\pi \im p)$ generated  in the first
braiding.  Note that this Aharonov-Bohm phase is {\em not} cancelled 
by the one implied by the matter coupling~(\ref{maco}) for this process.
To be specific,  this Aharonov-Bohm phase becomes
$\exp(\im S_{\mbox{\scriptsize matter}})=\exp(-4\pi \im p)$ corresponding to  the 
second braiding in figure~\ref{dreip} where the charge $2p$ is exchanged
with the flux $N$ in a clockwise fashion. The last two braidings do not 
contribute. Upon demanding the 
total topological action 
$\exp(\im S_{\mbox{\scriptsize CSI}}+\im S_{\mbox{\scriptsize matter}})= \exp(-2\pi \im p)$
to be trivial, we finally rederive the fact that the Chern-Simons parameter 
$p$ has to be  integral.
To conclude, the 3-cocycle condition~(\ref{repetee}) is necessary and 
sufficient for a consistent implementation of Dirac monopoles
in a $\Z_N$ Chern-Simons gauge theory.

\sectiona{$U(1) \times U(1)$ Chern-Simons theory} 
\label{typeII}

The Chern-Simons terms~(\ref{CSt2}) of type~II  
establish pairwise couplings between 
the different $U(1)$ gauge fields $A^{(i)}_\kappa$ of a gauge group $U(1)^k$.
Here, we discuss the simplest example of such a Chern-Simons 
theory of type~II, namely
that with gauge group $U(1) \times U(1)$ spontaneously broken down to 
the product of two cyclic groups 
$\Z_{N^{(1)}} \times \Z_{N^{(2)}}$.
The generalization of the following analysis to $k>2$ is straightforward.

The spontaneously broken planar 
$U(1) \times U(1)$ Chern-Simons theory we consider is of the 
specific form  
\bea        \label{action12}
S &=&  \int d\,^3x \;  ({\cal L}_{\mbox{\scriptsize YMH}} + 
{\cal L}_{\mbox{\scriptsize matter}}
+ {\cal L}_{\mbox{\scriptsize CSII}})  \\
{\cal L}_{\mbox{\scriptsize YMH}} &=& 
\sum_{i=1}^2\{-\frac{1}{4}F^{(i)\kappa\nu} F^{(i)}_{\kappa\nu} + 
({\cal D}^\kappa \Phi^{(i)})^*{\cal D}_\kappa \Phi^{(i)} - V(|\Phi^{(i)}|)\} \\
{\cal L}_{\mbox{\scriptsize matter}} &=& -\sum_{i=1}^2 j^{(i)\kappa}A^{(i)}_{\kappa}  
\label{j12mat} \\
{\cal L}_{\mbox{\scriptsize CSII}} &=& \frac{\mu}{2} \epsilon^{\kappa\nu\tau} 
A^{(1)}_{\kappa} \partial_{\nu}A^{(2)}_{\tau},    \label{CSac12}
\eea 
where $A^{(1)}_\kappa$ and $A^{(2)}_\kappa$  denote 
the two different $U(1)$ gauge fields.
We assume that these gauge symmetries are realized with quantized charges,
that is, we are dealing with compact $U(1)$ gauge groups.
To keep the discussion general, however,
we allow for different fundamental charges for the two different
compact gauge groups $U(1)$. The fundamental charge
associated to the gauge field $A^{(1)}_\kappa$ is denoted by $e^{(1)}$, 
whereas $e^{(2)}$ is the fundamental charge for 
$A^{(2)}_\kappa$. 
The  two Higgs fields $\Phi^{(1)}$ and $\Phi^{(2)}$ are assumed to 
carry charge $N^{(1)}e^{(1)}$ 
and $N^{(2)}e^{(2)}$ respectively, that is,
${\cal D}_\kappa \Phi^{(i)}= 
(\partial_{\kappa}+\im N^{(i)}e^{(i)} A_{\kappa}^{(i)})\Phi^{(i)}$.
The charges introduced by the matter currents
$j^{(1)}$ and $j^{(2)}$ in~(\ref{j12mat}), in turn,  are quantized
as $q^{(1)}= n^{(1)} e^{(1)}$ and $q^{(2)}= n^{(2)} e^{(2)}$ respectively 
with $n^{(1)},n^{(2)} \in \Z$. 
For convenience, both Higgs fields are endowed
with the same (nonvanishing) vacuum expectation value $v$
\bea                         \label{poti}
V(|\Phi^{(i)}|) &=& \frac{\lambda}{4}(|\Phi^{(i)}|^2-v^2)^2  \qquad\qquad
 \lambda, v > 0 \qquad \mbox{and $i=1,2$.}
\eea 
In other words, both compact $U(1)$  gauge groups  
are spontaneously broken down 
at the same energy scale $M_H = v \sqrt{2\lambda}$.

We proceed along the line of argument in the previous section. 
That is, we start with an analysis of the unbroken phase 
and present the argument for  
the quantization~(\ref{quantmuij}) of the topological mass $\mu$
in the presence of Dirac monopoles in section~\ref{ub12}. 
In section~\ref{br12}, we then discuss 
the Chern-Simons Higgs screening mechanism in the broken phase and 
establish the Aharonov-Bohm interactions 
between the charges and magnetic fluxes in the spectrum. 
Finally, section~\ref{revii} contains  
a study of the type~II $\Z_{N^{(1)}} \times \Z_{N^{(2)}}$ Chern-Simons theory 
describing the long distance physics in the broken phase of this model.

\subsection{Unbroken phase with Dirac monopoles}
\label{ub12}

In this section, we turn to a brief discussion of the 
unbroken phase of the model~(\ref{action12}), that is, we set $v=0$ and 
$\mu \neq 0$. For more detailed studies of this unbroken Chern-Simons
theory, the interested reader is referred 
to~\cite{ezawa,hagen,kimko,wesolo,wilccs} and the references given there.

Variation of the action~(\ref{action12}) 
w.r.t.\ the gauge fields $A_{\kappa}^{(1)}$ 
and $A_{\kappa}^{(2)}$, respectively, 
gives rise to the following field equations 
\bea       \label{fe1}  \ba{lcl}             
\partial_\nu F^{(1) \;\nu\kappa} + 
\frac{\mu}{2}\epsilon^{\kappa\nu\tau} \partial_{\nu}A^{(2)}_{\tau} &=& 
j^{(1)\;\kappa}+j^{(1)\;\kappa}_H                                     \\
\partial_\nu F^{(2) \;\nu\kappa} +
\frac{\mu}{2}\epsilon^{\kappa\nu\tau} \partial_{\nu}A^{(1)}_{\tau} &=&
j^{(2)\;\kappa}+j^{(2)\;\kappa}_H. 
\ea 
\eea
with $j^{(i)}$ the two matter currents in~(\ref{j12mat}) and
$j^{(i)}_H$ the two Higgs currents in this model.
This coupled set of differential equations 
leads to Klein-Gordon equations for the 
dual field strengths $\tilde{F}^{(1)}$ and $\tilde{F}^{(2)}$ 
with mass $|\mu|/2$.
Thus the field strengths fall off exponentially and 
the Gauss' laws take the form
\bea    \label{g1}   \ba{lcl}
Q^{(1)} &=& q^{(1)} +q_H^{(1)} + \frac{\mu}{2} \phi^{(2)} \; = \; 0   \\
Q^{(2)} &=& q^{(2)} +q_H^{(2)} + \frac{\mu}{2} \phi^{(1)} \; = \; 0,
\ea
\eea
with 
\beas             \ba{ll}
Q^{(i)} \; = \; \int\! d\, ^2x\, \nabla \!\cdot\! \mbox{\bf E}^{(i)} \;=\; 0, 
& \qquad
\phi^{(i)} \; = \; \int \! d\,^2x \,\epsilon^{jk}\partial_j A^{(i) \; k} \\
q^{(i)} \; = \; \int\! d\, ^2x \, j^{(i) \; 0} ,   &
\qquad q_H^{(i)} \; = \; \int\! d\,^2x \, j^{(i) \; 0}_H. 
\ea
\eeas 
Hence, the Chern-Simons screening mechanism operating in this theory
attaches fluxes, which belong to one $U(1)$ gauge group, 
to the charges of the other~\cite{hagen,wilccs}.

The long range interactions that remain between the 
particles in the spectrum of this model 
are the topological Aharonov-Bohm interactions implied by 
the couplings~(\ref{j12mat}) and~(\ref{CSac12}).       
These can be summarized as
\bea           \label{ons12}
{\cal R}^2 \; |q^{(1)},q^{(2)}\rangle|q^{(1)'},q^{(2)'}\rangle 
&=&
e^{-\im (\frac{2 q^{(1)} q^{(2)'}}{\mu} \, + \, \frac{2 q^{(1)'} 
q^{(2)}}{\mu})}         \, 
|q^{(1)},q^{(2)}\rangle|q^{(1)'},q^{(2)'}\rangle  \\
{\cal R}  \; |q^{(1)},q^{(2)}\rangle|q^{(1)},q^{(2)}\rangle  
&=&
e^{-\im \frac{2 q^{(1)} q^{(2)}}{\mu}} \,
|q^{(1)},q^{(2)}\rangle|q^{(1)},q^{(2)}\rangle.   \label{ons12an}
\eea
From~(\ref{ons12an}), we then conclude that the only particles
endowed with a nontrivial spin are those that carry 
charges w.r.t.\ both $U(1)$ gauge groups. 
In other words, only these particles obey
anyon statistics. The other particles are bosons.

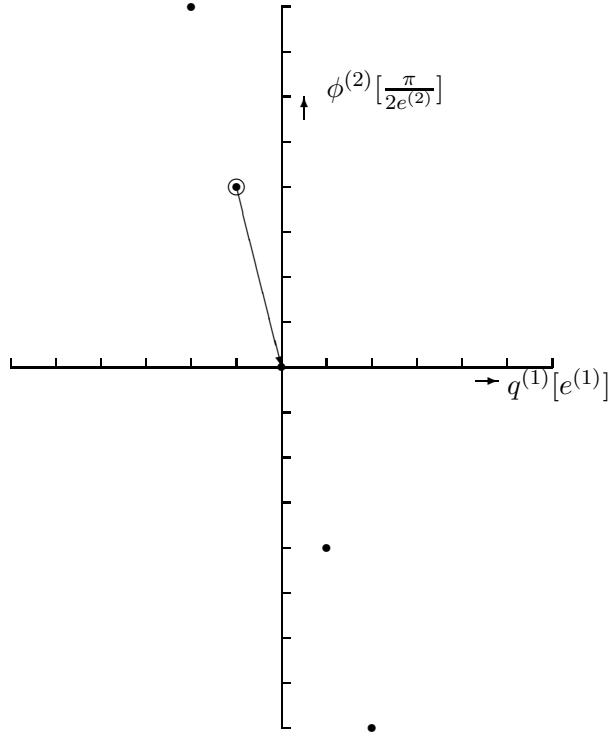
\begin{figure}[tbh] 
\begin{center}
\begin{picture}(125,160)(-60,-80)
\put(-60,0){\line(1,0){120}}
\put(0,-80){\line(0,1){160}}
\thinlines
\multiput(-60,0)(10,0){13}{\line(0,1){2}}
\multiput(0,-80)(0,10){17}{\line(1,0){2}}
\multiput(-20,80)(10,-40){5}{\circle*{2.0}}
%\multiput(-10,40)(10,-40){3}{\circle*{2.0}}
\put(-10,40){\circle{3.6}}
\put(-10,40){\vector(1,-4){10}} 
\put(5,55){\vector(0,1){5}}
\put(43,-3){\vector(1,0){5}}
\put(10,60){\small $\phi^{(2)}[\frac{\pi}{2e^{(2)}}]$}
\put(50,-6){\small$q^{(1)} [e^{(1)}]$}
\end{picture}
\end{center}
\vspace{0.5cm}
\caption{\sl Spectrum of unbroken  $U(1) \times U(1)$ 
Chern-Simons theory of type~II. 
We just depict the $q^{(1)}$ versus $\phi^{(2)}$ diagram. 
The topological mass $\mu$ is set to its minimal nontrivial value 
$\mu=\frac{e^{(1)} e^{(2)}}{\pi}$, i.e.\ $p=1$. 
The arrow represents the tunneling induced by a {\em charged} 
Dirac monopole/instanton $(2)$, which indicates that there are no stable 
particles in this theory for $p=1$.
The charge/flux diagram for $q^{(2)}$ versus $\phi^{(1)}$ 
is obtained from this one by the replacement $(1) \leftrightarrow (2)$.}
\label{u12}
\end{figure}

We proceed with the incorporation of Dirac monopoles/instantons in 
this compact Chern-Simons gauge theory.
There are two different species associated to the two compact 
$U(1)$ gauge groups. The magnetic charges carried by  these  
Dirac monopoles are quantized as
$g^{(i)} = \frac{2\pi m^{(i)}}{e^{(i)}}$ with $m^{(i)} \in \Z$ and 
$i=1,2$.
Given the coupling  between 
the two $U(1)$ gauge fields established by the 
Chern-Simons term~(\ref{CSac12}), 
the magnetic flux tunnelings induced by these monopoles
in one $U(1)$ gauge group are accompanied by charge tunnelings in the other.
Specifically, as indicated by the Gauss' laws~(\ref{g1}), 
the tunnelings associated with the two minimal Dirac monopoles become
\bea                             \label{inst1}
\mbox{instanton $(1)$ :} & & \left\{   \ba{lcllcl}
\Delta \phi^{(1)}  & = &  -\frac{2\pi}{e^{(1)}} \, , 
& \qquad \Delta \phi^{(2)}  &=&   0 \\
\Delta q^{(1)}  &=&  0 \, , 
& \qquad \Delta q^{(2)}  &=&   \mu \frac{\pi}{e^{(1)}} 
\ea \right.                                            \\
      &&  \nn \\                        
\label{inst2}
\mbox{instanton $(2)$ :} & & \left\{   \ba{lcllcl}
\Delta \phi^{(1)}  &=&   0 \, , 
& \qquad  \;  \Delta \phi^{(2)}  &=&  -\frac{2\pi}{e^{(2)}} \\
\Delta q^{(1)}  &=&  \mu \frac{\pi}{e^{(2)}} \, , 
& \qquad  \; \Delta q^{(2)}  &=&  0. 
\ea \right.
\eea
The presence of the Dirac monopole~(\ref{inst1}) implies 
quantization of the charges $q^{(1)}$ in multiples of~$e^{(1)}$. 
This can be seen by the following simple argument. 
The tunneling event induced by the monopole~(\ref{inst1})
should be invisible to the long range monodromies involving 
the various charges in the spectrum of this  theory.
Hence, from~(\ref{ons12}) we infer that  the Aharonov-Bohm phase
$\exp (-\im \frac{ 2 q^{(1)} \Delta q^{(2)}}{\mu})
=\exp (- \im \frac{2\pi q^{(1)}}{e^{(1)}})$  should be trivial. Therefore,
$q^{(1)}=n^{(1)} e^{(1)}$ with $n^{(1)} \in \Z$.
In a similar fashion, we see that the presence of the 
Dirac monopole~(\ref{inst2})
leads to quantization of $q^{(2)}$ in multiples of $e^{(2)}$.
Moreover, for consistency,
the tunneling events induced by the monopoles should 
respect these quantization rules for $q^{(1)}$ and $q^{(2)}$.
As follows from~(\ref{inst1}) and~(\ref{inst2}), this means that 
the topological mass is necessarily quantized as
\bea                 \label{mu12}
\mu &=& \frac{pe^{(1)}e^{(2)}}{\pi}   \qquad \qquad \mbox{with $p \in \Z$}.
\eea
It is easily verified that the consistency 
demand requiring the particles
connected by Dirac monopoles to give rise to the same spin factor
or quantum statistical parameter~(\ref{ons12an}),
does {\em not} lead to a further constraint on $\mu$ in this case.

To conclude, the spectrum of this unbroken 
$U(1) \times U(1)$ Chern-Simons theory, featuring Dirac monopoles,
can be presented as in~figure \ref{u12}. 
The modulo calculus for the charges $q^{(1)}$ and $q^{(2)}$
induced by the Dirac monopoles~(\ref{inst2}) and~(\ref{inst1}), respectively,
implies a compactification of the spectrum 
to $(p-1)^2$ different stable particles, with $p$ 
the integral Chern-Simons parameter in~(\ref{mu12}).

\subsection{Higgs phase}   \label{br12}

We now switch on  the  Higgs mechanism by setting $v \neq 0$.
At energies well below the symmetry breaking 
scale $M_H=v \sqrt{2\lambda}$
both  Higgs fields $\Phi^{(i)}$ are then completely condensed  
\bea 
 \Phi^{(i)}(x) & \longmapsto & v \exp (\im \sigma^{(i)}(x) )
\qquad \mbox{for $i=1,2.$}
\eea
Hence, the dynamics of the Chern-Simons Higgs medium in this model
is described by the effective action obtained 
from the following simplification in~(\ref{action12})
\bea  \label{ma12}
({\cal D}^{\kappa}\Phi^{(i)})^* {\cal D}_{\kappa}\Phi^{(i)}
-V(|\Phi^{(i)}|)   & \longmapsto &
\frac{M_A^{(i)\,2}}{2} \tilde A^{(i) \; \kappa}\tilde A^{(i)}_{\kappa}
\label{mhed12}                   \qquad  \qquad                  \\
\tilde{A}^{(i)}_{\kappa} & := & A^{(i)}_{\kappa} + 
\frac{1}{N^{(i)}e^{(i)}}\partial_{\kappa}
\sigma^{(i)}                           \qquad \qquad    \nn
\\
 M_A^{(i)} &=& N^{(i)}e^{(i)} v\sqrt{2}, \nn  \qquad \qquad
\eea
with $i=1,2$.   A derivation similar to the one for~(\ref{mass})
reveals that the two polarizations  $+$ and $-$ of the 
photon fields $\tilde{A}^{(i)}_\kappa$  acquire  
masses  $M^{(i)}_{\pm}$, which differ by the topological mass $|\mu|/2$.
We refrain from giving the explicit  expressions
of the masses $M^{(i)}_{\pm}$ in terms of  
$\mu$, $M_A^{(1)}$ and $M_A^{(2)}$.

In this broken phase, 
the Higgs currents $j_H^{(i)}$ appearing in the field 
equations~(\ref{fe1}) become 
screening currents $j_{\mbox{\scriptsize scr}}^{(i)}$ 
\bea 
j_H^{(i)} &\longmapsto& j_{\mbox{\scriptsize scr}}^{(i)} \; := \;
-M_A^{(i)\,2} \tilde{A}^{(i)}. 
\eea
In particular, the Gauss'  laws~(\ref{g1}) now take the form  
\bea    \label{g1higg}   \ba{lcl}
Q^{(1)} &=& q^{(1)} +q_{\mbox{\scriptsize scr}}^{(1)} + \frac{\mu}{2} \phi^{(2)} 
\; = \; 0    \\
Q^{(2)} &=& q^{(2)} +q_{\mbox{\scriptsize scr}}^{(2)} + \frac{\mu}{2} \phi^{(1)}
\; = \; 0,
\ea
\eea   
with 
\bea                             \label{occscre}
q_{\mbox{\scriptsize scr}}^{(i)} &:=&  \int \! d\,^2 x \, j^{(i)\,0}_{\mbox{\scriptsize scr}}.
\eea  
As we have seen in section~\ref{bp}, the emergence 
of the screening charges~(\ref{occscre}) is at the heart of the 
de-identification of charge and flux  occurring  
in the phase transition from the unbroken phase to 
the broken phase in a Chern-Simons gauge theory. 
They accompany the matter charges $q^{(i)}$ provided
by the currents $j^{(i)}$ as well as the magnetic vortices.

Let us first focus on the magnetic vortices in this model.  
There are two different species 
associated with the winding of the two different 
Higgs fields $\Phi^{(1)}$ and $\Phi^{(2)}$.
These vortices, which are of  characteristic size $1/M_H$,
carry the  quantized  magnetic fluxes 
\bea                       \label{fluqu12}
\phi^{(i)} &=& \frac{2 \pi a^{(i)}}{N^{(i)}e^{(i)}}  
\qquad  \mbox{with $a^{(i)}\in \Z$.}
\eea
As indicated by the Gauss' laws~(\ref{g1higg}), these vortices 
induce screening charges  in the Higgs medium
\beas        \ba{lcl}       
q_{\mbox{\scriptsize scr}}^{(1)} &=& -\frac{\mu}{2}\phi^{(2)}          \\
q_{\mbox{\scriptsize scr}}^{(2)} 
&=& -\frac{\mu}{2}\phi^{(1)},
\ea
\eeas 
which completely screen the Coulomb fields generated by 
their  magnetic fluxes.  
The screening charges do {\em not} couple to the Aharonov-Bohm interactions.
Therefore, the long range Aharonov-Bohm interactions among  the 
vortices implied by the Chern-Simons coupling~(\ref{CSac12})
are {\em not} screened
\bea           \label{onsflux12}
{\cal R}^2 \; |\phi^{(1)},\phi^{(2)}\rangle|\phi^{(1)'},\phi^{(2)'}\rangle 
&=&
e^{\im \frac{\mu}{2}(\phi^{(1)} \phi^{(2)'}  + \, \phi^{(1)'} \phi^{(2)})}         \, 
|\phi^{(1)},\phi^{(2)}\rangle|\phi^{(1)'},\phi^{(2)'}\rangle  \\
{\cal R}  \; |\phi^{(1)},\phi^{(2)}\rangle|\phi^{(1)},\phi^{(2)}\rangle  
&=&
e^{\im \frac{\mu}{2} \phi^{(1)} \phi^{(2)}} \,
|\phi^{(1)},\phi^{(2)}\rangle|\phi^{(1)},\phi^{(2)}\rangle.  \label{onsfl12an}
\eea    
Note that there are no Aharonov-Bohm phases generated 
among vortices of the same species. Thus there is only a nontrivial spin 
assigned to composites  carrying flux w.r.t.\ 
both  broken $U(1)$ gauge groups.

Finally, the matter charges 
$q^{(i)}$ provided by the currents $j^{(i)}$ induce the  screening charges 
$q_{\mbox{\scriptsize scr}}^{(i)} =-q^{(i)}$ in the Higgs medium, 
screening their Coulomb interactions, 
but not their Aharonov-Bohm interactions with the vortices.
The remaining long range interactions for these charges can 
then be summarized by
\bea                                 \label{qphi}
{\cal R}^2 \; |q^{(1)},q^{(2)}\rangle|\phi^{(1)},\phi^{(2)}\rangle 
&=&
e^{\im ( q^{(1)} \phi^{(1)}  +  \, q^{(2)} \phi^{(2)})}         \, 
|q^{(1)},q^{(2)}\rangle|\phi^{(1)},\phi^{(2)}\rangle, 
\eea 
as implied by the  matter coupling~(\ref{j12mat}).

\subsection{$\Z_{N^{(1)}} \times \Z_{N^{(2)}}$ Chern-Simons theory of type~II}
\label{revii}

The discussion in the previous section, in fact, pertains to
all values of the topological mass $\mu$.
Here, we again assume that the model features 
the Dirac monopoles~(\ref{inst1}) and~(\ref{inst2}), 
which implies the quantization~(\ref{mu12}) of $\mu$.
We will show that under these circumstances
the long distance physics of the Higgs phase
is described  by a $\Z_{N^{(1)}} \times \Z_{N^{(2)}}$ 
gauge theory with a  3-cocycle $\omega_{\mbox{\scriptsize II}}$ of type~II determined 
by the homomorphism~(\ref{homoII}).

Let us first recall  from the previous section that 
the spectrum of the $\Z_{N^{(1)}} \times \Z_{N^{(2)}}$ Chern-Simons 
Higgs phase consists of the matter charges $q^{(i)}=n^{(i)} e^{(i)}$, 
the quantized magnetic fluxes~(\ref{fluqu12}) 
and the dyonic combinations. We will label these 
particles as $\left( A,n^{(1)}  n^{(2)}\right)$ with $A := (a^{(1)}, a^{(2)})$
and $a^{(i)}, n^{(i)} \in \Z$.
Upon implementing~(\ref{mu12}), 
the Aharonov-Bohm interactions~(\ref{onsflux12}), (\ref{onsfl12an})
and~(\ref{qphi}) between these particles can then be recapitulated
as
\bea                             
{\cal R}^2 \; |A, n^{(1)}  n^{(2)} \rangle 
| A',n^{(1)'}  n^{(2)'} \rangle      &=&
\alpha' (A) \; \alpha (A') \; |A, n^{(1)}  n^{(2)} \rangle
|A',n^{(1)'}  n^{(2)'} \rangle      \qquad
                            \label{brz2}
\\                               \label{anz123jo}
{\cal R} \; |A, n^{(1)}  n^{(2)} \rangle|A, n^{(1)}  n^{(2)} \rangle &=&
\alpha (A) \;
|A, n^{(1)}  n^{(2)} \rangle|A, n^{(1)}  n^{(2)} \rangle
\\ \label{spin12}
T \; |A, n^{(1)}  n^{(2)} \rangle  &=& 
\alpha(A) \; |A, n^{(1)}  n^{(2)}\rangle ,
\eea 
with
\bea 
\alpha (A') &:=& \varepsilon_A(A') \; \Gamma^{n^{(1)}n^{(2)}} (A')    \\
\alpha' (A) &:=& \varepsilon_{A'} (A) \; \Gamma^{n^{(1)'}n^{(2)'}} (A).
\eea 
Here
\bea              \label{hrep12zon}
\Gamma^{n^{(1)} n^{(2)}} (A) 
&=&   \exp \left( \frac{2 \pi \im}{N^{(1)}} \, n^{(1)} a^{(1)} +
\frac{2 \pi \im}{N^{(2)}} \, n^{(2)} a^{(2)} \right),
\eea   
denotes an UIR of the group $\Z_{N^{(1)}} \times \Z_{N^{(2)}}$,
whereas the epsilon factors are identical to~(\ref{epii}), that is 
\bea
\varepsilon_A(A') &=& \exp \left( \frac{2\pi \im p}{N^{(1)}N^{(2)}} 
\, a^{(1)}a^{(2)'} \right) ,
\eea 
with $p$ the integral Chern-Simons parameter in~(\ref{mu12}).
Under these remaining long range 
Aharonov-Bohm interactions, the charge labels $n^{(i)}$ 
clearly become $\Z_{N^{(i)}}$ quantum numbers. Moreover, in the presence 
of the Dirac monopoles~(\ref{inst1}) and~(\ref{inst2})
the fluxes $a^{(i)}$ are conserved modulo $N^{(i)}$.
Specifically, in terms of the integral 
charge and flux quantum numbers $n^{(i)}$ and
$a^{(i)}$ the tunneling events corresponding to these 
minimal monopoles read
\bea                             \label{instb1}
\mbox{instanton $(1)$ :} & & \left\{   \ba{lcl}
a^{(1)} & \mapsto & a^{(1)} -N^{(1)}  \\
n^{(2)} & \mapsto & n^{(2)}  + p 
\ea \right.                                            \\
      &&  \nn \\                        
\label{instb2}
\mbox{instanton $(2)$ :} & & \left\{   \ba{lcl}
a^{(2)} & \mapsto &  a^{(2)} -N^{(2)} \\
n^{(1)} & \mapsto & n^{(1)} +p \, .
\ea \right.
\eea   
Here, we substituted~(\ref{mu12}) in~(\ref{inst1}) and~(\ref{inst2})
respectively. Hence, the decay of an unstable
flux corresponding to one residual cyclic gauge group is accompanied 
by the creation of the charge $p$ w.r.t.\ the other cyclic gauge group.
See also figure~\ref{figz42}.
It is again easily verified that these local 
tunneling events are invisible to the long range Aharonov-Bohm 
interactions~(\ref{brz2}) and that the particles connected by the 
monopoles exhibit the same spin factor~(\ref{spin12}).
The conclusion then becomes that the spectrum of a 
$\Z_{N^{(1)}} \times \Z_{N^{(2)}}$ Higgs phase corresponding to an
integral Chern-Simons parameter $p$ compactifies to
\bea                          \label{compsp12}
\left( A,n^{(1)}  n^{(2)}\right)    \qquad\qquad 
\mbox{with $A=(a^{(1)}, a^{(2)})$ and  $a^{(i)}, n^{(i)}
\in 0,1, \ldots, N^{(i)}-1$}, \qquad
\eea          
where the modulo calculus for the flux quantum numbers $a^{(i)}$ 
involves the charge jumps displayed in~(\ref{instb1}) and~(\ref{instb2}).

\begin{figure}[tbh] 
\begin{center}
\begin{picture}(85,80)(-15,-15)
\put(-5,-5){\dashbox(20,40)[t]{}}
\put(-15,0){\line(1,0){60}}
\put(0,-15){\line(0,1){60}}
\thinlines
\multiput(-10,-10)(0,10){6}{\multiput(0,0)(10,0){5}{\circle*{2.0}}}
\multiput(0,-10)(0,10){6}{\multiput(0,0)(20,0){3}{\circle*{2.0}}}
\put(0,40){\circle{3.6}}
\put(0,40){\vector(1,-4){10}} 
\put(-3,42){\vector(0,1){5}}
\put(43,-3){\vector(1,0){5}}
\put(-10,50){\small$\phi^{(2)}[\frac{\pi}{2e^{(2)}}]$}
\put(50,-6){\small$q^{(1)}[e^{(1)}]$}
\put(18,-24){(a)}
\end{picture}
\hspace{2cm}
\begin{picture}(85,80)(-15,-15)
\put(-5,-5){\dashbox(40,30)[t]{}}
\put(-15,0){\line(1,0){60}}
\put(0,-15){\line(0,1){60}}
\thinlines
%\multiput(-10,-10)(0,10){6}{\multiput(0,0)(20,0){3}{\circle*{2.0}}}
\multiput(-10,0)(0,20){3}{\multiput(0,0)(10,0){6}{\circle*{2.0}}}
\put(0,40){\circle{3.6}}
\put(0,40){\vector(1,-4){10}} 
\put(-3,42){\vector(0,1){5}}
\put(43,-3){\vector(1,0){5}}
\put(-10,50){\small$\phi^{(1)}[\frac{\pi}{e^{(1)}}]$}
\put(50,-6){\small$q^{(2)}[e^{(2)}]$}
\put(18,-24){(b)}
\end{picture}
\end{center}  
\vspace{0.5cm}
\caption{\sl The spectrum of a Higgs phase with  residual
gauge group $\Z_2 \times \Z_4$ and Chern-Simons action of type~II 
compactifies to the particles in the dashed boxes. 
We have displayed the flux $\phi^{(2)}$ 
versus the charge $q^{(1)}$ and  
the flux $\phi^{(1)}$ versus the charge $q^{(2)}$.
Here, the topological mass is assumed to take 
its minimal nontrivial value 
$\mu=\frac{e^{(1)}e^{(2)}}{\pi}$, that is, \ $p=1$. 
The arrows in figure~(a) and~(b) visualize the tunnelings corresponding 
to the Dirac monopole~(2)  and the monopole~(1)  respectively.}
\label{figz42} 
\end{figure}
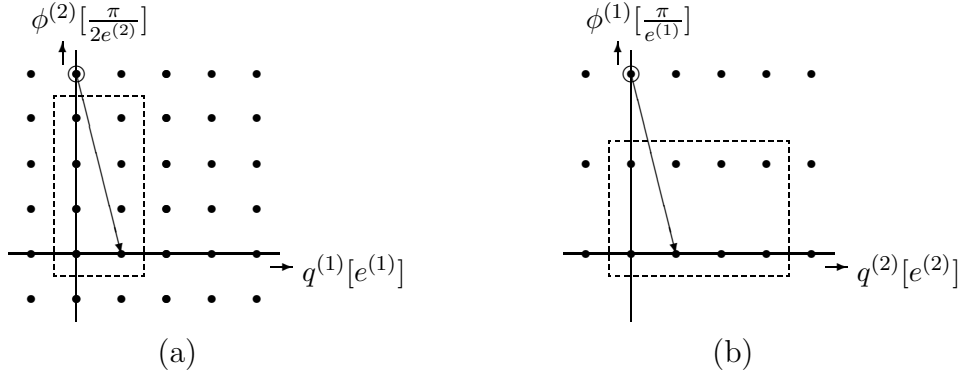

It is now readily checked that in accordance with the 
homomorphism~(\ref{homoII}) for this case, the 
$\Z_{N^{(1)}} \times \Z_{N^{(2)}}$ gauge theory 
labeled by the integral Chern-Simons parameter $p$ corresponds
to the 3-cocycle of type~II given in~(\ref{type2do}), which we repeat for
convenience
\bea                  \label{ronaldo}
\omega_{\mbox{\scriptsize II}} (A,B,C) &=&             
\exp \left( \frac{2 \pi \im p} {N^{(1)}N^{(2)}}  \;
a^{(1)}(b^{(2)} +c^{(2)} - [b^{(2)}+c^{(2)}]) \right).
\eea 
In other words, the spectrum~(\ref{compsp12}) with the topological 
interactions summarized in the expressions~(\ref{brz2}), (\ref{anz123jo}) 
and~(\ref{spin12}) 
is governed by the quasi-quantum 
double $D^{\omega_{\mbox{\tiny II}}} (\Z_{N^{(1)}} \times \Z_{N^{(2)}})$ 
with $\omega_{\mbox{\scriptsize II}}$ the 3-cocycle~(\ref{ronaldo}). In particular,
the fusion rules following from~(\ref{Ncoef}) 
\bea \label{CSfusion12}
\left( A,n^{(1)}  n^{(2)} \right) \times
\left( A',n^{(1)'}  n^{(2)'} \right)  &=&   
\left( [A+A'], n_{sum}^{(1)}  n_{sum}^{(2)} \right),
\eea
with
\beas 
[A+A'] &=& ([a^{(1)}+a^{(1)'}],  [a^{(2)}+a^{(2)'}]) \\
n_{sum}^{(1)}&=&[n^{(1)}+n^{(1)'}+\frac{p}{N^{(2)}}(a^{(2)}+a^{(2)'}-
[a^{(2)}+a^{(2)'}])]       \\
n_{sum}^{(2)}&=& [n^{(2)}+n^{(2)'}+\frac{p}{N^{(1)}}(a^{(1)}+a^{(1)'}-
[a^{(1)}+a^{(1)'}])],
\eeas
are again a direct reflection of the tunneling properties 
induced by the monopoles~(\ref{inst1}) and~(\ref{inst2}). 
Note that these `twisted' tunneling properties 
actually imply that the  complete spectrum~(\ref{compsp12}) of this theory 
is  just generated by the two fluxes $a^{(1)}=1$ 
and $a^{(2)}=1$, if the Chern-Simons parameter $p$ is set to $1$.

To conclude, at first sight the periodicity $\gcd (N^{(1)}, N^{(2)})$ 
in the Chern-Simons parameter $p$ as indicated by the mapping~(\ref{homoII}) 
is not completely obvious from the fusion rules~(\ref{CSfusion})
and the topological interactions~(\ref{brz2}), (\ref{anz123jo}) 
and~(\ref{spin12}). Here, we recall theorem~(\ref{theorem}), which was
the crucial ingredient in the proof that the 
3-cocycle~(\ref{ronaldo}) boils down to a 3-coboundary 
for $p=\gcd (N^{(1)}, N^{(2)})$. 
We then simply infer that setting 
the Chern-Simons parameter $p$ to $\gcd (N^{(1)}, N^{(2)})$ amounts 
to the automorphism 
\beas
\left( A,n^{(1)}  n^{(2)} \right) &\longmapsto& 
\left( A, [n^{(1)} +xa^{(2)}]  [n^{(2)}+ya^{(1)}] \right),
\eeas
of the spectrum~(\ref{compsp12}) for $p=0$,
where  $x$ and $y$ are the integers appearing in~(\ref{theorem}).
Hence, the theories for $p=0$ and $p=\gcd (N^{(1)}, N^{(2)})$ are the same 
up to a relabeling of the dyons.

\sectiona{$\Z_2 \times \Z_2 \times \Z_2$ Chern-Simons theory}
\label{typeIII}

Chern-Simons actions~(\ref{type3do}) of type~III occur 
for finite abelian gauge groups ${ H}$ corresponding to  direct products  
of three or more cyclic groups. 
As indicated by the homomorphism (\ref{homo}), 
such type~III Chern-Simons theories do {\em not} emerge
in spontaneously broken $U(1)^k$ Chern-Simons theories.
At present, it is not clear to us whether there actually 
exist symmetry breaking schemes
which give rise to 3-cocycles of type~III for a residual finite abelian gauge 
group in the Higgs phase. 
This point deserves further scrutiny especially 
since adding a type~III Chern-Simons action to 
an abelian discrete $H$ gauge theory has some drastic consequences.
It renders such a theory nonabelian. 
In general these type~III Chern-Simons theories are, in fact, 
dual versions of gauge theories 
featuring a finite {\em nonabelian} gauge group.

Here, we just focus on the simplest example 
of a type~III Chern-Simons theory, namely that with gauge 
group ${ H} \simeq \Z_2 \times \Z_2  \times \Z_2 $. 
The generalization to other abelian groups 
allowing for 3-cocycles of type~III is straightforward.
The outline is as follows. 
In  section~\ref{thes},  we will show 
that the incorporation of the 
type~III Chern-Simons action in a 
$\Z_2 \times \Z_2  \times \Z_2 $ gauge theory involves 
a `collapse' of the spectrum. 
Whereas the ordinary $\Z_2 \times \Z_2  \times \Z_2$ theory
features 64 different singlet particles, the spectrum 
just consists of 22 different particles in the presence of 
the 3-cocycle of type~III. 
Specifically, the dyon charges, 
which formed one dimensional UIR's of $\Z_2 \times \Z_2 \times \Z_2$, 
are  reorganized into two dimensional or doublet projective 
representations  of $\Z_2 \times \Z_2 \times \Z_2$.
This abelian gauge theory
then describes nonabelian topological interactions between these 
doublet dyons, which will be discussed in section~\ref{z23ab}. 
In section~\ref{emduality}, we finally establish that this theory
is a dual version of the ordinary discrete gauge theory with nonabelian
gauge group the dihedral group $D_4$. Furthermore, we show that 
upon adding a type~I Chern-Simons action, 
the theory actually 
becomes dual to the $\bar{D}_2$ gauge theory discussed 
in section~\ref{exampled2b}.

\subsection{Spectrum}
\label{thes}

The type~III  Chern-Simons action~(\ref{type3}) 
for the gauge group $\Z_2 \times \Z_2 \times \Z_2 $ takes the form
\bea                                  \label{trico}
\omega_{\mbox{\scriptsize III}} (A,B,C) &=& 
\exp \left( \pi \im \,  a^{(1)} b^{(2)} c^{(3)} \right),
\eea
where we have set the integral cocycle parameter to its nontrivial 
value, that is, $p_{\mbox{\scriptsize III}} = 1$.  
From the slant product~(\ref{c}) as applied to the 3-cocycle~(\ref{trico}), 
we infer that the 2-cocycle $c_A$, which enters the definition of the 
projective  $\Z_2 \times \Z_2 \times \Z_2 $ dyon charge
representations~(\ref{project}) for the magnetic flux $A$ 
in this Chern-Simons theory, reads 
\bea                     \label{cIII}
c_A (B, C) &=& \exp \left(
\pi \im \{ a^{(1)} b^{(2)} c^{(3)} +b^{(1)}c^{(2)}a^{(3)}
              -b^{(1)}a^{(2)}c^{(3)} \} \right).
\eea
For the trivial magnetic flux sector $A=0$, this 2-cocycle 
naturally vanishes, so the pure charges are given by the ordinary 
UIR's of $\Z_2 \times \Z_2 \times \Z_2$.
For the nontrivial magnetic flux sectors 
$A \neq 0$, the 2-cocycle  $c_A$ is nontrivial, that is,
it can not be decomposed as~(\ref{repphase}). 
Hence,  we are dealing with projective 
representations that can not be obtained from ordinary representations
by the inclusion of extra Aharonov-Bohm phases $\varepsilon$ 
as in~(\ref{rei}). 
An important result in projective representation theory now 
states that for a given finite group ${ H}$ 
the number of inequivalent irreducible projective 
representations~(\ref{project}) 
associated with a 2-cocycle $c$ equals the number of $c$-regular
classes in ${ H}$~\cite{karpov}. Here, an element $h \in { H}$ is called
c-regular iff $c(h,g)=c(g,h)$ for all $g \in { H}$. If $h$ 
is c-regular, so are all its conjugates. 
In our abelian example with the 2-cocycle $c_A$ for  $A \neq 0$, 
it is easily verified that there are only two $c_A$ 
regular classes in $\Z_2 \times \Z_2 \times \Z_2$, 
namely the trivial flux $0$ and $A$ itself.
Hence, there are only two inequivalent irreducible projective representations
associated with $c_A$. Just as for ordinary UIR's,
the sum of the squares of the dimensions 
of these  projective UIR's should equal the order 8 of 
the group $\Z_2 \times \Z_2 \times \Z_2$ 
and we find that both representations are two dimensional.
An explicit construction of these representations can be found 
in~\cite{karpov}.

Let us illustrate these general remarks by considering the effect
of the presence of the 2-cocycle $c_A$ for the particular
magnetic flux $A=100$. 
Substituting~(\ref{cIII}) in~(\ref{project}) yields the following 
set of defining relations for the generators 
of $\Z_2 \times \Z_2 \times \Z_2$ in 
the projective representation $\alpha$
\bea           \label{defre}   \ba{rcl}
\alpha(100)^2 &=& \alpha(010)^2 \; = \; \alpha(001)^2 \; = \; {\mbox{\bf 1}} \\
\alpha(100) \, \cdot \, \alpha(010) &=& 
\alpha(010) \, \cdot \, \alpha(100)  \\
\alpha(100) \, \cdot \, \alpha(001) &=& 
\alpha(001) \, \cdot \, \alpha(100)         \\
\alpha(010) \, \cdot \, \alpha(001) &=& 
-\alpha(001) \, \cdot \, \alpha(010).
\ea
\eea
In other words, the generators $\alpha(010)$ and 
$\alpha(001)$ become anti-commuting, which indicates that 
the  projective representation $\alpha$ is necessarily higher dimensional.
Specifically, the two inequivalent two dimensional 
projective UIR's associated to the 2-cocycle $c_{100}$ 
are given by~\cite{karpov}
\be        \label{matrixa}    
\alpha^1_{\pm} (100) =  \pm \left( \ba{rr} 1 & 0 \\ 0 & 1 \ea \right),
\qquad \alpha^1_{\pm} (010) = \left( \ba{rr} 0 & 1 \\ 1 & 0 \ea \right),
\qquad \alpha^1_{\pm} (001) = \left( \ba{rr} 1 & 0 \\ 0 & -1 \ea \right).
\ee
Here, the subscript $+$ and $-$ labels the two inequivalent representations,
whereas the superscript $1$ refers to 
the fact that $A=100$ denotes the nontrivial magnetic 
flux associated to the first gauge group $\Z_2$ in the 
product $\Z_2 \times \Z_2 \times \Z_2$. 
In passing, we note that the set of matrices~(\ref{matrixa}) 
generates the two dimensional UIR of the dihedral point group $D_4$.

It is instructive to examine the projective representations 
in~(\ref{matrixa}) a little closer.
In an ordinary $\Z_2 \times \Z_2 \times \Z_2$ gauge theory, 
the three global  $\Z_2$ symmetry generators commute with each other and 
with the flux projection operators.
Thus the total internal Hilbert space of this gauge theory allows for 
a basis of mutual eigenvectors $|A,n^{(1)}n^{(2)}n^{(3)}\rangle$, 
where the labels $n^{(i)} \in 0,1$ 
denote the $\Z_2$ representations and $A \in \Z_2 \times \Z_2 \times \Z_2$
the different magnetic fluxes. In other words, the spectrum 
consists of 64 different particles each carrying a one dimensional internal
Hilbert space labeled by a flux and a charge.
Upon introducing the type~III Chern-Simons action~(\ref{trico})
in this abelian discrete gauge theory, the global $\Z_2$ symmetry generators 
cease to commute with each other as we have seen explicitly for the flux 
sector $A=100$ in~(\ref{defre}). 
In this sector, the eigenvectors of the two non-commuting $\Z_2$ generators
are rearranged into an irreducible doublet representation. We can, however,
still diagonalize the generators in this doublet representation 
separately to uncover the $\Z_2$ eigenvalues $1$ and $-1$.
Hence, the $\Z_2$ charge quantum numbers $n^{(i)} \in 0,1$ 
remain unaltered in the presence of a Chern-Simons action 
of type~III.

The analysis is completely similar for the other flux sectors. 
First of all, the two 2-dimensional 
projective dyon charge representations $\alpha^2_{\pm}$ associated with
the magnetic flux $A=010$ follow from a cyclic 
permutation of the set of matrices in~(\ref{matrixa}), such that
the diagonal matrix $\pm {\mbox{\bf 1}}$ ends up at the second position,
that is,   $\alpha^2_{\pm} (010)=\pm {\mbox{\bf 1}}$. The two projective 
representations $\alpha^3_{\pm}$ for $A=001$ are then defined 
by the cyclic permutation of the matrices in~(\ref{matrixa})
fixed by $\alpha^3_{\pm}(001) =\pm {\mbox{\bf 1}}$.
To proceed, the two 2-dimensional 
projective representations $\beta_{\pm}^1$
for the flux $A=011$ are determined by 
\be        \label{matrixb}
\beta^1_{\pm} (100) =   \left( \ba{rr} 1 & 0 \\ 0 & -1 \ea \right),
\qquad \beta^1_{\pm} (010) =  \left( \ba{rr} 0 & 1 \\ 1 & 0 \ea \right),
\qquad \beta^1_{\pm} (001) = \pm \left( \ba{rr} 0 & 1 \\ 1 & 0 \ea \right).
\ee
Here, the subscript $+$ and $-$ again labels the two inequivalent 
representations, while the superscript now reflects the fact that $A=011$ 
corresponds to a trivial flux w.r.t.\ 
to the first gauge group $\Z_2$ in the product 
$\Z_2 \times \Z_2 \times \Z_2$. 
The two representations $\beta_{\pm}^2$ associated to the flux $A=101$
are defined by the same set of matrices~(\ref{matrixb}) 
moved one step to the right with cyclic boundary conditions, 
whereas the representations $\beta_{\pm}^3$ for $A=110$ are given 
by the same set moved two steps to the right with cyclic boundary conditions.
Finally, the two inequivalent dyon charge representations $\gamma_{\pm}$ 
for the magnetic flux $A=111$ are generated by  
the Pauli matrices  
\be        \label{matrixg}
\gamma_{\pm} (100) =   \pm \left(  \ba{rr} 0 & 1 \\ 1 & 0 \ea \right),
\; \gamma_{\pm} (010) = \pm \left( \ba{rr} 0 & -\im \\ \im & 0 \ea \right),
\; \gamma_{\pm} (001) = \pm \left( \ba{rr} 1 & 0 \\ 0 & -1 \ea \right).
\ee
In contrast with the sets of matrices contained in~(\ref{matrixa}) 
and~(\ref{matrixb}) which generate the 2-dimensional representation
of the dihedral group $D_4$, the two sets in~(\ref{matrixg}) 
generate the two dimensional UIR's of the truncated pure braid group
$P(3,4)$ displayed in the character table~\ref{char} 
of appendix~\ref{trubra}.

The complete spectrum of this $\Z_2 \times \Z_2 \times \Z_2$ Chern-Simons 
theory of type~III can now be summarized as
\be             \label{spectz23}
\ba{cc}
\mbox{particle}                &\qquad   \exp (2\pi \im  s)    \\
           & \\
(0,n^{(1)}n^{(2)}n^{(3)})  &\qquad  1    \\
(100, \alpha^1_{\pm}), \, 
(010, \alpha^2_{\pm}), \,
(001, \alpha^3_{\pm})      &\qquad   \pm 1   \\
(011,\beta^1_{\pm}), \,
(101,\beta^2_{\pm}), \, 
(110,\beta^3_{\pm})        &\qquad    \pm 1  \\
(111, \gamma_{\pm})        &\qquad    \mp \im, 
\ea
\ee
where the spin factors for the particles 
are obtained from the action of the flux of the particle on its  
own dyon charge as indicated by expression~(\ref{anp}). 
Hence, there are 7 nontrivial pure charges $(0,n^{(1)}n^{(2)}n^{(3)})$
labeled by the ordinary nontrivial one dimensional
$\Z_2 \times \Z_2 \times \Z_2$ representations~(\ref{hrepz}).
The trivial representation naturally corresponds to the vacuum.
In addition, there are 14 dyons carrying a nontrivial 
abelian magnetic flux and a doublet charge.
The conclusion then becomes that the introduction of a Chern-Simons action 
of type~III leads  to a compactification or `collapse' of the spectrum.
Whereas an ordinary $\Z_2 \times \Z_2 \times \Z_2$ gauge theory
features 64 different singlet particles, 
we only have 22 distinct particles in the presence 
of  a type~III Chern-Simons action~(\ref{trico}). 
To be specific,  the  singlet dyon charges
are rearranged into doublets so that the squares 
of the dimensions of the internal Hilbert spaces for the particles in the 
spectrum still add up to the order of the quasi-quantum double 
$D^{\omega_{\mbox{\tiny III}}} (\Z_2 \times \Z_2 \times \Z_2)$,
that is, $8^2= 8\cdot 1^2 + 14 \cdot 2^2$.
Let us close with the remark
that this collapse of the spectrum can also be seen 
directly by evaluating the Dijkgraaf-Witten invariant 
for the 3-torus $S^1 \times S^1\times S^1$
with the 3-cocycle~(\ref{trico}). 
See section~\ref{dijwit} in this connection.

\subsection{Nonabelian topological interactions}
\label{z23ab}

Here, we highlight the {\em non}abelian 
nature of the topological interactions
in this type~III Chern-Simons theory with {\em abelian}
gauge group $\Z_2 \times \Z_2 \times \Z_2$.

Let us start by considering  
the Aharonov-Bohm scattering experiment, depicted 
in figure~\ref{abverpres} of appendix~\ref{ahboverl},
in which the incoming projectile now is 
the dyon $(100, \alpha_+^1)$, while 
the dyon $(011, \beta^1_+)$  plays the role of the scatterer. 
We choose the following flux/charge eigenbasis  
for the four dimensional internal Hilbert space 
describing this 2-dyon system 
\bea        \label{cbasis}
e_{\uparrow \uparrow} 
&=& |100, \left( \ba{c} 1 \\ 0  \ea \right) \rangle \ot
        |011, \left( \ba{c} 1 \\ 0  \ea \right) \rangle     \\
e_{\downarrow \uparrow} 
&=& |100, \left( \ba{c} 0 \\ 1  \ea \right) \rangle \ot
        |011, \left( \ba{c} 1 \\ 0  \ea \right) \rangle  \nn  \\
e_{\uparrow \downarrow} 
&=& |100, \left( \ba{c} 1 \\ 0  \ea \right) \rangle \ot
        |011, \left( \ba{c} 0 \\ 1  \ea \right) \rangle  \nn  \\
e_{\downarrow\downarrow} 
&=& |100, \left( \ba{c} 0 \\ 1  \ea \right) \rangle \ot  
        |011, \left( \ba{c} 0 \\ 1  \ea \right) \rangle.     \nn
\eea
From~(\ref{braidaction}),~(\ref{matrixa}) and~(\ref{matrixb}), we 
then infer that the monodromy matrix takes the following block 
diagonal form in this basis
\bea                                   \label{alles}
{\cal R}^2 &=&  \left( \ba{rrrr} 0 & 1 & & \\ -1 & 0 & &  \\
                               & & 0 & -1 \\ 
                               & & 1 & 0 \ea \right),
\eea  
where we used 
\beas
\alpha_+^1 (011) \; = \; 
c_{100}^{-1} (010,001)\; \alpha_+^1(010) \, \cdot \, \alpha_+^1 (001)
\; = \; 
        \left( \ba{rr} 0 & 1  \\ -1 & 0  \ea \right),
\eeas
which follows from~(\ref{project}) and~(\ref{matrixa}).
The monodromy matrix~(\ref{alles})  reveals that the magnetic
flux $A=011$ acts as an Alice flux on the doublet dyon charge $\alpha_+^1$. 
Specifically, upon a parallel transport of the dyon $(100, \alpha_+^1)$
around  the dyon $(011, \beta^1_+)$, it returns with the orientation 
($\uparrow$ or $\downarrow$) of its 
charge $\alpha_+^1$ flipped ($\downarrow$ or $\uparrow$).
Furthermore, the orientation of the doublet dyon charge $\beta^1_+$ 
is unaffected by this process as witnessed by the block diagonal 
form of the monodromy matrix. 
Note that~(\ref{alles}) is, in fact, 
identical to the monodromy matrix~(\ref{allesd2}) in section~\ref{abd2b}
for a system of a pure doublet charge $\chi$ and a pure doublet flux
$\sigma_2^+$ in a $\bar{D}_2$ gauge theory.
In other words, this Aharonov-Bohm scattering problem 
is equivalent to the one discussed in section~\ref{abd2b} and leads 
to the same  cross sections, which we 
repeat for convenience 
\bea             
\frac{{\rm d} \sigma_+}{{\rm d} \theta}   &=&
\frac{1+\sin{(\theta/2)}}{8 \pi p \sin^2{(\theta/2)}}   \label{c+}  \\
\frac{{\rm d} \sigma_-}{{\rm d} \theta} &=& 
\frac{1-\sin{(\theta/2)}}{8 \pi p \sin^2{(\theta/2)}}    \label{c-}  \\
\frac{{\rm d} \sigma}{{\rm d}\theta} &=& 
\frac{{\rm d}\sigma_-}{{\rm d}\theta}  + 
\frac{{\rm d}\sigma_+}{{\rm d}\theta} \; = \; 
 \frac{1}{4 \pi p \sin^2{(\theta/2)}} \; ,             \label{singvac}
\eea
with $\theta$ the scattering angle and $p$ the momentum of the incoming 
projectiles $(100, \alpha_+^1)$. 
In this case, the multi-valued exclusive 
cross section~(\ref{c+}) is measured by 
a detector which only signals scattered dyons $(100, \alpha_+^1)$ with 
the same charge orientation as the incoming beam of  
projectiles.  A device just detecting 
dyons $(100, \alpha_+^1)$ with charge orientation opposite to the 
charge orientation of the projectiles, in turn,  measures the 
multi-valued charge flip cross section~(\ref{c-}).
Finally, Verlinde's single-valued inclusive
cross section~(\ref{singvac}) for this case  
is measured by a detector, which signals scattered 
dyons $(100, \alpha_+^1)$  irrespective of the orientation of their
charge.

The fusion rules for the particles in the spectrum~(\ref{spectz23})
are easily obtained  from expression~(\ref{Ncoef}). 
We refrain from presenting the complete set and confine ourselves 
to the fusion rules that will enter the discussion later on.
First of all, the pure charges naturally add modulo $2$ 
\bea                                       \label{chesf}
(0,n^{(1)}n^{(2)}n^{(3)}) \times (0,n^{(1)'}n^{(2)'}n^{(3)'})
\:=\: (0,[n^{(1)}+n^{(1)'}] [n^{(2)}+n^{(2)'}][n^{(3)} + n^{(3)'}]).  \qquad 
\eea 
The same holds for the magnetic fluxes of the dyons, whereas
the composition rules for the dyon charges are less trivial
\bea 
(100, \alpha^1_{\pm}) \times (100, \alpha^1_{\pm})
    &=&(0) + (0,010) + (0,001) + (0,011)   \label{1al} \\
(011, \beta^1_{\pm}) \times (011, \beta^1_{\pm})
    &=&(0) + (0,100) + (0,111) + (0,011)   \label{1bet} \\
(010, \alpha^2_{\pm}) \times (001, \alpha^3_{\pm})    \label{hetv}
    &=& (011, \beta^1_{+}) + (011, \beta^1_{-})    \\
(100, \alpha^1_{\pm}) \times (011, \beta^1_{\pm})
  &=& (111, \gamma_{+}) + (111, \gamma_{-}),    \label{hetvo}
\eea
where $(0)$ denotes the vacuum. The occurrence 
of the vacuum in the fusion rules~(\ref{1al}) 
and~(\ref{1bet}), respectively, then indicates that the dyons 
$(100, \alpha^1_{\pm})$  and $(011, \beta^1_{\pm})$ are their own
anti-particles. In fact, this observation is valid for all particles in the 
spectrum.

The fusion rule~(\ref{1al}) shows that a  pair of dyons 
$(100, \alpha^1_{+})$ can carry three different types of nontrivial
Cheshire charge, which is also the case for a pair 
of dyons $(011, \beta^1_{+})$, as expressed by~(\ref{1bet}).  
The nondiagonal form of the 
matrix~(\ref{alles}) implies that these two different 
pairs exchange Cheshire charges in the  monodromy 
process~depicted in figure~\ref{ccatd2} of 
section~\ref{chesd2b} 
for a pair of doublet charges $\chi$ and a pair of 
fluxes $\sigma_a^+$ in a $\bar{D}_2$ gauge theory.
Suppose that  a certain timeslice sees the creation
of a $(100, \alpha^1_+)$ 
dyon/anti-dyon pair and a $(011, \beta^2_+)$ 
dyon/anti-dyon pair from the vacuum. 
Hence, both pairs carry a trivial Cheshire charge at this stage, 
that is, both pairs are in the vacuum channel $(0)$ of their fusion rule.
After a monodromy involving a dyon in the pair 
$(100, \alpha^1_+)$  and a dyon in the pair $(011, \beta^1_+)$,
both pairs carry Cheshire charge $(0, 011)$, which become localized 
charges upon fusing the members of the pairs. 
As follows from the rule~(\ref{chesf}), these localized charges
annihilate each other when they are brought together. Hence,
global charge is naturally conserved in this process.
To be explicit, in terms of the associated internal quantum states, 
this process reads
\bea   \label{uitz}
|0 \rangle 
 & \longmapsto & \frac{1}{2} 
\{ |100, \left( \ba{c} 1 \\ 0  \ea \right) \rangle 
|100, \left( \ba{c} 1 \\ 0  \ea \right) \rangle
\; + \; |100, \left( \ba{c} 0 \\ 1  \ea \right) \rangle
|100, \left( \ba{c} 0 \\ 1  \ea \right) \rangle \} \ot   
\qquad \label{eennul} \\  & & 
\ot 
\{ |011, \left( \ba{c} 1 \\ 0  \ea \right) \rangle
|011, \left( \ba{c} 0 \\ 1  \ea \right) \rangle  \; + \;
|011, \left( \ba{c} 0 \\ 1  \ea \right) \rangle
|011, \left( \ba{c} 1 \\ 1  \ea \right) \rangle  \}  \label{tweenul} 
\qquad   \\
  & \stackrel{{\mbox{\scriptsize \bf 1}} \ot {\cal R}^2 \ot 
{\mbox{\scriptsize\bf 1}}}{\longmapsto} &
\frac{1}{2}
\{ |100, \left( \ba{c} 1 \\ 0  \ea \right) \rangle
|100, \left( \ba{c} 0 \\ 1  \ea \right) \rangle
\; - \; |100, \left( \ba{c} 0 \\ 1  \ea \right) \rangle
|100, \left( \ba{c} 1 \\ 0  \ea \right) \rangle\}  
\ot  \qquad  \nn \\  & & 
\ot
\{ |011, \left( \ba{c} 0 \\ 1  \ea \right) \rangle 
|011, \left( \ba{c} 1 \\ 0  \ea \right) \rangle  \; - \; 
|011, \left( \ba{c} 1 \\ 0  \ea \right) \rangle 
|011, \left( \ba{c} 0 \\ 1  \ea \right)\rangle \}  \nn 
\qquad \\
& \longmapsto & |0,011 \rangle \ot |0, 011 \rangle \nn \\
  & \longmapsto & |0 \rangle.            \nn
\eea 
The quasi-quantum double 
$D^{\omega_{\mbox{\tiny III}}} (\Z_2 \times \Z_2 \times \Z_2)$
acts on the two particle state~(\ref{eennul})
for the dyons $(100, \alpha^1_+)$ through
the comultiplication~(\ref{coalgebra}) with the 2-cocycle~(\ref{cIII}).
From the action of the flux projection operators, we 
formally obtain that this state carries trivial total flux.
The global symmetry transformations, which act by means of 
the matrices~(\ref{matrixa}), 
then leave  this two particle state invariant. In other words, this state
indeed carries trivial total charge.
In a similar fashion, we infer 
that the two particle state~(\ref{tweenul}) for
the dyons $(011, \beta^1_{+})$ corresponds to trivial total flux 
and charge. After the monodromy, 
which involves the matrix~(\ref{alles}), both 
two particle states then carry the global charge $(0,011)$. 
Note that this exchange of Cheshire charge is again accompanied 
by an exchange of quantum statistics (see the discussion at the end of 
section~\ref{d2bqst}). The two particle states~(\ref{eennul})
and~(\ref{tweenul}) are bosonic in accordance with the trivial
spin~(\ref{spectz23}) assigned to the dyons $(100, \alpha^1_+)$ and 
$(011, \beta^1_{+})$ respectively. Both two particle states
emerging after the monodromy, in turn, are fermionic.

We conclude this section with a concise analysis of  
the truncated braid group representations that may occur in this theory. 
To start with, the only identical particle configurations 
that obey braid statistics are those that consist either of 
the dyons $(111, \gamma_{+})$ or of the 
dyons $(111, \gamma_{-})$. It is easily verified that 
the braid operators for such systems are of order $4$.
Thus the  internal Hilbert spaces for systems of $n$ of 
these dyons decomposes into UIR's of the truncated braid 
group $B(n,4)$. The one dimensional UIR's that may occur in 
this decomposition correspond to semion statistics and the higher
dimensional UIR's to nonabelian braid statistics. 
All other identical particle systems realize permutation statistics.
Specifically, the pure charges are bosons, whereas the remaining 
dyons in general may obey bose, fermi or parastatistics. 
Furthermore, the maximal order of the monodromy 
operator for distinguishable particles in this theory is 4. 
Thus the   distinguishable particles configurations
are ruled by the pure braid group $P(n,8)$ and its subgroups.  
Let us just focus on  a system containing
the three dyons $(100, \alpha^1_{+})$, $(010, \alpha^2_{+})$ and 
$(001, \alpha^3_{+})$.
From the fusion rules~(\ref{hetv}) and~(\ref{hetvo}), 
we obtain that under the action of the quasi-quantum double 
$D^{\omega_{\mbox{\tiny III}}} (\Z_2 \times \Z_2 \times \Z_2)$, 
the internal Hilbert space for this three particle system decomposes into
the following direct sum of irreducible representations
\bea         \label{ein}
(100, \alpha^1_{+}) \times (010, \alpha^2_{+}) \times (001, \alpha^3_{+})
    &=& 2 \; (111, \gamma_{+}) + 2 \; (111, \gamma_{-}).
\eea 
The occurrence of two pairs of equivalent fusion channels now 
implies that 2-dimensional irreducible representations 
of the pure braid group are conceivable for this system. 
This indeed turns out to be the case.
The monodromy operators for this system are of order 2.
Hence, the associated truncated pure braid group is 
$P(3,4) \subset P(3,8)$, which
has been discussed in appendix~\ref{trubra}. 
A straightforward calculation then reveals that the 
$P(3,4)$ representation carried by the internal Hilbert space of 
this system breaks up into the following irreducible pieces
\bea                \label{eindt}
\Lambda_{P(3,4)} &=& 2 \; \Omega_8 + 2 \; \Omega_9,
\eea
where~$\Omega_8$ and $\Omega_9$ denote the two dimensional UIR's
contained in the character table~\ref{char} of appendix~\ref{trubra}.
Finally, from~(\ref{ein}) and~(\ref{eindt}), we infer that under the action
of  the direct product 
$D^{\omega_{\mbox{\tiny III}}} 
(\Z_2 \times \Z_2 \times \Z_2) \times P(3,4)$
the internal Hilbert space decomposes into the following irreducible
subspaces
\bea 
\left( (111, \gamma_{+}), \Omega_8 \right) \: + \: 
\left( (111, \gamma_{-}) , \Omega_9 \right),
\eea 
where $\left( (111, \gamma_{+}), \Omega_8 \right)$ and 
$\left( (111, \gamma_{-}) , \Omega_9 \right)$ both label 
a four dimensional representation.

\subsection{Electric/magnetic duality}
\label{emduality}

The analysis of the previous sections actually  revealed some 
striking similarities between the type~III Chern-Simons theory with 
gauge group $\Z_2 \times \Z_2 \times \Z_2$ 
and  the $\bar{D}_2$ gauge theory discussed in 
section~\ref{exampled2b}. 
To start with, the orders of these gauge groups
are the same $|\Z_2 \times \Z_2 \times \Z_2|=|\bar{D}_2|=8$.
Moreover, the spectrum of both 
theories consists  of $8$  singlet particles 
and $14$ doublet particles, which adds up to a 
total number of 22 distinct particles. 
Also, the charge conjugation operation 
acts trivially (${\cal C}={\mbox{\bf 1}}$)
on these spectra, that is, the particles in both theories appear 
as their own anti-particle. 
Finally, the truncated braid groups that govern the topological interactions 
in these discrete gauge theories are similar.
Hence, it seems that these theories are dual w.r.t.\ each other. 
As it stands, however, that is not the case. 
This becomes clear upon comparing 
the spins assigned to the particles in the different theories, as  
displayed in~(\ref{spind2})  and~(\ref{spectz23}).
The $\bar{D}_2$ theory features three
particles corresponding to a spin factor $\im$
and three particles to a spin factor $-\im$, 
whereas the spectrum of the $\Z_2 \times \Z_2 \times \Z_2$ contains
just one particle with spin factor $\im$ and one with $-\im$.
In other words, the modular $T$ matrices  associated 
to these models are different.
It can be verified that the modular $S$ matrices,
which classify the monodromy properties of the 
particles in these theories, are also distinct.

Let us now recall from~(\ref{conj3e}), that 
the full set of Chern-Simons actions for the 
gauge group  $\Z_2 \times \Z_2 \times \Z_2$
consists of three nontrivial 3-cocycles of type~I, three of type~II,
one of type~III and products thereof. It then turns out that the 
$\Z_2 \times \Z_2 \times \Z_2$ Chern-Simons theories corresponding to 
the product of the 3-cocycle of type~III and either one of the three 
3-cocycles of type~I are actually
dual to a $\bar{D}_2$ gauge theory. 
Here, we just explicitly show  this duality for the 
$\Z_2 \times \Z_2 \times \Z_2$ Chern-Simons theory defined by
\bea                                  
\omega_{\mbox{\scriptsize{I+III}}} (A,B,C) &=& 
\exp \left( 
\frac{\pi \im }{2} a^{(1)}(b^{(1)} +c^{(1)} -[b^{(1)}+c^{(1)}]) 
+ \pi \im \,  a^{(1)} b^{(2)} c^{(3)} \right). 
\qquad
\eea
In other words, the total Chern-Simons action is the 
product of the 3-cocycle~(\ref{trico}) of type~III and the nontrivial
3-cocycle~(\ref{type1}) of type~I for the first $\Z_2$ gauge group
in $\Z_2 \times \Z_2 \times \Z_2$.
As indicated by~(\ref{anp}), the introduction of this type~I 
3-cocycle, in particular, involves the assignment of  an additional 
imaginary spin factor $\im$ to those dyons in the 
spectrum~(\ref{spectz23}) that 
carry nontrivial flux w.r.t.\  the first $\Z_2$ gauge group 
of the product $\Z_2 \times \Z_2 \times \Z_2$.
The spin factors of the other particles are unaffected.
Consequently, the spin factors associated to the different 
particles in this theory become
\be             \label{spz23de}
\ba{cc}
\mbox{particle}                &\qquad   \exp (2\pi \im  s)    \\
           & \\
(0,n^{(1)}n^{(2)}n^{(3)})  &\qquad  1    \\
(011,\beta^1_{\pm}), \, 
(010, \alpha^2_{\pm}), \,
(001, \alpha^3_{\pm})      &\qquad   \pm 1   \\
(100, \alpha^1_{\pm}), \,
(101,\beta^2_{\pm}), \, 
(110,\beta^3_{\pm})        &\qquad    \pm \im  \\
(111, \gamma_{\pm})        &\qquad    \pm 1.
\ea
\ee      
Note that the spin structure of this spectrum indeed corresponds
to that of the $\bar{D}_2$ gauge theory exhibited in~(\ref{spind2}).
Moreover, it is readily checked that the modular 
$S$ matrix~(\ref{fusion}) for this 
$\Z_2 \times \Z_2 \times \Z_2$ Chern-Simons theory is equivalent to that
for the $\bar{D}_2$ theory given in table~\ref{modsd2}. 
In other words, these two theories are dual.
To be explicit,  the duality transformation
\bea
\bar{D}_2   &\longleftrightarrow &
\{\Z_2 \times \Z_2 \times \Z_2, \omega_{\mbox{\scriptsize{I+III}}} \},
\eea 
\bea            \label{duality}
\ba{rclrcl}
1                 &\longleftrightarrow &   (0),            &
\qquad \chi              &\longleftrightarrow &   (111, \gamma_+)      \\
J_1               &\longleftrightarrow &   (0,011),        &
\qquad \bar{\chi}        &\longleftrightarrow &   (111, \gamma_-)       \\
J_2               &\longleftrightarrow &   (0,101),        &
\qquad \sigma^{\pm}_1    &\longleftrightarrow &   (011, \beta^1_{\pm})  \\
J_3               &\longleftrightarrow &   (0,110),        &
\qquad \sigma^{\pm}_2    &\longleftrightarrow &   (010, \alpha^2_{\pm}) \\
\bar{1}           &\longleftrightarrow &   (0,100),        &
\qquad \sigma^{\pm}_3    &\longleftrightarrow &   (001, \alpha^3_{\pm}) \\
\bar{J}_1         &\longleftrightarrow &   (0,111),        &
\qquad \tau^{\pm}_1      &\longleftrightarrow &   (100, \alpha^1_{\pm}) \\
\bar{J}_2         &\longleftrightarrow &   (0,001),        &
\qquad \tau^{\pm}_2      &\longleftrightarrow &   (101, \beta^2_{\pm})  \\
\bar{J}_3         &\longleftrightarrow &   (0,010),        &
\qquad \tau^{\pm}_3      &\longleftrightarrow &   (110, \beta^3_{\pm}),
\ea   
\eea
corresponds to an invariance of the modular matrices 
\bea
S_{\bar{D}_2}  &=&   S_{\mbox{\scriptsize \bf Z}_2 \times 
\mbox{\scriptsize \bf Z}_2 \times \mbox{\scriptsize \bf Z}_2} \\
T_{\bar{D}_2}  &=&   T_{\mbox{\scriptsize \bf Z}_2 \times 
\mbox{\scriptsize \bf Z}_2 \times \mbox{\scriptsize \bf Z}_2}
\eea 
which implies that both models describe the same topological interactions.
Note that the duality 
transformation~(\ref{duality}) establishes an 
interchange of electric and magnetic quantum numbers.
Specifically, 
the nonabelian $\bar{D}_2$ magnetic flux doublets are mapped 
into the $\Z_2 \times \Z_2 \times \Z_2$  
doublet dyon charges, while the $\Z_4$ singlet dyon charges associated
to these $\bar{D}_2$ doublet fluxes are sent into 
the abelian $\Z_2 \times \Z_2 \times \Z_2$  magnetic fluxes. 
Hence, we are in fact dealing with 
an example of nonabelian electric/magnetic duality. 
Here, it should be remarked though that the interchange
of electric and magnetic quantum numbers does not extend to 
the pure singlet charges. That is,  
as expressed by~(\ref{duality}),
the pure $\bar{D}_2$ singlet charges $J_1$, $J_2$ and $J_3$ 
are related to  pure $\Z_2 \times \Z_2 \times \Z_2$  charges.

In a similar fashion, we obtain duality between the $\bar{D}_2$ theory
and the $\Z_2 \times \Z_2 \times \Z_2$ gauge theory with Chern-Simons 
action being the product of the 3-cocycle of type~III and either one of 
the other two 3-cocycles of type~I. The duality transformation for these 
cases  simply amounts to a natural permutation 
of the particles in~(\ref{duality}).  
Finally, we note that duality with the $\bar{D}_2$ theory also emerges 
for the $\Z_2 \times \Z_2 \times \Z_2$ gauge theory featuring the 
Chern-Simons action $\omega_{\mbox{\scriptsize I+I+I+III}}$ 
but is lost for the case
$\omega_{\mbox{\scriptsize I+I+III}}$. 
Here, $\omega_{\mbox{\scriptsize  I+I+I+III}}$ denotes
the product of the three distinct 3-cocycles of type~I 
and the 3-cocycle of type~III, while
$\omega_{\mbox{\scriptsize I+I+III}}$ stands for a product of two distinct 
 3-cocycles of type~I
and the 3-cocycle of type~III.
It is easily verified that the spin structure of the spectrum
for the latter theory
does not match with that~(\ref{spind2}) of $\bar{D}_2$.
To be specific, the spectrum for $\omega_{\mbox{\scriptsize  I+I+III}}$
contains  five dyons with spin factor $\im$ and five with $-\im$.

A complete discussion, which is beyond the scope of this thesis,
not only involves $\bar{D}_2$, but also the other nonabelian 
gauge group of order 8, namely the dihedral group $D_4$, 
and the finite set of Chern-Simons actions for these two 
nonabelian gauge groups. Let us just remark that the 
$\Z_2 \times \Z_2 \times \Z_2$ gauge theory corresponding 
to the type~III Chern-Simons action~(\ref{trico}) itself
is dual to the ordinary $D_4$ gauge theory 
\bea     \label{d4dual}
D_4   &\longleftrightarrow &
\{\Z_2 \times \Z_2 \times \Z_2, \omega_{\mbox{\scriptsize III}} \},  
\eea 
which is in accordance with our earlier observation that 
the sets of matrices~(\ref{matrixa}) and~(\ref{matrixb}) 
associated to the dyon charges $\alpha^i_{\pm}$ 
and $\beta^i_{\pm}$, respectively,
generate the two dimensional UIR of $D_4$.
Furthermore, the incorporation of either one of the three 3-cocycles
of type~III does not destruct this duality
\bea
D_4   &\longleftrightarrow &
\{\Z_2 \times \Z_2 \times \Z_2, \omega_{\mbox{\scriptsize II+III}} \},
\eea 
where the duality transformation between the two spectra again 
boils down to a permutation of that associated with~(\ref{d4dual}).

The extension of this nonabelian duality to higher order abelian groups 
which allow for Chern-Simons actions of type~III is left for future work.
An interesting question in this respect is whether the 
nonabelian dual gauge groups are restricted to the dihedral and double 
dihedral series or also involve  other nonabelian finite groups.

\sectiona{Dijkgraaf-Witten invariants}
\label{dijwit}

In~\cite{diwi}, Dijkgraaf and Witten defined a topological invariant
for a compact, closed oriented  three
manifold ${\cal M}$ in terms of a 3-cocycle
$\omega \in H^3({ H}, U(1))$  for a finite group $H$.
They represented this invariant as the partition function
$Z({\cal M})$ of a lattice gauge theory with gauge group $H$ and 
Chern-Simons action $\omega$. It was shown explicitly 
that $Z(M)$ is indeed a combinatorial invariant of the manifold ${\cal M}$.
With the same data (a finite group ${ H}$ and a 3-cocycle~$\omega$), 
Altschuler and Coste~\cite{altsc1,altsc}
subsequently constructed a surgery invariant ${\cal F}({\cal M})$ from a
surgery presentation of the manifold ${\cal M}$. 
They conjectured that up to normalization these two invariants are the same
\bea 
{\cal F}({\cal M)} &=& \frac{Z({\cal M})}{Z(S^3)},  \label{cj}
\eea
with $Z(S^3)= 1/|{ H}|$. Altschuler and Coste verified their conjecture
for lens spaces, using the 3-cocycles of type~I for  cyclic groups
${ H} \simeq \Z_N$.
In this section, this analysis will be extended with some numerical results
using the 3-cocycles of type~II and of type~III, 
which  were not treated in~\cite{altsc1,altsc}.
In addition, we will evaluate the Dijkgraaf-Witten invariant for the 
3-torus ${\cal M} = S^1 \times S^1 \times S^1$ associated with 
the three types of 3-cocycles for  
${ H} \simeq \Z_2 \times \Z_2 \times \Z_2$. 

The Dijkgraaf-Witten invariant for the 
lens space $L(p,q)$ associated with an abelian finite 
group $H$ and 3-cocycle $\omega$ 
takes the following form~\cite{altsc1,diwi}  
\bea
Z(L(p,q)) &=& \frac{1}{|{ H}|} \sum_{\{A \in { H}|[A^p]=0\}} \;\; 
\prod_{j=1}^{p-1}  \omega (A, A^j, A^n),
\eea
with $n$ the inverse of $q$ mod $p$.
The surgery invariant constructed by Altschuler and Coste,
on the other hand, reads~\cite{altsc}
\bea                  \label{dl3}
{\cal F} (L(rs-1,s)) &=& \frac{1}{|{ H}|^2} 
\sum_{\{A,B,C \in { H}|[A^{rs-1}]=0\}} 
                           c_A^*(A^{-r},B) \;
                          c_{A^{-r}}^*(A,C) \\
                          & &\times  
\prod_{m=1}^r \omega^* (A,A^{-m} \cdot B, A)
          \prod_{n=0}^{s-1} \omega^* (A^{-r},A^{1-nr}
                                     \cdot C, A^{-r}),   \nn
\eea
where $*$ stands for complex conjugation,
$r,s$ denote positive integers and $c_A$ is 
the 2-cocycle defined in~(\ref{c}).
Note that formula~(\ref{dl3}) differs 
by an overall complex conjugation from the one given in~\cite{altsc},
where the orientation of the manifold was treated incorrectly.
The conjecture~(\ref{cj}) of Altschuler and Coste now states that these two 
invariants are equivalent
\bea
{\cal F}(L(p,q)) &=& \frac{Z(L(p,q))}{Z(S^3)} \, .
\eea

With the help of Mathematica, we numerically verified this 
conjecture 
for several lens spaces with the three different types of 
3-cocycles given in section~\ref{aa3c}, and did  {\em not}
find any counter-examples.
For the  3-cocycles~(\ref{type1}) of type~I for ${ H} \simeq \Z_5$, 
for instance, these numerical evaluations confirmed the 
fact that the Dijkgraaf-Witten invariant can distinguish the lens spaces 
$L(5,1)$ and $L(5,2)$, which are homeomorphic but of different
homotopy type~\cite{altsc, altsc1}   
\bea
Z(L(5,1)) &=& \left\{ \ba{ll}
               1& \mbox{for  $p_{\mbox{\scriptsize I}} =0$} \\
              \frac{1}{\sqrt{5}} &  \mbox{for  $p_{\mbox{\scriptsize I}}=1,4$} \\
             -\frac{1}{\sqrt{5}} &  \mbox{for  $p_{\mbox{\scriptsize I}} =2,3$} \ea \right.\\
Z(L(5,2)) &=& \left\{ \ba{cl}
                 1& \mbox{for  $p_{\mbox{\scriptsize I}} =0$} \\
                 -\frac{1}{\sqrt{5}} & \mbox{for  $p_{\mbox{\scriptsize I}}=1,4$} \\
                 \frac{1}{\sqrt{5}} &  \mbox{for  $p_{\mbox{\scriptsize I}} =2,3$}. \ea \right.
\eea
This nice property of the Dijkgraaf-Witten invariant
is lost for 3-cocycles of type~II and~III.
Specifically, for ${ H} \simeq \Z_5 \times \Z_5$ 
and a 3-cocycle~(\ref{type2}) of type~II,
one arrives at
\bea
Z(L(5,1)) \; = \; Z(L(5,2)) \; = \; \left\{ \ba{ll}
               1 & \mbox{for  $p_{\mbox{\scriptsize II}} = 0$} \\
            \frac{1}{5} &  \mbox{for  $p_{\mbox{\scriptsize II}}
 = 1,\dots, 4$,} \ea \right.
\eea
while for ${ H} \simeq \Z_5 \times \Z_5 \times \Z_5$  
and a 3-cocycle~(\ref{type3})  
of  type~III  the situation becomes completely trivial
\bea
Z(L(5,1)) \; = \; Z(L(5,2)) \; = \; 1   
\qquad \mbox{for  $p_{\mbox{\scriptsize III}} = 0,1, \ldots, 4$.} 
\eea
To proceed, the result for the nontrivial 3-cocycle of type~I
for ${ H} \simeq \Z_2$ 
\bea
Z(L(p,1)) &=& \left\{ \ba{ll}
                 \frac{1}{2} & \mbox{for odd  $p$} \\
                \frac{1}{2}(1+(-1)^{p/2}) & \mbox{for even $p$,} \ea \right. 
\eea
established in~\cite{diwi},
generalizes in the  following manner to the nontrivial 3-cocycles 
of type~II and~III for ${ H} \simeq \Z_2 \times \Z_2$ and 
${ H} \simeq \Z_2 \times \Z_2 \times \Z_2$ respectively
\bea
Z(L(p,1)) &=&  \left\{ \ba{ll}
               \frac{1}{4} & \mbox{for odd  $p$} \\
               \frac{1}{4}(3+(-1)^{p/2}) & \mbox{for even $p$} \ea 
\right.\\
Z(L(p,1))&=&  \left\{ \ba{ll}
            \frac{1}{8} &      \mbox{for odd  $p$} \\
         \frac{1}{8}(7+(-1)^{p/2}) & \mbox{for even $p$.} \ea \right.
\eea

Finally, the Dijkgraaf-Witten
invariant for the 3-torus $S^1 \times S^1 \times S^1$
is of particular interest, since it 
counts the number of particles in the spectrum of 
a discrete ${ H}$ Chern-Simons gauge theory~\cite{diwi}.
For abelian groups ${ H}$ it takes the form 
\bea
Z(S^1 \times S^1 \times S^1) &=&\frac{1}{|{ H}|} \sum_{A,B,C \in { H}} 
W(A,B,C),
\eea
with
\bea 
W(A,B,C) &=& \frac{\omega(A,B,C) \; \omega(B,C,A) \; \omega(C,A,B)}{
\omega(A,C,B) \; \omega(B,A,C) \; \omega(C,B,A)} \, .
\eea 
It is not difficult to check that for the three different types 
of 3-cocycles for the direct product group 
${ H} \simeq \Z_2 \times \Z_2 \times \Z_2$, the invariant yields
\bea
Z(S^1 \times S^1 \times S^1) &=&  \left\{ \ba{cl}
               64 & \mbox{for type~I and~II} \\
               22 & \mbox{for type~III}, \ea \right.
\eea 
expressing the collapse of the spectrum we found for 
3-cocycles of type~III in section~\ref{typeIII}.

\aanhangsel 
\sectiona{ Cohomological derivations}
\label{gc}

This appendix provides some background to
the group cohomological results entering the discussion 
in this chapter.   The outline is as follows.
We start by establishing  the isomorphism~(\ref{clasi}).
Next  we  turn to the content of the 
cohomology groups that play a role in  
abelian discrete $H$ Chern-Simons gauge theories.
Specifically, we will derive the 
following identities for ${ H} \simeq \Z_N^k$
\bea   \left\{    \ba{lcl}         \label{conj1}         
H^1(\Z_N^k,U(1)) & \simeq & \Z_N^k      \\
H^2(\Z_N^k,U(1)) & \simeq & \Z_N^{\frac{1}{2}k(k-1)}    \\
H^3(\Z_N^k,U(1)) & \simeq & \Z_N^{k+\frac{1}{2}k(k-1)+\frac{1}{3!}k(k-1)(k-2)}, 
\ea
\right.
\eea    
and subsequently generalize these results to abelian groups ${ H}$
being direct products of cyclic groups possibly of 
different order. Finally, we will show that the content
of the cohomology group, which classifies 
the Chern-Simons actions for the compact $U(1)^k$ gauge group, reads
\bea                    \label{uki}
H^4(B(U(1)^k), \Z) &\simeq& \Z^{ k + \frac{1}{2}k(k-1)}.
\eea 
In passing, we stress that we will consider the cohomology and
abelian groups in additive rather than  multiplicative form.
This turns out to be more convenient for the manipulations planned 
in this appendix. In this additive presentation,   
a  direct product of $k$ cyclic factors $\Z_N$, for example,
then becomes the  direct sum denoted by  $\Z_N^k := \oplus_{i=1}^k \Z_N$.

Our first objective is to prove the isomorphism (\ref{clasi}).
This will be done using  
the universal coefficients theorem (see for example~\cite{rotman}),
which 
relates cohomology groups with different coefficients. 
We will need the universal coefficients theorem in the specific form
\bea                         \label{uni}
H^n(X, {\mbox{\bf B}}) &\simeq& H^n(X, \Z) \otimes {\mbox{\bf B}}
 \oplus \mbox{Tor} (H^{n+1}(X,\Z),{\mbox{\bf B}}),
\eea
relating the cohomology of some topological
space $X$ with coefficients in some abelian group ${\mbox{\bf B}}$
and the cohomology of $X$ with integer coefficients $\Z$.
Here, $\ot$ stands for the symmetric tensor product and  
$\mbox{Tor}(\, . \, \, ,\, . \,)$ for the torsion product.
The symmetric tensor product ${\mbox{\bf A}} \otimes {\mbox{\bf B}}$ 
(over $\Z$) for abelian groups 
${\mbox{\bf A}}$ and ${\mbox{\bf B}}$, is the abelian group of all 
ordered pairs $a \otimes b$
($a \in {\mbox{\bf A}}$ and $b \in {\mbox{\bf B}}$) with relations~\cite{rotman} 
\beas
(a+a') \ot b &=& a \ot b + a' \ot b   \\
a \ot (b+b') &=& a \ot b + a \ot b'  \\
m(a \ot b )  &=& ma \ot b \; =\; a \ot mb \qquad \forall \, m \in \Z.
\eeas 
It is not difficult to check that these relations imply the 
following identifications
\bea      
\Z_N \otimes \Z_M & \simeq &  \Z_{{\gcd}(N,M)}  \label{begin} \\
\Z_N \otimes \Z &\simeq& \Z_N  \\
\Z_N \otimes U(1) &\simeq& 0  \label{nec0}   \\
\Z \otimes U(1) &\simeq& U(1)         \\
\Z \otimes \Z  &\simeq& \Z \label{nas},     
\eea
with ${\gcd}(N,M)$ being the greatest common divisor of $N$ and $M$.
An important property of the symmetric
tensor product $\otimes$ is that it is distributive 
\bea                                     \label{dizzy}
(\oplus_{i} {\mbox{\bf A}}_i)\otimes {\mbox{\bf B}} &\simeq& \oplus_{i} ({\mbox{\bf A}}_{i} 
\otimes {\mbox{\bf B}}).
\eea
The definition of  the torsion product $\mbox{Tor}(\, . \, \, ,\, . \,)$
can be found in any textbook on algebraic topology, 
for our purposes the following properties suffice~\cite{rotman}.
Let ${\mbox{\bf A}}$ and ${\mbox{\bf B}}$ again be abelian groups, then
\beas
\mbox{Tor} ({\mbox{\bf A}},{\mbox{\bf B}}) &\simeq& \mbox{Tor}({\mbox{\bf B}},{\mbox{\bf A}})   \\
\mbox{Tor} (\Z_N, {\mbox{\bf B}}) &\simeq& {\mbox{\bf B}}[N] \; \simeq \; 
\{ b \in {\, \mbox{\bf B}}\,| \, N b =0 \,\}, 
\eeas
so in particular
\bea    
\mbox{Tor} (\Z_N, \Z_M) &\simeq& \Z_{{\gcd}(N,M)}  \label{difor} \\
\mbox{Tor} (\Z_N, U(1)) &\simeq& \Z_N   \label{nec}         \\
\mbox{Tor} ({\mbox{\bf A}}, \Z) &\simeq& 0    \qquad \qquad \forall {\mbox{\bf A}}.  
\label{torp}
\eea
The last identity follows from  the fact
that the group of integers $\Z$
is torsion free, that is, it does not contain elements of finite order.
Just as the symmetric tensor product, the torsion product is  distributive 
\bea                               \label{eind}
\mbox{Tor} (\oplus_{i} {\mbox{\bf A}}_{i}, {\mbox{\bf B}}) &\simeq& \oplus_{i} 
\mbox{Tor} ( {\mbox{\bf A}}_{i}, {\mbox{\bf B}}).
\eea

The proof of the isomorphism~(\ref{clasi}) now goes as follows. 
First we note that  for finite
groups ${ H}$ all cohomology in fixed degree $n>0$ is finite.
With this knowledge, the universal coefficients theorem~(\ref{uni}) 
directly gives the desired result
\bea                                   
 H^n({ H}, U(1)) & \simeq &  H^n({ H}, \Z) \ot U(1) \oplus
               \mbox{Tor} (H^{n+1}({ H},\Z), U(1))    \nn      \\
             & \simeq &  H^{n+1}({ H}, \Z)  \qquad \mbox{for $n>0$}. 
\label{ole}
\eea
In the last step we used the distributive property of the tensor product and 
the torsion product together with  the identities~(\ref{nec0}) 
and~(\ref{nec}).

We now turn to the derivation of the identities 
in~(\ref{conj1}). Our starting point will be 
the standard result  (e.g.~\cite{spanier})
\bea
H^n(\Z_N,\Z) &\simeq&  \left\{ \ba{ll}
                 \Z_N & \mbox{if $n$ is even} \label{nec3} \\
                 0 & \mbox{if $n$ is odd} \\
                 \Z &  \mbox{if $n=0$,} \ea \right.
\eea 
which together with~(\ref{ole}) immediately imply that
the identities in~(\ref{conj1})  are valid for $k=1$.
The extension to $k>1$  involves the so-called
K\"{u}nneth formula (see for example~\cite{rotman})
\bea                  \label{kun}
  H^n(X \times Y, \Z) \; \simeq 
 \sum_{i+j=n} H^i(X, \Z) \otimes H^j(Y, \Z) \oplus \!\!\!
 \sum_{p+q=n+1} \!\!\! \mbox{Tor} (H^{p}(X,\Z), H^{q}(Y,\Z)),  \qquad \;\;
\eea
which states that the cohomology of a direct product space is completely
determined in terms of the cohomology of its factors.
With the ingredients~(\ref{nec3}) and~(\ref{kun}),
the identities~(\ref{conj1}) can now be proven by induction.
To lighten the notation a bit,
we will omit explicit mention of 
the coefficients of the cohomology groups,
if the  integers $\Z$ are meant, that is, 
$H^n(\Z_N^k) := H^n(\Z_N^k, \Z)$.  Let us start with  the 
trivial cohomology group $H^0(\Z_N^k)$.
Upon using the K\"{u}nneth formula~(\ref{kun}), 
the property~(\ref{torp}) of the torsion product
and the  result~(\ref{nec3}), we easily infer
\bea                   \label{eros}
H^0(\Z_N^k) \; \simeq \; H^0(\Z_N^{k-1}) \ot H^0(\Z_N^k)  
             \; \simeq \; H^0(\Z_N^{k-1}) \ot \Z  \; \simeq \; \Z,  
\eea
where the last isomorphism follows by induction.
To be explicit, as indicated by~(\ref{nec3}) this isomorphism 
obviously holds for $k=1$. 
If we subsequently assume that this isomorphism is valid 
for some fixed $k$, we obtain with~(\ref{nas}) that it also holds for $k+1$.
To proceed, in a similar fashion, we arrive at 
\be                   
H^1(\Z_N^k) \; \simeq \; H^1(\Z_N^{k-1}) \ot H^0(\Z_N) 
         \; \simeq \; H^1(\Z_N^{k-1}) \; \simeq \; 0.
\ee
These results enter the following derivation starting  from the K\"unneth
formula~(\ref{kun}) 
\bea                         \label{era}
H^2(\Z_N^k) &\simeq& H^0(\Z_N^{k-1}) \ot H^2(\Z_N) \oplus
                      H^2(\Z_N^{k-1}) \ot H^0(\Z_N)  \\
             &\simeq& \Z_N \oplus H^2(\Z_N^{k-1}) \; \simeq \; \Z_N^k.   \nn
\eea 
Here we used the distributive property~(\ref{dizzy}) of the
tensor product and again induction 
to establish  the last isomorphism.
We continue with 
\bea                                  \label{era1}
H^3(\Z_N^k) &\simeq&  H^3(\Z_N^{k-1}) \ot H^0(\Z_N) \oplus 
                       \mbox{Tor}(H^2(\Z_N^{k-1}), H^2(\Z_N)) \\
             &\simeq&  H^3(\Z_N^{k-1}) \oplus \Z_N^{k-1} \; \simeq \;
                       \Z_N^{\frac{1}{2}k(k-1)}. \nn
\eea
Finally, using the previous results and  induction, we obtain
\bea                      \label{eureka}
H^4(\Z_N^k) &\simeq& H^0(\Z_N^{k-1})\ot H^4(\Z_N) \oplus 
                      H^2(\Z_N^{k-1}) \ot H^2(\Z_N)  \oplus     \\
            &  &    H^4(\Z_N^{k-1})\ot H^0(\Z_N) 
                  \oplus \mbox{Tor}(H^3(\Z_N^{k-1}), H^2(\Z_N))  \nn \\
            &\simeq& H^4(\Z_N) \oplus H^2(\Z_N^{k-1})\oplus  H^4(\Z_N^{k-1})
                    \oplus \mbox{Tor}(H^3(\Z_N^{k-1}), H^2(\Z_N))  \nn \\
            &\simeq &  \Z_N \oplus \Z_N^{k-1} \oplus H^4(\Z_N^{k-1})
                 \oplus \Z_N^{\frac{1}{2}(k-1)(k-2)}     \nn \\
            &\simeq & \Z_N^{k+\frac{1}{2}(k-1)(k-2)} \oplus H^4(\Z_N^{k-1}) 
                     \nn \\
     &\simeq &  \Z_N^{k+\frac{1}{2}k(k-1)+\frac{1}{3!}k(k-1)(k-2)} .   \nn
\eea
To conclude, the results~(\ref{era}),~(\ref{era1}) and~(\ref{eureka}) 
together with~(\ref{ole}) lead to the identities~(\ref{conj1}).

This derivation, at the same time,  gives a nice insight into the 
structure of the terms that build 
up the cohomology group $H^4(\Z_N^k)\simeq H^3(\Z_N^k, U(1))$.
We can, in fact, distinguish three types of terms that contribute here.
By induction, we find that there are  $k$ terms  
of the form  $H^4(\Z_N)$. 
These are the terms that label the 3-cocycles of type~I 
exhibited in~(\ref{type1}). 
By a similar argument, we infer that there are 
$\frac{1}{2} k(k-1)$ terms of the  form  
$ H^2(\Z_N^{k-1})$. These terms 
label the 3-cocycles of type~II displayed in~(\ref{type2}). 
Finally, the $\frac{1}{3!}k(k-1)(k-2)$ terms we are left with are 
entirely due to torsion products
and label the 3-cocycles of type~III in~(\ref{type3}).

The generalization of the above results 
to abelian groups ${ H}$, which are 
direct products of cyclic groups possibly of different order, is now
straightforward. The picture that the 3-cocycles divide 
into three different types remains unaltered. 
If the direct product ${ H}$
consists of $k$ cyclic factors, then there are again $k$ different 
3-cocycles of type~I, $\frac{1}{2} 
k(k-1)$ different 3-cocycles of type~II and $\frac{1}{3!}k(k-1)(k-2)$ 
different 3-cocycles of type~III.
The only distinction is that through~(\ref{begin}) and~(\ref{difor})
the greatest common divisors of the orders of the different cyclic factors 
constituting the direct product group $H$ enter the scene 
for 3-cocycles of type~II and~III.
This is best illustrated by considering the direct product
group $H \simeq \Z_N \times \Z_M \times 
\Z_K$, which is the simplest example where all three types of 3-cocycles
appear. The derivation~(\ref{eros})-(\ref{eureka}) for this 
particular case leads to the following content of the 
relevant cohomology groups 
\bea   \left\{    \ba{lcl}         \label{conj1do}         
H^1(\Z_N \times \Z_M \times \Z_K, U(1)) 
& \simeq & \Z_N \oplus \Z_M \oplus \Z_K \\
H^2(\Z_N \times \Z_M \times \Z_K,U(1)) & \simeq & 
\Z_{{\gcd}(N,M)} \oplus \Z_{{\gcd}(N,K)} \oplus \Z_{{\gcd}(M,K)}
\\
H^3(\Z_N \times \Z_M \times \Z_K,U(1)) & \simeq & 
\Z_N \oplus \Z_M \oplus \Z_K \oplus  \\ & & 
\Z_{{\gcd}(N,M)} \oplus \Z_{{\gcd}(N,K)} 
\oplus \Z_{{\gcd}(M,K)} \oplus \\
& & \Z_{{\gcd}(N,M,K)}.
\ea
\right.
\eea   
The 3-cocycles of type~I labeled by the terms $\Z_N$, $\Z_M$
and $\Z_K$ are of the form~(\ref{type1do}), whereas the explicit 
the 3-cocycles of type~II labeled by the terms
$\Z_{{\gcd}(N,M)}$, $\Z_{{\gcd}(N,K)}$ and $\Z_{{\gcd}(M,K)}$ 
take the form~(\ref{type2do}). The explicit realization of the 
3-cocycles of type~III corresponding to the term $\Z_{{\gcd}(N,M,K)}$
can be found in~(\ref{type3do}).

We would like to conclude
this appendix by establishing the content~(\ref{uki}) of 
the cohomology group  $H^4(B(U(1)^k))$.
The standard result (see for instance~\cite{diwi})
\bea
H^n(BU(1)) &\simeq&  \left\{ \ba{ll}
                 \Z & \mbox{if $n=0$ or $n$ even} \label{necu1} \\
                 0 & \mbox{otherwise,}
                 \ea \right.
\eea 
generated by the first Chern class of degree 2,
indicates that~(\ref{uki}) holds for $k=1$.
For $k > 1$, we may again appeal to the K\"unneth
formula, because the classifying space of the product group 
$U(1)^k$ is the same as the 
product of the classifying spaces of the factors, that is,
$B(U(1)^k) = B(U(1)^{k-1}) \times BU(1)$ 
(see for instance~\cite{novikov}, page 132).   
The derivation of the result~(\ref{uki}) then becomes 
similar to the one given for the finite abelian group $\Z_N^k$.
Since the group $\Z$ is torsion free, however, the terms 
due to  torsion products vanish in this case.
The terms that persist are the following.
First of all, there are $k$ terms of the form $H^4(BU(1))\simeq \Z$. These
label the different Chern-Simons actions of type~I displayed in~(\ref{CSt1}).
In addition, there are $\frac{1}{2} k(k-1)$ terms 
of the form $H^2(BU(1)) \simeq \Z$, which label
the Chern-Simons actions of type~II given in~(\ref{CSt2}).

\sectie

\chapter{Nonabelian discrete Chern-Simons theories}
\label{chap4}

This final chapter contains a 
concise discussion of Chern-Simons theories
with a finite nonabelian gauge group $H$. In fact, we confine ourselves to 
stating some salient results. 
A more detailed discussion will be presented elsewhere. 
The outline is as follows. In section~\ref{quaqudo}, we start with a
brief discussion of  the quasi-quantum double $\DW$ associated with
a nonabelian discrete $H$ Chern-Simons theory defined 
by a 3-cocycle $\omega \in H^3(H,U(1))$. 
We then turn to an explicit example in 
section~\ref{csaliceel}, namely the Chern-Simons theories 
with dihedral gauge group $H \simeq D_N$. 
Such theories, which can be seen as discrete versions 
of Chern-Simons Alice electrodynamics,  may for example occur
in spontaneously broken $SO(3)$ or $SU(3)$ Chern-Simons gauge theories.
We present the explicit realization of the 
Chern-Simons actions  $\omega \in H^3(D_{2N+1},U(1))$
for the odd dihedral groups $D_{2N+1}$ 
and subsequently elaborate on the related  fusion rules. 
Section~\ref{dubbel} contains a 
similar treatment of the Chern-Simons theories
corresponding to a double dihedral group $\bar{D}_N$, 
which may appear as the long distance remnant of 
a spontaneously broken $SU(2)$ Chern-Simons gauge 
theory. 
Finally, in appendix~\ref{laap} we have gathered some cohomological results 
which enter the discussion in the main text.

\sectiona{$\DW$ for nonabelian $H$}  \label{quaqudo}

In this section, we briefly recall the structure of the quasi-quantum double 
$\DW$ related to a nonabelian discrete $H$ gauge theory
endowed with a nontrivial Chern-Simons action $\omega \in H^3({ H}, U(1))$.
For a more detailed discussion, the reader is referred to the original paper
by Dijkgraaf, Pasquier and Roche~\cite{dpr}. 
See also~\cite{altsc1,spm1}.

We will cling to the notation established in 
section~\ref{thequdo} in the discussion of the  quantum 
double $D(H)$ for a nonabelian finite group $H$.
The deformation of $D(H)$ into the quasi-quantum double $\DW$ 
by means of a 3-cocycle $\omega \in H^3({ H}, U(1))$
can then be summarized as~\cite{dpr}
\bea 
\varphi &=& \sum_{g,h,k}\,\omega^{-1}(g,h,k)\;  
{\mbox{P}}_g \otimes {\mbox{P}}_h \otimes {\mbox{P}}_{k}, \label{isomna}  \\
{\mbox{P}}_g \, x \cdot {\mbox{P}}_{h} \, y &=& 
\delta_{g, x h x^{-1}} \; {\mbox{P}}_g \, xy  \;\;\theta_g(x,y) 
\label{naalgco}   \\
 \Delta(\,{\mbox{P}}_g \, x \,) &=& 
\sum_{h \cdot k=g} {\mbox{P}}_{h} \, x \ot {\mbox{P}}_{k} \, {y}
\;\; \gamma_x(h,k) \label{nacoalgco}\, , \; 
\eea
and 
\bea \label{quasicoasna}
({\mbox{id}} \ot \Delta) \, \Delta( \, {\mbox{P}}_g \, x \, ) &=& 
\varphi\cdot (\Delta \ot {\mbox{id}}) \, \Delta( \, {\mbox{P}}_g \, x \, ) 
\cdot\varphi^{-1},
\eea
with $g,h,k,x,y \in H$. 
The cochains $\theta$ and $\gamma$ appearing in the 
multiplication~(\ref{naalgco}) and 
comultiplication~(\ref{nacoalgco}) 
are defined as 
\bea  
\label{theta}
\theta_g(x,y) &=& i_g \omega (x,y) \; = \; 
\frac{\omega(g,x,y)\;\omega(x,y,(xy)^{-1}gxy)}
{\omega(x,x^{-1}gx,y)}  \\ 
\gamma_x(h,k)&=& \tilde{\im}_x\omega(h,k) \; = \; 
\frac{\omega(h,k,x)\;\omega(x,x^{-1}hx,x^{-1}kx)}{\omega(h,x,x^{-1}kx)}\, .
\label{gamma}
\eea
Here, $i_g$ is the nonabelian version of the slant product~(\ref{ig}) 
(see also appendix~\ref{laap}), while $ \tilde{\im}_x$ stands for 
a different inner product which lowers the degree of cochains by 1. 
Note that the inner product $ \tilde{\im}$ coincides with the 
slant product $i$ for abelian $H$, so that $\theta=\gamma$ 
in accordance with expression~(\ref{algebra}) and~(\ref{coalgebra}) of
section~\ref{symalg}.
To proceed, by repeated use of the  3-cocycle condition
\bea
\label{pentagonna}
\delta\omega(g,h,k,l) &=& 
\frac{\omega(g,h,k) \; \omega(g,hk,l)\;\omega(h,k,l)}{\omega
(gh,k,l)\;\omega(g,h,kl)} \;=\; 1,
\eea
the following relations are readily checked
\bea
\tilde{\delta} \theta_{g}(x,y,z) &:=& 
\frac{\theta_{x^{-1}gx}(y,z) \; \theta_{g}(x,yz)}{\theta_{g}(x,y) \; 
\theta_{g}(xy,z)   } \: =\; 1      \label{2cocycle}   \\ 
(\delta \gamma_x) \left(\frac{\omega}{{\rm ad}_x \omega} \right) (g,h,k)
&:=& \frac{\gamma_{x}(h,k) \; \gamma_{x}(g,hk)}{\gamma_x(g,h) \; 
\gamma_x(gh,k) } \frac{\omega(g,h,k)}{\omega(x^{-1}gx,x^{-1}hx,x^{-1}kx)}
\; = \; 1  \qquad  \;\; \;  \;\;
\label{adj} \\
\theta_{gh}(x,y) \; \gamma_{xy}(g,h) &=&
\theta_{g}(x,y) \; \theta_{h}(x,y) \; \gamma_{x}(g,h) \; 
\gamma_{y}(x^{-1}gx,x^{-1}hx) . \qquad
\label{raar}
\eea
Relation~(\ref{2cocycle}) implies  
that the multiplication~(\ref{naalgco}) is associative,   
relation~(\ref{adj}) that 
the comultiplication~(\ref{nacoalgco}) 
is quasi-coassociative~(\ref{quasicoasna}), whereas~(\ref{raar}) 
indicates that the comultiplication is an algebra morphism.

The operator $\tilde{\delta}$ is the so-called conjugated 
coboundary operator corresponding to the cohomology groups
$H^n(H,U(1)[H])$ defined in appendix~\ref{laap}. Here, we write 
$U(1)$ rather then ${\mbox{\bf R}}/{\Z}$ to emphasize that we are dealing with 
the multiplicative presentation.
The conjugated 2-cocycle condition in~(\ref{2cocycle}) then indicates 
that $\theta \in H^2(H,U(1)[H])$.  
Although nontrivial 2-cocycles $\theta$ are in principle also concievable, 
we restrict ourselves for convenience to nonabelian discrete $H$ 
Chern-Simons theories corresponding to 3-cocycles $\omega \in H^3(H,U(1))$
for which $\theta$ is trivial.
That is, $\theta$ boils down to a 2-coboundary
\be \label{jrep}
\theta_g(x,y) \; = \; 
\tilde{\delta} \varepsilon_g(x,y) \; = \;
\frac{\varepsilon_{x^{-1}gx}(y) \; \varepsilon_g(x)}{\varepsilon_g(xy)} \;\; ,
\ee
where $\varepsilon$ denotes a 1-cochain.  
The dyon charges in such a Chern-Simons
theory form trivial projective centralizer representations. 
Specifically,  the action of the quasi-quantum double $\DW$ 
on the internal Hilbert space $V^A_\alpha$ corresponding to a 
particle $(\,^A C, \alpha \,)$ is given as in~(\ref{13}) albeit
deformed by the cochain $\varepsilon$ following from~(\ref{jrep})
\bea \label{13na}                                              
\Pi^A_{\alpha}(\, {\mbox{P}}_{h} \, {g}\, ) 
\; |\,^A\!h_i,\,^{\alpha}\!v_j \rangle &=&
\delta_{h,g\,^A\!h_i \, g^{-1}}\;\; \varepsilon_h(g)\; 
|\, g\,^A\!h_i \, g^{-1},\,{\alpha}(\tilde{g})_{mj}\,^{\alpha}v_m \rangle,
\eea
with $\tilde{g} := \,^A\!x_k^{-1}\, g \,\, ^A\!x_i$ and 
$\,^A\!x_k$  defined through $\,^A\!h_k := g\,^A\!h_i \, g^{-1}$.
To avoid confusion, we stress that unlike the conventions set 
in section~\ref{symalg} for the abelian case, the epsilon factors 
are not absorbed in the definition of $\alpha$.
In other words, here $\alpha$ denotes an {\em ordinary} 
unitary irreducible representation 
of the centralizer associated to the conjugacy class $^A C$. 
It is then easily verified with~(\ref{naalgco}) and~(\ref{jrep}) 
that~(\ref{13na}) indeed defines an ordinary representation 
of the quasi-quantum double.

The definition~(\ref{cruijff}) 
of the universal $R$-matrix remains unaltered in the presence of a 
3-cocycle. From~(\ref{13na}), we then infer that 
action of the braid operator~(\ref{braidact}) on the  two particle 
internal Hilbert space $V^A_{\alpha}\ot V^B_{\beta}$ is given by
\bea                 \label{braidactep}
\lefteqn{ {\cal R} \; |\,^A\!h_i,\,^{\alpha}\!v_j \rangle 
|\,^B\!h_m,\,^{\beta}\!v_n \rangle \;= }    \\
&  & \qquad \qquad \varepsilon_{^A\!h_i \, ^B\!h_m \,^A\!h_i^{-1}}(\,^A\!h_i  ) \;
|\,^A\!h_i \, ^B\!h_m \,^A\!h_i^{-1},\,\beta(\,^A\!\tilde{h}_i\,)_{ln} \,
^{\beta}\!v_l \rangle
|\,^A\!h_i,\,^{\alpha}\!v_j \rangle \, .   \nn
\eea
Hence, the 1-cochains $\varepsilon$ 
describe additional Aharonov-Bohm interactions 
among the nonabelian magnetic fluxes. 
Further, the relations~(\ref{theta}) and~(\ref{gamma}) imply
that the quasitriangularity conditions~(\ref{grcom}), (\ref{tria2}) 
and~(\ref{tria2}) are met. The braid operators~(\ref{braidactep}) 
then satisfy the quasi-Yang-Baxter equation~(\ref{qujaba}).
It should be emphasized that due to the occurrence 
of flux metamorphosis for nonabelian fluxes, the 
isomorphisms~(\ref{intermi}) do not drop out of 
the truncated braid group representations~(\ref{brareco}) 
realized by the multi-particle systems in this nonabelian
discrete Chern-Simons theory.
This in contrast with the abelian case discussed in section~\ref{symalg}.

To conclude, the fusion rules are determined by 
\bea          \label{Ncoemet}
N^{AB\gamma}_{\alpha\beta C} &=& \frac{1}{|{ H}|} \sum_{h,g}
    \,   \mbox{tr} \left( \Pi^A_{\alpha} \ot \Pi^B_{\beta}
                      (\Delta (\, {\mbox{P}}_h \, g \,)  ) \right)   \;
         \mbox{tr} \left( \Pi^C_{\gamma} (\, {\mbox{P}}_h \, g \,) \right)^*, 
\eea 
whereas the modular matrices take the form
\bea                                \label{fusionna}
S^{AB}_{\alpha\beta} &:=& 
\frac{1}{|{ H}|} \, 
\mbox{tr} \; {\cal R}^{-2 \; AB}_{\; \; \; \; \; \alpha\beta}       \\
&=& \frac{1}{|{ H}|} \, 
\sum_{\stackrel{\,^A\!h_i\in\,^A\!C\,,^B\!h_j\in\,^B\!C}{[\,^A\!h_i,
\,^B\!h_j]=e}} \!\!\!\!
\mbox{tr}  \left( \alpha(\,^A\!x_i^{-1}\,^B\!h_j\,^A\!x_i) \right)^*
\; \mbox{tr}  \left( \beta(\,^B\!x_j^{-1}\,^A\!h_i\, ^B\!x_j) \right)^*
\; \sigma(^A\!h_i|^B\!h_j)  \;\;\;
\nn \\
T^{AB}_{\alpha\beta} &:=&  
\delta_{\alpha,\beta} \, \delta^{A,B} \; \exp(2\pi \im s_{(A,\alpha)})
\;=\; \delta_{\alpha,\beta} \, \delta^{A,B}  
\, \frac{1}{d_\alpha} 
\mbox{tr} \left(  \alpha(^A\!h_1) \right) \; \varepsilon_{^A\!h_1} (^A\!h_1) 
\, .
\label{modultspi}
\eea
Here, the symmetric cochain $\sigma$ in~(\ref{fusionna}) 
is defined as $\sigma(g|h) := \varepsilon_g^*(h)\varepsilon_h^* (g)$.

\sectiona{Discrete Chern-Simons Alice electrodynamics}
\label{csaliceel}

The dihedral groups $D_N$,  which are of the semi-direct product form
\bea                        \label{alicegroup}
D_N  &\simeq&  \Z_2  \ltimes  \Z_N \, ,
\eea 
constitute an infinite series of nonabelian subgroups of $SO(3)$.
In fact, the dihedral groups can be seen as discrete versions of the 
group $\Z_2 \ltimes U(1)$, which is the gauge group of 
Alice electrodynamics~\cite{schwarz,alice,preskra,alicetop}.  
This section contains an analysis of the Chern-Simons theories 
with a dihedral gauge group, which may 
serve as a stepping stone for an eventual study 
of Alice electrodynamics endowed with a Chern-Simons action.

\begin{table}[h]
\begin{center}
\begin{tabular}[tbh]{ccccc} \hline  \\[-4mm]
$ H $ & $\qquad H^3(H,U(1)) $ & $\qquad H^2(H,U(1))$ & $\qquad H^1(H,U(1))$
\\ \hl \\[-4mm]
$D_{2N+1}$ & $\qquad \Z_{2N+1} \times \Z_2$ & $\qquad 0$ &  $\qquad \Z_2$ 
\\ 
$D_{2N}$ & $\qquad \Z_{2N} \times \Z_2 \times \Z_2$ & 
$\qquad \Z_2$ & $\qquad \Z_2 \times \Z_2$
\\[1mm] \hl
\end{tabular}
\end{center} 
\caption{\sl Cohomology groups for the odd and even dihedral
groups $D_{2N+1}$ and $D_{2N}$.}
\label{comtab}  
\end{table}

The dihedral series naturally falls into the  even and odd dihedral 
groups  $D_{2N}$ and $D_{2N+1}$ (with $N \geq 1$) respectively.  
At the group level, the main distinction between the odd and even
groups is that the latter have a nontrivial 
centre $\Z_2 \subset \Z_{2N} \subset D_{2N}$. 
We have gathered the related cohomology groups in 
table~\ref{comtab}.  
The first cohomology group $H^1(H,U(1))$ corresponds 
to the algebra of 1-dimensional UIR's of the finite group $H$.
For the odd dihedral groups there are two 1-dimensional UIR's,
while there are four 1-dimensional UIR's for even dihedral groups.
Also, as indicated by the second cohomology 
groups displayed in table~\ref{comtab}, the 2-cocycles for odd 
dihedral groups  are trivial. In contrast, the even dihedral groups 
allow for a nontrivial 2-cocycle and consequently 
nontrivial projective representations. It now 
follows from Shapiro's lemma~(\ref{shapiro}) that the associated 
conjugated second cohomology groups  become
\bea 
H^2 (D_{2N},U(1)[D_{2N}]) &\simeq& \oplus_A  \: H^{2} (^A N, U(1)) 
\; \simeq \;  \Z_2^4     \label{dihedrap}   \\
%                          &\simeq&  
%\left( \oplus_{i=1}^2 H^{2} (D_{2N}, U(1))\right) 
%\left( \oplus_{j=1}^2 H^{2} (\Z_2 \times \Z_2, U(1))\right) 
%\left(\oplus_{k=1}^{N-1} H^{2} (\Z_{2N}, U(1)) \right) \\
H^2 (D_{2N+1},U(1)[D_{2N+1}]) &\simeq&  0 \, .  \label{dihedrap2}  
\eea 
In establishing the result~(\ref{dihedrap}), we used
the obvious fact that the centralizer of 
the two centre elements is the full group $D_{2N}$,  
which contributes two $\Z_2$ terms.
Further, there are also two conjugacy classes 
with centralizer $\Z_2 \times \Z_2$ which lead to 
the other two $\Z_2$ terms as follows from~(\ref{conj2e}). 
The remaining conjugacy classes of $D_{2N}$ 
have centralizer $\Z_{2N}$ with trivial second 
cohomology group as indicated by~(\ref{conj2e}).
In a similar fashion, we infer that~(\ref{dihedrap2}) vanishes,
which implies that the conjugated 2-cocycle $\theta$ obtained from 
the 3-cocycles $\omega \in H^3(D_{2N+1}, U(1))$ through the 
slant product~(\ref{theta}) are always trivial. That is, they boil down 
to a coboundary~(\ref{jrep}). Hence, the dyon charges in $D_{2N+1}$ 
Chern-Simons theories form trivial projective centralizer representations. 
Since~(\ref{dihedrap}) is nontrivial this is not necessarily true for 
$D_{2N}$ Chern-Simons theories.
To proceed, the third cohomology group of an  
odd dihedral group is generated by just one element of order  $4N+2$
\bea
H^3 (D_{2N+1},U(1)) &\simeq&  \Z_{2N+1} \times \Z_2 \; \simeq \; 
\Z_{4N+2} \, . \label{h3} 
\eea
The minimal set of  generators for the 
third cohomology group of the even dihedral groups, on the other hand, 
consists of three (commuting) elements: two of order 2 and one of order $2N$.
Let us now recall from section~\ref{topclas}, 
that a discrete $H$  gauge theory ($H \subset SO(3)$)
may occur as the long distance remnant of 
a spontaneously broken $SO(3)$ gauge theory featuring
the $\Z_2$ Dirac monopoles with magnetic charge $g=\frac{2\pi}{e}$.
(Otherwise, the residual gauge group becomes 
the lift $\bar{H} \subset SU(2)$). Alternatively, such a discrete $H$
gauge theory may arise from a $SU(3)$ gauge theory, either directly
$SU(3) \rightarrow H$ or as the final phase of the hierarchy
$SU(3) \rightarrow SO(3) \rightarrow H$, where the
$\Z_2$ monopoles in the intermediate $SO(3)$ phase
are of the regular 't Hooft-Polyakov type.
The Chern-Simons actions for $SU(3)$ and $SO(3)$ 
are classified~\cite{diwi} by the integers: $H^4(BSU(3),\Z) \simeq \Z$ and 
$H^4(BSO(3),\Z) \simeq \Z$. As these cohomology groups are generated by one 
element and $H^3 (D_{2N},U(1))$ by three, we immediately 
conclude that the natural homomorphisms 
$H^4(BSU(3),\Z) \to H^3 (D_{2N},U(1))$ and $H^4(BSO(3),\Z) \to H^3 
(D_{2N},U(1))$, induced by the 
inclusions $D_{2N} \subset SU(3)$ and $D_{2N} \subset SO(3)$ respectively
(see~(\ref{homo})), are not onto. Thus only a subset of the conceivable
$D_{2N}$ Chern-Simons theories occur in spontaneously broken $SU(3)$ and 
$SO(3)$ Chern-Simons gauge theories. Since~(\ref{h3}) 
is generated by just one element, this reasoning does not apply to the 
odd dihedral groups. It is then expected
that the full set of $D_{2N+1}$ Chern-Simons theories may appear 
in spontaneously broken $SO(3)$ or $SU(3)$ Chern-Simons theories. 
Future work should point 
out whether this is indeed the case. We will not dwell
any further on these embeddings. In the following, 
we simply give the explicit realization of the complete set of independent
3-cocycles for the odd dihedral groups 
and discuss the associated Chern-Simons theories. 
In passing, we mention that the realization of the 3-cocycles for the even 
dihedral groups is currently under investigation. 
An interesting question in this respect is whether 
there are 3-cocycles that under the slant product~(\ref{theta})
map into a nontrivial element of the conjugated 
cohomology group~(\ref{dihedrap}).  
That is, whether there exist $D_{2N}$ Chern-Simons theories
featuring dyon charges which correspond to nontrivial projective 
$\Z_2 \times \Z_2$  representations.

Let us start by setting some notational conventions.
The two generators $X$ and $R$ of the odd dihedral group $D_{2N+1}$ 
are subject to the conditions
\bea    \label{feyt}
R^{2N+1} \; = \; e, \qquad X^2 \; = \; e, \qquad
  XR \; = \; R^{-1} X,
\eea
with $e$ the unit element of $D_{2N+1}$.
We will label  the elements of $D_{2N+1}$ by the 2-tuples
\bea    \label{defelemd}
(A,a) & := & X^A R^a  \qquad \qquad \mbox{with $A \in 0,1$ and               
$a \in -N, -N+1,\ldots,N$.} \qquad
\eea
Hence, the capital $A$ represents an element of the $\Z_2$ subgroup 
of $D_{2N+1}$ and the lower-case letter $a$ an 
element of the $\Z_{2N+1}$ subgroup. 
From~(\ref{feyt}) and~(\ref{defelemd}), we then 
infer that the multiplication law becomes 
\bea                  \label{adddodd}
(A,a) \cdot (B,b) &=& ([A+B],[(-)^B a + b])\, ,
\eea
with the abbrevation $(-) := (-1)$. 
Here, the rectangular brackets appearing in the first entry 
naturally indicate modulo $2$ calculus such that the sum lies 
in the range $0,1$ and those for the second entry
modulo $2N+1$ calculus in the range $-N,\ldots,N$. 
%It is now easily verified that the inverse of the elements 
%and conjugation take the form
%\bea
%(A,a)^{-1} &=& (A,[-(-)^A a])  \\
%(B,b) \cdot (A,a) \cdot (B,b)^{-1} &=& (A, [(-)^B (a -2Ab)]).
%\eea

With the conventions established above, 
an explicit realization of the set of 
$4N+2$ independent 3-cocycles corresponding 
to the elements of~(\ref{h3}) can be presented as 
\bea                                      \label{nonabco}
\lefteqn{\omega\left( (A,a),(B,b),(C,c) \right) \;= }\\  
& & \exp \left( \frac{2\pi \im p}{(2N+1)^2} 
\{ (-)^{B+C} a 
\left( (-)^Cb+c-[(-)^Cb+c] \right) +\frac{(2N+1)^2}{2}ABC\} \right) . 
\nn  \;\;\;
\eea
Here, the integral Chern-Simons parameter $p$ characterizing the different 
3-cocycles naturally exhibits the periodicity $4N+2$. We 
choose the range $p \in 0, 1, \ldots,4N+1$ for convenience. 
Furthermore, the rectangular brackets denote modulo $2N+1$ 
calculus in the range $-N, \ldots,N$. 
With the multiplication law~(\ref{adddodd}), it is 
then easily verified that~(\ref{nonabco}) indeed satisfies the 3-cocycle 
condition~(\ref{pentagonna}).
Let us also note that the last term in~(\ref{nonabco}), in fact, 
constitutes the usual 3-cocycle~(\ref{type1}) for the 
subgroup $\Z_2 \subset D_{2N+1}$ albeit written in a compact form.
To continue, the conjugated 2-cocycle $\theta$ appearing in 
the multiplication~(\ref{naalgco}) for the quasi-quantum double
$D^\omega ( D_{2N+1})$ simply follows from a substitution of~(\ref{nonabco}) 
in the slant product~(\ref{theta}). The cochain $\gamma$ entering 
the definition of the comultiplication~(\ref{nacoalgco}) 
is obtained by plugging~(\ref{nonabco}) into~(\ref{gamma}).
%From~(\ref{nonabco}) we can determine $\theta$ as
%\bea                              \label{deo}
%\lefteqn{\theta_{(A,a)}((B,b),(C,c)) \; = } \\
%& & \exp \left( \frac{2\pi \im p}{(2N+1)^2} 
%          \{ (-)^{B+C}a \left((-)^C b +c -[(-)^C b +c] \right) +  \right. 
%\nn \\
%& & \qquad  \mbox{} -(-)^{A+C}b \left((-)^C[(-)^Ba +2Ab] +2Ac 
%               -[(-)^C((-)^Ba +2Ab) +2Ac] \right) + \nn \\ 
%& &  \qquad   \left. \mbox{}      +\frac{(2N+1)^2}{2}ABC \} 
%\right),\nn
%\eea
%while  the comultiplication cochain $\gamma$ takes the form
%\bea
%\lefteqn{\gamma_{(A,a)}((B,b),(C,c)) \; = } \\ 
%& & \exp \left( \frac{2\pi \im p}{(2N+1)^2}  
%   \{ (-)^{C+A}b \left( (-)^A c + 2Ca-[(-)^Ac+2Ca] \right) + \right. \nn \\
%& & \qquad \mbox{}  +(-)^{B+C}a \left( (-)^C[(-)^Ab +2Ba]+[(-)^A c +2Ca] 
%+ \right. \nn \\
%& & \qquad \left. \mbox{}  
%-[(-)^C((-)^Ab + 2Ba) + (-)^A c +2Ca] \right) 
% \left. +\frac{(2N+1)^2}{2} ABC \} \right) . \nn
%\eea   
It may now be checked explicitly that the 
cochain
\bea
\varepsilon_{(A,a)}((B,b))   &=& \exp  \left( \frac{2 \pi \im p}{(2N+1)^2}
                          \{  b \, 
[(-)^B a +2Ab] - Ab^2 +\frac{(2N+1)^2}{4}AB \} \right), \;\;\; \qquad  
\label{eppiek}
\eea
solves relation~(\ref{jrep}) for this case. Here, the rectangular 
brackets again indicate modulo $2N+1$ calculus.
In a similar fashion as for the abelian case discussed in the previous 
chapter, the exponents of these 
additional Aharonov-Bohm phases implied by the Chern-Simons 
action~(\ref{nonabco}) can be interpreted 
as the inner product of the nonabelian fluxes $(A,a)$ and $(B,b)$. 
In particular, upon restricting~(\ref{eppiek}) to 
the fluxes associated to the subgroups $\Z_{2N+1}$ and $\Z_2$
respectively, we arrive at
\begin{eqnfourarray}       \label{ep22}
\varepsilon_{(0,a)}((0,b)) &=& 
\exp \left( \frac{2\pi \im p}{(2N+1)^2} \,  ab \right)
&$\qquad\qquad  -N\leq a,b \leq N$ \\
\varepsilon_{(A,0)}((B,0)) &=& 
\exp \left( \frac{2\pi \im p}{4} \, AB \right)
&$\qquad\qquad  A,B \in 0,1$,
\end{eqnfourarray}
which is exactly the result~(\ref{vareen}) for the abelian fluxes 
in a  $\Z_{2N+1}$ and a $\Z_2$ Chern-Simons theory.

\begin{table}[h] 
\begin{center}
\begin{tabular}[t]{lll} \hline   \\[-4mm]
$ \mbox{Conjugacy class}      $ & & $\str \mbox{Centralizer}       $\\ 
\hl \\[-4mm]  
$ ^0 C=\{ (0,0) \} \qquad              
$& &$ \str D_{2N+1}$\\ 
$ ^a C=\{ (0,a), (0,-a) \} \qquad     
$& &$ \str \Z_{2N+1} \simeq \{ (0,b) \}_{b=-N}^{N}$\\ 
$ ^X C=\{ (1,b)\}_{b=-N}^{N}  \qquad   
$& &$ \str \Z_2 \simeq \{ (0,0), (1,-N) \} $  \\[1mm] \hl
\end{tabular}
\end{center} 
\caption{\sl Conjugacy classes of the odd dihedral group 
$D_{2N+1}$ and their centralizers. 
Here, the label $a$ takes values in the range $1,2,\ldots, N$.}  
\label{tabcon}  
\end{table}
\begin{table}[h] 
\begin{center}
\begin{tabular}[t]{lcccc} \hline \\[-4mm]
$\str D_{2N+1}\qquad $&$ ^0 C $&$ ^a C                       $&$ ^X C $\\ 
\hl \\[-4mm]  
$ \Lambda^+ \str    $&$  1   $&$  1                         $&$  1   $\\ 
$ \Lambda^- \str    $&$  1   $&$  1                         $&$ -1   $\\ 
$ \Lambda^n \str    $&$  2   $&$ 2\cos\left( \frac{2\pi na}{2N+1}\right)   
$&$  0   $\\[1mm]  \hl
\end{tabular} \qquad \qquad
\end{center}
\caption{\sl Character table of $D_{2N+1}$. The representation 
label $n$ and the conjugacy class label $a$ both take values in the range 
$1,2, \ldots, N$.}
\label{tabchar}
\end{table}

The spectrum of a $D_{2N+1}$ gauge theory consists of
$(2N^2+2N+4)$ distinct particles.  
As indicated by the table~\ref{tabcon}, there are 
$N+1$ different particles carrying  nontrivial pure magnetic flux.
To start with, the conjugacy class $^X C$ labels the Alice flux
which in this discrete version of Alice electrodynamics  may take 
$2N+1$ different disguises being the different elements of $^X C$.
The $N$ different conjugacy classes $^a C$ then 
describe magnetic flux  doublets. That is, these
consist of a nontrivial $\Z_{2N+1}$ flux $(0,a)$ 
and the associated anti-flux $(0,-a)$, which transform 
into each other under conjugation by the elements of the conjugacy 
class $^X C$. In other words, when a $\Z_{2N+1}$ 
flux $(0,a)$ encircles an Alice flux $\in$$\, ^X C$ it returns as 
its anti-flux $(0,-a)$.
Furthermore, as follows from the character table~\ref{tabchar},
this theory features $N+1$ different nontrivial charges: 
one singlet charge $\Lambda^-$ and $N$ doublet charges $\Lambda^n$.
In fact, the doublet charges $\Lambda^n$ consist of a 
nontrivial $\Z_{2N+1}$ charge paired with its anti-charge, 
which transform into each other
under the action of the Alice fluxes $\in$$\,^X C$.
The trivial $D_{2N+1}$ representation $\Lambda^+$ denotes the vacuum. 
The remaining  $2N^2+1$ particles in the spectrum are dyons.
As displayed in table~\ref{tabcon}, the centralizer related to the 
conjugacy classes $^a C$ is the cyclic group $\Z_{2N+1}$.
Hence, there are $2N^2$ distinct dyons carrying a magnetic doublet flux and
nontrivial $\Z_{2N+1}$ charge. Finally, there is just one dyon 
associated to the Alice flux $^X C$, namely that with nontrivial
$\Z_2$ charge. Henceforth, these particles will be denoted as 
\bea             \ba{rclrcl}          \label{vespe}
(0,+)    &:=&  (\, ^0 C, \, \Lambda^+ \, )
& \qquad \qquad  (a,l)    &:=&  (\,  ^a C , \, \Gamma^l \, )  \\   
(0,-)    &:=&  (\, ^0 C, \,\Lambda^- \, )     
& \qquad \qquad  (X,+)    &:=& (\, ^X C,\, \Gamma^+ \, )   \\
(0,n)    &:=& (\, ^0 C,\,\Lambda^n \, )     
&\qquad \qquad  (X,-)    &:=& (\, ^X C,\, \Gamma^- \, ) \, .
\ea
\eea 
Here the flux label $a$ and the pure charge label 
$n$ run from $1$ to $N$, the dyon charge label 
$l$ takes values in the range $0,1,\ldots,2N$ and $\Gamma^+$ denotes 
the trivial and $\Gamma^-$ the nontrivial $\Z_2$ representation.

The spin factors assigned to the particles~(\ref{vespe})
depend on the Chern-Simons action~(\ref{nonabco}) 
added to this $D_{2N+1}$ gauge theory.
From~(\ref{modultspi}) and~(\ref{eppiek}), we obtain
\bea             \label{spd2n2}
\ba{cc}
\mbox{particle}                &\qquad   \exp (2\pi \im  s)    \\
           & \\
(0,+), \, (0,-), \, (0,n)   &\qquad     1    \\
(a, l)                      &\qquad   
\exp \left( \frac{2\pi \im}{2N+1}(al +\frac{p}{2N+1} \, aa) \right)    \\
(X,\pm)                     &\qquad \pm  \im^p \, ,
\ea          
\eea    
with $p$ the integral Chern-Simons parameter labeling the 
independent  3-cocycles~(\ref{nonabco}).

Let us turn to the fusion algebra~(\ref{Ncoemet}) for this 
$D_{2N+1}$ Chern-Simons theory.
First of all, the fusion rules for the pure charges are of course 
unaffected by the presence of a nontrivial Chern-Simons 
action and simply follow from the character table~\ref{tabchar}
\bea         
(0,\pm)\times(0,\pm) &=&  (0,+)   \\
(0,\pm )\times(0,\mp)&=& (0,-)     \\
(0,\pm) \times (0,n) &=&  (0,n)    \\
(0,n) \times (0,n')  &=&  \left\{ \ba{ll} (0,+)+(0,-) &
                              \;\; \mbox{if $n=n'$} \\
                            (0,|n-n'|)  &
                              \;\; \mbox{otherwise} \ea \right. \qquad \\
                     & &  \mbox{}      
                          + \left\{ \ba{ll} (0,n+n') & 
                              \;\; \mbox{if $n+n' \leq N$} \\
                            (0,2N+1-n-n') & 
                              \;\; \mbox{otherwise.} \ea \right. \qquad \nn
\eea 
The fusion rules for these pure charges with the other particles 
in the spectrum~(\ref{spd2n2}) then read
\bea
(0,\pm) \times (a,l)   &=& (a,l)       \\  
(0,n) \times (a,l)     &=& (a,[l+n]) + (a,[l-n])  \label{rect2N} \\ 
(0,\pm) \times (X,\pm) &=& (X,\pm)   \\
(0,\mp) \times (X,\pm) &=& (X,\mp)   \\
(0,n)\times(X,\pm) &=& (X,+) + (X,-) \,  ,
\eea                                    
where the rectangular brackets in~(\ref{rect2N}) 
indicate modulo $2N+1$ calculus such that the sum always lies in the range 
$0,1,\ldots, 2N$. The fusion rule~(\ref{rect2N}), in fact, expresses
that the $\Z_{2N+1}$ charge/anti-charge paired in the doublet $(0,n)$ simply 
add/subtract with the $\Z_{2N+1}$ charge $l$ of the dyon $(a,l)$. 
To proceed, the presence of a nontrivial Chern-Simons action~(\ref{nonabco})
affects fusion processes among particles carrying a doublet flux. 
Specifically, the fusion rules for particles 
carrying the same doublet flux become
\bea
(a,l)\times(a,l') &=& \left\{ \ba{ll} (0,+) +(0,-) & 
                              \qquad \mbox{if $l=l'$} \\   
                                      (0,|l-l'|) & 
                              \qquad \mbox{if $0 < |l-l'| \leq N$} \\
                                     (0,2N+1-|l-l'|) & 
                              \qquad \mbox{if $N< |l-l'| \leq 2N$} \ea \right.  \\
             & & \mbox{} +\left\{ \ba{ll}
                   (2a,[l+l']) &  \mbox{if $2a\leq N$}\\
                   (2N+1-2a, [-l-l'-2p]) & \mbox{otherwise,} \ea \right. \nn 
\eea
while particles carrying different doublet flux $a \neq a'$ amalgamate as
\bea 
(a,l)\times(a',l')  &=& \left\{ \ba{ll} 
                 (a-a',[l-l']) & \qquad \qquad \mbox{if $a-a'>0$} \\
                 (a'-a,[l'-l]) & \qquad \qquad \mbox{otherwise} \ea \right. \\
              & & \mbox{} +\left\{ \ba{ll}
                   (a+a',[l+l']) &  \mbox{if $a+a'\leq N$}\\
                   (2N+1-(a+a'), [-l-l'-2p]) & \mbox{otherwise,} \ea \right.
\qquad \nn 
\eea
with $p$ the integral Chern-Simons parameter. As in the $\Z_N$ Chern-Simons 
theory discussed in section~\ref{rev}, the twist 
in these fusion rules simply reflect the fact that the flux 
tunneling $\Delta a = -(2N+1)$ induced
by a minimal  monopole/instanton 
is accompanied by a charge jump $\Delta l = 2p$ in the presence
of a Chern-Simons action.
Furthermore, fusing a particle which carries a doublet flux 
with a particle carrying Alice flux yields
\bea 
(a,l)\times(X,\pm) &=& (X,+)+(X,-) \, .
\eea  
To conclude, the fusion rules for particles both carrying Alice flux read
\bea
(X,\pm)\times(X,\pm) &=& (0,+) + \sum_{n=1}^N \; (0,n) + 
                \sum_{a=1}^N \sum_{l=0}^{2N} \; (a,l)  \\
(X,\pm)\times(X,\mp) &=& (0,-) + \sum_{n=1}^N \; (0,n) + 
                \sum_{a=1}^N \sum_{l=0}^{2N} \; (a,l) \, ,
\eea
which in particular express conservation of $\Z_2$  charge.

Some remarks concerning this fusion algebra are pertinent. 
First of all, the 3-cocycle related to 
the $\Z_2$ subgroup of $D_{2N+1}$, i.e.\ the last term in~(\ref{nonabco}),
has no effect on the fusion rules. The charge jump $2p$ accompanying the flux
tunneling $\Delta A =-2$ induced by a minimal monopole is absorbed
by the modulo 2 calculus for the $\Z_2$ charges.
In other words, the periodicity of the fusion algebra
in the integral Chern-Simons parameter $p$ is half of that 
of the 3-cocycle~(\ref{nonabco}), that is, there are only $2N+1$ 
different sets of fusion rules.
Finally, a characteristic feature of these $D_{2N+1}$ 
Chern-Simons theories is that all particles 
are their own anti-particle as indicated by the occurrence 
of the vacuum in the fusion rules for identical particles.

\sectiona{$\bar{D}_N$ Chern-Simons theory}
\label{dubbel}

As a general result for finite subgroups $H$ of $SU(2)$, we 
have (see appendix~\ref{laap}) 
\bea                      \label{generaal}
H^3(H,U(1)) & \simeq & \Z_{|H|} \, .
\eea 
In other words, the number of independent 3-cocycles for a finite group 
$H \subset SU(2)$ coincides with the order $|H|$ of $H$.  
Moreover, the complete set of discrete $H$ Chern-Simons 
theories corresponding to these 3-cocycles may 
appear in a spontaneously broken $SU(2)$ 
Chern-Simons gauge theory $SU(2) \rightarrow H$. That is, 
the natural homomorphism or restriction~(\ref{homo}) 
induced by the inclusion $H \subset SU(2)$ is surjective
\bea 
H^4(BSU(2), \Z) \;\simeq \; \Z 
  &\longrightarrow&  H^3(H,U(1)) \; \simeq \; \Z_{|H|}   \\
p &\longmapsto & p \qquad  \mbox{mod $|H|$.} \nn
\eea
Thus the  integral $SU(2)$ Chern-Simons parameter $p$ becomes 
periodic (period $|H|$) in a 
broken phase with residual finite gauge group $H$.
Further, the 2-cocycles for finite subgroups $H$ of $SU(2)$ 
are trivial
\bea  \label{geennar}
H^2(H,U(1)) &\simeq& 0.
\eea 
This result, a proof of which is contained in appendix~\ref{laap},
implies that the conjugated second cohomology group also vanishes
\bea   \label{thetatriv}
H^2 (H, U(1) [H]) 
&\simeq&  \oplus_A  \: H^{2} (^A N, U(1)) \; \simeq \; 0 \,. 
\eea 
Here, the first isomorphism is due to Shapiro's lemma~(\ref{shapiro}) 
whereas~(\ref{geennar}), which 
naturally indicates that $H^{2} (^A N, U(1)) \simeq  0$ for the centralizers
$^A N \subset H \subset SU(2)$ related to the conjugacy classes $^A C$
of $H$, subsequently accounts for the last isomorphism. 
From~(\ref{thetatriv}), we then
conclude that the conjugated 2-cocycle $\theta \in H^2 (H, U(1)[H])$, 
following from the 3-cocycle $\omega \in H^3 (H, U(1))$ through the slant 
product~(\ref{theta}), boils down to  a coboundary~(\ref{jrep}). 
In short, the dyon charges 
in Chern-Simons theories with gauge group a finite subgroup
$H$ of $SU(2)$ form trivial projective centralizer representations.

In this section, we focus on the Chern-Simons theories with gauge group 
the double dihedral group $\bar{D}_N \subset SU(2)$. 
As the order of $\bar{D}_N$
equals $4N$, we infer from~(\ref{generaal}) 
\bea      \label{co3}
H^3(\bar{D}_N,U(1)) &\simeq & \Z_{4N} \, ,
\eea 
This result can actually also be found in~\cite{cartan}, where it was
derived by means of a complete resolution for $\bar{D}_N$.
The explicit realization of the 3-cocycles related to the different 
elements of this cohomology group
involves some notational conventions which we establish first.

The double dihedral group $\bar{D}_N$ can be presented 
by two generators $R$ and $X$ subject to the relations
\bea    \label{fey}
R^{2N} \; = \; e, \qquad X^2 \; = \; R^N, \qquad  XR \; = \; R^{-1} X.
\eea
Here, $e$ is the unit element of $\bar{D}_N$. 
We will denote the elements of $\bar{D}_N$  by the 2-tuples
\bea    \label{defelem}
(A,a) & := & X^A R^a  \qquad \qquad \mbox{with $A \in 0,1$ and               
$a \in -N+1, -N+2, \ldots,N$.   } \qquad
\eea
So for instance $e=(0,0)$. As follows from~(\ref{fey}) 
and~(\ref{defelem}), the multiplication law then reads
\bea     \label{multdn}
(A,a) \cdot (B,b) &=& ([A+B], [(-)^B a+b+NAB]),
\eea 
where the rectangular brackets for the first entry of the 2-tuple
indicate  modulo 2 calculus such that the sum lies in the range $0,1$,
while those for the second entry imply
modulo $2N$ calculus such that the sum lies in the range $-N+1,\ldots,N$. 
%With~(\ref{multdn}) it is now readily verified that 
%that the inversion operation and conjugation take the form
%\bea
%(A,a)^{-1} &=& (A,[-(-)^A a+NA]) \\
%(B,b) \cdot (A,a) \cdot (B,b)^{-1} &=& (A, [(-)^B (a - 2Ab)]).
%\label{conju}
%\eea

In this additive presentation of $\bar{D}_N$, 
the 3-cocycles corresponding to the even elements of~(\ref{co3})
are of the form
\bea    \label{oom}
\lefteqn{\omega((A,a),(B,b),(C,c)) \; =} \\
& & \exp \left( \frac{2\pi \im p}{2N^2}
                 \{ (-)^{B+C}a \left( (-)^C b + c- [(-)^C b + c +NBC] \right) 
                 - \frac{N^2}{2} ABC \} \right), \qquad  \nn
\eea
where the integral and periodic parameter $p$ labeling  the 
independent 3-cocycles takes values in the range 
$0,1,\ldots, 2N-1$. The rectangular brackets indicate modulo $2N$
calculus in the range $-N+1,\ldots, N$.  
With the multiplication rule~(\ref{multdn}),
it is readily checked that~(\ref{oom}) indeed satisfies 
the relation~(\ref{pentagonna}).   In passing, we remark that
the 3-cocycles related to the odd elements of~(\ref{co3}), i.e.\
$p \rightarrow p/2$, are currently  under investigation.
The conjugated 2-cocycle $\theta$ deforming the multiplication,
and the cochain $\gamma$ entering the comultiplication 
of the quasi-quantum double $D^\omega (\bar{D}_N)$, 
follow from substituting~(\ref{oom}) in the expressions~(\ref{theta}) 
and~(\ref{gamma}) respectively. 
%\bea                              \label{de}
%\lefteqn{\theta_{(K,k)}((L,l),(M,m)) \; =}   \\
%& & \exp \left( \frac{2\pi \im p}{2N^2}  
%    \{ (-)^{L+M}k \left( (-)^M l +m -[(-)^M l +m+NLM] \right) + \right. \nn\\
%& & \qquad \mbox{} -(-)^{K+M}l \left( (-)^M[(-)^Lk +2Kl] +2Km 
%-[(-)^M((-)^Lk +2Kl) +2Km]\right) 
%\nn \\
%& &  \left. \mbox{} -\frac{N^2}{2}KLM \} \right) ,
%\nn \qquad 
%\eea
%while  the comultiplication cochain $\gamma$ takes the form
%\bea
%\lefteqn{\gamma_{(M,m)}((K,k),(L,l)) \;=}    \\
%& & \exp \left( \frac{2\pi \im p}{2N^2}
%\{ (-)^{L+M}k \left((-)^M l + 2Lm-[(-)^Ml+2Lm] \right) + \right.   \nn \\
%& & \mbox{} +(-)^{K+L}m \left( (-)^L[(-)^Mk +2Km]+[(-)^M l +2Lm] \, + 
%\right. \nn \\  
%& &  \left.  \mbox{}  -[(-)^L((-)^Mk + 2Km) + (-)^M l +2Lm +NKL] \right)
% \left. -\frac{N^2}{2} KLM \} \right) . \nn 
%\eea
From relation~(\ref{jrep}), we then obtain that the cochain 
$\varepsilon$ associated to the 3-cocycle~(\ref{oom}) becomes
\bea       \label{eppo}
\varepsilon_{(A,a)}((B,b)) \, = \,
\exp \left( \frac{2\pi \im p}{2N^2} 
                   \{ -(-)^Ab[-(-)^A\{(-)^Ba +2Ab\}]-Ab^2+\frac{N^2}{4}AB
                   \} \right).      \;   \qquad
\eea
The rectangular brackets again imply modulo $2N$ calculus in the range
$-N+1,\ldots,N$. 

\begin{table}[htb] 
\begin{center}
\begin{tabular}[t]{lcl} \hline     \\[-4mm]
$ \mbox{Conjugacy class}                      $ & & $\str 
\mbox{Centralizer}       $\\ \hl   \\[-4mm]
$ ^0  C=\{(0,0)\} \str                   $& &$ \bar{D}_N$\\ 
$ ^N  C=\{(0,N)\} \str                 $& &$ \bar{D}_N$\\ 
$ ^a  C=\{(0,a),(0,-a)\}\str              
$& &$ \Z_{2N} \simeq \{(0,0),(0,1),\ldots,(0,[2N-1])\}$\\ 
$ ^X C=\{(1,0),(1,[2]),\ldots,(1,[2N-2])\}       
$& &$ \str \Z_4 \simeq\{(0,0),(1,0),(0,N),(1,N)\} $\\ 
$ ^{\bar{X}}  C=\{(1,1),(1,[3]),\dots,(1,[2N-1])\}
$& &$ \str \Z_4 \simeq \{(0,0),(1,1),(0,N),(1,-N+1)\} $\\[1mm] \hl
\end{tabular}
\end{center} 
\caption{\sl Conjugacy classes of the double dihedral 
group $\bar{D}_N$ and the associated centralizers. 
Here, the label $a$ takes values 
in the range $1,\ldots,N-1$, whereas the rectangular brackets indicate 
modulo $2N$ calculus in the range $-N+1, -N+2, \ldots, N$.}
\label{tabce}  
\end{table} 
\begin{table}[htb] 
\begin{center}
\begin{tabular}[t]{lccccc} \hline      \\[-4mm]
$\str \bar{D}_N\qquad$&$ ^0 C $&$ ^a C$&$ ^N C  $&$ ^X C  $&$ ^{\bar{X}} C
$\\ \hl \\[-4mm]
$ \Gamma^{++} \str$&$ 1 $&$ 1      $&$ 1  $&$ 1  $&$ 1  $\\ 
$ \Gamma^{+-} \str$&$ 1 $&$ 1      $&$ 1  $&$-1  $&$-1  $\\ 
$ \Gamma^{-+} \str$&$ 1 $&$ (-1)^a $&$ (-1)^N  $&$ \im^N  $&$ -\im^N $\\ 
$ \Gamma^{--} \str$&$ 1 $&$ (-1)^a $&$ (-1)^N  $&$-\im^N  $&$ \im^N $\\ 
$ \Gamma^n\str$&$ 2 $&$ 2\cos{\left(\frac{\pi na}{N}\right)} 
$&$ 2\cos{(\pi n)} $
&$ 0  $&$ 0  $\\[1mm]   \hl
\end{tabular} \qquad \qquad
\end{center}
\caption{\sl Character table of $\bar{D}_N$. 
The representation label $n$ and the conjugacy 
class label $a$  both take values in the range $1,2,\ldots,N-1$.}
\label{charis}
\end{table}

Let us now establish the spectrum of a $\bar{D}_N$ gauge theory. 
As follows from the character 
table~\ref{charis}, such a theory 
features $N+2$ nontrivial charges.
That is, three singlet charges labeled 
by $\Gamma^{+-}, \Gamma^{-+}, \Gamma^{--}$ and $N-1$ doublet 
charges $\Gamma^n$. The trivial $\bar{D}_N$ 
representation $\Gamma^{++}$ denotes the vacuum.
Furthermore, the elements of $\bar{D}_N$ are divided 
into $N+3$ conjugacy classes. These are displayed in table~\ref{tabce}
together with their centralizers. The conjugacy class $^N C$ contains the 
nontrivial centre element  $(0,N)$. In other words,
the associated centralizer is the full group $\bar{D}_N$. Hence,
there are $N+3$ different particles with the 
singlet flux $(0,N)$, namely the pure flux $(0,N)$ itself 
and a total number of $N+2$ dyons carrying this flux 
and a nontrivial $\bar{D}_N$ charge. The spectrum 
also contains $N-1$ different  doublet fluxes 
labeled by the conjugacy classes $^a C$.
The related dyons carry a nontrivial $\Z_{2N}$ centralizer charge
$\Gamma^l$  (with $l \in 0,1,\ldots, 2N-1$) as defined in~(\ref{znorrep}).
Finally, the conjugacy classes $^X C$ and $ ^{\bar{X}} C$ both consist 
of $N-1$ elements and have centralizer  $\Z_4$. 
We will denote the corresponding  
$\Z_4$ charges as $\Gamma^\sigma$ with $\sigma \in 0,1,2,3$.
To conclude, the complete spectrum consists of a total number 
of $2N^2+14$ distinct particles, which will be labeled as 
\bea   \label{spectredN}       \ba{rclrcl}
(0,rs)   &:=& ( \,  ^0 C, \, \Gamma^{rs} \, )

& \qquad \qquad (N,rs)   &:=& 
(\, ^N C, \,  \Gamma^{rs} \, )   \\

(0,n)   &:=& ( \, ^0 C, \, \Gamma^{n} \, )        

& \qquad \qquad (N,n)   &:=& 
( \, ^N C, \, \Gamma^{n} \, )             \\

(X, \sigma) &:=& ( \, ^X C , \, {\Gamma^\sigma}\, )

& \qquad \qquad ( \, {\bar{X}}, \, \sigma \,) &:=& 
( \, ^{\bar{X}} C , \, {\Gamma^\sigma} \, )  \\

(a,l) &:=& (\,^a C, \, \Gamma^{l}\,) \, .   &  && 
\ea
\eea
Here,  $r$ and $s$ label the singlet $\bar{D}_N$ charges, that is 
$r,s\in +, -$, whereas the doublet charge label $n$ and the doublet flux
label $a$ both take values in the range $1,\ldots,N-1$.

The introduction of the Chern-Simons action~(\ref{oom}) in this $\bar{D}_N$ 
gauge theory affects the spin factors assigned to the particles 
in the spectrum~(\ref{oom}). 
Specifically, from~(\ref{modultspi}) and~(\ref{eppo}) we obtain
\bea             \label{spd2bn2}
\ba{cc}
\mbox{particle}                &\qquad   \exp (2\pi \im  s)    \\
           & \\
(0,rs), \, (0,n)     &\qquad     1             \\
(N,+ \pm)                  &\qquad     (-1)^{p}    \\
(N,- \pm)                  &\qquad     (-1)^{N+p}  \\
(N,n)                   &\qquad     (-1)^{n+p}     \\
(a, l)                  &\qquad   
\exp \left( \frac{2\pi \im}{2N}(a l +\frac{p}{N} \, aa ) \right)    \\
(X,\sigma)              &\qquad  \im^{\sigma +p}  \\
(\bar{X},\sigma)         & \qquad \im^{\sigma+p} \, ,
\ea          
\eea  
with $p$ the integral Chern-Simons parameter in~(\ref{oom}).

We close this section by enumerating 
the fusion rules~(\ref{Ncoemet}) for this $\bar{D}_N$ Chern-Simons theory.
To start with, the pure charges amalgamate as 
\bea
(0,rs)\times(0,r's')&=&\left\{ \ba{ll} 
        \left(0,(r \cdot r')((-)^N s \cdot s')\right)&\mbox{if $r=r'=-$}  \\
        (0,(r \cdot r')(s \cdot s')) & \mbox{otherwise} \ea \right.    
                 \label{eend} \\
(0,rs)\times(0,n)  &=& \left(0,|n-\frac{1}{2}(1-r)N| \right) 
\label{defactor}    \\
(0,n)\times(0,n')&=& \left\{ \ba{ll} 
                   (0,++)+(0,+-) & \mbox{if $n=n'$}  \\
                   (0,|n-n'|) & \mbox{otherwise} \ea \right.  \\
               & & +\left\{ \ba{ll} 
                   (0,n+n')       & \mbox{if $n+n'<N$}  \\
                   (0,-+)+(0,--) & \mbox{if $n+n'=N$}  \\
                   (0,2N-n-n') & \mbox{if $n+n'>N$,} \ea \right.    \nn
\eea
where the factor $(1-r)$ appearing in~(\ref{defactor}) by 
definition equals $2$ if $r=-$ and $0$ if $r=+$. 
The fusion rules $(0,\alpha)\times(N,\beta)$, and $(N,\alpha)\times(N,\beta)$,
where $\alpha$ and $\beta$  label the set of  $\bar{D}_N$   representations,
follow from the above results and the class algebra.
To proceed, 
the composition rules for the pure charges with the other particles read
\bea
(0,rs)\times(a,l)      &=&    \left(a,[l+\frac{1}{2}(1-r)N]\right) 
\label{z2nrect}   \\
(0,n)\times(a,l)       &=&    \left(a,[l+n]\right)+ \left(a,[l-n]\right)  
\label{z2nrect1} \\
(0,rs)\times(X,\sigma)
&=& \left(X,[\sigma+(1-s)+\frac{1}{2}(1-r)N]\right)  \\
(0,rs)\times(\bar{X},\sigma)
&=&\left(\bar{X},[\sigma+(1-s)+\frac{1}{2}(1-r)(N+2)]\right)  \\  
(0,n)\times(X/\bar{X},\sigma)
&=&\left(X/\bar{X},[\sigma+n]\right) + 
\left(X/\bar{X},[\sigma+n+2]\right).
\eea
Here, the rectangular brackets for the $\Z_{2N}$ charges in~(\ref{z2nrect})
and~(\ref{z2nrect1}) naturally denote modulo $2N$ calculus, while
those for the $\Z_4$ charges in the other rules denote modulo 4 calculus.
The presence of the 
Chern-Simons action~(\ref{oom}) affects fusion 
of a particle $(N,rs)$ or $(N,n)$ with 
a particle carrying a doublet flux 
\bea
(N,rs)\times(a,l)&=&(N-a,[-l-2p+\frac{1}{2}(1-r)N]) \label{sok} \qquad \\
(N,n)\times(a,l)&=&(N-a,[-l+n-2p])+(N-a,[-l-n-2p])  \qquad \label{sok2} \\
(N,rs)\times(X,\sigma)&=&
\left\{ \ba{ll} 
(X,[-\sigma+(1-s)+\frac{1}{2}(1-r)N]) & \mbox{for even $N$}  \qquad \\
(\bar{X},[-\sigma+(1-s)+\frac{1}{2}(1-r)(N+2)]) & \mbox{for odd $N$} \ea 
\right.   \qquad  \\ 
(N,rs)\times(\bar{X},\sigma)&=&
\left\{ \ba{ll} 
(\bar{X},[-\sigma+(1-s)+\frac{1}{2}(1-r)(N+2)]) & \mbox{for even $N$}  
\qquad \\
(X,[-\sigma+(1-s)+\frac{1}{2}(1-r)N]) & \mbox{for odd $N$} \ea 
\right.  \qquad  \\ 
(N,n)\times(X/\bar{X},\sigma)&=&
\left\{ \ba{ll} 
(X/\bar{X},[\sigma+n]) + (X/\bar{X},[\sigma+n+2]) & \mbox{for even $N$}  
\qquad  \\
(\bar{X}/X,[\sigma+n]) + (\bar{X}/X,[\sigma+n+2]) & 
\mbox{for odd $N$,}    \ea 
\right.                                                        
\eea
with $p$ in~(\ref{sok}) and~(\ref{sok2}) the integral Chern-Simons parameter.
The fusion rules for two particles both carrying a doublet flux 
also depend on the Chern-Simons parameter
\bea                                 \label{coq}
(a,l)\times(a',l') &=& \left\{ \ba{ll} 
                   (a+a',[l+l']) & \mbox{if $a+a'<N$}  \\
                   (2N-a-a',[-l-l'-4p]) & \mbox{if $a+a'>N$} \ea \right.  \\
               & & \mbox{} + \left\{ \ba{ll} 
 \left(N,++\right)+\left(N,+-\right) & \mbox{if $a+a'=N$ and $[l+l'+2p]=0$} \nn \\  
 \left(N,[l+l'+2p]\right) & \mbox{if $a+a'=N$}    \\
                          &  \mbox{and $0<[l+l'+2p]<N$} \\  
 \left(N,-+\right)+\left(N,--\right) & \mbox{if $a+a'=N$ and $[l+l'+2p]=N$} \\                    
 \left(N,2N-[l+l'+2p]\right) & \mbox{if $a+a'=N$} \\
                                     &\mbox{and $N<[l+l'+2p]<2N$} \\  
                     \ea \right.  \\
                  & & \mbox{} + \left\{ \ba{ll} 
                   \left(a-a',[l-l']\right)       
                     & \qquad \mbox{if $a-a'>0$}  \nn \\
                   \left(a'-a,[l'-l]\right) 
                     & \qquad \mbox{if $a-a'<0$} \ea \right. \\
               & & \mbox{} + \left\{ \ba{ll} 
     \left(0,++\right)+\left(0,+-\right) & \qquad \mbox{if $a=a'$ and $l-l'=0$} \nn \\  
     \left(0,|l-l'|\right) & 
     \qquad \mbox{if $a=a'$ and $0<|l-l'|<N$} \\  
     \left(0,-+\right)+\left(0,--\right) & 
     \qquad \mbox{if $a=a'$ and $|l-l'|=N$} \\                    
     \left(0,2N-|l-l'|\right) & \qquad \mbox{if $a=a'$ and $N<|l-l'|<2N$.} \\  
                     \ea \right.  
\eea
Further
\bea
(a,l)\times(X/\bar{X},\sigma)&=&
\left\{ \ba{ll} 
(X/\bar{X},[\sigma+l]) + (X/\bar{X},[\sigma+l+2]) & \mbox{if $a$ is even}  \\
(\bar{X}/X,[\sigma+l]) + (\bar{X}/X,[\sigma+l+2]) & \mbox{if $a$ is odd.}  \ea
\right.  \qquad                                                      
\eea
For the remaining fusion rules, it is again important to make the distinction 
between even and odd $N$. For even  $N$, we have
\bea
(X,\sigma)\times(X,\sigma')&=&\delta_{[\sigma-\sigma'],0}\; (0,++) +
                         \delta_{[\sigma-\sigma'],2}\; (0,+-)    \\
           & & \mbox{} +\delta_{[\sigma-\sigma'+N],0}\; (0,-+) +
                         \delta_{[\sigma-\sigma'+N],2}\; (0,--) \nn \\
           && \mbox{} +\delta_{[\sigma-\sigma'],{\rm even }}
              \sum_{n=1}^{\frac{1}{2}(N-2)} (0,2n)  + 
             \delta_{[\sigma-\sigma'],{\rm odd }}
              \sum_{n=0}^{\frac{1}{2}(N-2)} (0,2n+1)  \nn \\
           & &\mbox{} +\sum_{a=1}^{\frac{1}{2}(N-2)}\sum_{l=0}^{N-1}
               \{\delta_{\sigma+\sigma',{\rm even }} \; (2a,2l)+
              \delta_{\sigma+\sigma',{\rm odd }} \; (2a,2l+1)\} \nn \\
           & & \mbox{} +\delta_{[\sigma+\sigma'],0} \;(N,++) +
                         \delta_{[\sigma+\sigma'],2}\; (N,+-)    \nn \\
           & & \mbox{} +\delta_{[\sigma+\sigma'+N],0}\; (N,-+) +
                         \delta_{[\sigma+\sigma'+N],2}\; (N,--) \nn \\
           & & \mbox{} +\delta_{[\sigma+\sigma'],{\rm even }}
              \sum_{n=1}^{\frac{1}{2}(N-2)} (N,2n) 
           + \delta_{[\sigma+\sigma'],{\rm odd }}
              \sum_{n=0}^{\frac{1}{2}(N-2)} (N,2n+1)  \nn \\
(X,\sigma)\times(\bar{X},\sigma')&=&   \!\!\!
            \sum_{a=0}^{\frac{1}{2}(N-2)}\sum_{l=0}^{N-1}\{
             \delta_{\sigma+\sigma',{\rm even }} \; (2a+1,2l)   
                       +
              \delta_{\sigma+\sigma',{\rm odd }} \; (2a+1,2l+1)\} 
\;\;\;\;\;\;  \qquad \\
(\bar{X},\sigma)\times(\bar{X},\sigma') &=& 
                \delta_{[\sigma-\sigma'],0}\; (0,++) +
                         \delta_{[\sigma-\sigma'],2}\; (0,+-)    \\
           & & \mbox{} +\delta_{[\sigma-\sigma'+N],2}\; (0,-+) +
                         \delta_{[\sigma-\sigma'+N],0}\; (0,--) \nn \\
           & & \mbox{} + \delta_{[\sigma-\sigma'],{\rm even }}
              \sum_{n=1}^{\frac{1}{2}(N-2)} (0,2n)  + 
             \delta_{[\sigma-\sigma'],{\rm odd }}
              \sum_{n=0}^{\frac{1}{2}(N-2)} (0,2n+1)  \nn \\
           & & \mbox{} + \sum_{a=1}^{\frac{1}{2}(N-2)}\sum_{l=0}^{N-1}
                \{\delta_{\sigma+\sigma',{\rm even }} \; (2a,2l)+
              \delta_{\sigma+\sigma',{\rm odd }} \; (2a,2l+1)\} \nn \\
           & & \mbox{} +  \delta_{[\sigma+\sigma'],0} \;(N,++) +
                         \delta_{[\sigma+\sigma'],2}\; (N,+-)    \nn \\
           & & \mbox{} +   \delta_{[\sigma+\sigma'+N],2}\; (N,-+) +
                         \delta_{[\sigma+\sigma'+N],0}\; (N,--) \nn \\
           & & \mbox{} +  \delta_{[\sigma+\sigma'],{\rm even }}
              \sum_{n=1}^{\frac{1}{2}(N-2)} (N,2n) 
           + \delta_{[\sigma+\sigma'],{\rm odd }}
              \sum_{n=0}^{\frac{1}{2}(N-2)} (N,2n+1),  \nn    
\eea
with $\delta$ the kronecker delta function.
For odd $N$, we then arrive at 
\bea
(X,\sigma)\times(X,\sigma')&=&\delta_{[\sigma+\sigma'],0} \;(N,++) +
                         \delta_{[\sigma+\sigma'],2}\; (N,+-)     \\
           & & \mbox{} +   \delta_{[\sigma+\sigma'+N],2}\; (N,-+) +
                         \delta_{[\sigma+\sigma'+N],0}\; (N,--) \nn \\
           & & \mbox{} +   \delta_{[\sigma+\sigma'],{\rm even }} 
              \sum_{n=1}^{\frac{1}{2}(N-1)} (N,2n) 
           + \delta_{[\sigma+\sigma'],{\rm odd }} 
              \sum_{n=0}^{\frac{1}{2}(N-3)} (N,2n+1)  \nn  \\
           & & \mbox{} +  \sum_{a=0}^{\frac{1}{2}(N-3)} \sum_{l=0}^{N-1}
               \{\delta_{\sigma+\sigma',{\rm even }} \; (2a+1,2l)+
              \delta_{\sigma+\sigma',{\rm odd }} \;  (2a+1,2l+1)\} \nn \\
(X,\sigma)\times(\bar{X},\sigma')&=&\delta_{[\sigma-\sigma'],0}\; (0,++) +
                         \delta_{[\sigma-\sigma'],2}\; (0,+-)    \\
           & & \mbox{} +   \delta_{[\sigma-\sigma'+N],2}\; (0,-+) +
                         \delta_{[\sigma-\sigma'+N],0}\; (0,--) \nn \\
           & & \mbox{} +   \delta_{[\sigma-\sigma'],{\rm even }}
              \sum_{n=1}^{\frac{1}{2}(N-1)} (0,2n)  + 
             \delta_{[\sigma-\sigma'],{\rm odd }}
              \sum_{n=0}^{\frac{1}{2}(N-3)} (0,2n+1)  \nn \\
           & & \mbox{} +  \sum_{a=1}^{\frac{1}{2}(N-1)} \sum_{l=0}^{N-1}
               \{\delta_{\sigma+\sigma',{\rm even }} \; (2a,2l)+
              \delta_{\sigma+\sigma',{\rm odd }} \; (2a,2l+1)\} \nn  \\
(\bar{X},\sigma)\times(\bar{X},\sigma')&=& \delta_{[\sigma+\sigma'],0} \;(N,++) +
                         \delta_{[\sigma+\sigma'],2}\; (N,+-)     \\
           & & \mbox{} +  \delta_{[\sigma+\sigma'+N],0}\; (N,-+) +
                         \delta_{[\sigma+\sigma'+N],2}\; (N,--) \nn \\
           & & \mbox{} +  \delta_{[\sigma+\sigma'],{\rm even }}
              \sum_{n=1}^{\frac{1}{2}(N-1)} (N,2n) 
           + \delta_{[\sigma+\sigma'],{\rm odd }}
              \sum_{n=0}^{\frac{1}{2}(N-3)} (N,2n+1)  \nn  \\
           & & \mbox{} +  \sum_{a=0}^{\frac{1}{2}(N-3)} \sum_{l=0}^{N-1}
               \{\delta_{\sigma+\sigma',{\rm even }} \;  (2a+1,2l)+
              \delta_{\sigma+\sigma',{\rm odd }} \; (2a+1,2l+1)\}. \nn
\eea           
For $N=2$ and $p=0$, this fusion algebra naturally 
coincides with that given in 
section~\ref{chesd2b} for an ordinary $\bar{D}_2$ gauge theory.
Finally, the fusion rules given above show
that for even $N$ the charge conjugation operation is 
trivial (${\cal C}=\mbox{\bf 1}$), whereas for odd $N$ this is not the case.
Specifically, for odd $N$ all particles are their own anti-particle
except the singlet charges $(0,-+), (0,--)$ and the 
singlet dyons $(N,-+), (N,--)$ which form pairs under 
under charge conjugation as implied by~(\ref{eend}).

\aanhangsel

\sectiona{Conjugated cohomology}
\label{laap}

In this appendix, we briefly review the notion
of conjugated cohomology as it appears in the structure 
of the quasi-quantum double  $\DW$ for a nonabelian 
finite group $H$. We then recall the relation with  
ordinary cohomology.
Finally, we give a proof of the results~(\ref{generaal}) 
and~(\ref{geennar}) for finite subgroups $H$ of $SU(2)$.
It should be stressed that in contrast with the main text, the cohomology 
will be presented in additive rather then multiplicative form.
For convenience, we will also omit explicit mentioning of the coefficients 
for the cohomology groups if the integers are meant.
So $H^n(BG) := H^n(BG, \Z)$.

Let $\Z[H]=\{\sum_{h\in H} a_h h| a_h \in \Z \}$
be the group algebra for a finite group $H$. Hence, addition of the  
elements of $\Z[H]$ corresponds to that in $\Z$ and  
multiplication is defined by that in $H$.  
A so-called $H$-module (or $\Z[H]$-module)  
is an abelian group $A$ on which $H$ acts. That is, 
there exists a homomorphism from $H$ into the group 
of automorphisms of $A$.  For every $H$-module $A$, we then have homology 
and cohomology groups $H_n(H,A)$ and $H^n(H,A)$ 
respectively.   The latter are defined as follows~\cite{cartan,serre}.
Let $C^n(H,A)$ with $n \geq 0$ be the collection of $n$-cochains
$c: H^n \to A$ and $d_A$ the homomorphism 
$d_A: C^n(H,A) \to C^{n+1}(H,A)$ given by 
\bea         \label{concoh}
\lefteqn{ d_A c \, (h_1,\ldots ,h_{n+1}) \; := } \\
& & h_1 c \, (h_2,\dots ,h_{n+1})
+(-)^{n+1}c \, (h_1,\ldots ,h_n) + 
\sum _{i=1}^n (-)^i c \, (h_1,\ldots ,h_i \cdot h_{i+1},\ldots, h_{n+1}) \, .
\;\; \nn  
\eea
Here, the cochain $h_1 c$ is determined by the definition of the  
action of $H$ on the module $A$. 
The homomorphism~(\ref{concoh}) is a coboundary operator, i.e.\ $d_Ad_A=0$,
and $H^n(H,A)=(\mbox{ker}\; d_A)/ (\mbox{im} \; d_A)$ 
is the  cohomology of  $(C^{\bullet}(H,A),d_A)$ in degree $n$.
The ordinary cohomology is obtained by the  trivial action of 
$H$ on $A$. In particular, note that 
$d_A$, for $A={\mbox{\bf R}}/\Z$ with trivial action of $H$, 
is the additive version of the coboundary operator $\delta$ defined in 
expression~(\ref{coboundop}) of section~\ref{fincoh}.
The conjugated cohomology of $H$ now corresponds to the module 
$A := {\mbox{\bf R}}/\Z[H]=\{\sum_{h\in H} a_h h|a_h \in {\mbox{\bf R}}/\Z \}$, 
where the elements $g \in H$ act through conjugation, i.e.\ 
$g (\sum _{h\in H} a_h h) := \sum _{h\in H} a_h \, ghg^{-1}$.  
The related coboundary operator $d_{{\mbox{\bf R}}/\Z[H]}$ 
then becomes the additive version of the operator $\tilde\delta$ 
given in relations~(\ref{2cocycle}) and~(\ref{jrep}). Here, the conjugated
$n$-cochains $\alpha _h$ correspond to mappings 
$H^n \to {\mbox{\bf R}}/\Z[H]$: 
$(h_1,\dots ,h_n) \mapsto \sum _{h\in H} \alpha _h (h_1,\dots ,h_n) h$.

Let $u\in H_1(H,{\mbox{\bf R}}/\Z[H])$ now be the $1$-cycle 
\bea
u &:=& \sum _{h\in H} h\otimes h \,.
\eea
The slant product of $u$  with the $n$-cochains 
$c \in C^n (H, {\mbox{\bf R}}/\Z)$ then defines the mapping~\cite{spanier}   
\bea                                   \label{slano}
i: \; C^n (H, {\mbox{\bf R}}/\Z)
&\longrightarrow& C^{n-1}(H, {\mbox{\bf R}}/\Z[H]) \\
   c &\longmapsto&  ic := \sum_{h\in H} i_h c \, h  \,,  \nn
\eea
given by 
\bea 
\!\! \lefteqn{i_h \, c \, (h_1,\ldots ,h_{n-1}) \; := } \\
& & \!\!\!\!\!\!\!\! (-)^{n-1} c\, (h_2,\dots ,h_{n+1})
+\sum _{i=1}^{n-1} (-)^{n-1+i} 
c \, (h_1,\ldots ,h_i,(h_1 \cdots h_i)^{-1} h h_1 \cdots h_i, h_{i+1} \, , 
\ldots, h_{n-1}) .                   \nn
\eea 
It is easily  verified that 
\bea
d_{{\mbox{\scriptsize \bf R}}/\mbox{\scriptsize \bf Z}[H]} ( ic )  
&=& i (d_{{\mbox{\scriptsize \bf R}}/\mbox{\scriptsize \bf Z}} c) \, .
\eea 
Therefore, if $c$ is a $n$-cocycle, then $ic$ is a conjugated $n-1$
cocycle. In other words, the slant product~(\ref{slano}) defines 
an homomorphism from the ordinary cohomology groups 
of $H$ into the conjugated cohomology groups
\bea
i: \; H^n (H,{\mbox{\bf R}}/\Z) & \longrightarrow &  H^{n-1}(H,{\mbox{\bf R}}/\Z[H]) \, ,
\eea
which lowers the degree by one.

Under the action of $H$ (given as conjugation), the module 
${\mbox{\bf R}}/ \Z[H]$  naturally decomposes into a direct sum of submodules
\bea
{\mbox{\bf R}}/\Z[H] &\simeq& \oplus_A \: {\mbox{\bf R}}/ \Z[^A C] \,,
\eea
where  $A$ labels the set of conjugacy classes $^A C$ of $H$. 
The corresponding conjugated cohomology groups decompose accordingly
\bea         \label{with}
H^n (H, {\mbox{\bf R}}/ \Z[H]) &\simeq&  \oplus_A \: H^n (H, {\mbox{\bf R}}/\Z[^A C]) \, .
\eea
We may now use Shapiro's lemma 
(see for instance~\cite{evens} and~\cite{serre} page 117) 
which for this case states  
\bea                \label{sha}
H^n (H, {\mbox{\bf R}}/\Z[^A C]) &\simeq&  H^n (^A N, {\mbox{\bf R}}/\Z) \, ,
\eea
with $^A N$ the centralizer associated to the conjugacy class $^A C$.
With~(\ref{with}) and~(\ref{sha}) we then arrive at 
\bea   \label{shapiro}
H^n (H, {\mbox{\bf R}}/\Z[H]) &\simeq& \oplus_A  \: H^{n} 
(^A N,{\mbox{\bf R}}/\Z) \, ,
\eea
which expresses the fact that the conjugated cohomology of a finite 
group $H$ is completely determined by the ordinary cohomology of 
its centralizers.

We finally turn to a proof of the 
results~(\ref{generaal}) and~(\ref{geennar}) which upon passing to
integer coefficients (see relation~(\ref{ole}) in appendix~\ref{gc}) 
take the form 
\bea                          
H^4(H) &\simeq& {\mbox{\bf Z}}_{|H|} \label{sut} \\
H^3(H) &\simeq& 0,             \label{sut3}
\eea   
with $H$ a finite subgroup of $SU(2)$ and $|H|$ its order. 
We will appeal  to Leray's spectral sequences which are treated
in almost every textbook on algebraic topology. 
In particular, an exposition aimed at physicists 
can be found in reference~\cite{bott}.
Let $ESU(2)$ now be a contractible space characterized by 
a free action of $SU(2)$. 
Of course, every subgroup $H$ of $SU(2)$ then also acts freely on $ESU(2)$.
The classifying spaces $BSU(2)$ and $BH$ are constructed from $ESU(2)$ 
by dividing out the action of $SU(2)$ and $H$ respectively, that is,
$BSU(2)=ESU(2)/SU(2)$ and $BH = ESU(2)/H$.
We now have a fiber bundle 
\bea
\Pi: \; BH &\rightarrow& BSU(2) \, ,
\eea
with fiber $SU(2)/H$ and simply connected base 
space $BSU(2)$. Leray's theorem (e.g.~\cite{bott}) then states 
that there exists a spectral sequence $\{E_r,d_r\}$ with 
nilpotent operator
\bea
d_r: \;  E_r^{p,q} &\rightarrow&  E_r^{p+r,q-r+1} \, ,
\eea 
and $E_2$ term
\bea         \label{sympa}
E_2^{p,q} &\simeq& H^p (BSU(2), H^q(SU(2)/H)) \, ,
\eea  
which converges to the cohomology of the total space $BH$
\bea              \label{feiv}
H^n (BH) &\simeq&  \oplus_{p+q=n} E_{\infty}^{p,q} \, .
\eea
The cohomology of $BSU(2)$ is known to be a polynomial ring 
$\Z [e]$ in the universal Euler class $e$ of degree 4
\bea             \label{ophou}
H^n(BSU(2)) &\simeq&  \left\{ \ba{ll}
                 \Z & \mbox{if $n=0$ or a multiple of 4}  \\
                 0 & \mbox{otherwise.} \ea \right.
\eea
Since $H^*(BSU(2))$ contains no torsion, 
the universal coefficients theorem~(\ref{uni}) applied to~(\ref{sympa})
yields                                        
\bea                 \label{e2}
E^{p,q}_2 
&\simeq& H^p(BSU(2))  \otimes H^q(SU(2)/H) \, . \qquad
\eea
From~(\ref{ophou}), (\ref{e2}) and the fact that 
the cohomology of degree larger then 
the dimension of the space under consideration vanishes, 
we conclude that $E^{p,q}_2 =0$ unless $p=0,4,8,\ldots $ 
and $q=0,1,2,3$. 
The next step is to construct the terms $E_r$ for $r>2$
and to check for which $r$  this sequence becomes stationary.
We have 
\bea
E_{r+1}^{p,q} &\simeq& 
   (\mbox{ker}\; d_r \{E_r^{p,q} \rightarrow E_r^{p+r,q-r+1}\}) /  
   (\mbox{im}\; d_r \{E_r^{p-r,q+r-1}\rightarrow E_r^{p,q}\}) \, .
\eea
If we apply this to~(\ref{e2}) and iterate this process, 
we simply arrive at $E_2\simeq E_3 \simeq E_4$ due to all the zeros. 
The term $E_5$ is slightly 
different, but from here the sequence becomes stationary:
$E_5 \simeq E_6 \simeq \ldots \simeq E_{\infty}$.
Hence, with~(\ref{feiv}) we infer 
$H^n (BH) \simeq \oplus_{p+q=n} E_5^{p,q}$. It
is easily seen that $E_5^{p,q}$ only differs from $E_2^{p,q}$
for $p$ a multiple of 4 and $q=0,3$. This implies
\bea       \label{expra}
H^3(H) &\simeq& E_5^{0,3} \; \simeq  \; 
            \mbox{ker} \; d_4 \{E_2^{0,3} \rightarrow E_2^{4,0}\}   \\
H^4(H) &\simeq& E_5^{4,0} \; \simeq \;  E_2^{4,0}/ 
       \mbox{im} \; d_4 \{E_2^{0,3} \rightarrow E_2^{4,0}\} \,,
\label{exprb}
\eea
where we used the fact~\cite{milnor} 
that the cohomology for a finite group $H$ 
is the same as that for its classifying space $BH$, i.e.\ 
$H^n(BH) \simeq H^n(H)$. 
To proceed, $E_2^{0,3} \simeq H^3(SU(2)/H)$ 
and $E_2^{4,0} \simeq H^4(BSU(2))$ as indicated by~(\ref{e2}). 
In other words, the expressions~(\ref{expra}) and~(\ref{exprb}) 
state that $H^3(H)$ is the kernel and $H^4(H)$ the cokernel of  
the  homomorphism 
\bea         \label{hom}
d_4: \; H^3(SU(2)/H) &\rightarrow& H^4(BSU(2)) \, .
\eea
This mapping is the composition of the isomorphism~\cite{borel}
$H^3(SU(2))\simeq H^4(BSU(2))$ and the homomorphism
\bea \label{nathom}
H^3(SU(2)/H) &\to& H^3(SU(2)) \, ,
\eea 
induced by the projection $\pi :SU(2)\to SU(2)/H$.
Both $H^3(SU(2)/H)$ and $H^3(SU(2))$ are isomorphic to $\Z$ (generated by 
the fundamental class) and the natural homomorphism~(\ref{nathom}) 
is simply  multiplication by the degree $|H|$ of the projection $\pi$.
The kernel of this homomorphism is trivial, so $H^3(H) \simeq 0$,
while $H^4(H) \simeq \Z/|H|\Z \simeq \Z_{|H|}$. This proves our claim.

\sectie

\chapter*{Samenvatting}
\addcontentsline{toc}{chapter}{Samenvatting}
\markboth{SAMENVATTING}{SAMENVATTING}

%\selectlanguage{dutch}

De vertaling van de titel van dit proefschrift luidt:
`Topologische wisselwerkingen in gebroken ijktheorie\"{e}n'.

Fase-overgangen waarin de symmetrie van een systeem vermindert,
komen in de natuur veelvuldig voor.
Een voorbeeld waar we in het dagelijks leven 
geregeld mee worden geconfronteerd is de overgang van water naar ijs.
In dit geval breekt de translatie- en rotatie-symmetrie 
van water naar de discrete symmetrie van het ijskristal.
De structuur van het ijskristal is echter 
niet noodzakelijkerwijs volkomen regelmatig.
Tijdens de vorming van het ijs verschilt de orientatie van het kristal 
in veschillende gebieden in de ruimte, wat soms aanlieding geeft tot
`weeffouten', ook wel defecten genoemd.

Dergelijke symmetrie-brekende fase-overgangen en de daarbij optredende 
defecten worden in verschillende disciplines van de natuurkunde bestudeerd.
Volgens de standaard oerknal-theorie in de cosmologie 
bijvoorbeeld is het heelal in een vroeg
stadium afgekoeld door middel van een reeks van symmetrie-brekende
fase-overgangen. In een aantrekkelijk maar nog steeds 
speculatief scenario hebben de hierbij optredende punt-, lijn- 
en andersoortige defecten een belangrijke rol gespeeld in 
de vorming van de melkwegstelsels en andere grote schaalstructuren 
in het huidige heelal.
Een fraai voorbeeld van een defect 
bestudeerd in de deeltjesfysica is de 't Hooft-Polyakov monopool waar we mee te maken krijgen
in elk groot unificatie-model waarin een compacte ijkgroep is gebroken 
naar een ondergroep die de electromagnetische groep $U(1)$ bevat.
Een ander voorbeeld van een defect is de magnetische 
fluxbuis in type~II supergeleiders waarin de 
electromagnetische  ijkgroep $U(1)$ is gebroken naar de eindige
ondergroep $\Z_2$. Verschillende soorten defecten komen 
voor in de vele fasen van supervloeibaar helium-3. 
Als laatste voorbeeld in de natuurkunde van de gecondenseerde materie 
noemen we hier de overgang van de wanordelijke
fase naar de geordende fase in vloeibare kristallen resulterend in  
punt-, lijndefecten en textuur.

In dit proefschrift beschouwen we 2+1 dimensionale modellen waarin een 
continue ijkgroep is gebroken naar een eindige ondergroep $H$.
De defecten die in deze modellen verschijnen zijn deeltjesachtige objecten
met een magnetische flux gekarakteriseerd door een element van de 
residuele ijkgroep $H$. Verder bestaat het spectrum van 
een dergelijk model uit `electrische' puntladingen, die 
overeenkomen met de verschillende irreducibele representaties 
van $H$, en zogenaamde dyonen, i.e.\ samenstellingen van de 
voorgenoemde magnetische- en electrische deeltjes. 
De ijkvelden in deze modellen zijn massief, dus de krachten tussen deze 
deeltjes zijn van eindige dracht.
Toch kunnen deeltjes elkaar over willekeurig grote afstanden be\"{i}nvloeden.
Een electrische puntlading dat rond een magnetische deeltje
beweegt, ondervindt in het algemeen een Aharonov-Bohm effect: 
het keert terug in een andere hoedanigheid.
In het geval van een niet-abelse ongebroken eindige ijkgroep $H$
doet zich het opzienbarende verschijnsel van flux metamorfose voor:
de fluxen van  om elkaar heen bewegende magnetisch deeltjes 
kunnen veranderen.  
Deze Aharonov-Bohm effecten vormen de topologische 
wisselwerkingen waar de titel van dit proefschrift naar verwijst.
De toevoeging topologisch slaat op het feit dat deze wisselwerkingen
onafhankelijk zijn van de afstand tussen de deeltjes 
en alleen afhangen van het aantal keren dat de deeltjes elkaar omcirkelen.

Hoofstuk~\ref{chap2} bevat een uitvoerige behandeling  van 
deze spontaan gebroken ijktheorie\"{e}n in drie-dimensionale ruimtetijd.
We laten zien dat de optredende deeltjes systemen 
in een quantummechanische  beschrijving vlecht-statistiek realizeren.
De dyonen die voorkomen in het geval van een abelse 
residuele eindige ijkgroep $H$ hebben fractionele spin en 
gedragen zich niet als bosonen of fermionen maar als anyonen: 
een georienteerde verwisseling van twee identieke dyonen 
in het vlak geeft aanleiding tot een
fase-factor $\exp(\im \Theta) \neq \pm 1$ in de golffunctie.
De deeltjes die voorkomen in het geval van een niet-abelse eindige 
groep $H$ vormen in het algemeen niet-abelse generalizaties van anyonen.
Verder bespreken we ondermeer de 
spin-statistiek connectie en de fusieregels voor deze deeltjes, 
de cross-secties voor de Aharonov-Bohm verstrooiings-experimenten 
met deze deeltjes, en het intrigerende fenomeen van Cheshire lading.
Een belangrijk resultaat in dit hoofsdtuk
is de identificatie van de quantumgroep $D(H)$
gerelateerd aan een theorie met een residuele eindige ijkgroup $H$.

In de hoofdstukken~\ref{chap3} en~\ref{chap4} bestuderen we de gevolgen
van de aanwezigheid van een zogenaamde Chern-Simons term in deze modellen.

\chapter*{Dankwoord}
\addcontentsline{toc}{chapter}{Dankwoord}
%\selectlanguage{dutch}

Tot slot wil ik van de gelegenheid gebruik maken
om een aantal mensen te bedanken die direct of indirect hebben 
bijgedragen aan het tot stand komen van dit proefschrift.

Allereerst gaat mijn dank uit naar mijn promotor Sander Bais
voor de inspanningen die het hebben mogelijk gemaakt dat ik in Amsterdam
heb kunnen promoveren en bovenal voor zijn stimulerende begeleiding.
Sander, de vele inspirende discussies, die 
soms tot diep in de avond voortduurden,
hebben de afgelopen vier jaar 
voor mij tot een boeiende en leerzame tijd gemaakt.

Zonder de samenwerking met mijn co-auteurs 
zou dit proefschrift nooit in deze vorm zijn verschenen.
Ik wil Peter van Driel bedanken voor een intensieve 
en vruchtbare samenwerking.
I also had the privilige to work with Alosha Morozov.
Alosha, I would like to thank you for an inspiring collaboration and your 
warm hospitality during my stay in Moscow and Alushta.

I am deeply indebted to Danny Birmingham for many fruitful
and illuminating discussions and in particular 
for a careful reading of this manuscript. 

Verder ben ik dank verschuldigd aan 
Robbert Dijkgraaf en Eduard Looijenga voor verschillende 
verhelderende gesprekken en belangwekkende suggesties.

Alle collega's die deel uitmaken of de laatste jaren deel hebben uitgemaakt 
van het instituut voor theoretische 
fysica, met name Alec Maassen van den Brink, Per John,
Nathalie Muller, Tjark Tjin en Alain Verberkmoes, wil ik bedanken 
voor hun hulp, stimulerende discussies en voor de aangename sfeer.

Mijn moeder en Hans bedank ik voor hun  morele steun en zorg.
Mijn dierbare Kalli dank ik voor haar liefde,
en in het bijzonder voor haar aanmoedigingen gedurende 
de laatste maanden.

\end{document}